\documentclass{article}
\usepackage{emulateapj,timesfonts}
\usepackage{float,epsfig,psfig}
\usepackage{pstricks}
   \def\apj#1#2#3#4{\par #4 19#3, {ApJ,\/} {#1}, #2 }
   \def\pasp#1#2#3#4{\par #4 19#3, {PASP,\/} {#1}, #2 }
   
   \def\apjs#1#2#3#4{\par #4 19#3, {ApJ (Supplement Series),\/} { #1}, #2 }
   
   \def\pasj#1#2#3#4{\par #4 19#3, {PASJ,\/} { #1},#2 }

   \def\aa#1#2#3#4{\par #4 19#3, {A\&A,\/} { #1}, #2 }
   
   \def\aas#1#2#3#4{\par #4 19#3, {A\&A (Supplement Series),\/} { #1}, #2 }
   
   \def\mnras#1#2#3#4{\par #4 19#3, {MNRAS,\/} { #1}, #2 }

   \def\apjl#1#2#3#4{\par #4 19#3, {ApJ (Letters),\/} { #1}, #2 }

   \def\nature#1#2#3#4{\par #4 19#3, {Nature,\/} { #1}, #2 }
   
   \def\asr#1#2#3#4{\par #4 19#3, {\em Adv. Space Res.,\/} { #1}, #2 }

   \def\BIB {\par}

\def\hea4{{\it HEAO~A4}}
\def\heaoa2{{\it HEAO~A2}}
\def\heao1{{\it HEAO~1}}

\def\amin{$^\prime$}

\def\eg{{\it e.g.}~}
\def\ie{{\it i.e.}~}

\def\h0{$H_{\rm o}=50$~km~s$^{-1}$~Mpc$^{-1}$}
\def\q0{$q_{\rm o}$}

\def\msun     {$M_{\odot}$}
\def\lsun     {$L_{\odot}$}

\def\lsel     {$10^{11}L_{\odot}$}

\def\etal    {{et~al.}~}

\def\cms3  {~{cm$^{-3}$}}

\begin{document}

\submitted{Submitted to ApJ 2000, August 31. Accepted 2001, March 6. In
  press 2001, July 1, v 555.}
\title{{Temperature and Heavy Element Abundance Profiles of Cool Clusters of
    Galaxies from ASCA}}
\author{A.~Finoguenov$^{1,3}$, M.~Arnaud$^2$ and L.P.~David$^3$ }
\affil{\vspace*{0.5cm} {$^1$ Max-Planck-Institut f\"ur extraterrestrische
    Physik, Giessenbachstra\ss e, 85748 Garching, Germany}\\
{$^2$ Service d'Astrophysique, CEA Saclay, 91191 Gif sur Yvette Cedex, France}\\
{$^3$ Smithsonian Astrophysical Observatory, 60 Garden st., MS 2, Cambridge,
  MA 02138, USA}}
% {$^4$ Space Research Institute, Profsoyuznaya 84/32 117810 Moscow, Russia}\\
% {$^5$ Alexander von Humboldt Fellow}}
\submitted{}

\begin{abstract}
  
We perform a spatially resolved X-ray spectroscopic study of a set of 18
relaxed clusters of galaxies with gas temperatures below 4 keV.  Spectral
analysis was done using ASCA/SIS data coupled with the spatial
information contained in ROSAT/PSPC and Einstein/IPC observations.  We
derive the temperature profiles using single-temperature fits and also
correct for the presence of cold gas at the cluster centers. For all of the
clusters in the sample, we derive Si and Fe abundance profiles. For a
few of the clusters, we also derive Ne and S abundance profiles. We present
a comparison of the elemental abundances derived at similar overdensities
as well as element mass-to-light ratios.  We conclude that the
preferential accretion of low entropy, low abundance gas into the
potentials of groups and cold clusters can explain most of the observed
trends in metallicity.  In addition, we discuss the importance of energy
input from SNe II on cluster scaling relations and on the relation between
the observed scatter in the retainment of SN Ia products with differences
between the epoch of cluster formation.
  
\end{abstract}


\section{Introduction}

%\doublespace

\begin{table*}
{
\begin{center}
\footnotesize
\tabcaption{\centerline{\footnotesize
X-ray and optical quantities of the sample}
\label{tab:opt}}

\begin{tabular}{lrrrrrrccccccccc}
\hline
\hline
Name  &D   &\amin & $L_B$ & $R_{mes}$ & $L_{B_{cD}}$& $R_c$ &$\alpha$&$\beta$&$r_{c}$& $\nwarrow$&$kT_{unc}$& $\delta kT_{unc}$&$kT_e$& $\delta kT_e$ & $R_{v}$\\
      &Mpc & kpc &\lsel & Mpc       & \lsel       &  Mpc  &     &  & kpc  & & keV     & keV & keV & keV          & Mpc     \\
\hline                                                                    
 A2197E & 177 & 49 & 4.25$^{1   }$ & 0.29 &  2.7 &   0.38  &  0.89   &0.35 & 17 & & 1.097 & 0.07 & 1.097 & 0.07 & 1.291 \\
   A400 & 145 & 40 & 21.$^{2   }$  & 3.   &  0.81 &  0.08  &  0.83   &0.56 & 180 &SE \ \ W & 1.83 & 0.16 & 1.83 & 0.16 & 1.67 \\
   A194 & 107 & 30 & 5.1$^{3   }$  & 0.8  &  0.00 &  0.267 &  0.925  &0.60 &  26 &NE \ \  SW& 2.12 & 0.43 & 2.12 & 0.43 & 1.79 \\
   A262 &  97 & 27 & 10.$^{4   }$  & 0.82 &  0.55 &  0.02  &  0.59   &0.46 &  58 &SE$\Rightarrow$W & 2.18 & 0.21 & 2.26 & 0.23 & 1.85 \\
  MKW4S & 171 & 47 & 6.3$^{\dag}$  & 0.5  &  2.62 &  0.25  &  1.     &0.51 & 124 &SE & 2.13 & 0.64 & 2.29 & 0.26 & 1.87 \\
   A539 & 174 & 48 & 38.$^{5   }$  & 2.   &  0.79 &  0.04  &  0.78   &0.69 & 250 & E  & 2.81 & 0.34 & 2.81 & 0.34 & 2.07 \\
   AWM4 & 195 & 57 & 6.4$^{\dag}$  & 0.5  &  1.09 &  0.25  &  1.     &0.62 & 110 &  & 2.92 & 0.39 & 2.92 & 0.39 & 2.11 \\
   MKW9 & 240 & 65 & 6.7$^{\dag}$  & 0.5  &  1.46 &  0.25  &  1.     &0.52 &  54 & E \ \  W & 2.66 & 0.57 & 2.92 & 0.43 & 2.11 \\
  A2197 & 183 & 50 & 3.0$^{1   }$  & 0.31 &  1.90 &  0.185 &  0.94   &0.43 &  53 &  & 2.21 & 0.49 & 3.05 & 0.82 & 2.15 \\
  A2634 & 189 & 52 & 70.$^{6   }$  & 1.5  &  1.57 &  0.324 &  0.47   &0.69 & 448 &SW & 3.06 & 0.35 & 3.06 & 0.35 & 2.16 \\
  A4038 & 171 & 47 & 15.$^{7   }$  & 1.2  &  1.04 &  0.043 &  0.86   &0.61 & 160 &NW$\Rightarrow$E & 3.31 & 0.25 & 3.31 & 0.
25 & 2.24 \\
 2A0335 & 211 & 57 & 7.3$^{\dag}$  & 0.5  &  0.71 &  0.25  &  1.     &0.65 &  80 &  & 3.07 & 0.27 & 3.40 & 0.39 & 2.27 \\
  HCG94 & 253 & 68 & 7.0$^{\dag}$  & 0.5  &  0.29 &  0.25  &  1.     &0.48 &  75 &NE \ \  SE& 2.94 & 0.45 & 3.45 & 0.50 & 2.29 \\
  A2052 & 210 & 57 & 26.$^{6   }$  & 1.5  &  1.10 &  0.139  &  0.7    &0.64 & 100 &  & 3.24 & 0.32 & 3.46 & 0.39 & 2.29 \\
   A779 & 136 & 38 & 13.1$^{8   }$ & 2.9  &  2.11 &  0.525  &  0.95   &0.34 &  54 & N \ \  S & 2.97 & 0.39 & 3.56 & 0.53 & 2.33 \\
  MKW3S & 273 & 73 & 9.4$^{\dag}$  & 0.5  &  1.67 &  0.25  &  1.     &0.71 & 300 &SE$\Rightarrow$NW& 3.45 & 0.47 & 3.79 & 0.53 & 2.40 \\
    CEN &  63 & 18 & 48.$^{9   }$ & 1.0  &  1.40 &  0.230 &  0.88   &0.44 &  28 &N  & 3.56 & 0.28 & 3.82 & 0.29 & 2.41 \\
  A2063 & 214 & 58 & 12.$^{6   }$ & 0.5  &  1.91 &  0.253 &  0.94   &0.69 & 220 & N$\Rightarrow$E & 3.83 & 0.39 & 3.86 & 0.24 & 2.42 \\
\hline
\end{tabular}
\end{center}
\hspace*{2.6cm} {$^{\dag}$} \hspace*{0.3cm}{\footnotesize Interpolated value
using Eq.1 from FDP.}  
\hspace*{0.6cm} $^{1}$ \hspace*{0.3cm}{\footnotesize present work} 
\hspace*{0.6cm} $^{2}$ \hspace*{0.3cm}{\footnotesize Arnaud \etal 1992} 
\hspace*{0.6cm} $^{3}$ \hspace*{0.3cm}{\footnotesize Nikogossyan \etal 1999} 
\hspace*{0.6cm} $^{4}$ \hspace*{0.3cm}{\footnotesize Sakai \etal 1994} 
\hspace*{0.6cm} $^{5}$ \hspace*{0.3cm}{\footnotesize Ostriker \etal 1988}
\hspace*{0.6cm} $^{6}$ \hspace*{0.3cm}{\footnotesize Cirimele \etal 1998} 
\hspace*{0.6cm} $^{7}$ \hspace*{0.3cm}{\footnotesize Green \etal 1990} 
\hspace*{0.6cm} $^{8}$ \hspace*{0.3cm}{\footnotesize Trevese \etal 1996 } 
\hspace*{0.6cm} $^{9}$ \hspace*{0.3cm}{\footnotesize Jerjen \& Tammann 1997} 
}
\end{table*}

The intracluster medium (ICM) has long been a subject of research for
analyzing elements produced and lost by galaxies. By determining the total
mass in metals produced by the stellar population of a cluster, one can
place constraints on the population of massive stars, or equivalently, the
Initial Mass Function (IMF), and the integrated Type Ia supernovae (SNe Ia)
rate.  In addition, the abundance and distribution of heavy elements in the
ICM are sensitive to the processes of hierarchical clustering and therefore
provide a unique tool for accessing the details of how present-day clusters
evolve.

In Finoguenov, David \& Ponman (2000, hereafter FDP), we presented an
extensive study of the distribution of heavy elements in groups and clusters
and discussed the implications regarding the chemical enrichment of the
intracluster medium. For clusters with the best photon statistics, we found
that the Fe abundance decreases significantly with radius, while the
Si abundance is either flat or decreases slightly.  This results in an
increasing Si/Fe ratio with radius and implies a radially increasing
predominance of Type II SNe enrichment in clusters. In rich clusters, the
overall Fe mass from SN~II requires $N_{SN II}/N_{SN Ia}\sim 20$ within
$0.2r_{180}$ ($r_{180}$ is the radius at which the mean cluster density
is 180 times the critical density), substantially 
different from the balance in our Galaxy
($7:1$). SN~Ia products show a strong central concentration in most
clusters, suggesting a release mechanism with a strong density dependence
(e.g., ram pressure stripping of infalling gas rich galaxies). Moreover, the
Iron Mass-to-Light Ratio (IMLR) for SN~Ia decreases with radius in most
clusters. The dependence of [Si/Fe] vs [Fe/H] reveals a smaller role of
SN~II in the enrichment of groups compared to clusters, suggesting that
SN~II products were only weakly captured by the shallower potential wells of
groups due to the possible high entropy of preheated gas.

In this {\it Paper}, we extend the approach developed in FDP to systems with
properties intermediate between cool groups and hot clusters.  Given the
temperature range of these systems, the ASCA/SIS sensitivity is sufficient
to constrain the temperature distribution and determine the radial profiles
of Fe and Si.  For clusters with good photon statistics we
can also constrain the distribution of Ne and S.  We present
new spectroscopic results for A262, A2197 (subclusters E and W), A539,
MKW3S, MKW4S, MKW9, AWM4, HCG94, A779, A400, A2052, A2634, A4038 (Klemola
44), 2A0335+096, A2063, and A194. We also include an analysis of the
Centaurus cluster to compare with the method of Ikebe \etal (1999).

This paper is organized as follows: in Sec. {\it\ref{sec:data}} we describe
our analysis technique; temperature profiles are discussed in
{\it\ref{sec:te}}; in {\it\ref{sec:ab}} we present the radial abundance
profiles of Fe, Si, Ne and S; in {\it\ref{sec:vir}} we analyze the elemental
abundances derived at similar overdensities and their corresponding
mass-to-light ratios with a further discussion of the roles of SN II in
{\it\ref{sec:infall} \& \ref{sec:en-sn}} and SN Ia in
{\it\ref{sec:sn1a}}. Our main results are summarized in
{\it\ref{sec:sum}}. We assume {\h0, $q=0.5$} throughout the paper.

\section{Data Reduction}\label{sec:data}

A detailed description of the ASCA observatory as well as the SIS detectors
can be found in Tanaka, Inoue \& Holt (1994) and Burke \etal (1991).  All
observations are screened using FTOOLS version 4.2 with standard screening
criteria. The effect of the broad ASCA PSF is treated as described in
Finoguenov \etal (1999) including the geometrical projection of the
three-dimensional distribution of X-ray emitting gas. Our minimization
routines are based on the $\chi^2$ criterion.  No energy binning is done,
but a special error calculation is introduced as in Churazov \etal (1996) to
properly account for small number statistics. Model fits to ROSAT (Truemper
1983) surface brightness profiles are used as input to the ASCA data
modeling.  The details of our minimization procedure for ASCA spectral
analysis are described in Finoguenov and Ponman (1999). We adapted the XSPEC
analysis package to perform the actual fitting and error estimation.  The
spatially resolved spectral characteristics are quoted as the best fit
solution plus an estimate of the 90 \% confidence area of possible parameter
variation based on the regularization technique (Press \etal 1992;
Finoguenov and Ponman 1999). To study the systematic errors related to the
spatially resolved spectroscopic analysis of the ASCA data, we follow the
approach described in FDP. For all ROSAT imaging analysis we use the
software described in Snowden \etal (1994) and references therein.

We use the MEKAL plasma code (Mewe \etal 1985, Mewe and Kaastra 1995,
Liedahl \etal 1995) in all of our spectral analysis.  All abundances are
given relative to the solar values in Anders \& Grevesse (1989).  The
abundances of He and C are fixed to their solar value.  The remaining
elements are combined into five groups for fitting: Ne; Mg; Si; S and Ar;
and Ca, Fe, and Ni. We restrict our analysis to the energy range 0.8--7.0
keV to avoid the large systematic uncertainties at low energies, which
prevent us from determining the O abundance.  We do not report the Mg
abundance due to the proximity of the Mg K lines with the poorly understood
4-2 transition lines of iron, which are strongest at temperatures of 2 to
4~keV (see Fabian \etal 1994 and Mushotzky \etal 1996).

Table \ref{tab:opt} contains the optical and X-ray properties of our sample.
Column (1) identifies the system, (2) adopted luminosity distance, (3)
corresponding scale length, (4) total blue light along with a reference, (5)
corresponding radius of the optical measurement, (6) B luminosity of the cD
galaxy from NED, (7) core radius of the galaxy distribution, and (8) the slope
$\alpha$ of the modified King profile $(1+(r/R_c)^2)^{-3\alpha/2}$.
Columns (9--10) give the results of the cluster surface brightness fitting
outside the central region using a $\beta$ model. Since the outer region
chosen for the spatial analysis only corresponds to a part of the cluster,
due to \eg\ number of CCDs read-out, in col. (11) we denote the position
of the observed region relative to the cluster center with NW$\Rightarrow$E
denoting an area extending from the North-West to the East in a
counterclockwise direction.  Column (12) gives the best fit cluster
emission-weighted temperature, (13) the corresponding 90\% error, (14) the
best fit cluster emission-weighted hotter temperature in a two-temperature
fit (data for the colder component are listed in Tab.\ref{tab:cold}), (15)
the corresponding 90\% error, and (16) an estimate of the virial radius of
the system ($r_{180}=1.23 T_{keV}^{0.5}h_{50}^{-1}$ Mpc; Evrard \etal 1996)
using $kT_e$. Estimates of the virial radii using X-ray observations carried
out in Finoguenov, Reiprich and Boehringer (2001) show that the
actual virial radii are 20\% lower than Evrard's value.

\subsection{Optical data reduction}

A modified King profile was assumed for the galaxy distribution. The core
radius and slope were taken from the literature when available (the
references are given in Table 1), otherwise we assumed average values of
$R_c=0.25$ Mpc and $\alpha=1$. For A779, A2197(E,W) and Centaurus, we fitted
the cumulative galaxy distribution derived from the galaxy catalog of
Tr\`evese et al (1997), Dixon, Godwin \& Peach (1989) and Jerjen \& Dressler
(1997), respectively.  Only galaxies brighter than the catalog completeness
limit were considered. The corresponding background counts were estimated as
described in Tr\`evese, Cirimele \& Appodia (1996) for A779 and from the deep
counts of the ESO-Sculptor Redshift Survey (Arnouts \etal 1997) for A2197.
They are negligible in the later case.  No background correction was applied
for Centaurus; cluster members and background objects are distinguished in
the catalog based on morphological criteria and we only considered definite
cluster members. A2197 is a double cluster, as revealed by the X-ray
morphology; one subcluster is centered on NGC 6160 (A2197W) and the other on
NGC 6173 (A2197E). To limit confusion between the two components, we only
considered the central region of each subcluster within $6'$ ($\sim 300$
kpc) from its central galaxy. The galaxy distribution parameters of A4038
were derived from a fit to the surface density of galaxies published in
Green, Godwin \& Peach (1990, Table 4).

The total blue light of each cluster within a given projected radius is
listed in Table 1. Some processing was required to obtain a set of
homogeneous values from the published data (see the references in Table
\ref{tab:opt}). The published luminosities of A194 and A262 were corrected for
the faint end of the luminosity function (LF). We used a Schechter
luminosity function with $M_{\rm B}^*=-20.6$ and $\alpha=-1.25$ and
converted the published limiting magnitudes to limiting luminosities taking
into account interstellar absorption. The z-correction is negligible in our
redshift range. The luminosities of A2052, A2063 and A2634 within 1.5 Mpc
were computed using a Schechter LF and a modified King galaxy distribution
with the parameters published in Cirimele et al (1997, 1998).  The
luminosity of A779 within $2.85\ {\rm Mpc}$ was computed from the
luminosity function determined by Tr\`evese, Cirimele \& Appodia (1996) in
that region.  The normalization of the LF was deduced from the published
galaxy counts within the completeness limit. For these four clusters and
A400 the luminosity in the V-band was converted into a B-band 
luminosity assuming
$L_{\rm V} = 1.3 L_{\rm B}$. The luminosity of the Centaurus cluster was
estimated using the catalog of Jerjen \& Dressler (1997). We summed the
luminosities of all possible cluster members within 1.5 Mpc from the cluster
center and down to the catalog completeness limit ($M_{\rm B}=-15.3$). The
error induced by the uncertainty on cluster membership is small. The
luminosity is only $20\%$ less if only definitive cluster members are
considered. The correction for the faint end of the LF, estimated from the
Schechter LF given in Jerjen \& Tammann (1997), was found negligible. A
similar procedure was performed for A2197W and A2197E using the galaxy
catalog of Dixon, Godwin \& Peach (1989), down to its completeness
limit. The background correction, estimated statistically from the deep
counts in each magnitude bin (Arnouts \etal 1997), is negligible (at the few
$\%$ level) as well as the correction for incompleteness.

\subsection{Details of X-ray data reduction}

Although the analysis technique was intended to be similar for all of the
clusters, some corrections were individually made, which we describe in this
section.

A194, A2063, AWM4, and A539 have bright point-like sources in the field of
view. In these cases we extract the spectra of point sources with both
ROSAT/PSPC and ASCA/SIS, subtract the cluster background, analyze the
spectra and then based on the ASCA response matrix estimate the
contribution from these point sources to the regions selected for analysis
of the diffuse cluster emission. For A194 the contribution of the point
sources was severe at the center and we omit the central region of A194 from
further discussion.

The 1996 observation of A400 was performed in 4-CCD mode which has significant
SIS1 calibration problems, so we omit the SIS1 data from further analysis.

The observation of MKW9 has a high background, which is noticeable at the
edges of the detector (due to warm CCDs as pointed out by Buote 2000). This
background dominates the spectrum below 1 keV. We analyze and subtract the
excess background and only fit the source spectrum in the 0.9--3.5 keV
energy band. Thus, the Fe abundance for this cluster is derived from the
L-shell line complex. Buote (2000) claims that the derived Fe abundance in
MKW9 increases from 0.4 to 0.7 solar, when adding a second temperature
component. We find that this conclusion depends on whether the high-energy
part of the spectrum is included in the spectral fitting, and does not
depend on the inclusion of the second temperature component as advocated by
Buote (2000). While we attribute this effect to uncertainties in the
subtraction of the background, our error bars allow for both Fe abundance
values. The spatial modeling of MKW9 was taken from an Einstein/IPC
observation.

The spectral analysis of the MKW4S data reveals significant residuals at high
energies which we attribute to an enhanced SIS background. Since the
spectral data on this cluster are of poor quality at energies above 4 keV,
this effect leads to a spurious rise in the derived temperature accompanied
by an unacceptable large $\chi^2$. To overcome these difficulties, we
restricted the spectral analysis to the 0.7--3.5 keV band. Thus, the Fe
abundance for this cluster is derived from measurements of the L-shell
line complex only.

For several clusters (MKW4S, AWM4, MKW9, A4038, HCG94, and MKW3S), whose
observations have obvious calibration problems, we add a 10\% systematic
error at the 68\% confidence level below 1.5 keV to the spectra. After the
corrections mentioned in this section we are able to reduce the $\chi^2$
values of the fit to acceptable values.  A spatially resolved spectroscopic
analysis of Zw1615+35 was not feasible due to the presence of 2 X-ray bright
sources in the field.

From the ASCA observations of the Centaurus cluster, we selected the longest
observation of the cluster center done in 1-CCD mode to analyze the emission
within central 7\amin.  Other observations are used to analyze the cluster
emission in the 7\amin--30\amin\ radius region. We also excluded the Eastern
part of the cluster which may have an infalling subgroup (Churazov \etal
1999).

We correct for the lower ASCA SIS normalizations relative to ASCA GIS
following Iwasawa, Fabian \& Nandra (1999). This increases our gas mass
estimates by approximately 10\%.

\section{Temperature Profiles}\label{sec:te}

\begin{figure*}

   \includegraphics[width=1.6in]{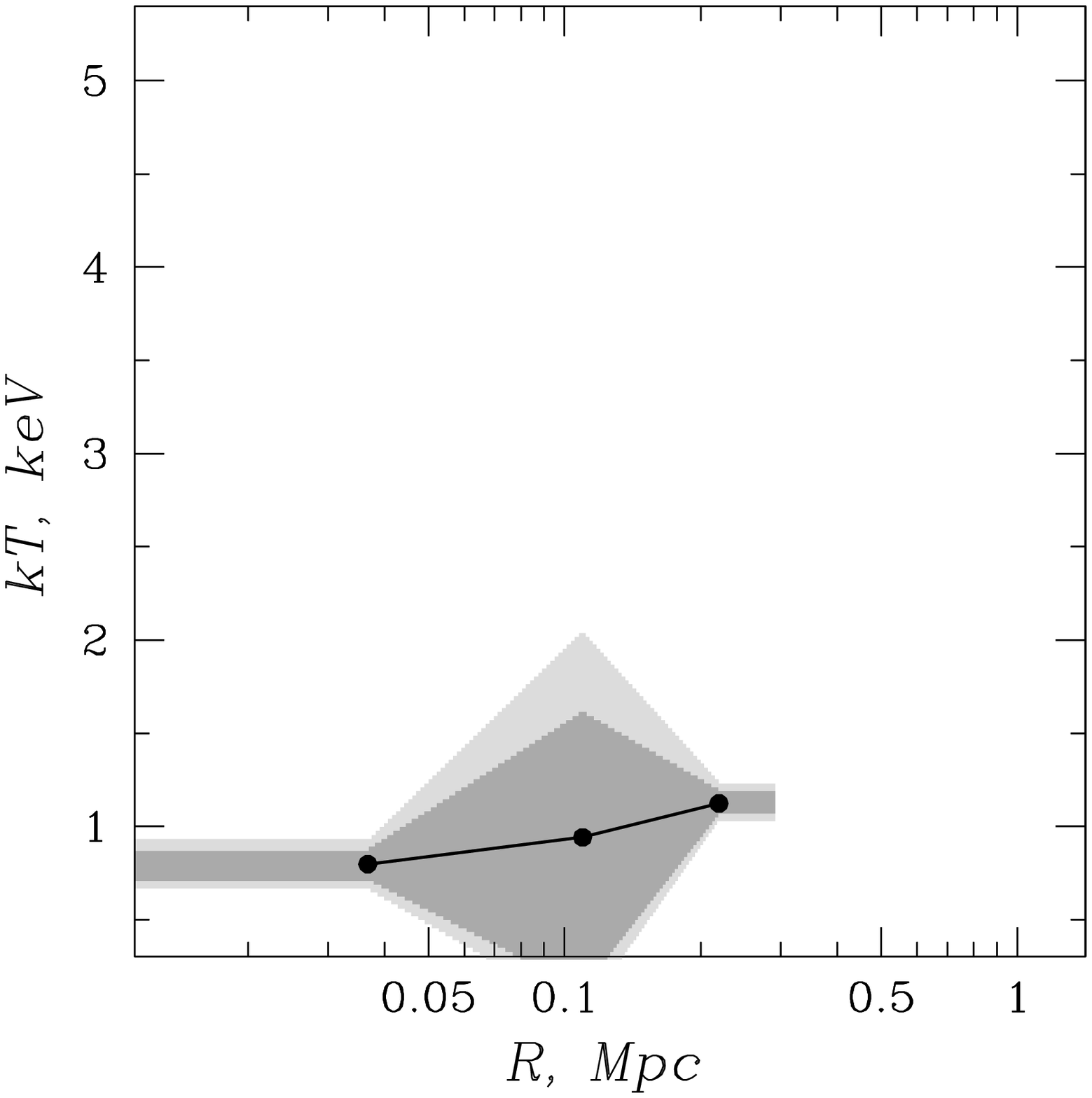} \hfill
   \includegraphics[width=1.6in]{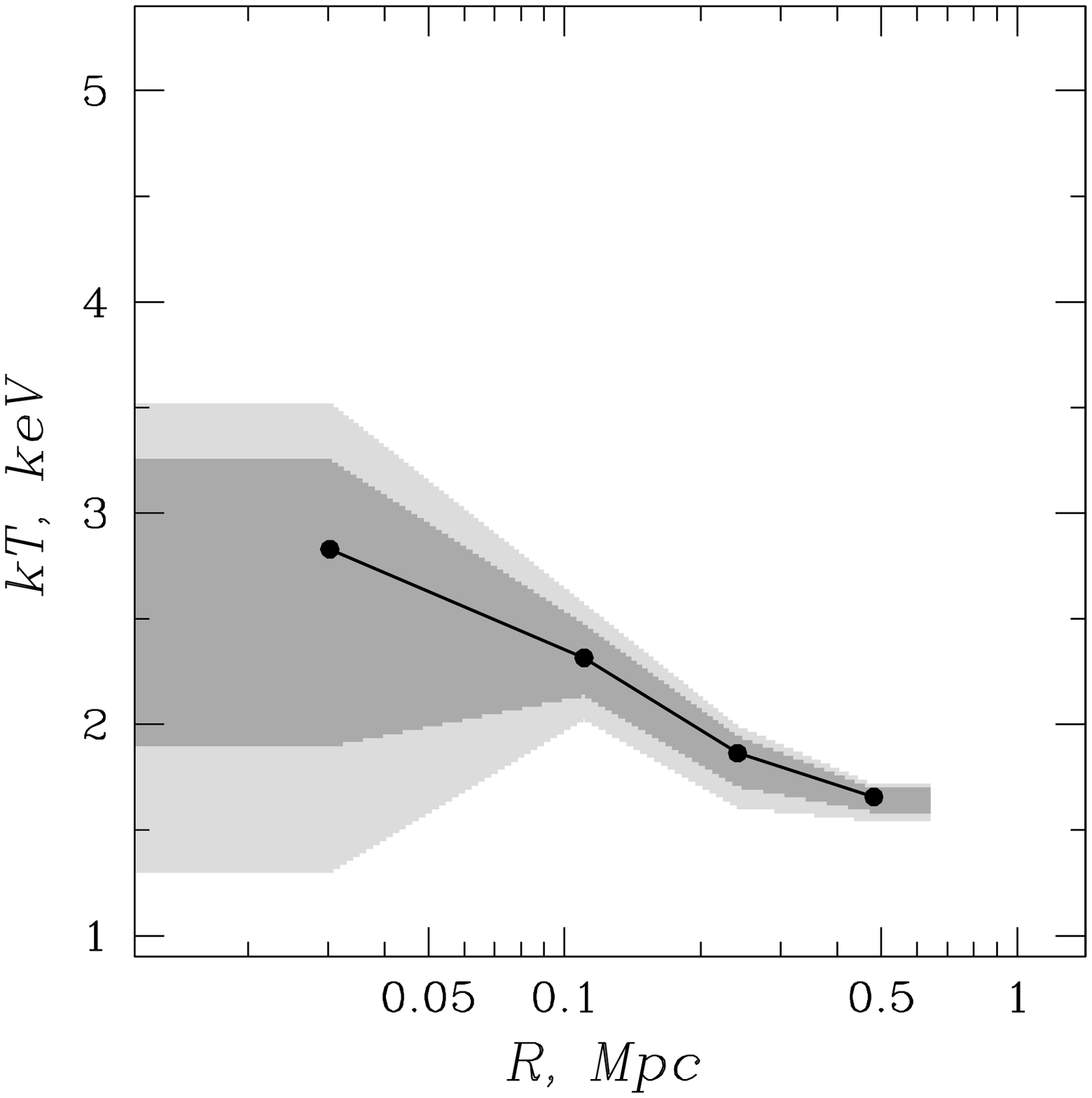} \hfill
   \includegraphics[width=1.6in]{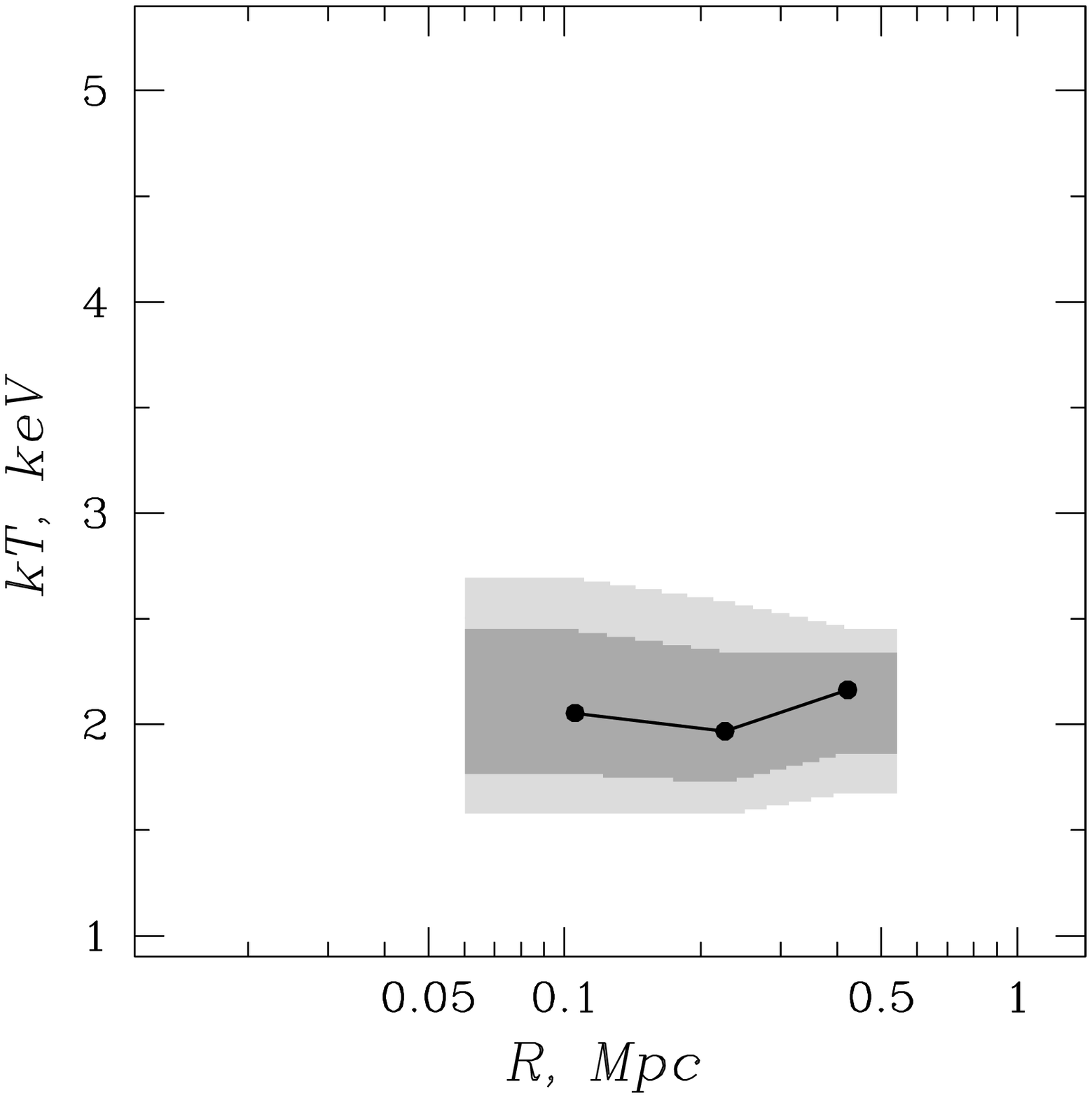} \hfill
   \includegraphics[width=1.6in]{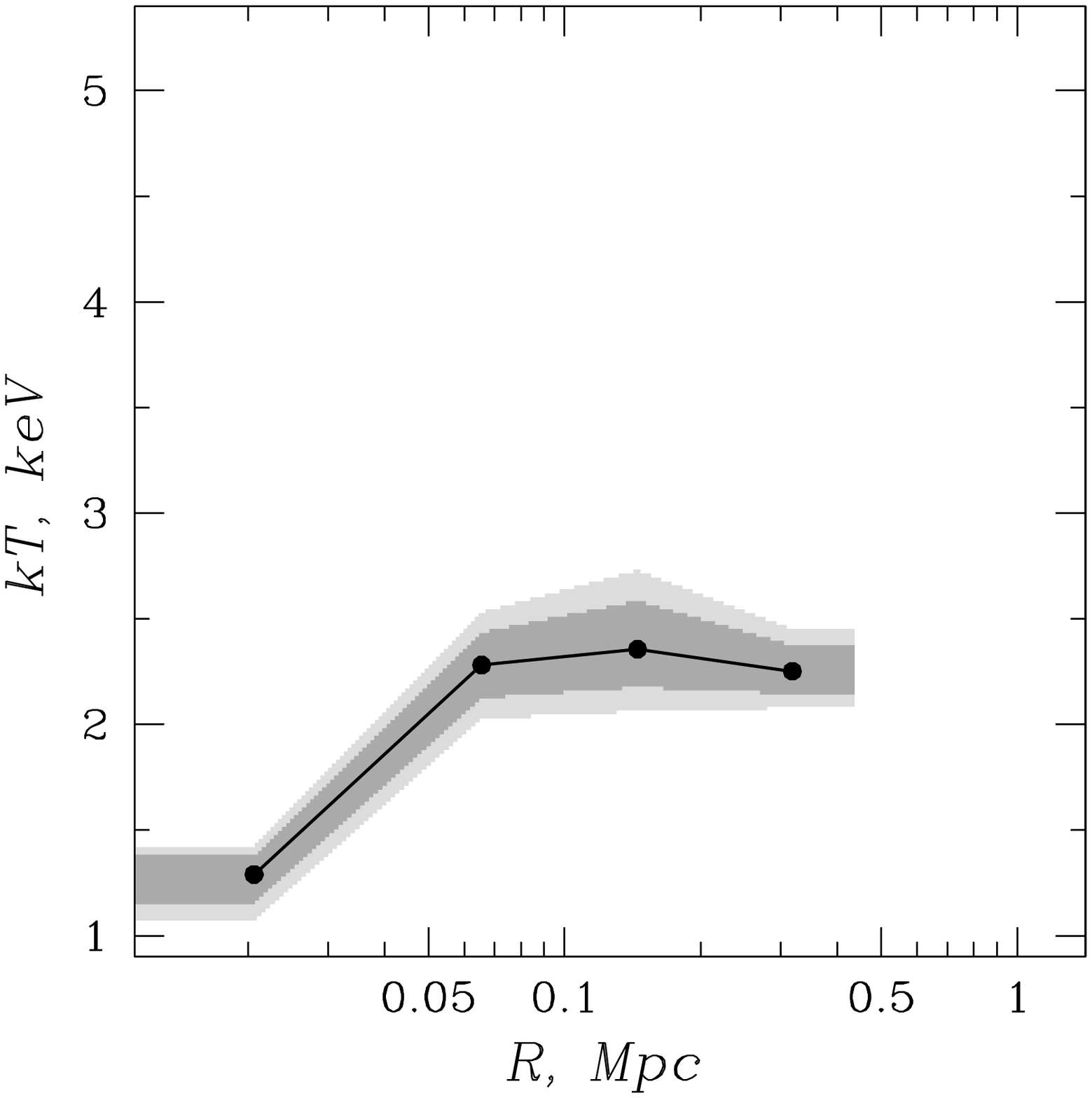} 

  \includegraphics[width=1.6in]{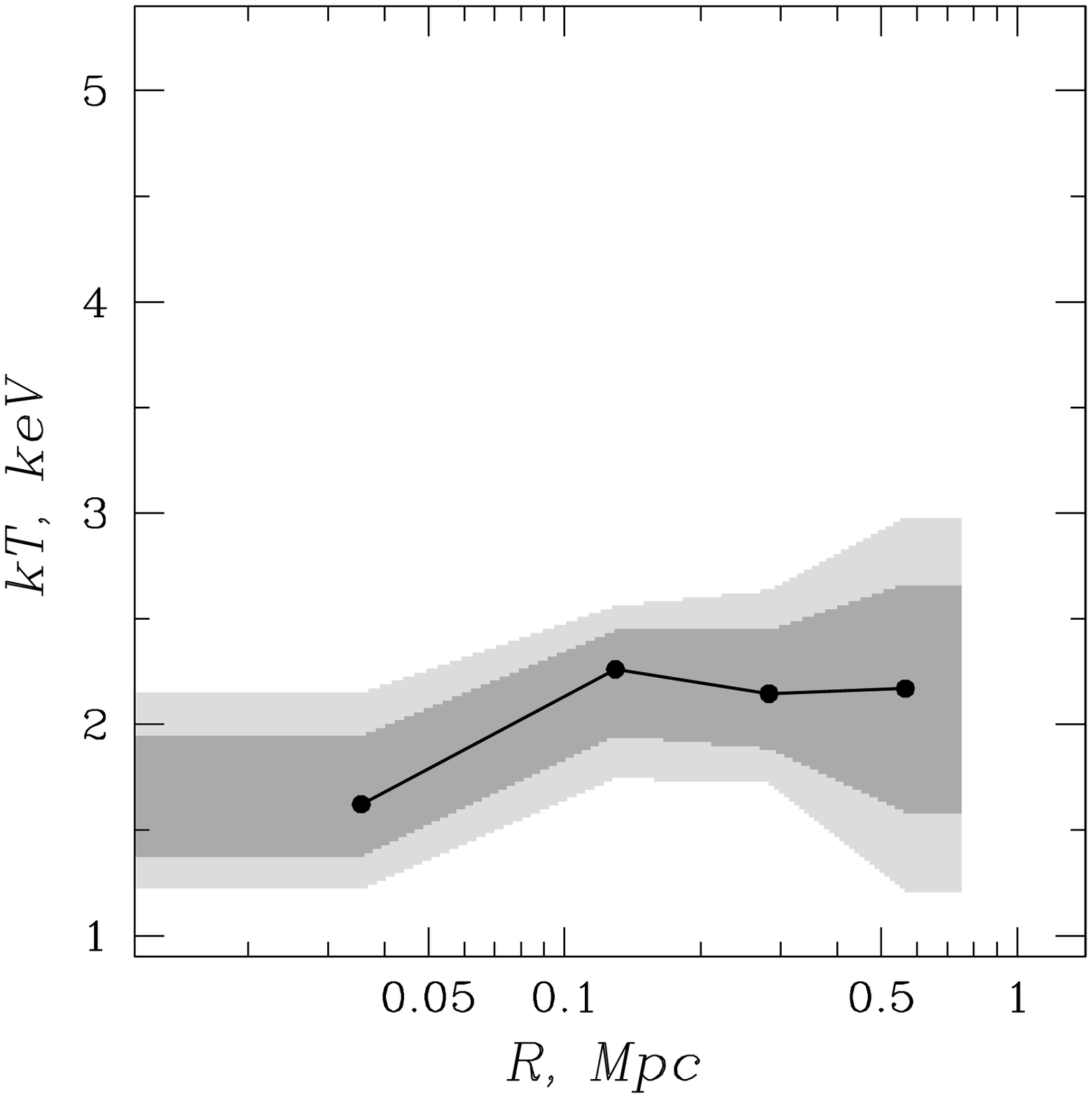} \hfill  
   \includegraphics[width=1.6in]{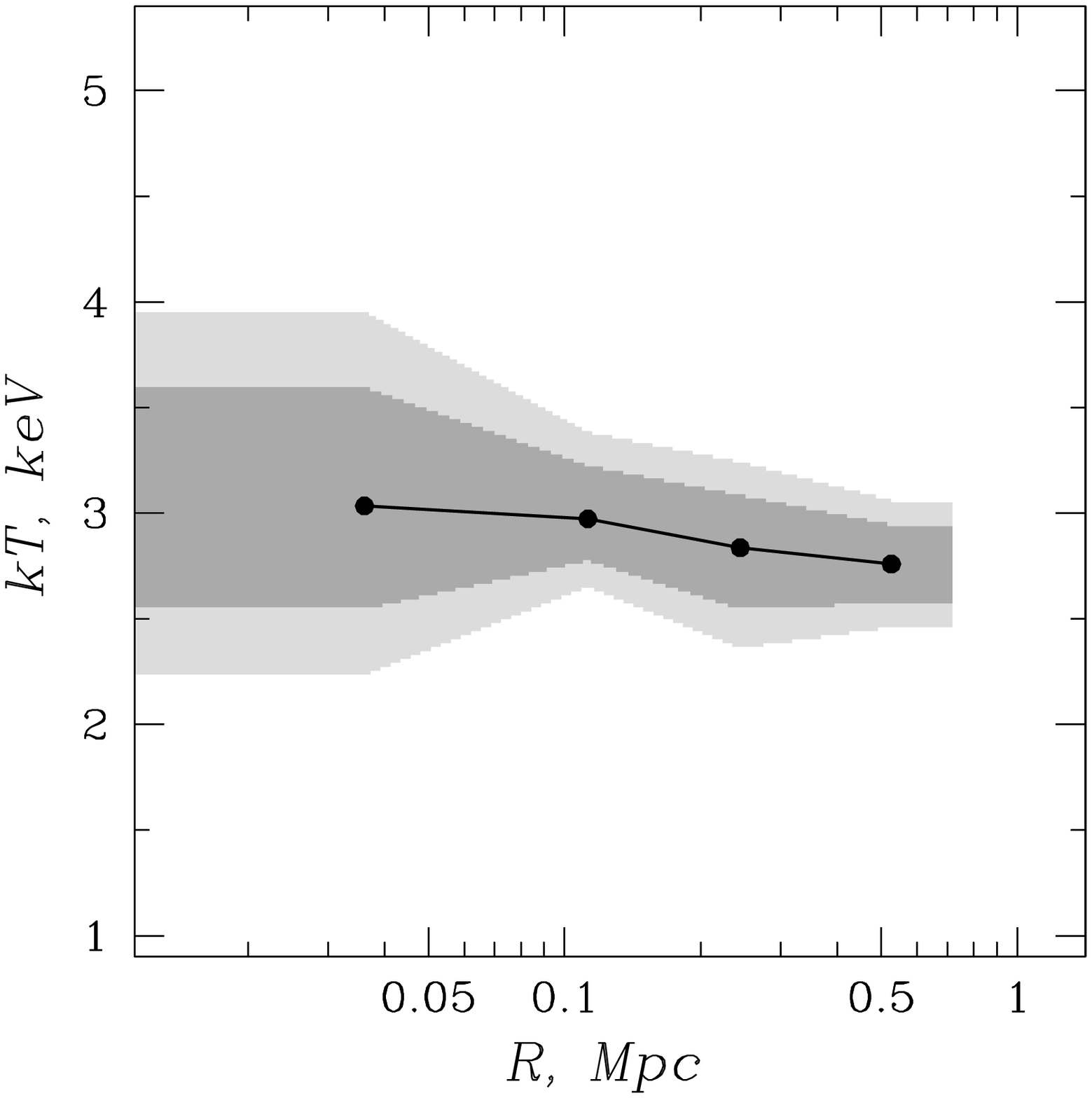} \hfill 
  \includegraphics[width=1.6in]{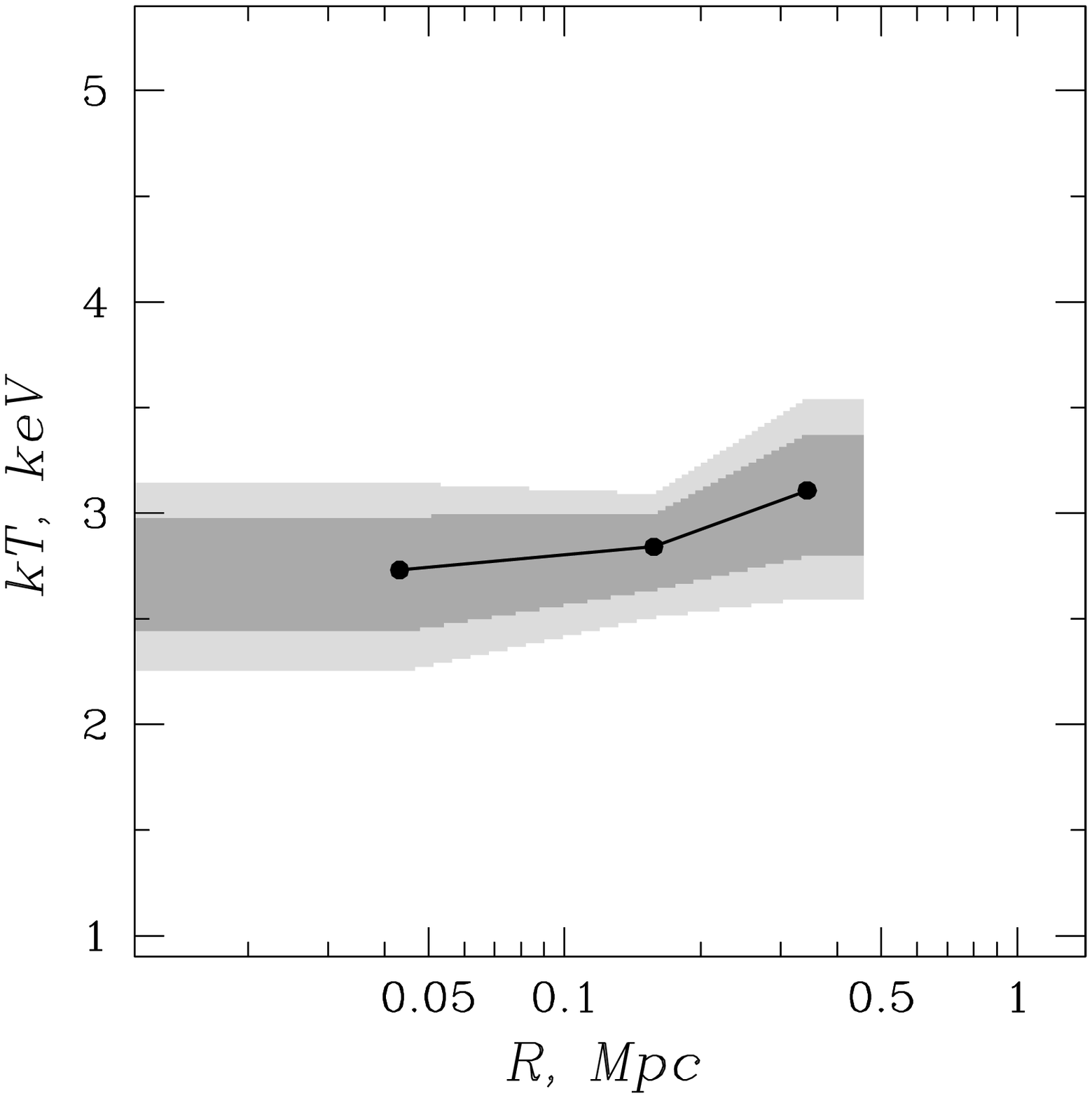} \hfill 
  \includegraphics[width=1.6in]{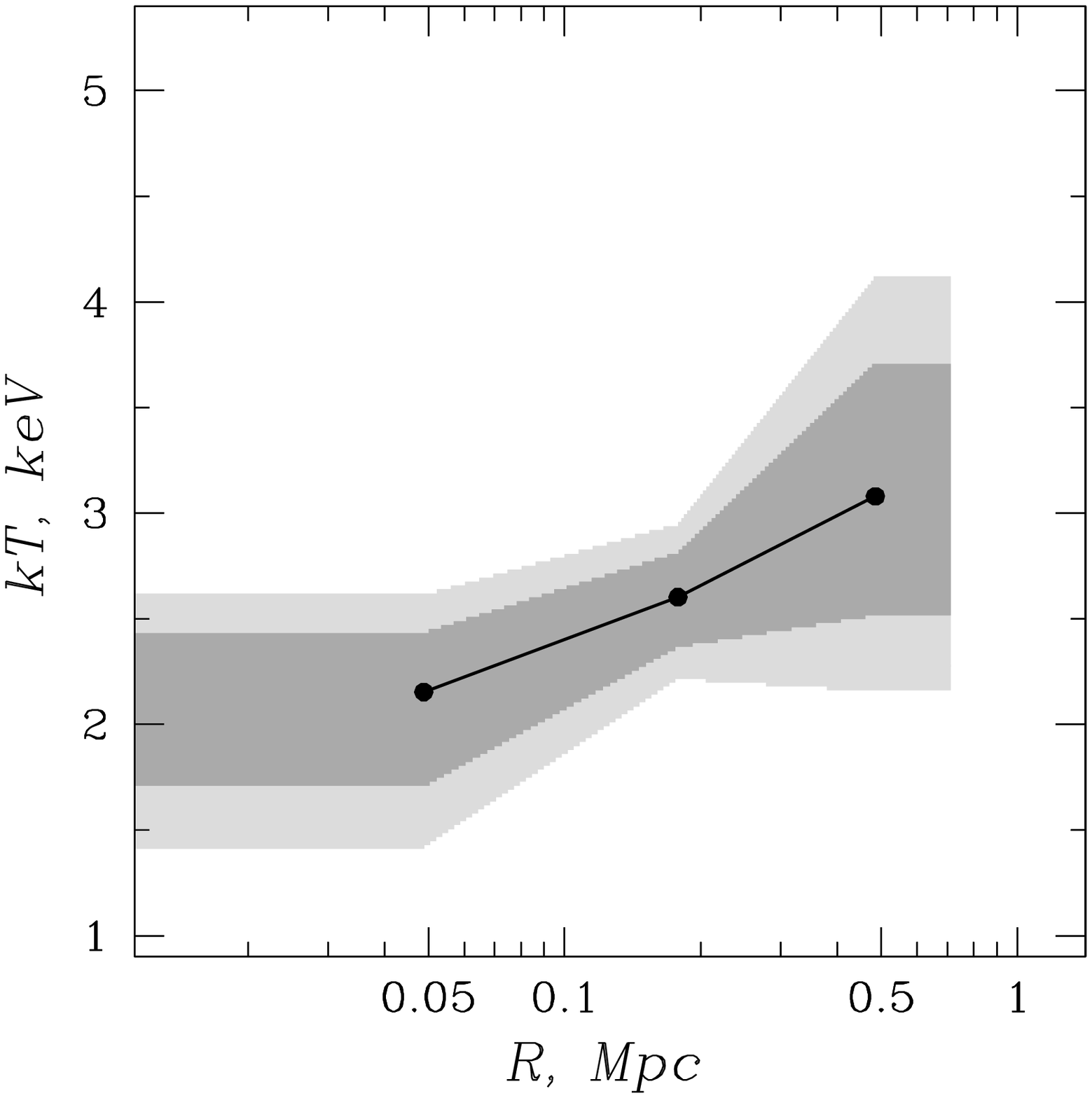}

   \includegraphics[width=1.6in]{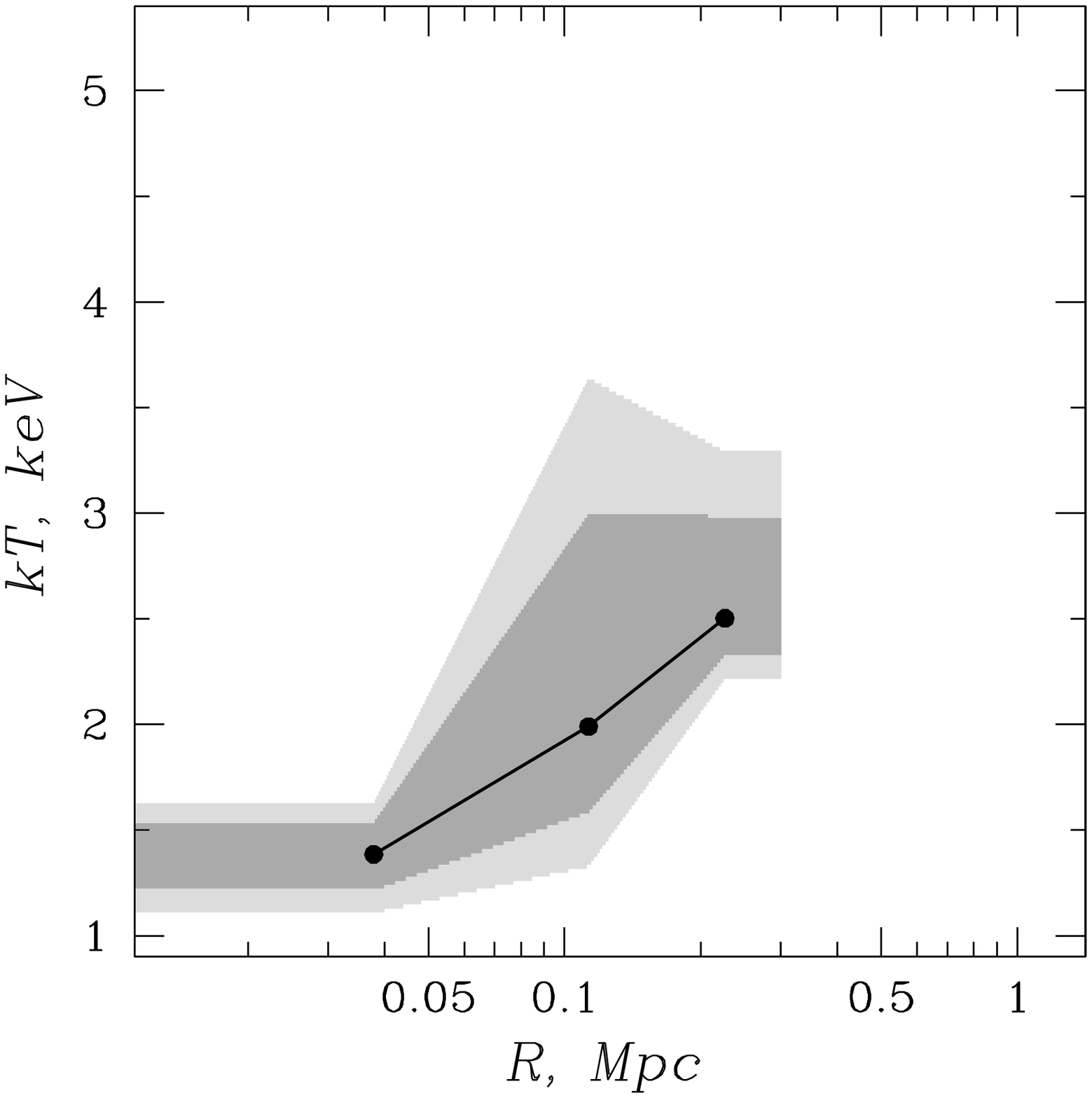}  \hfill 
  \includegraphics[width=1.6in]{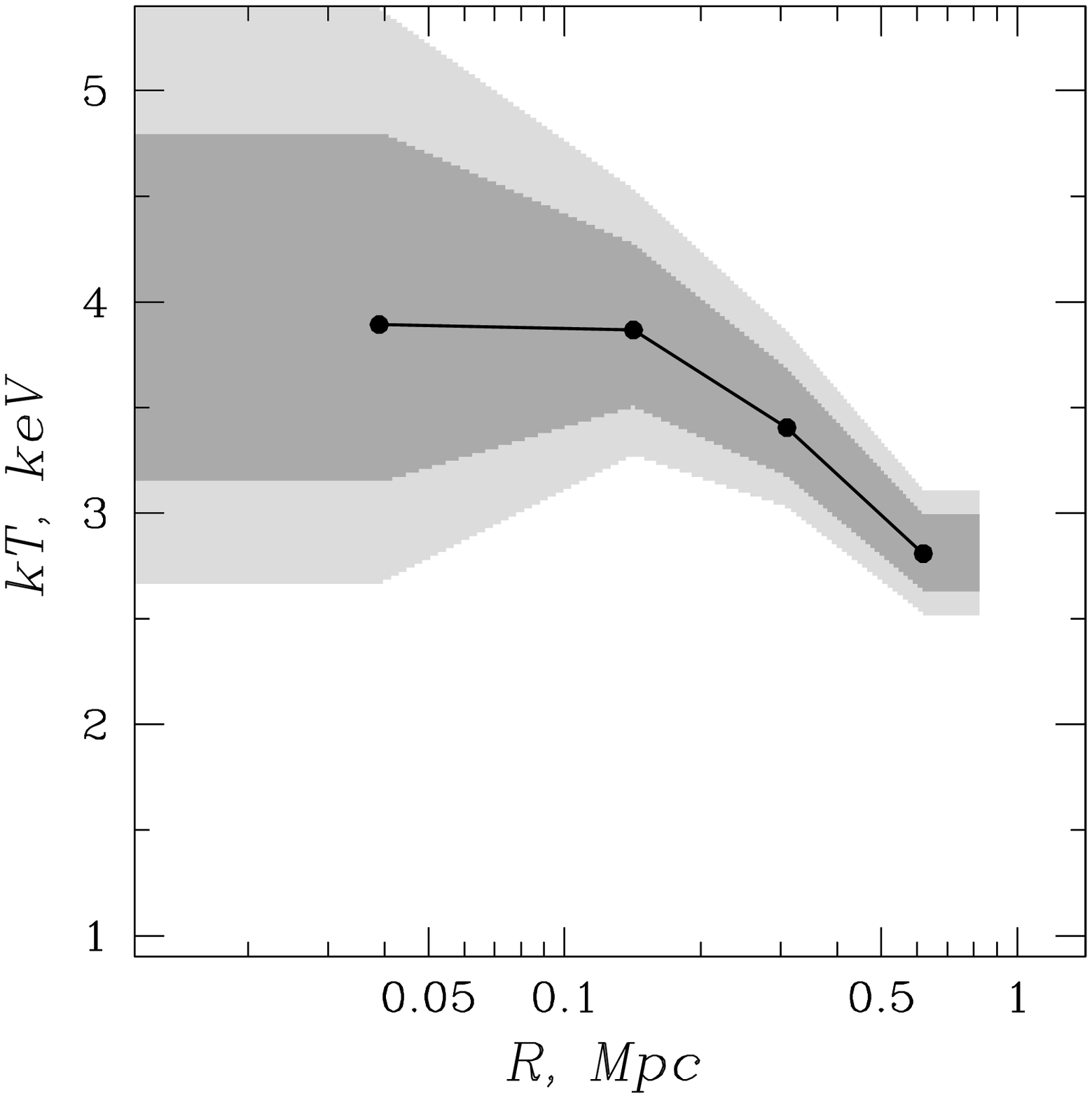} \hfill 
 \includegraphics[width=1.6in]{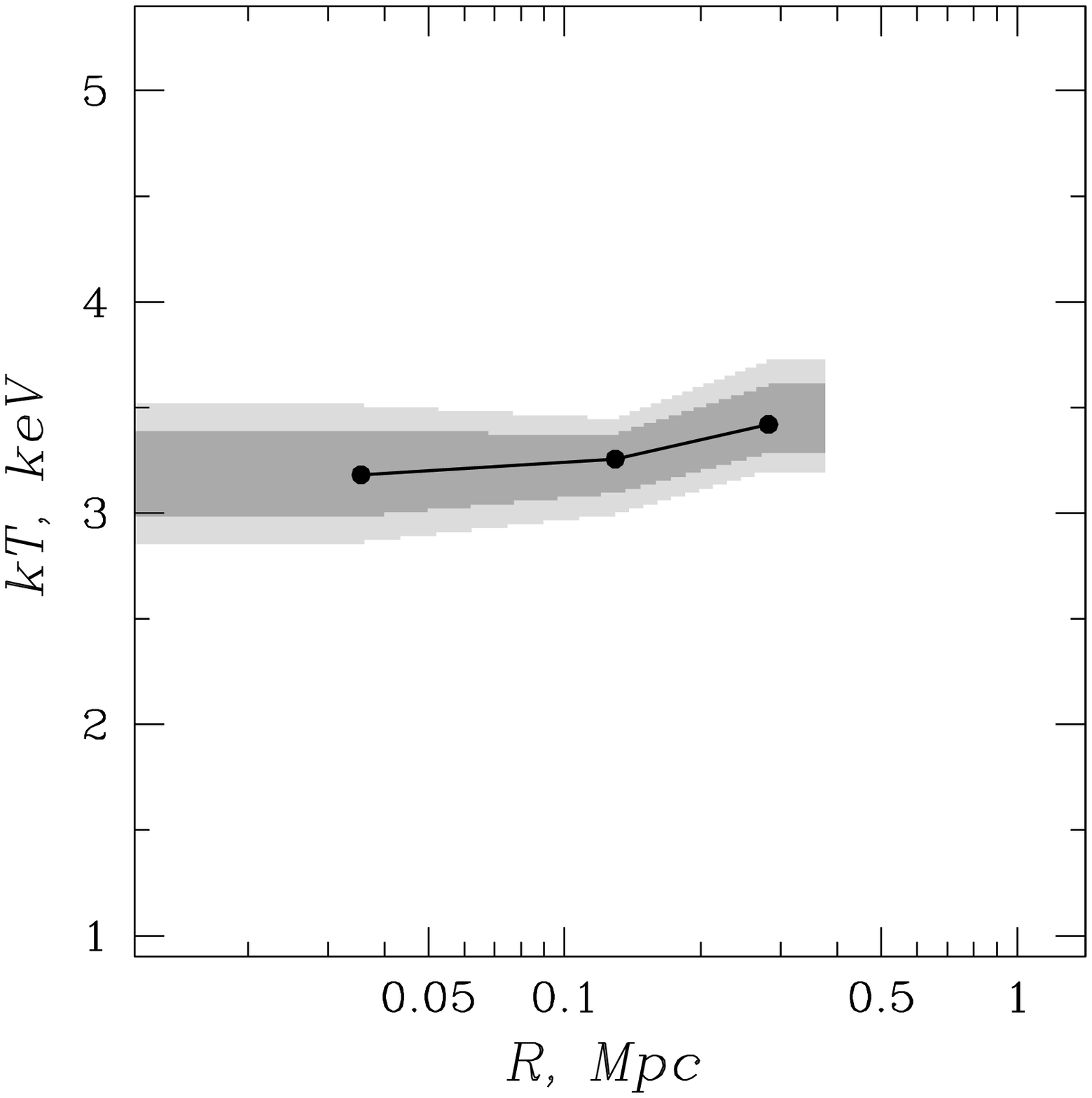} \hfill 
  \includegraphics[width=1.6in]{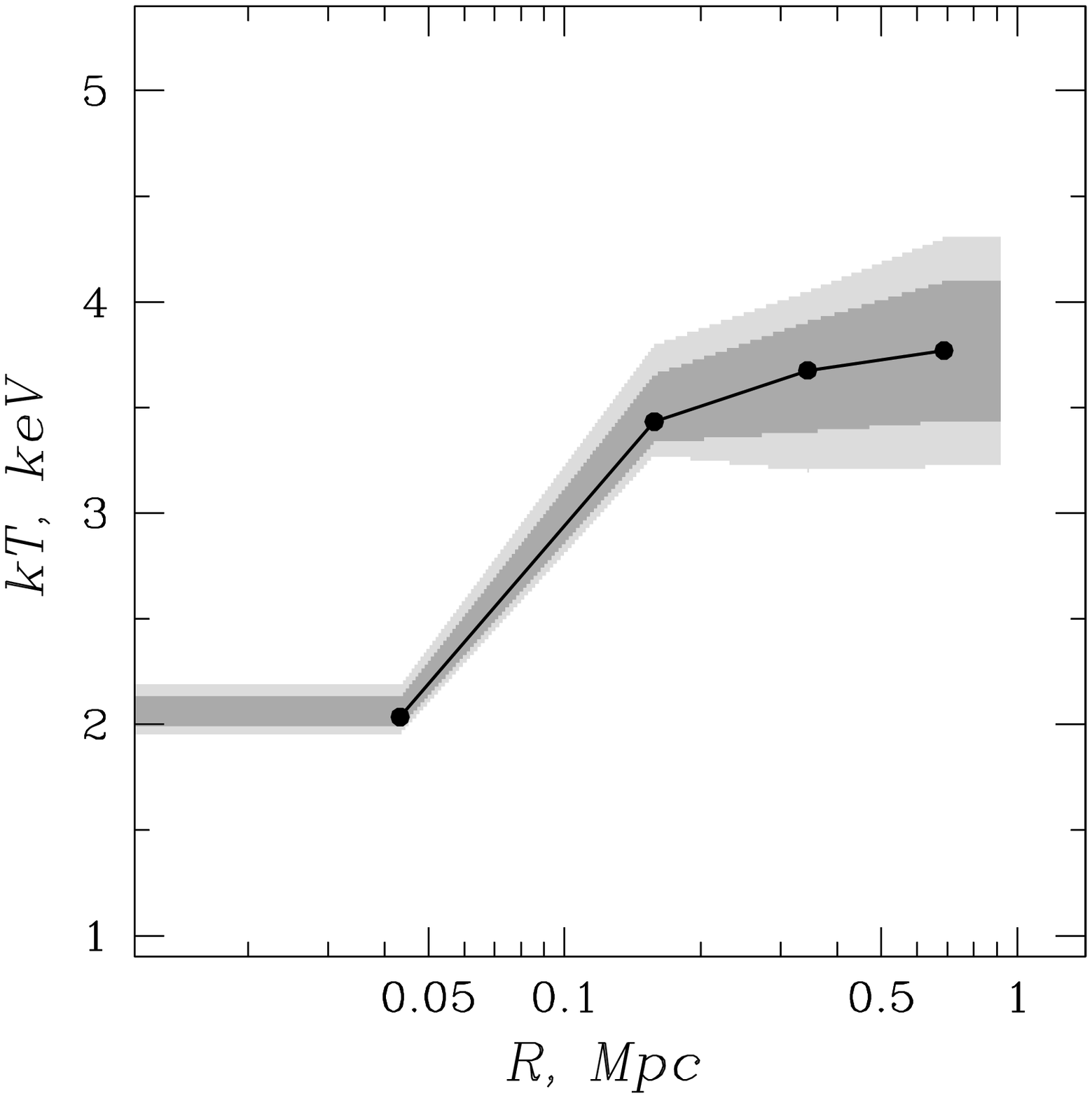} 

   \includegraphics[width=1.6in]{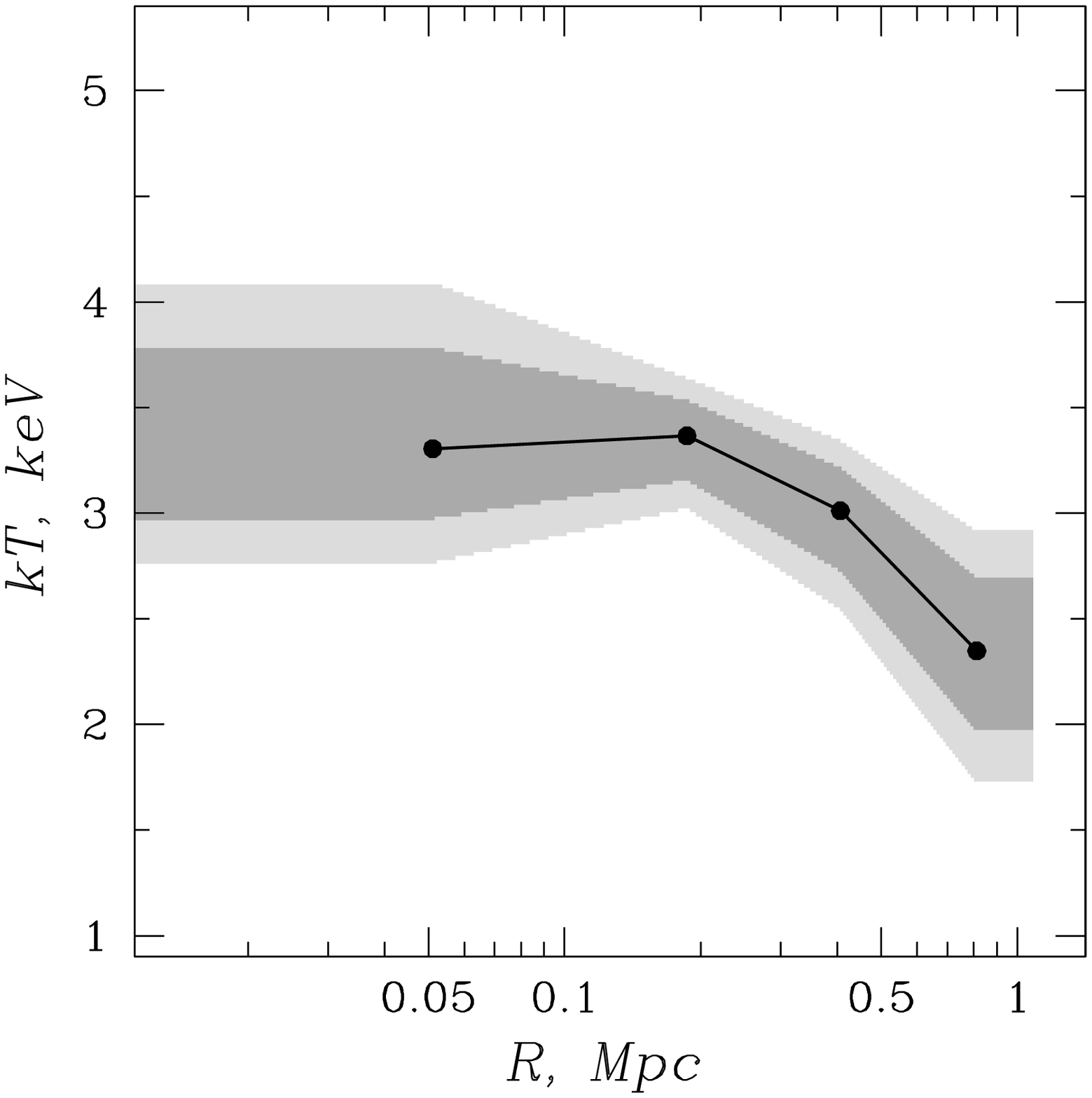} \hfill 
   \includegraphics[width=1.6in]{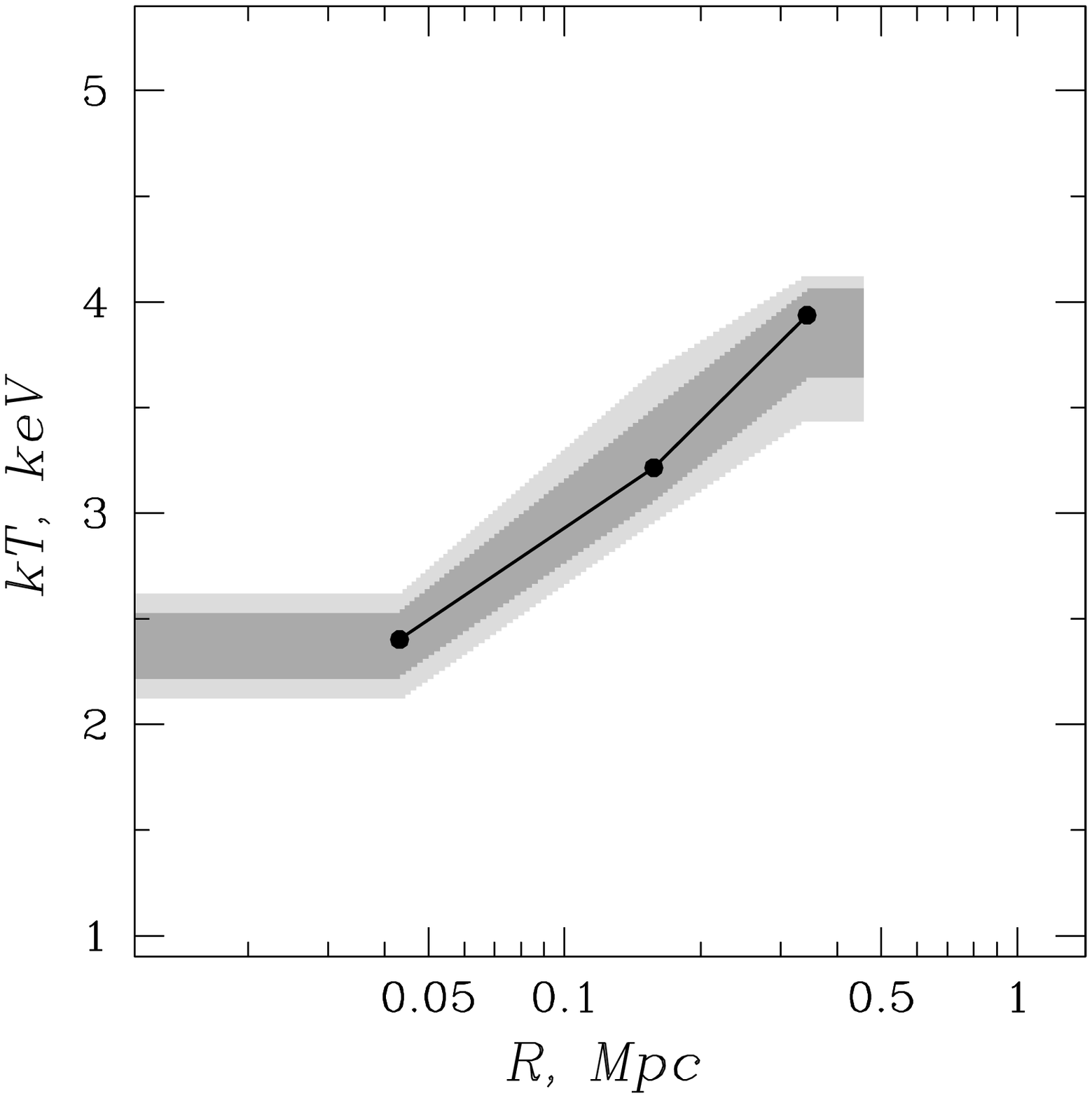} \hfill 
  \includegraphics[width=1.6in]{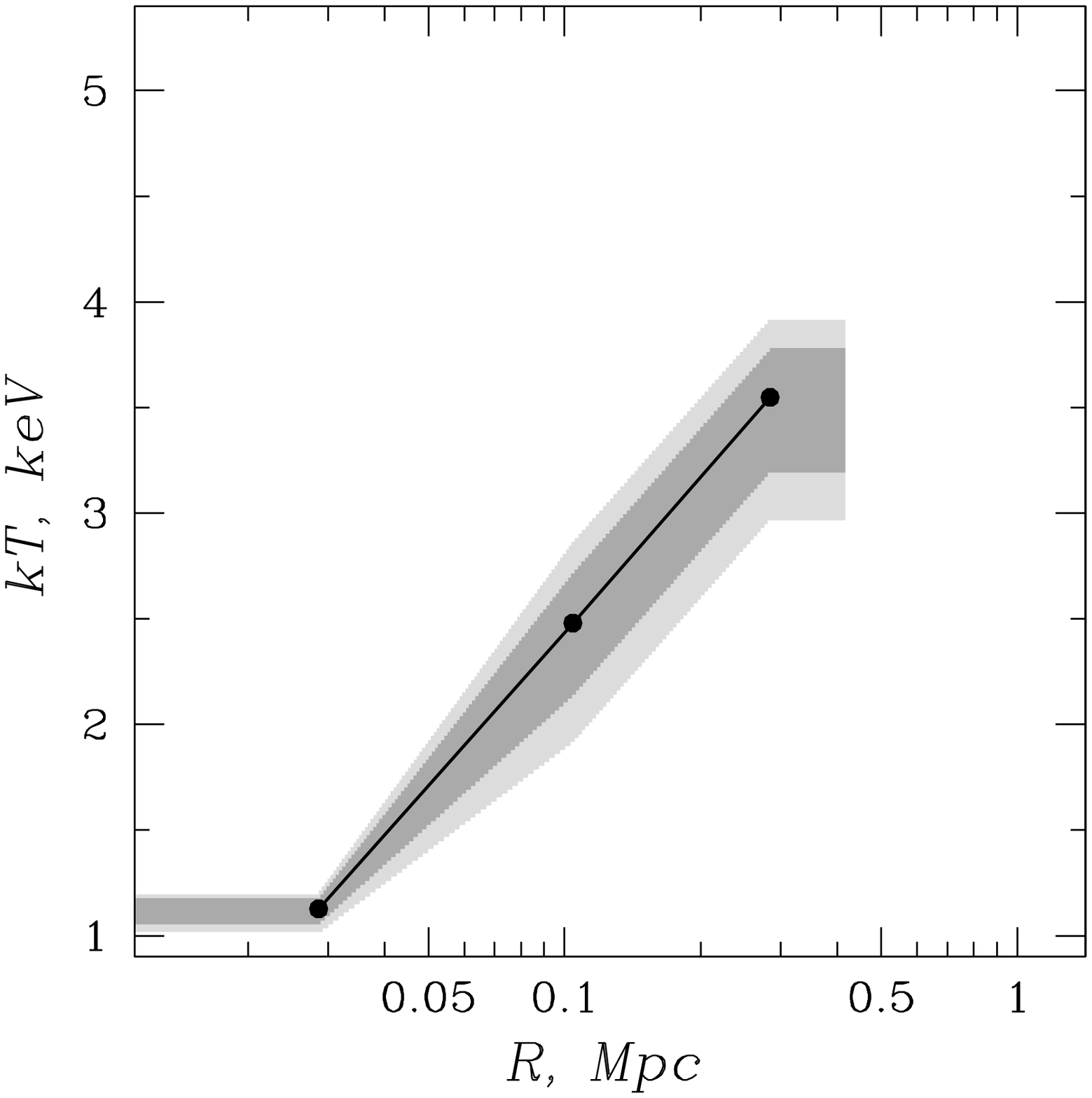} \hfill 
  \includegraphics[width=1.6in]{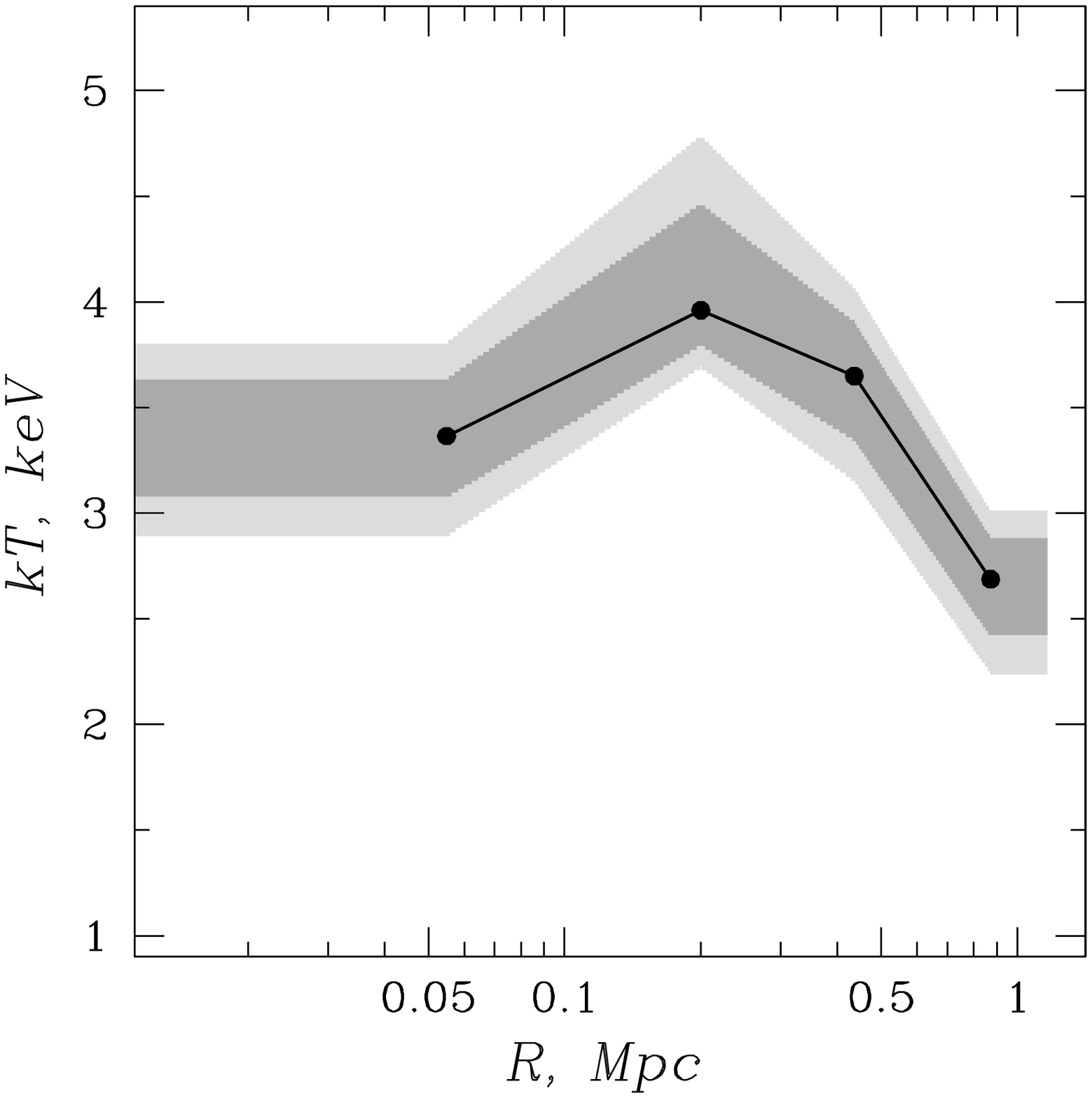} 

    \includegraphics[width=1.6in]{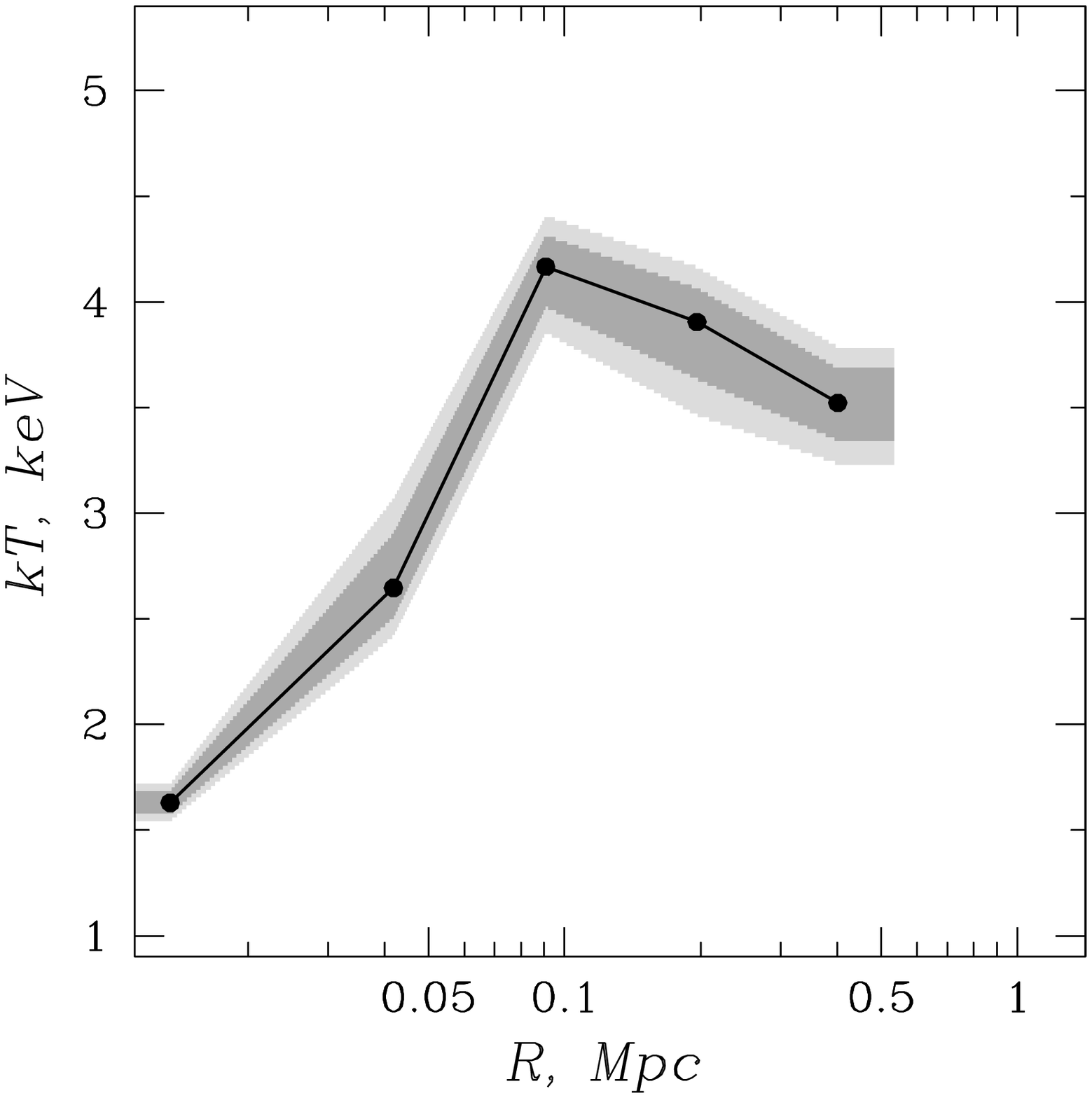}  \hfill 
  \includegraphics[width=1.6in]{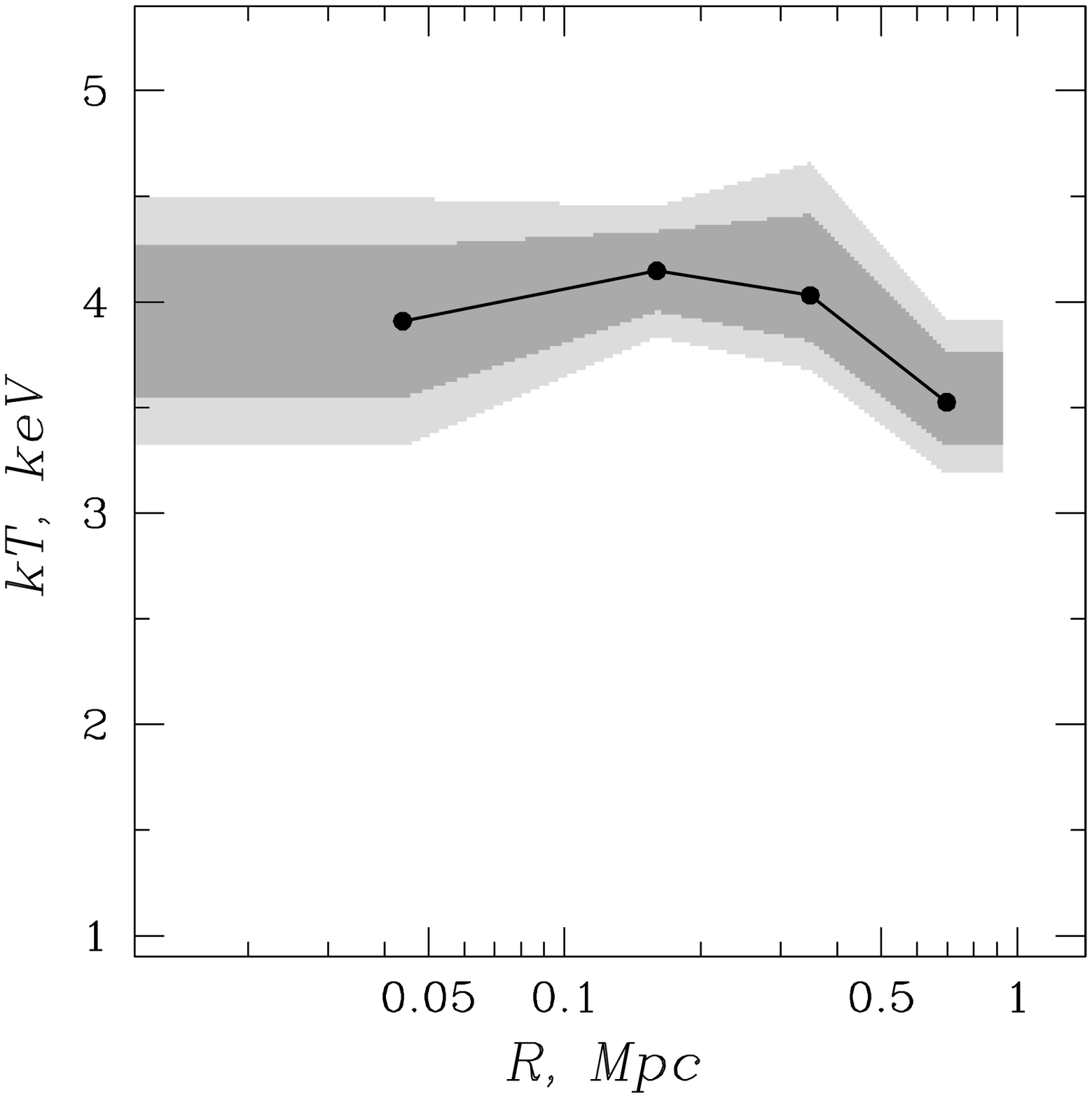}

\figcaption{Temperature profiles, derived using a simple one-temperature
model to fit the data.  The solid lines correspond to the best-fit with
filled circles indicating the spatial binning used in the analysis. Dark and
light shaded zones around the best fit curves denote the 68 and 90 per cent
confidence areas.  
\label{kt-fig}}
\vspace*{-20.1cm}

{\it \hspace*{2.5cm} A2197E \hspace*{3.9cm} A400 \hspace*{3.9cm} A194 \hspace*{3.9cm} A262}

\vspace*{3.55cm}

{\it \hspace*{2.6cm} MKW4S \hspace*{3.6cm} A539 \hspace*{3.8cm} AWM4 \hspace*{3.8cm} MKW9}

\vspace*{3.55cm}

{\it \hspace*{2.4cm} A2197W \hspace*{3.7cm} A2634 \hspace*{3.6cm} A4038 \hspace*{3.5cm} 2A0335}

\vspace*{3.55cm}

{\it \hspace*{2.6cm} HCG94 \hspace*{3.5cm} A2052 \hspace*{3.8cm} A779 \hspace*{3.5cm} MKW3S}

\vspace*{3.55cm}
{\it \hspace*{2.8cm} CEN \hspace*{13.3cm} A2063}

\vspace*{3.7cm}

\end{figure*}

\begin{figure*}

   \includegraphics[width=1.6in]{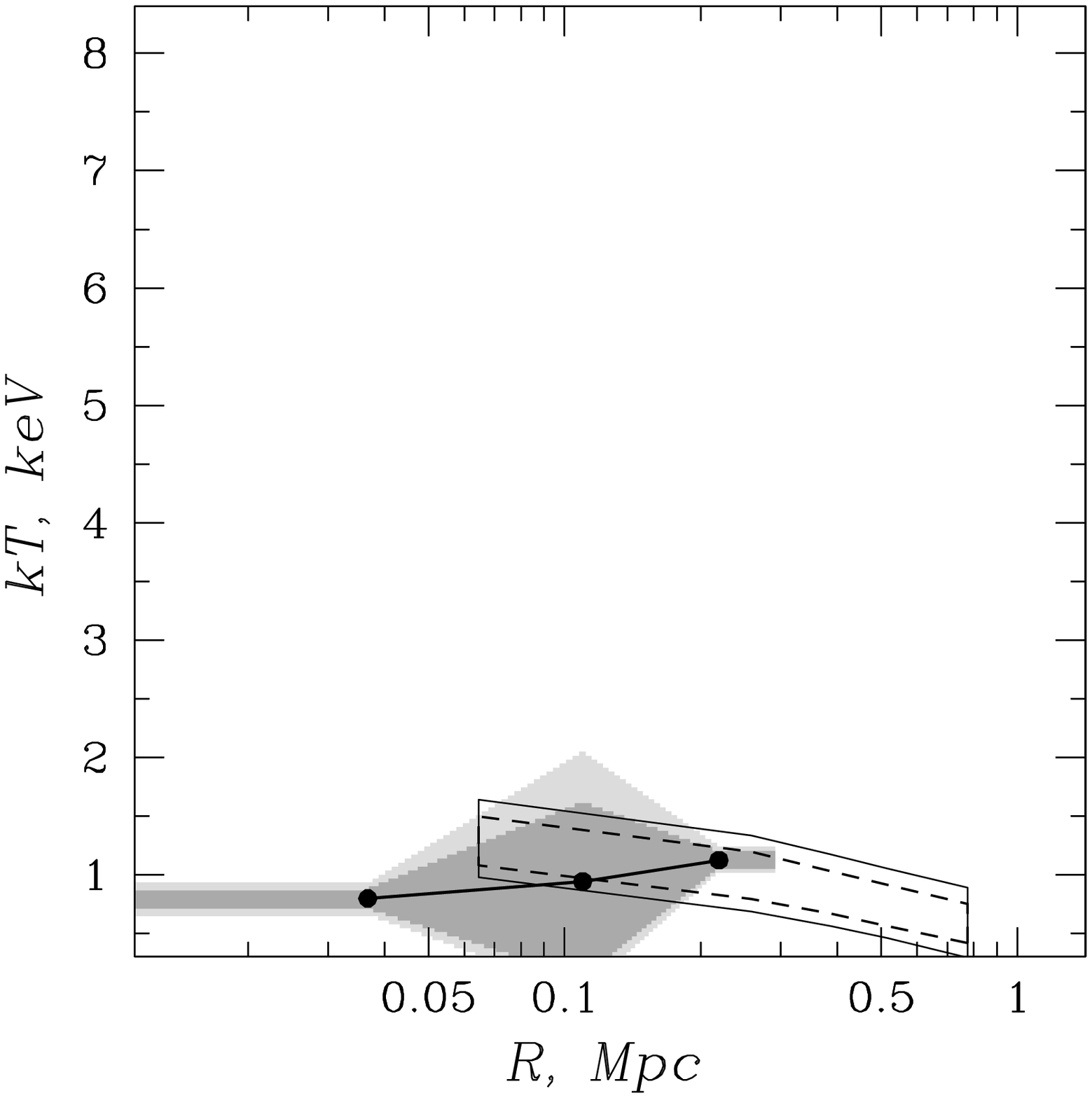} \hfill
   \includegraphics[width=1.6in]{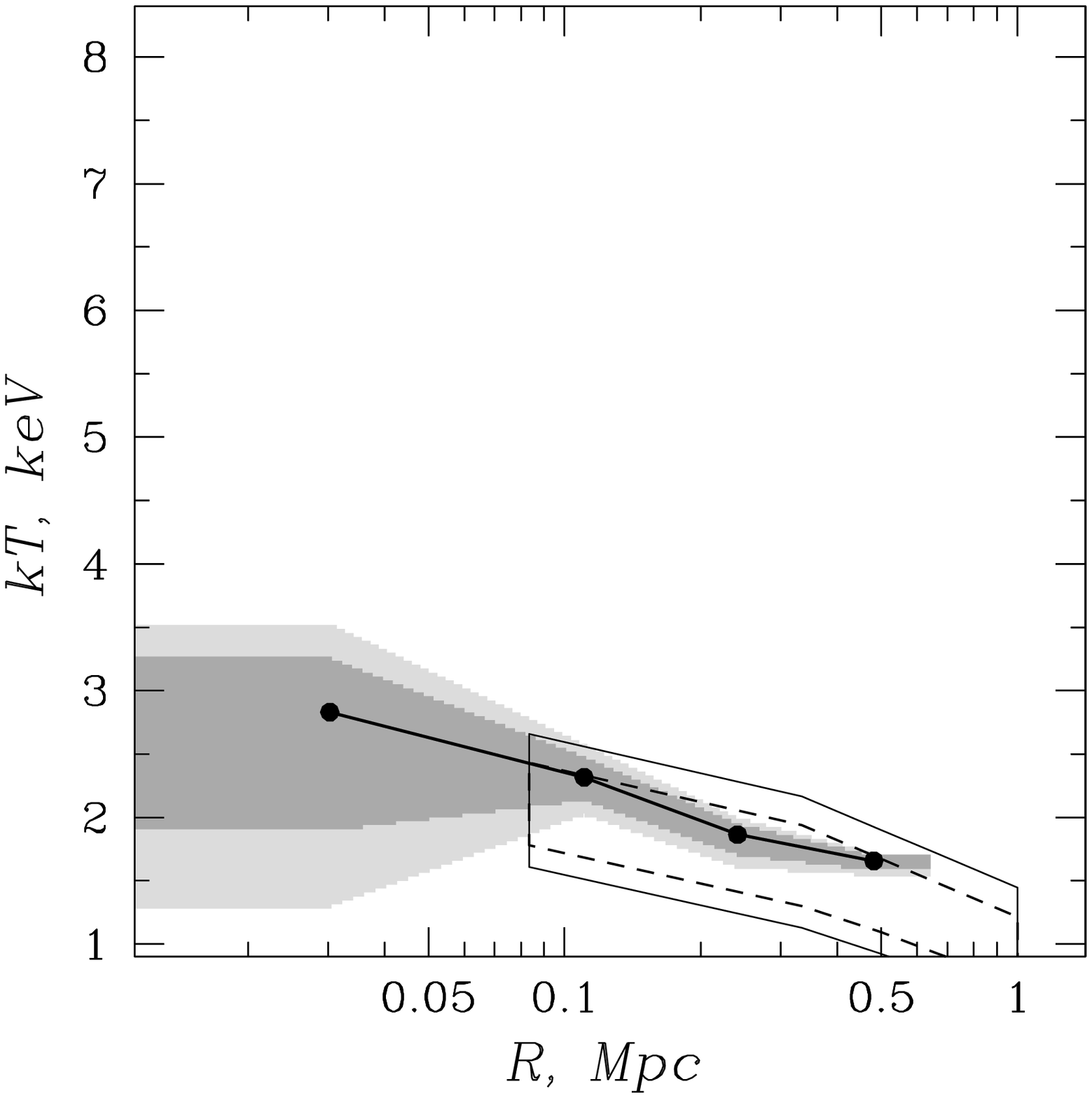} \hfill
   \includegraphics[width=1.6in]{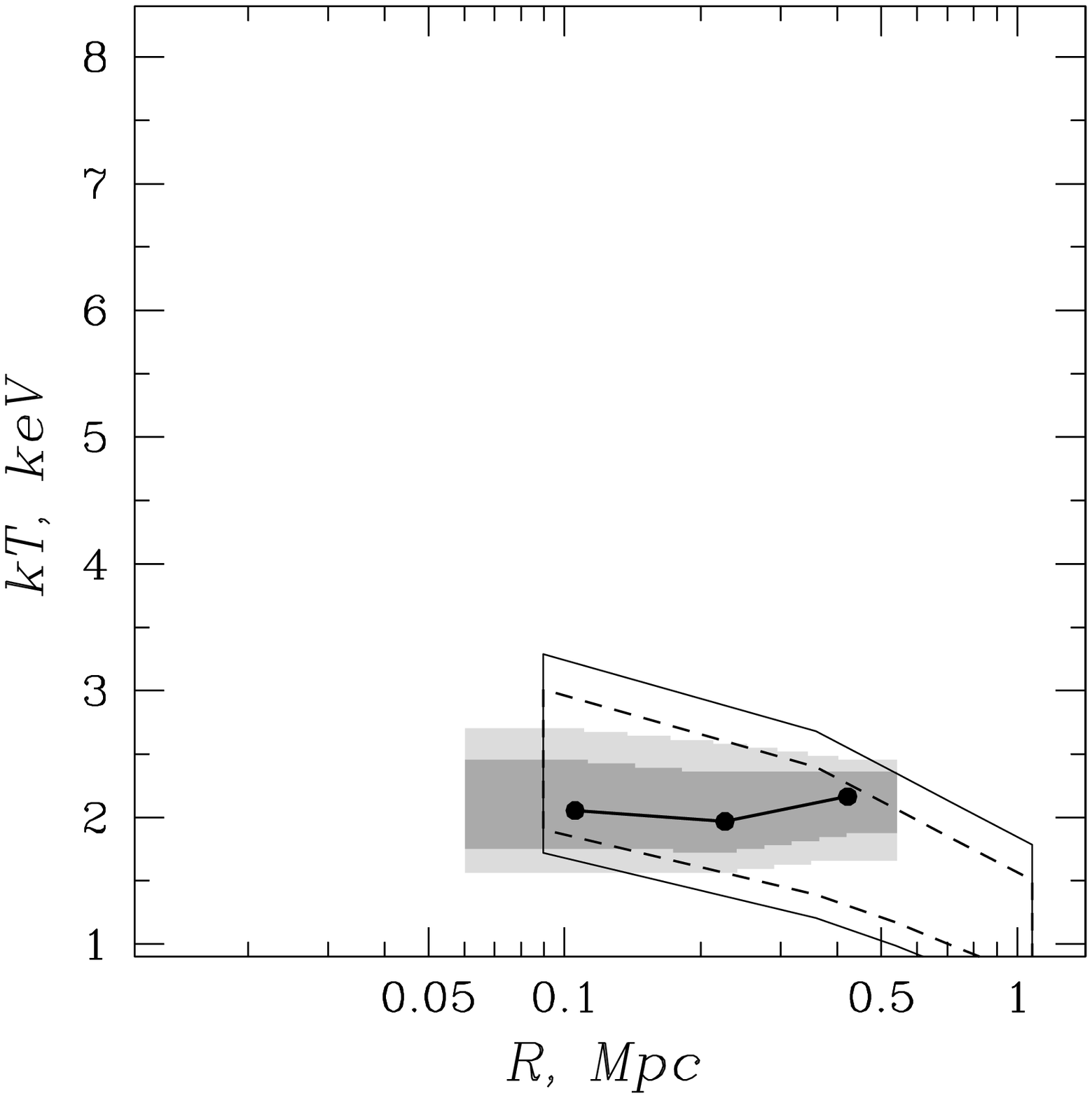} \hfill
   \includegraphics[width=1.6in]{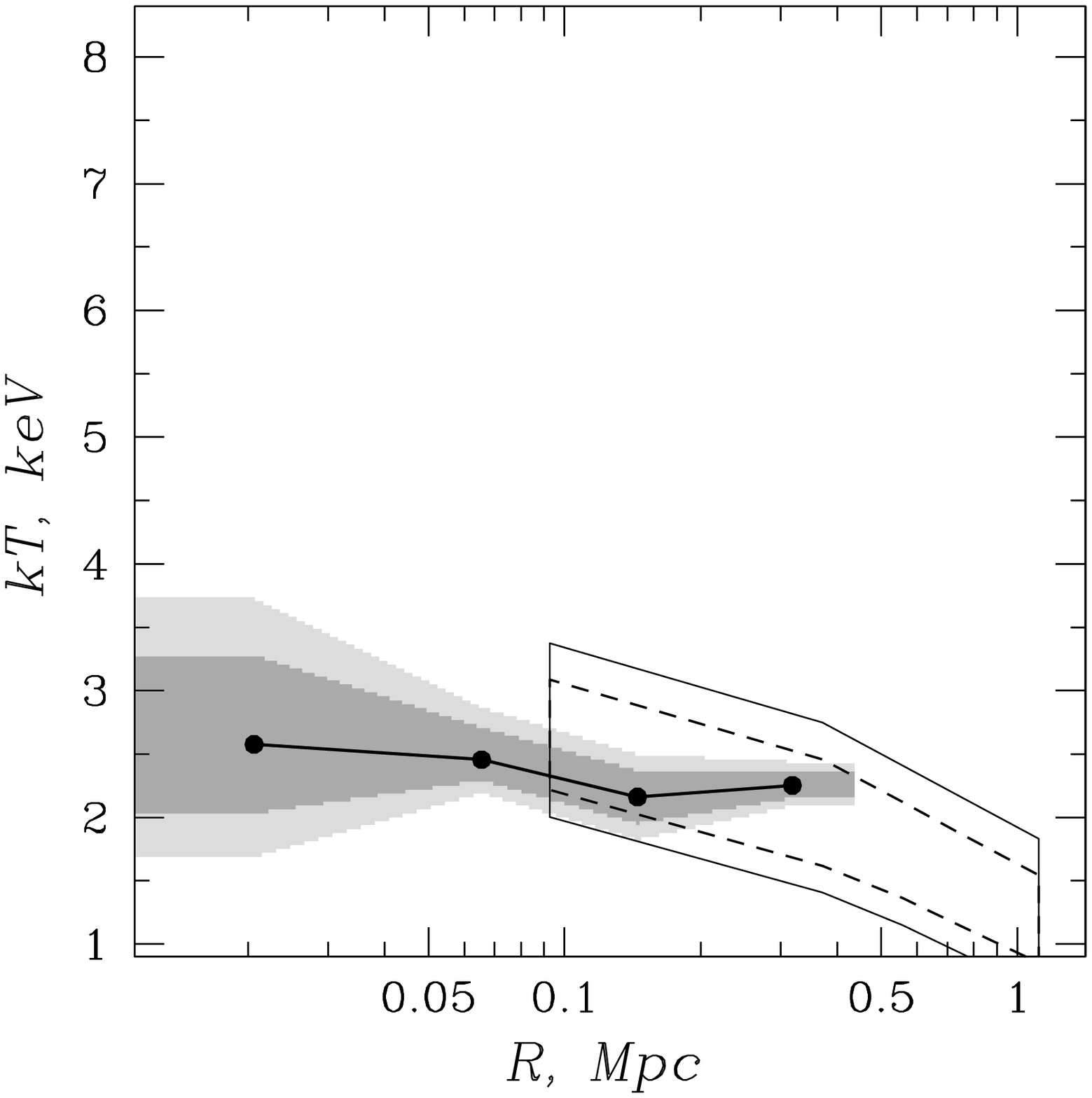} 

  \includegraphics[width=1.6in]{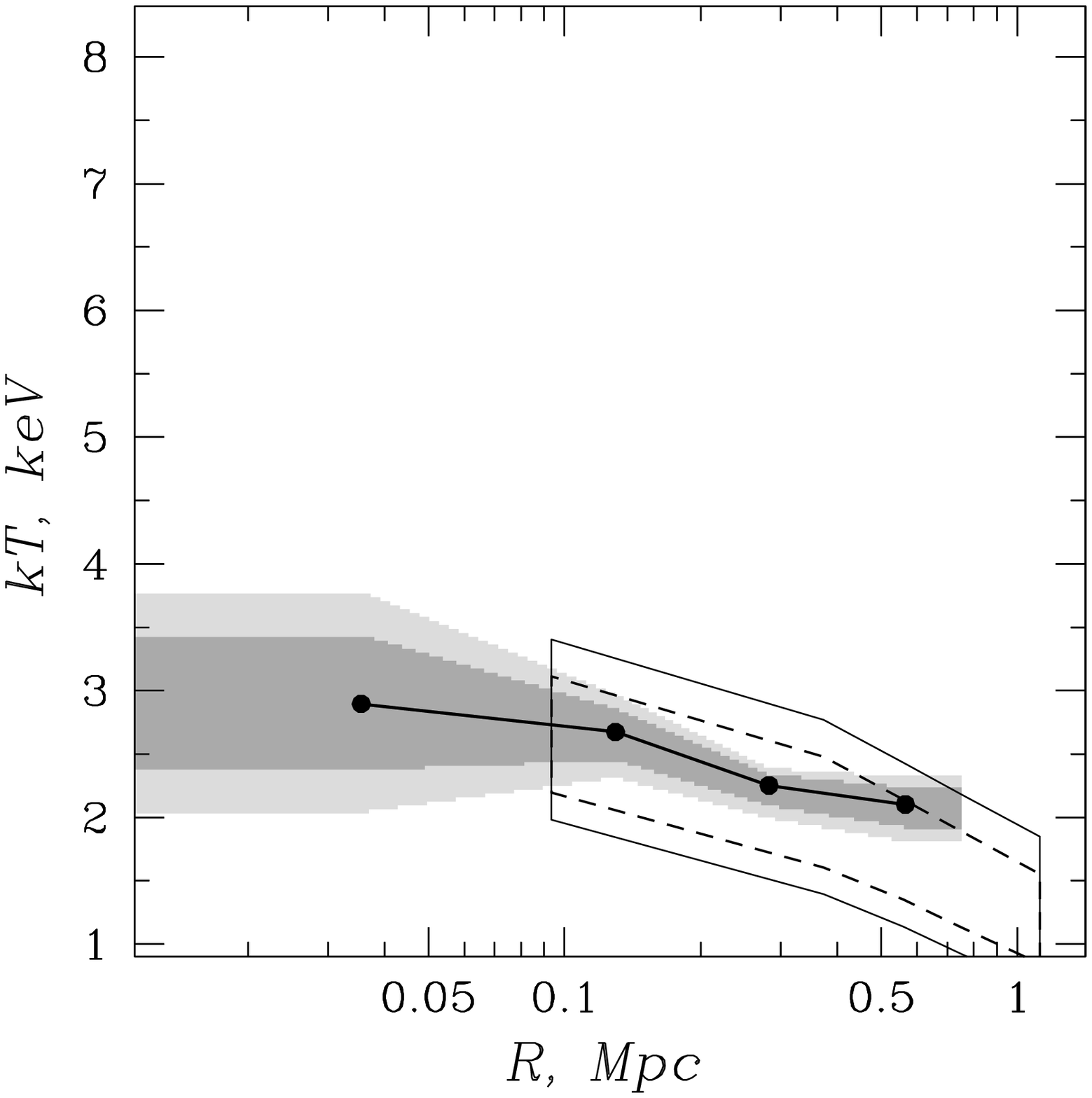} \hfill  
   \includegraphics[width=1.6in]{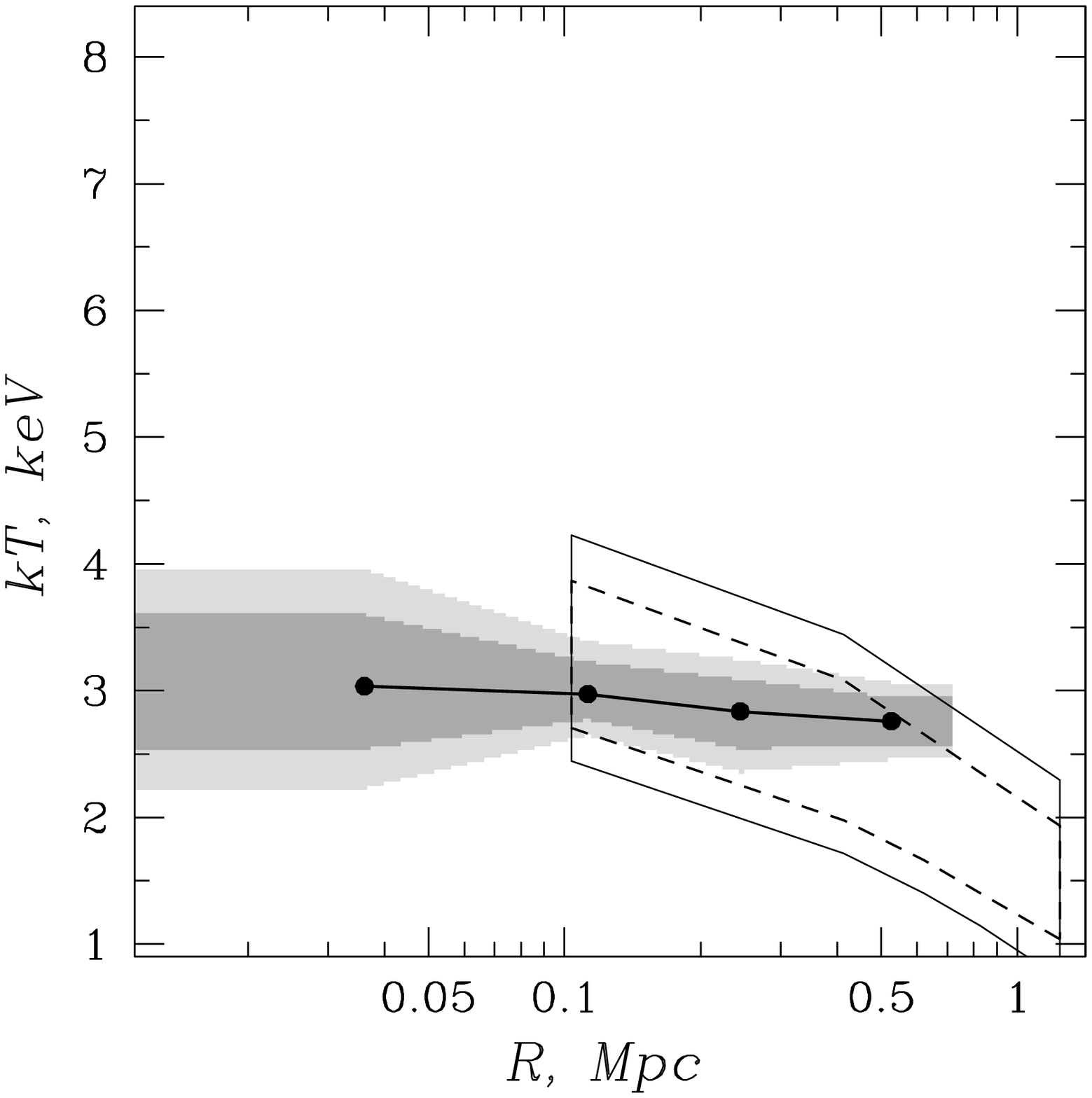} \hfill 
  \includegraphics[width=1.6in]{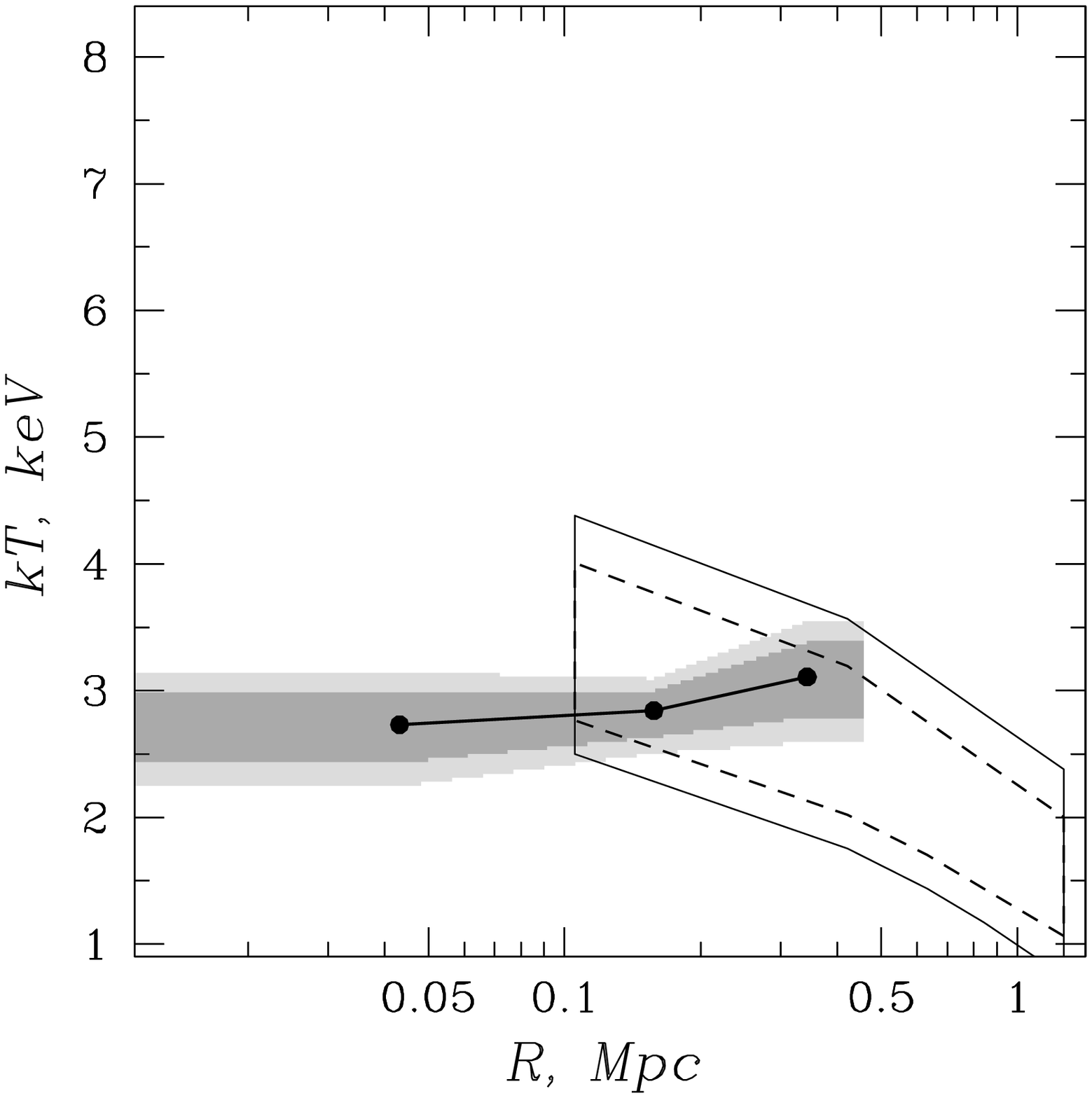} \hfill 
  \includegraphics[width=1.6in]{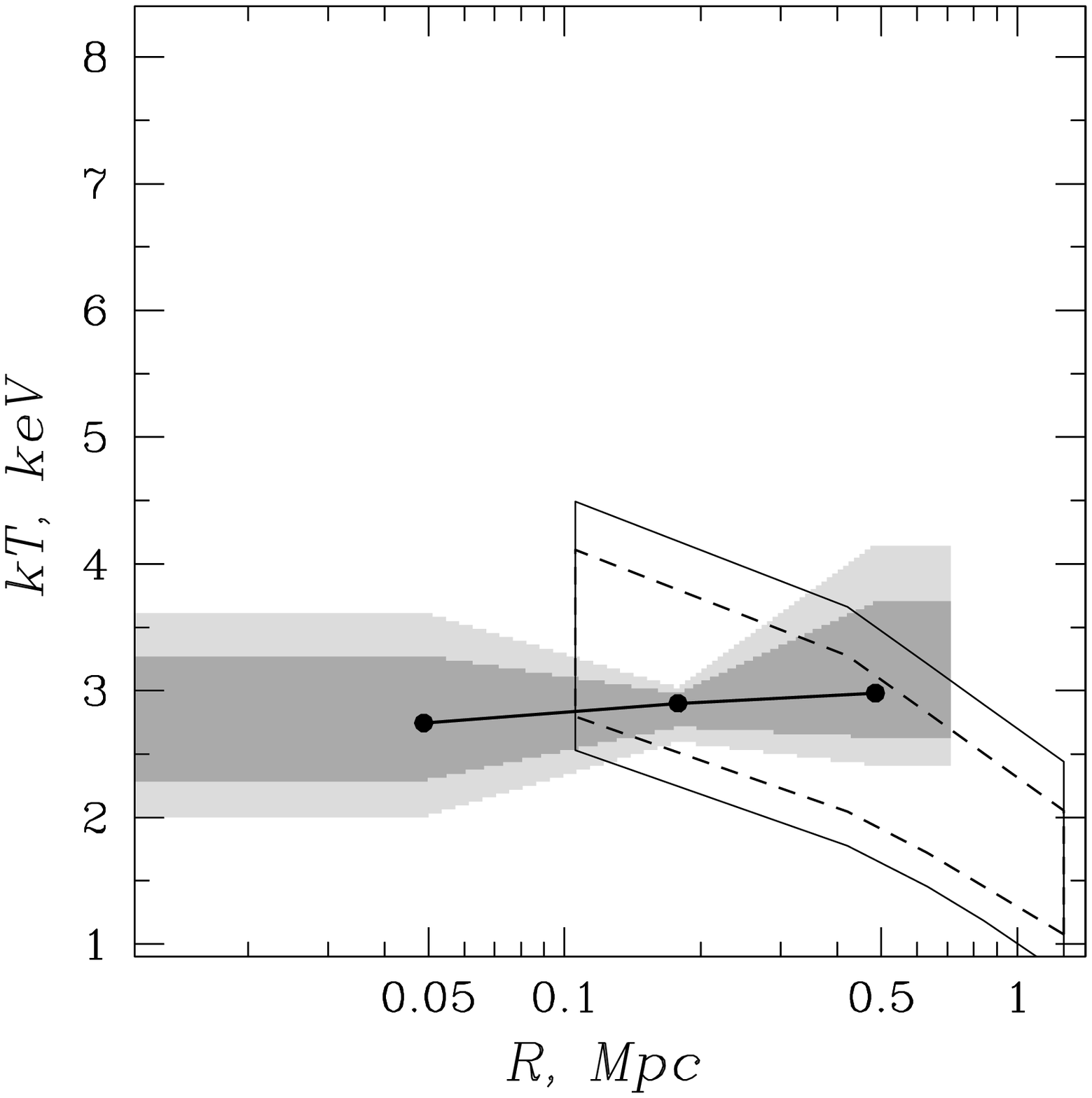}

   \includegraphics[width=1.6in]{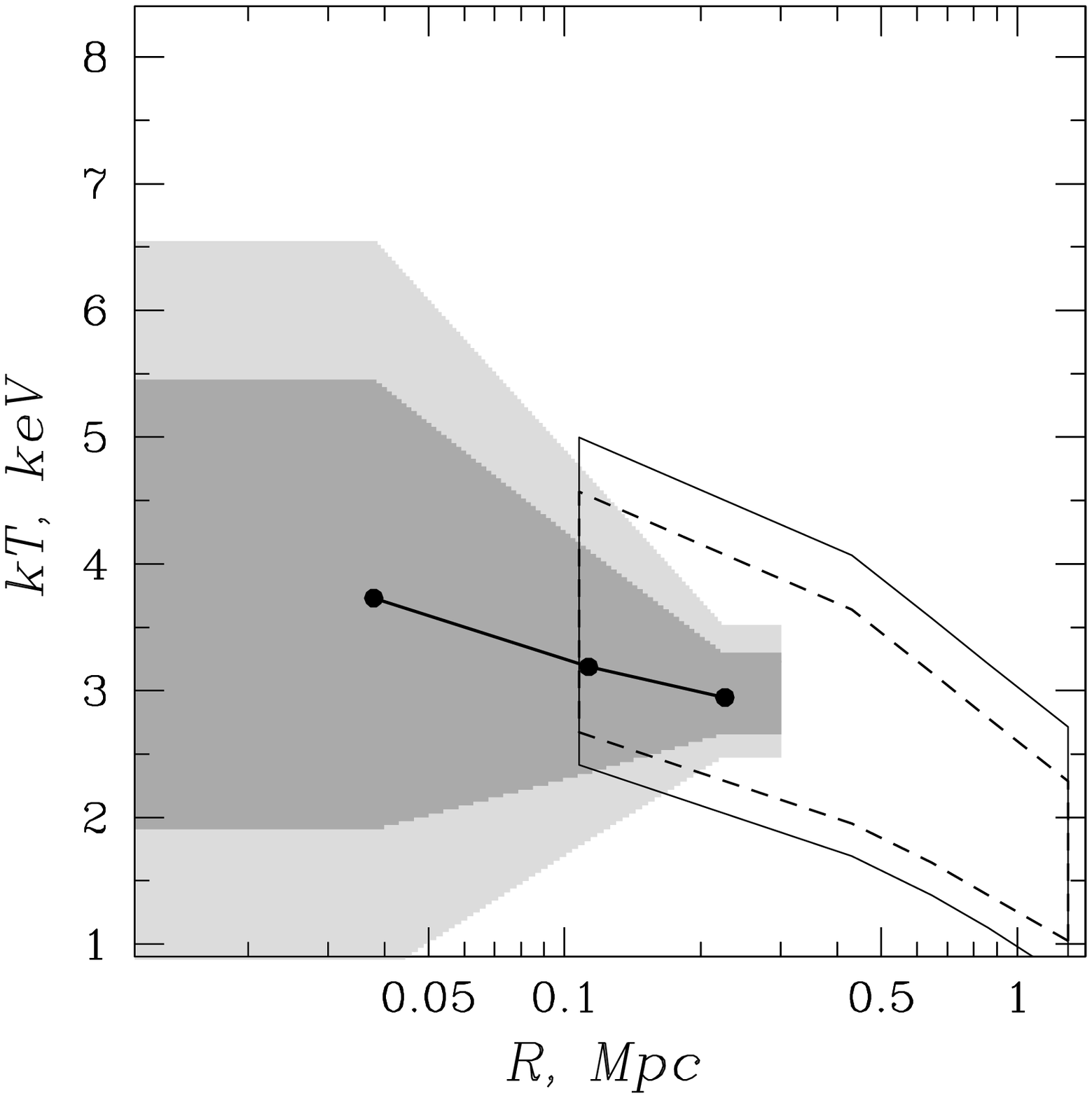}  \hfill 
  \includegraphics[width=1.6in]{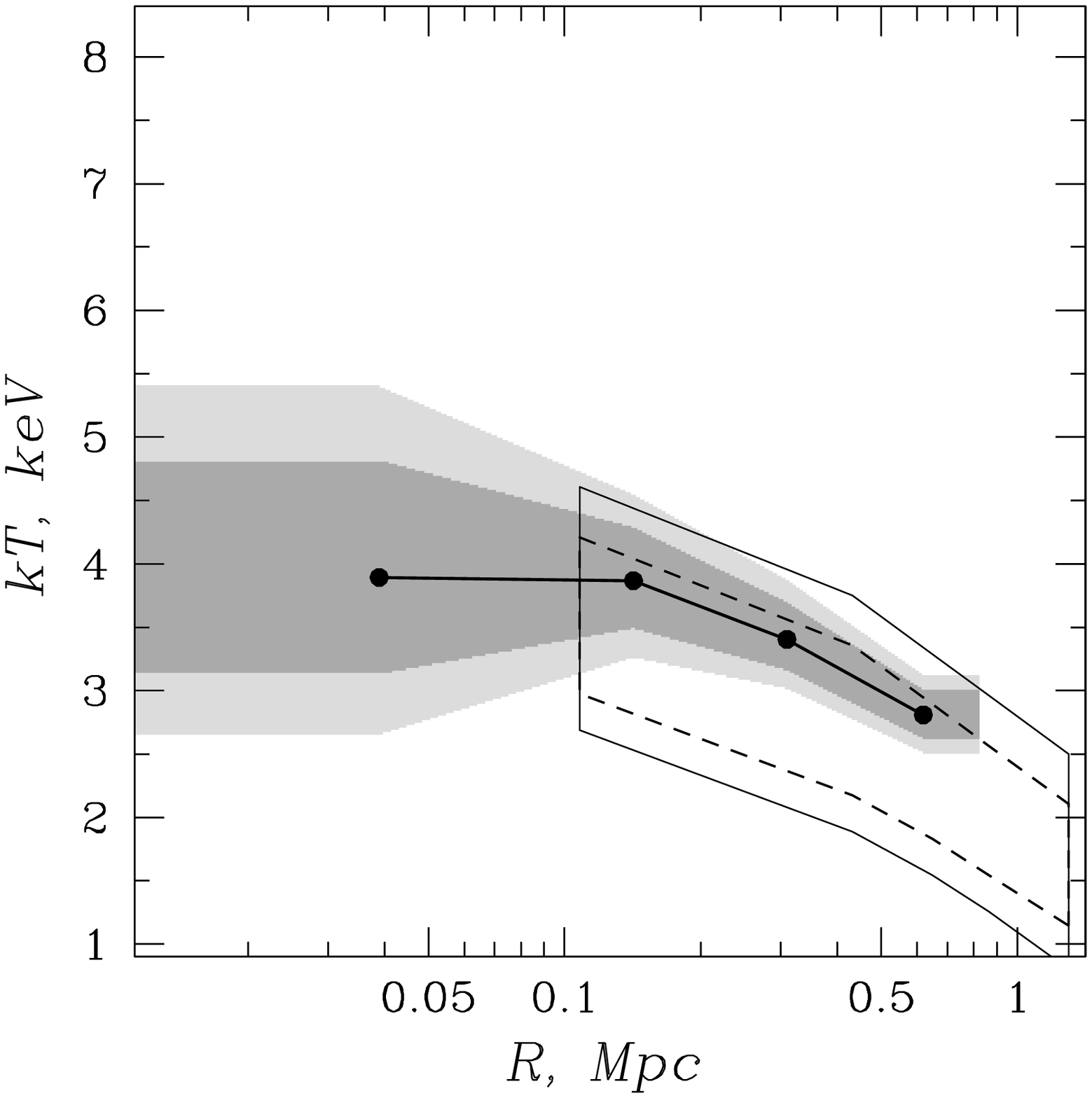} \hfill 
 \includegraphics[width=1.6in]{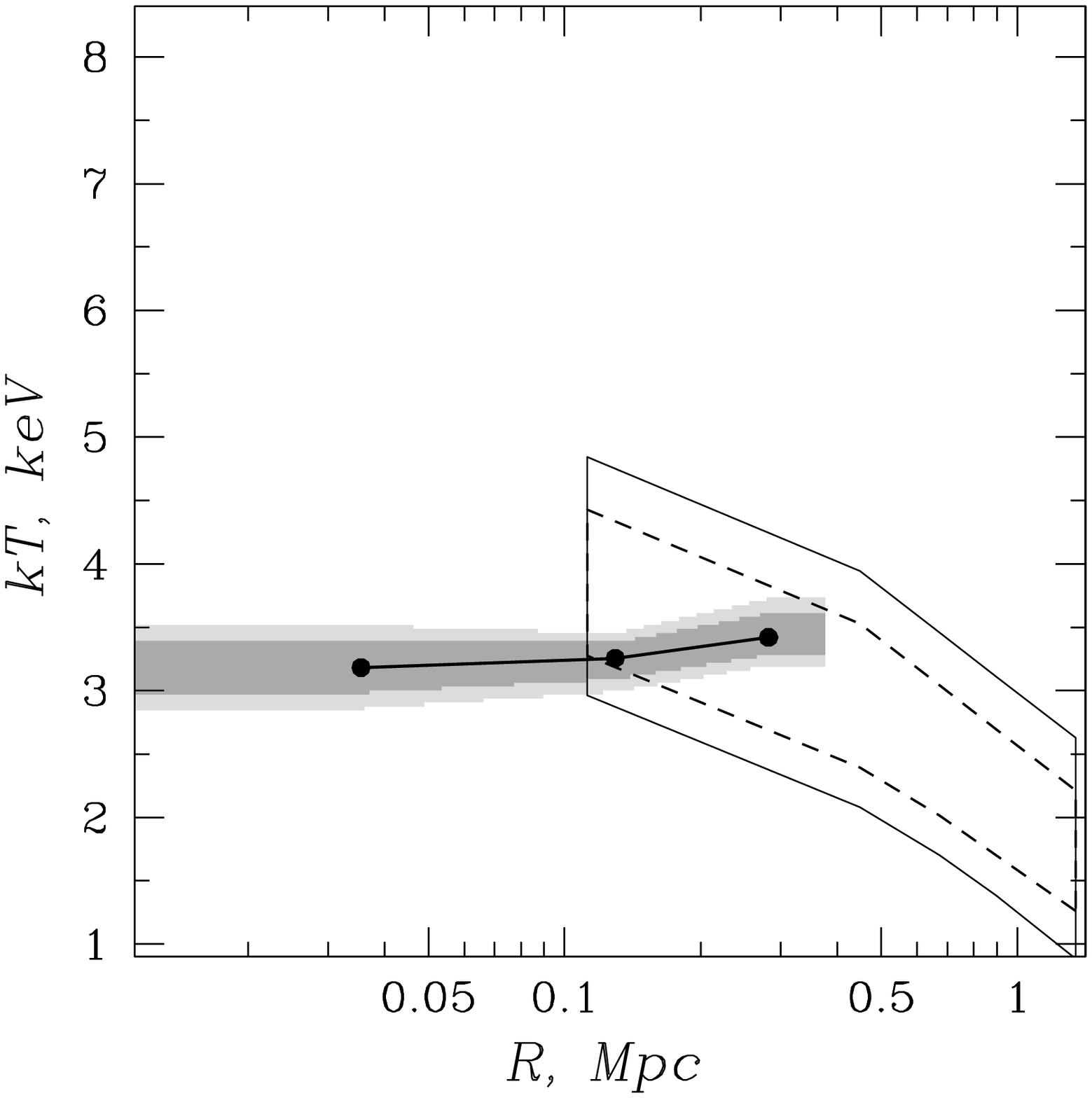} \hfill 
  \includegraphics[width=1.6in]{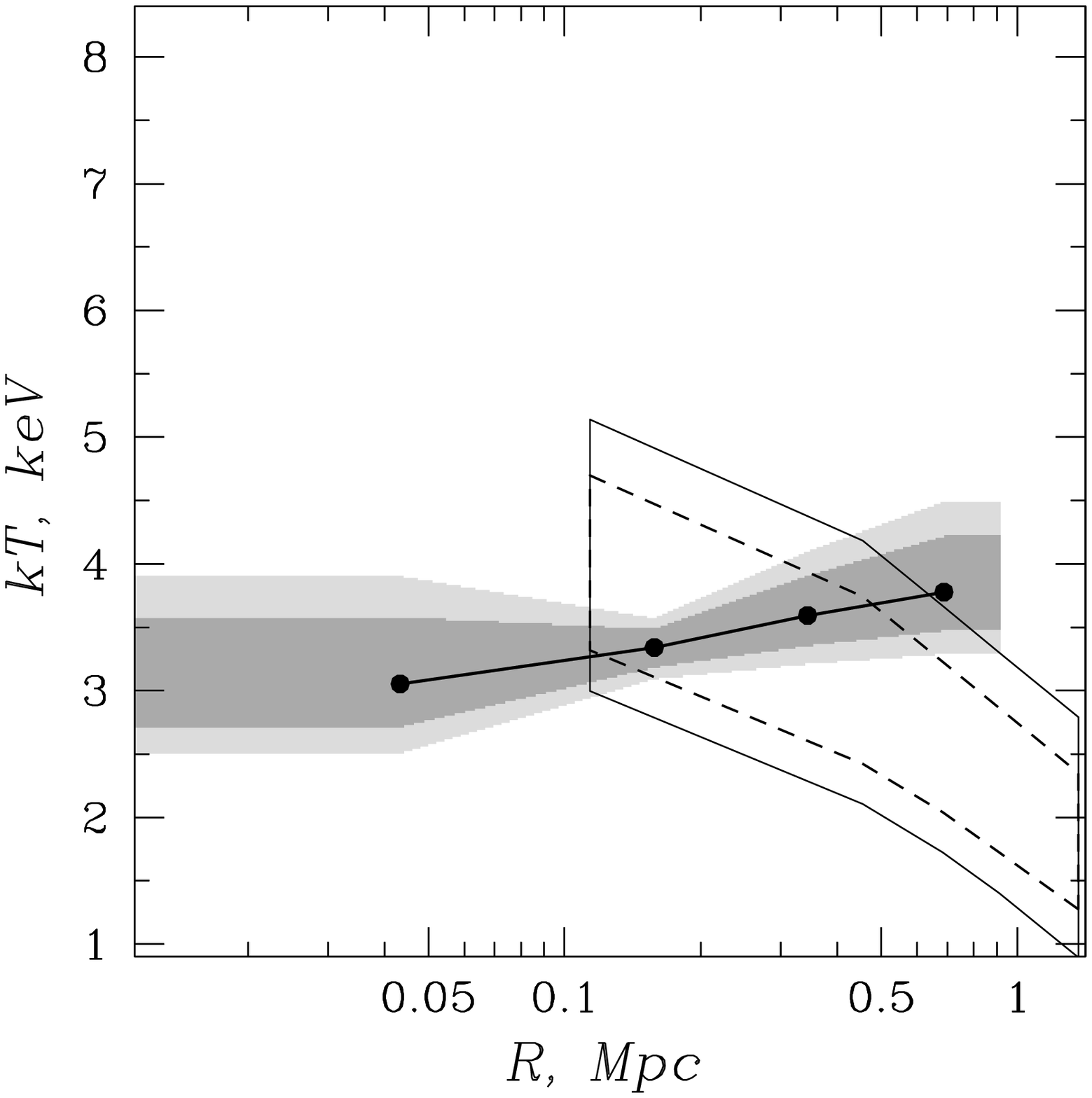} 

   \includegraphics[width=1.6in]{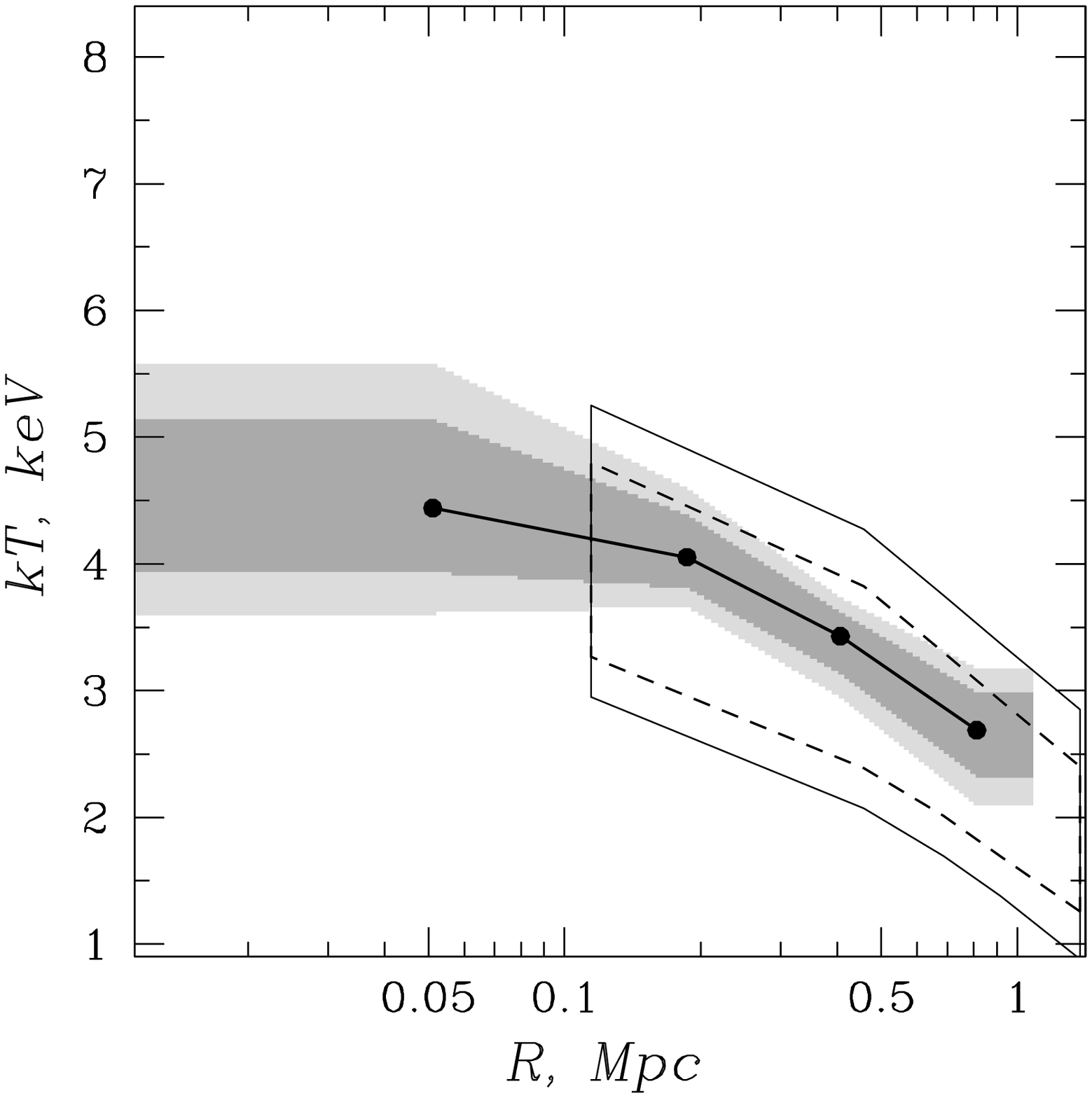} \hfill 
   \includegraphics[width=1.6in]{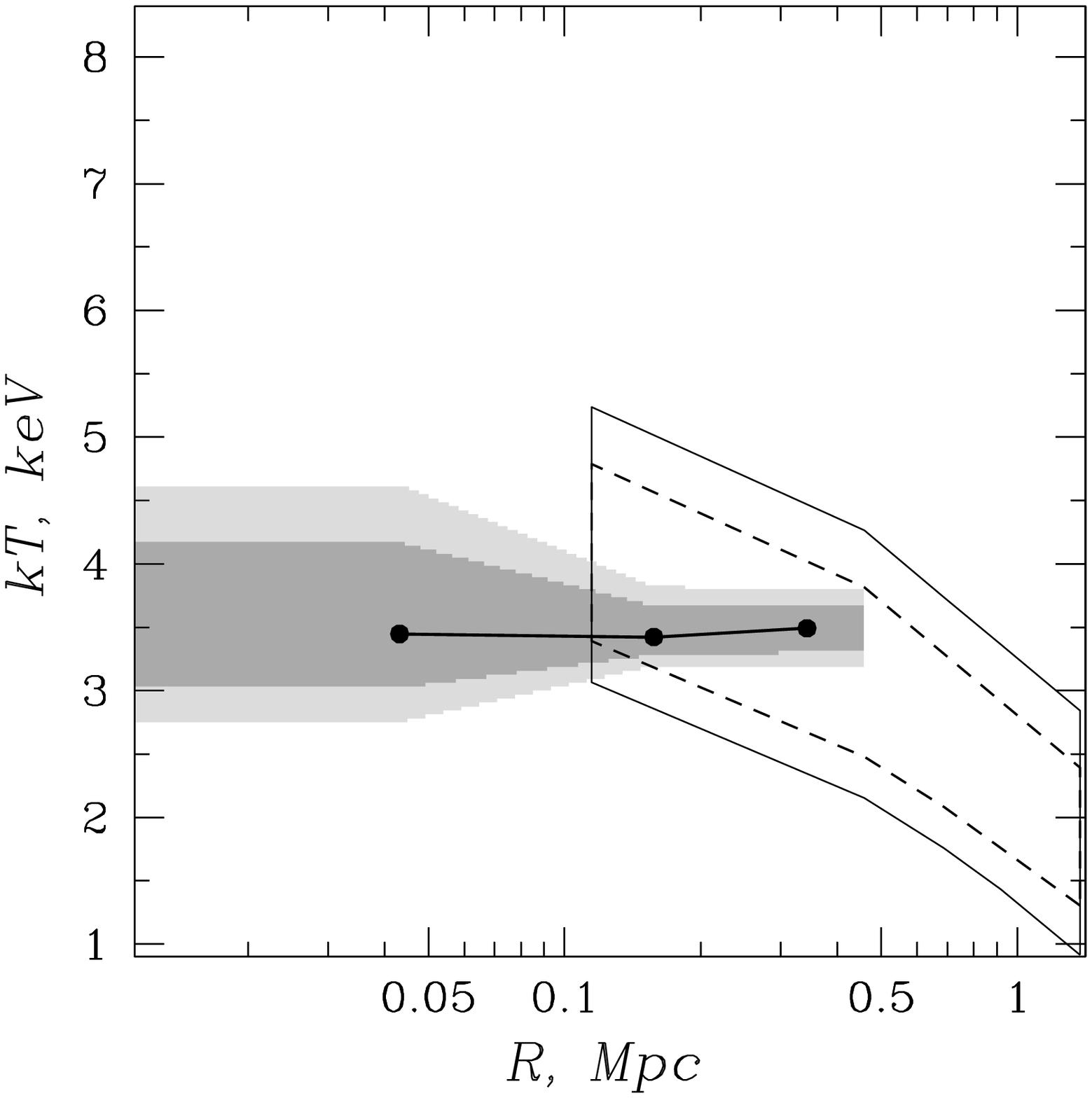} \hfill 
  \includegraphics[width=1.6in]{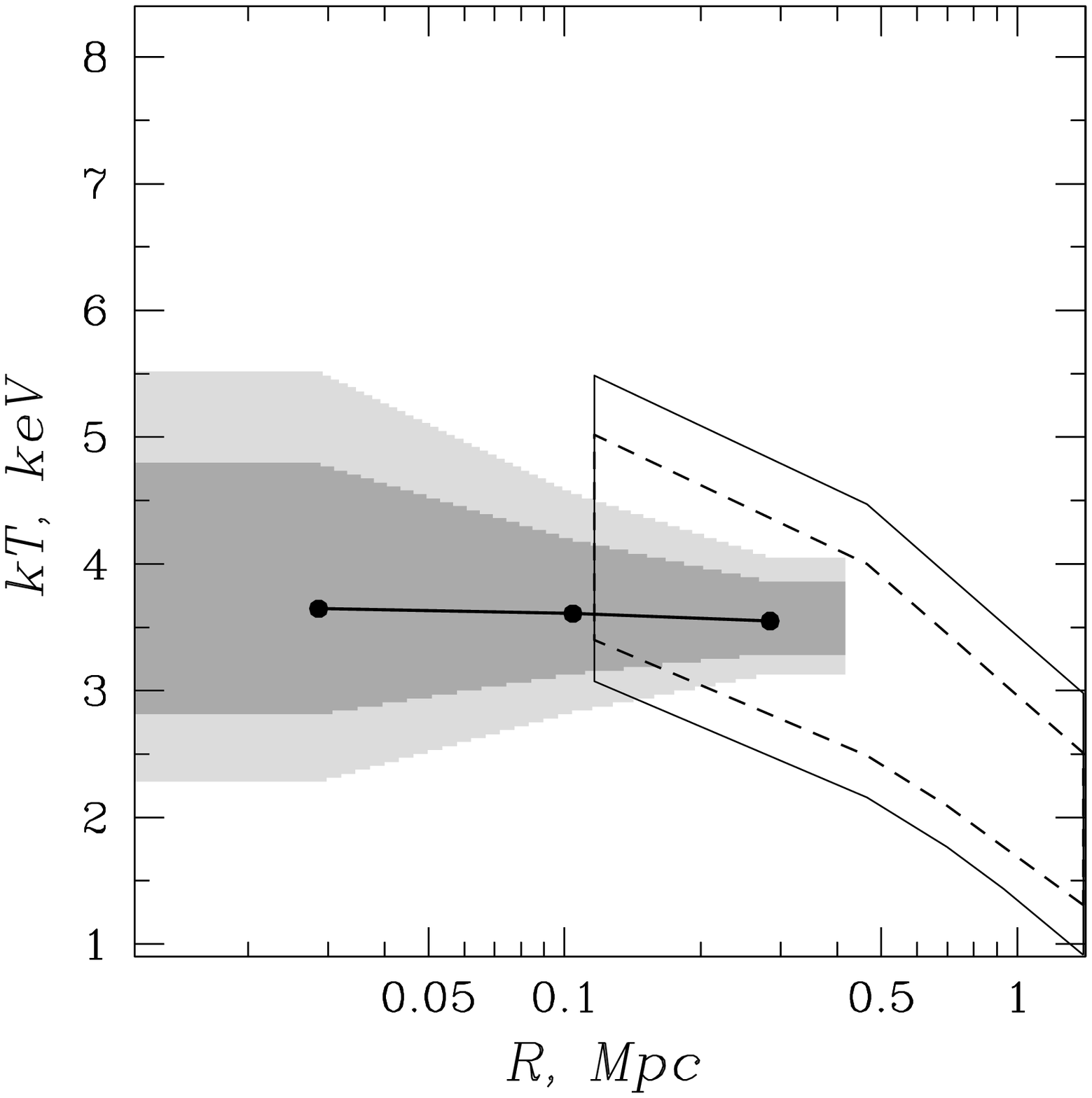} \hfill 
  \includegraphics[width=1.6in]{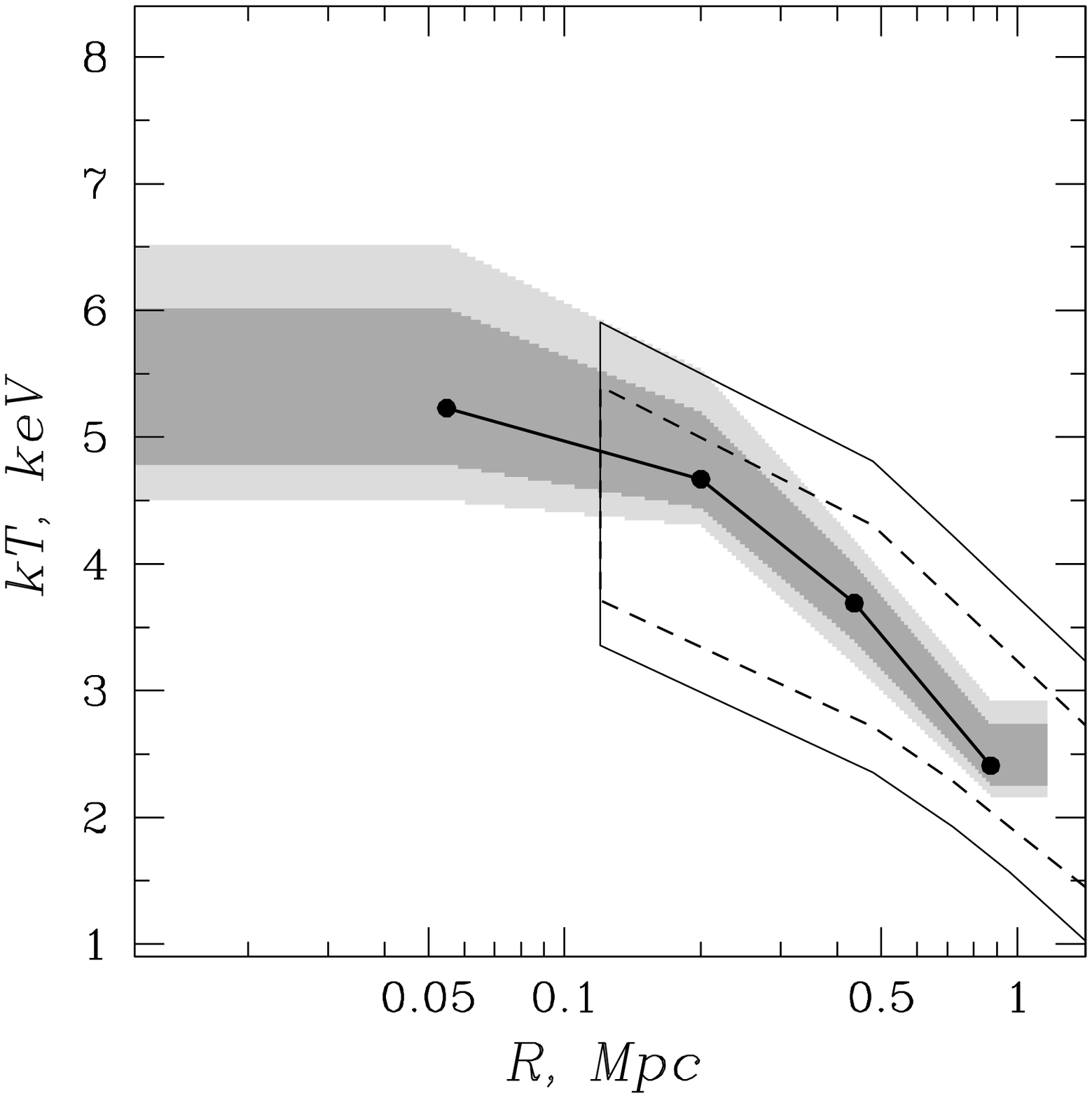} 

    \includegraphics[width=1.6in]{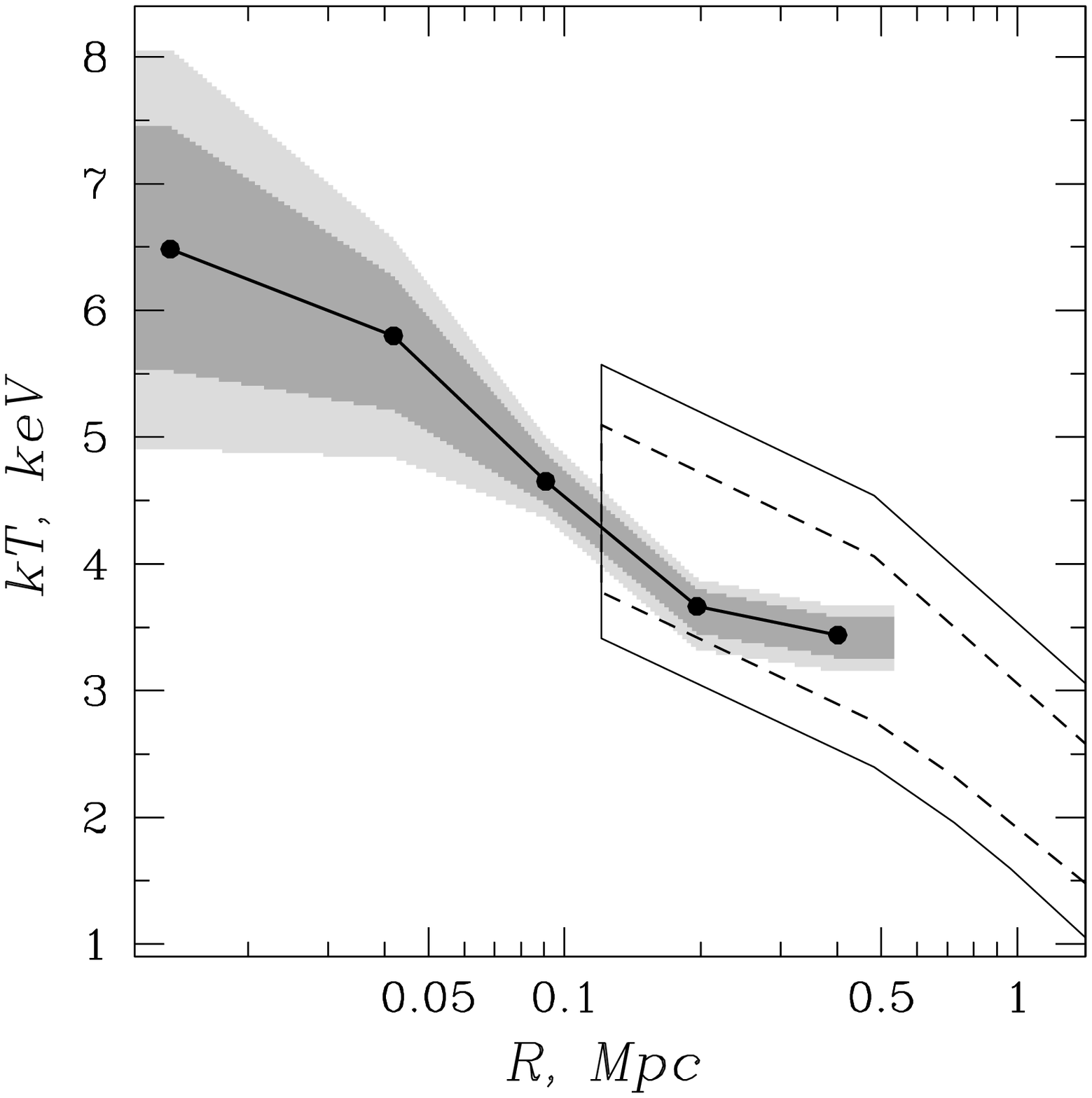}  \hfill 
  \includegraphics[width=1.6in]{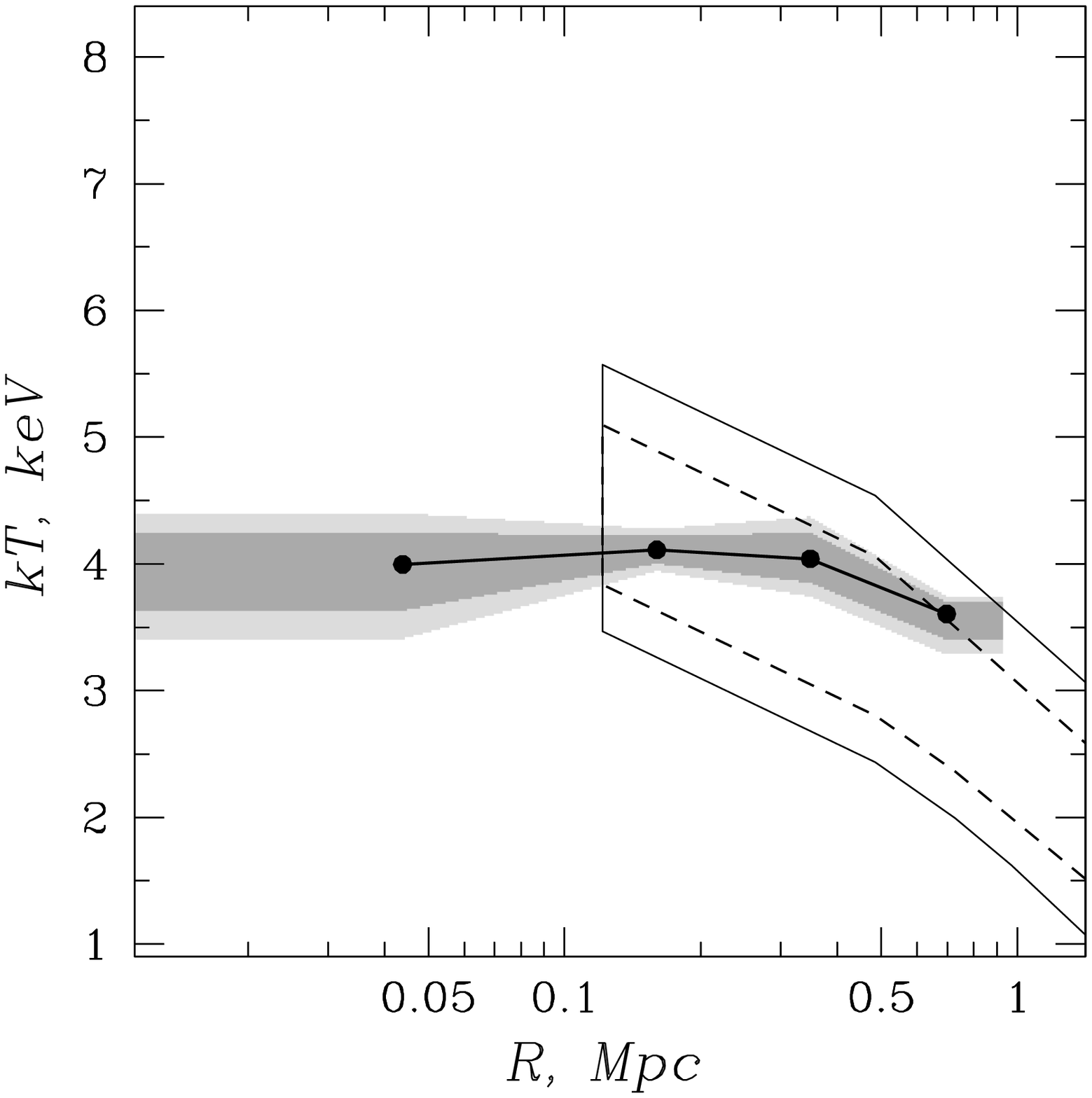}

\figcaption{Corrected temperature profiles, derived considering a two-phase
model for the center to fit the data.  The solid lines correspond to the
best-fit with filled circles indicating the spatial binning used in the
analysis. Dark and light shaded zones around the best fit curves denote the
68 and 90 per cent confidence areas. Contours denote the range of
temperatures found in Markevitch \etal (1998), scaled according to the
luminosity weighted temperature of the cluster, $kT_e$ (col. 14 in
Tab.\ref{tab:opt}) using virial units for radii (col. 16 in
Tab.\ref{tab:opt}).
\label{ktc-fig}}
\vspace*{-20.4cm}

{\it \hspace*{2.4cm} A2197E \hspace*{3.9cm} A400 \hspace*{3.9cm} A194 \hspace*{3.9cm} A262}

\vspace*{3.55cm}

{\it \hspace*{2.6cm} MKW4S \hspace*{3.6cm} A539 \hspace*{3.8cm} AWM4 \hspace*{3.8cm} MKW9}

\vspace*{3.55cm}

{\it \hspace*{2.4cm} A2197W \hspace*{3.7cm} A2634 \hspace*{3.6cm} A4038 \hspace*{3.5cm} 2A0335}

\vspace*{3.55cm}

{\it \hspace*{2.6cm} HCG94 \hspace*{3.5cm} A2052 \hspace*{3.8cm} A779 \hspace*{3.5cm} MKW3S}

\vspace*{3.55cm}
{\it \hspace*{2.8cm} CEN \hspace*{13.3cm} A2063}

\vspace*{3.7cm}

\end{figure*}

We present the derived temperature distributions using single-temperature
fits in Fig.\ref{kt-fig} and the temperature profiles of the hotter
component in the two-temperature fits in Fig.\ref{ktc-fig}. In the central
spatial bins of the clusters exhibiting a presence of a colder temperature
component, we also attempt the two-temperature fits. The single temperature
fits are used to calculate $kT_{unc}$ in Tab.\ref{tab:opt}, while
temperature of the hot component in the two-temperature fits is used to
calculate $kT_{e}$. One can see from the Table that there is no difference
for A400, A194 (however for this cluster the center is excluded anyway),
A539, AWM4, and A2634, while changes in the weighted temperature for other
cluster appear not to be very significant. In the single-temperature fits,
the temperature appears to drop significantly with radius in A400, A2634,
HCG94, in the outskirts of MKW3S, and also marginally in Cen. Adding a
second temperature in the fits adds A2063 and marginally MKW4S to the list
of clusters with declining temperature profiles. More evident is the removal
of rising temperature profiles in A262, 2A0335+096, A2052, A779, Cen and
marginally in MKW3S. The hot component is rather faint in the center and as
was shown in case of M87 may be a spatially distinct from cold component
(Finoguenov \& Jones 2000). However, the use of a two-temperature model
mimics the cooling-flow model of Johnstone \etal (1992, cf Buote 2000). To
quantify the characteristics of the cold component, we present in
Tab.\ref{tab:cold} the temperature, emissivity, and gas mass of this
component. The corresponding cooling rates can then be derived from these
values.
% Our attempts to fit the {\it mkcflow} model in XSPEC were not very
% successful.  
It was pointed out by Fukazawa (1997), that the cold component
of a two temperature fit in the range between 1 and 2 keV may reflect the
potential of the central cD rather than being an indication of a cooling
flow. While we cannot distinguish between these two possibilities with our
present sample, in the FDP sample our fits to A2029, A3112, and the A780
spectra resulted in temperatures for the cold components of $\sim$ 5, 3 and
3 keV, respectively.  It is therefore more likely that these clusters have
cooling flows.

%\begin{table*}
{
\begin{center}
\footnotesize
\tabcaption{\centerline{\footnotesize
Characteristics of the central cold component.$^{\flat}$}
\label{tab:cold}}

\begin{tabular}{lccc}
\hline
\hline
Name  &$kT_{c}$              & EM$^{\natural}$ & $M_{gas}$\\
      & keV                  &            & $10^{11}$ \msun\     \\
\hline                                                                    
   A262 & 1.259 (1.17:1.35)  &  2.747 (2.02:3.47) & 0.545 (0.47:0.61)\\
  MKW4S & 1.437 (1.33:1.48)  &  0.147 (0.09:0.17) & 0.504 (0.40:0.53)\\
   MKW9 & 1.600 (1.49:1.71)  &  0.257 (0.21:0.28) & 1.452 (1.31:1.51)\\
  A2197 & 1.178 (1.04:1.30)  &  0.117 (0.03:0.23) & 0.523 (0.29:0.74)\\
  A4038 & 1.507 (1.31:1.73)  &  3.908 (2.22:5.60) & 2.868 (2.16:3.43)\\
 2A0335 & 1.533 (1.44:1.67)  &  6.602 (4.72:8.27) & 5.508 (4.66:6.16)\\
  HCG94 & 1.518 (0.79:2.05)  &  0.411 (0.12:0.67) & 2.112 (1.12:2.71)\\
  A2052 & 1.583 (1.44:1.70)  &  5.784 (4.25:7.59) & 5.160 (4.43:5.92)\\
   A779 & 1.017 (0.97:1.10)  &  0.211 (0.12:0.29) & 0.347 (0.27:0.41)\\
  MKW3S & 1.898 (1.70:2.10)  &  3.934 (3.18:4.89) & 7.980 (7.18:8.89)\\
    CEN & 1.622 (1.54:1.71)  &  15.05 (13.3:17.2) & 1.213 (1.15:1.28)\\
  A2063 & 1.521 (1.19:1.87)  &  0.075 (0.00:0.18) & 0.617 (0.00:0.97)\\
\hline               
\end{tabular}
\end{center}

$^{\flat}$ \hspace*{0.3cm}{\footnotesize Errors
are quoted on the 68\% confidence level.}

$^{\natural}$ \hspace*{0.3cm}{\footnotesize The emission
measures are quoted in units of $10^{-17}\int n_e n_p dV/4\pi D^2$, i.e. $10^{-3}$
times the XSPEC units for {\it norm}. Correction for SIS vs GIS
normalizations of 1.2 has been made} 
}
%\end{table*}

\medskip

Most of the clusters in our sample have temperature profiles similar to the
universal temperature profile of Markevitch $\etal$ (1998). Only 2A0335+096
can be considered as discrepant, similar to findings of Kikuchi \etal
(1999). However, our conclusion regarding MKW3S favors the results of
Markevitch \etal (1998). In MKW3S we detect a lower temperature in the outer
annulus even with the single-temperature models. However, at the radius in
question, the SIS data do not cover the Northeastern part of the
cluster. The presence of high temperature gas in this region would account
for our discrepancies with Kikuchi \etal (1999). Some of the clusters in
this sample (A194, A539, AWM4, MKW9, A4038, A2052 and A779) suggest an
isothermal temperature distribution.  The implications of this result on the
M--T relation are further discussed in Finoguenov, Reiprich \& Boehringer
(2001). Modeling of ASCA data is especially complicated in the presence of
strong cooling flows (Markevitch 2000, private communication). Therefore,
our observation of the apparent temperature declines in the non-cooling flow
clusters A400 and A2634 is particularly important for confirmation of
temperature gradients.

The steep rise in the central gas temperature we detect in the Centaurus
cluster mimics the presence of the non-thermal component reported in Allen
\etal (1999).  Due to the limitations of the SIS data, i.e., the complex PSF
and low sensitivity at high energies, a more detailed study must await XMM
and Chandra observations.

Temperature and iron abundance profiles for the clusters reported in this
paper were also analyzed by White (2000) using GIS data.  A comparison of
our temperature profiles with those in White shows remarkable agreement for
single-temperature fitting (our Fig.\ref{kt-fig}), while our results on the iron
abundance supersedes in quality White (2000) results, because of the 
greater spectral resolution and sensitivity of the SIS compared to the 
GIS. For hotter clusters, the GIS iron abundance results have a similar 
quality with the 
SIS results of FDP. Given the differences in White's method for 
modeling of ASCA PSF effects (White \& Buote 2000), this agreement reassures
the robustness of ASCA measurements.

%\clearpage

\begin{figure*}

   \includegraphics[width=1.6in]{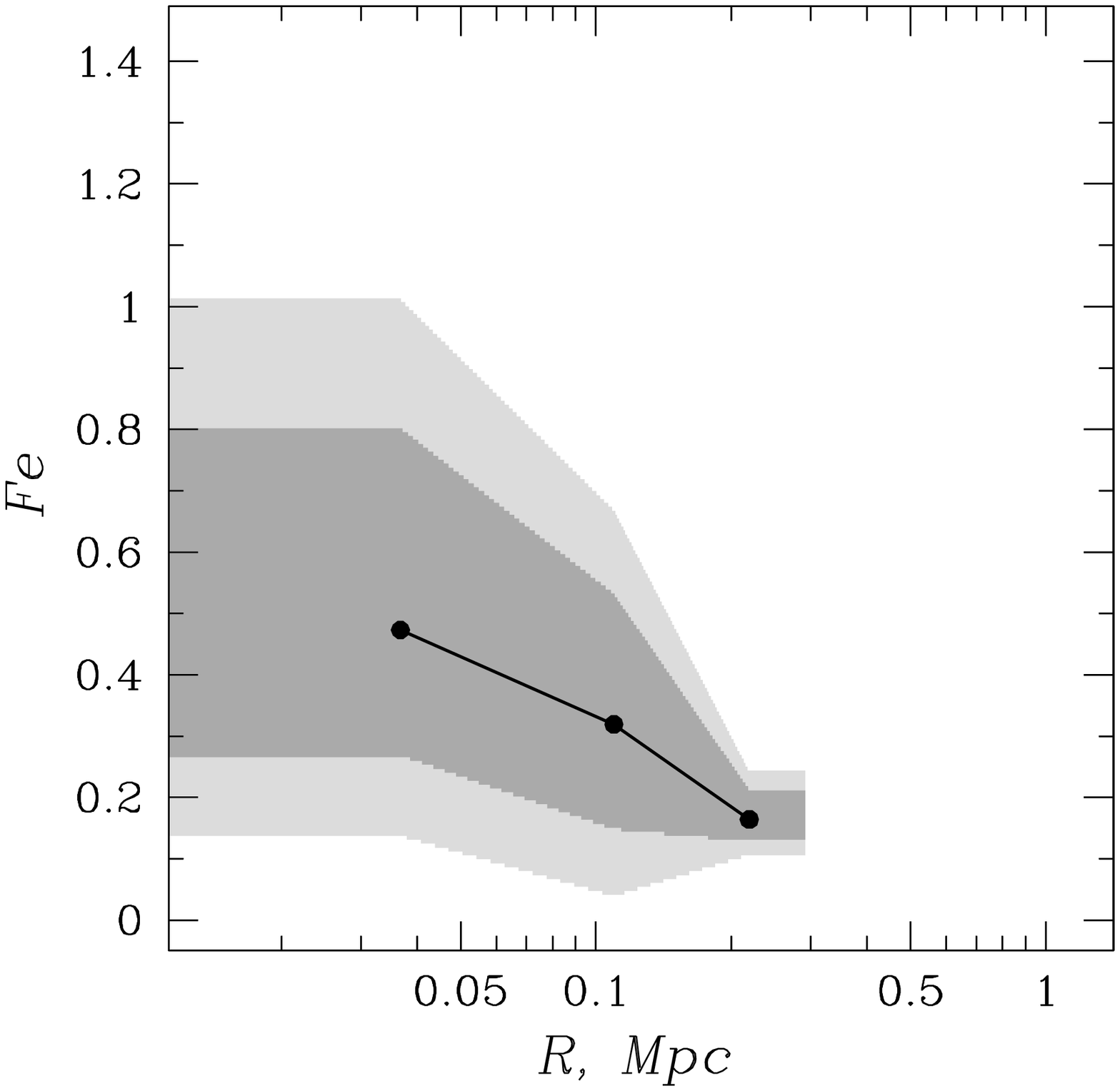} \hfill
   \includegraphics[width=1.6in]{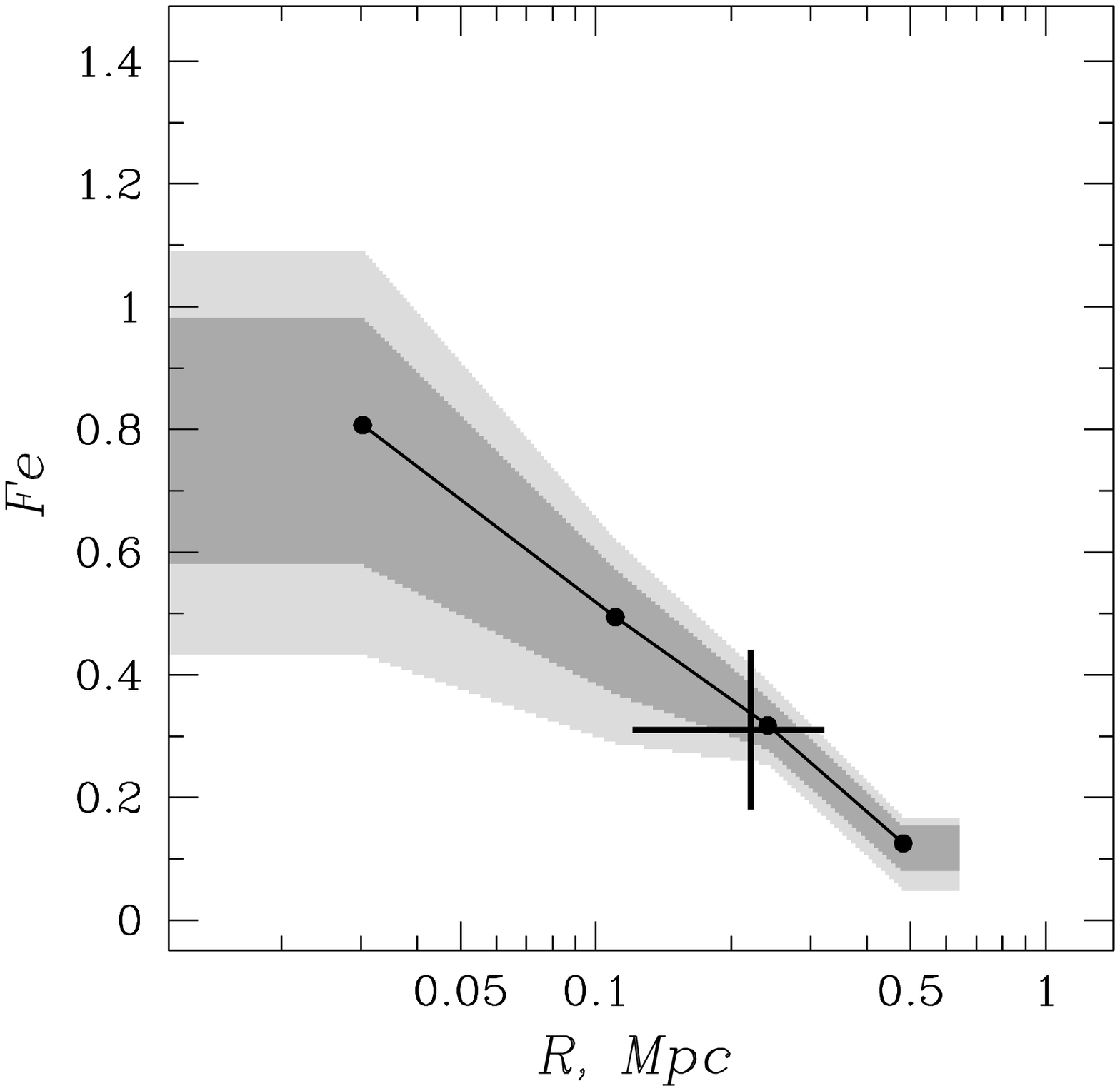} \hfill
   \includegraphics[width=1.6in]{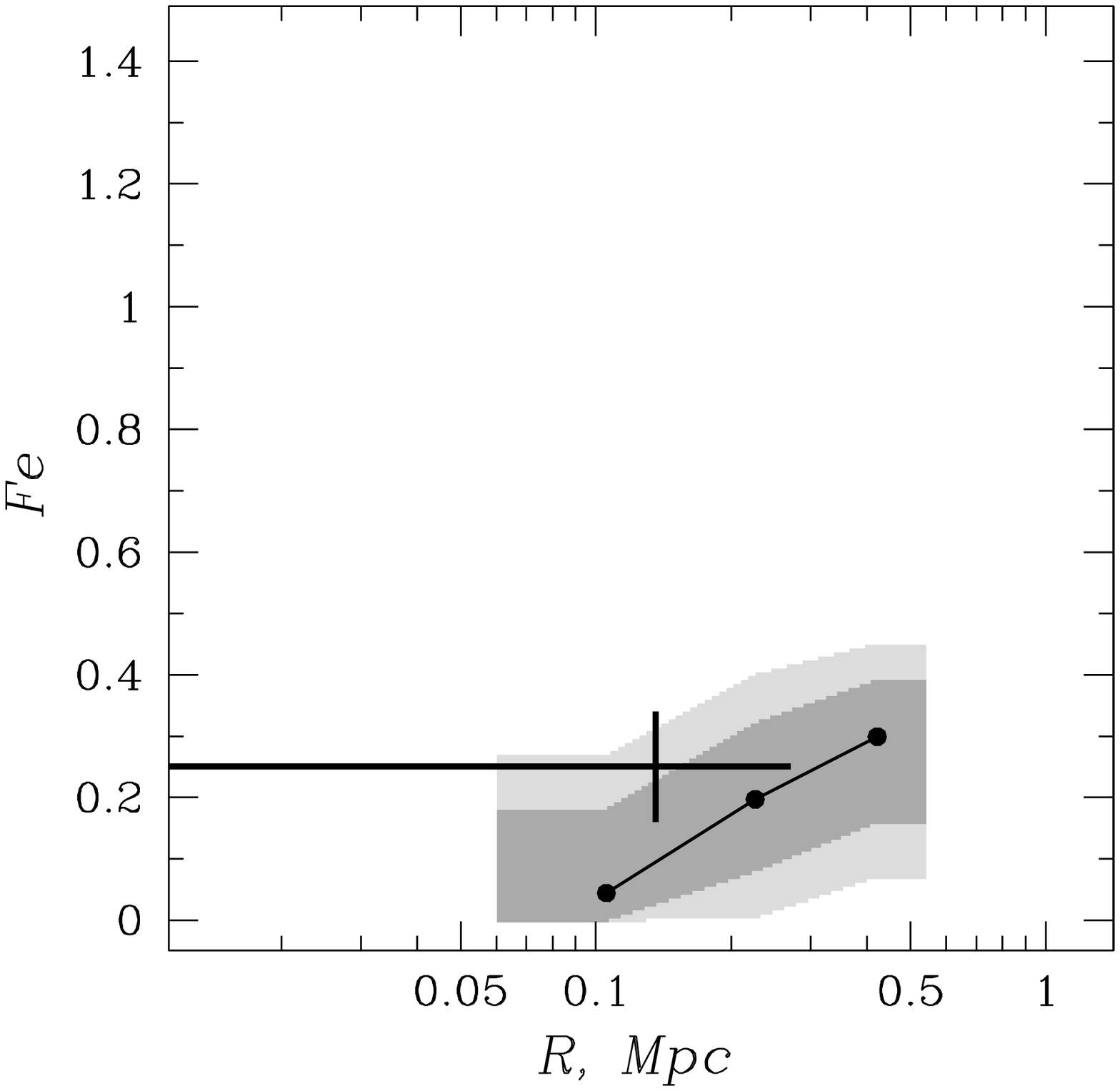} \hfill
   \includegraphics[width=1.6in]{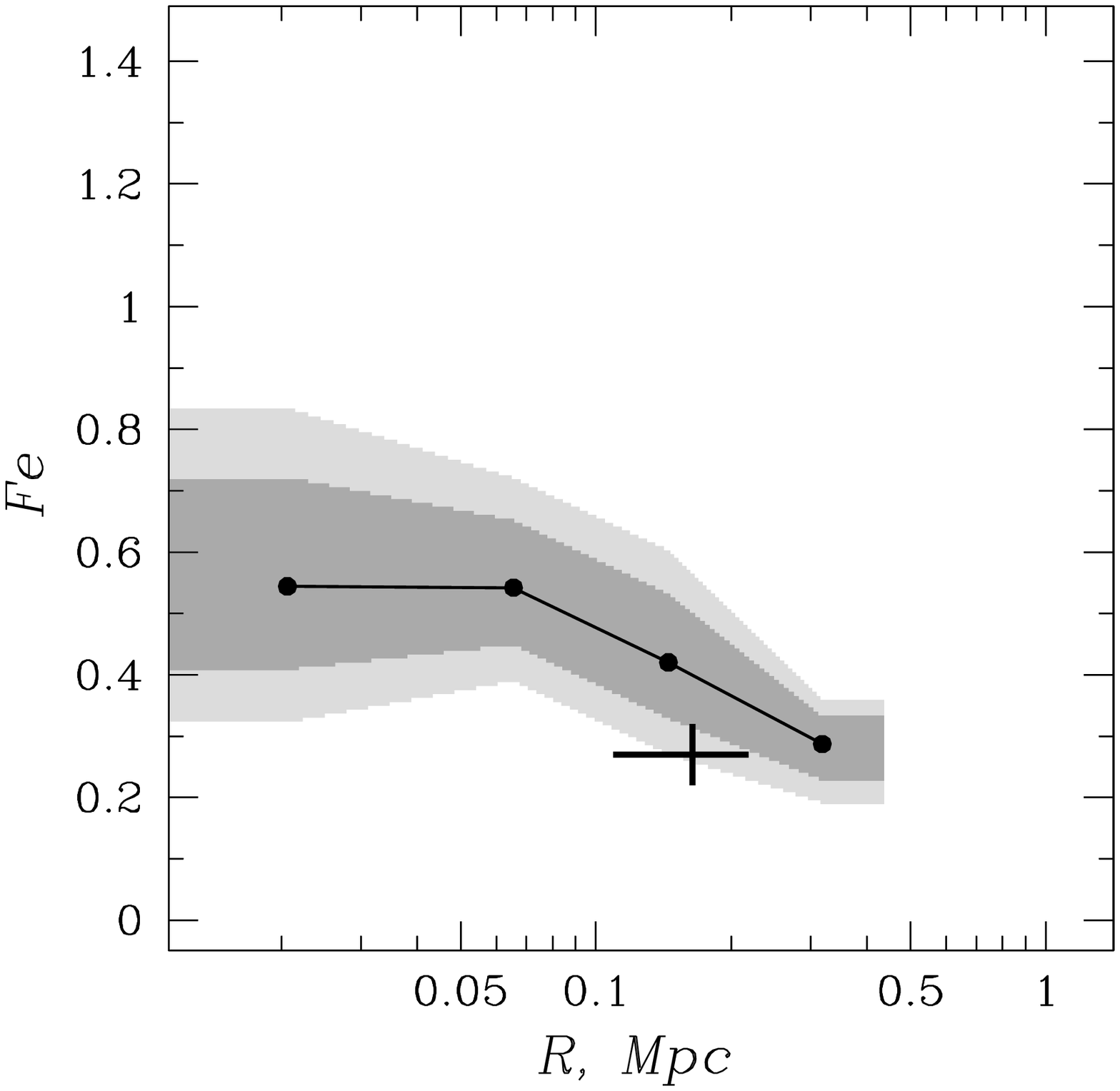} 

  \includegraphics[width=1.6in]{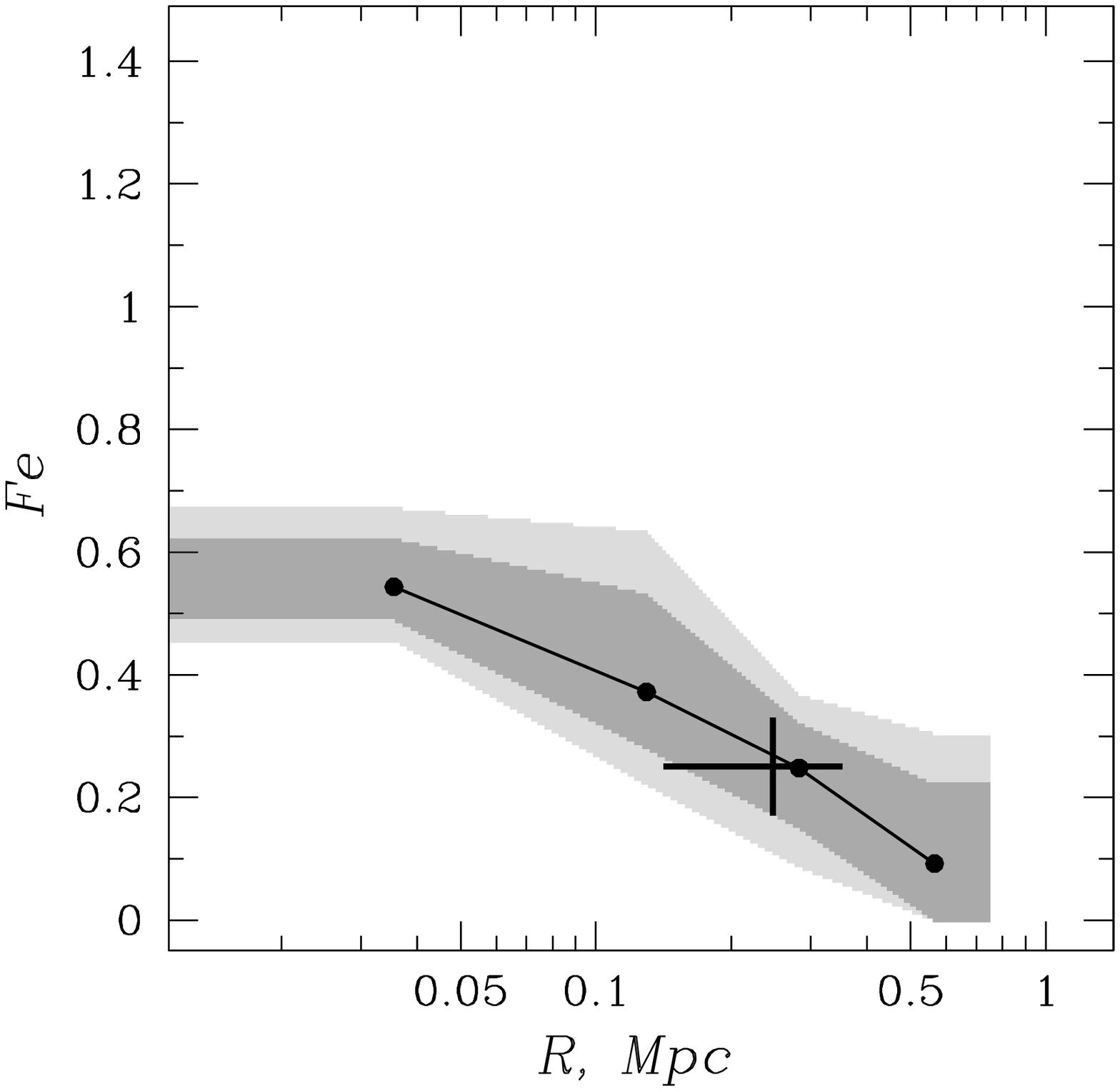} \hfill  
   \includegraphics[width=1.6in]{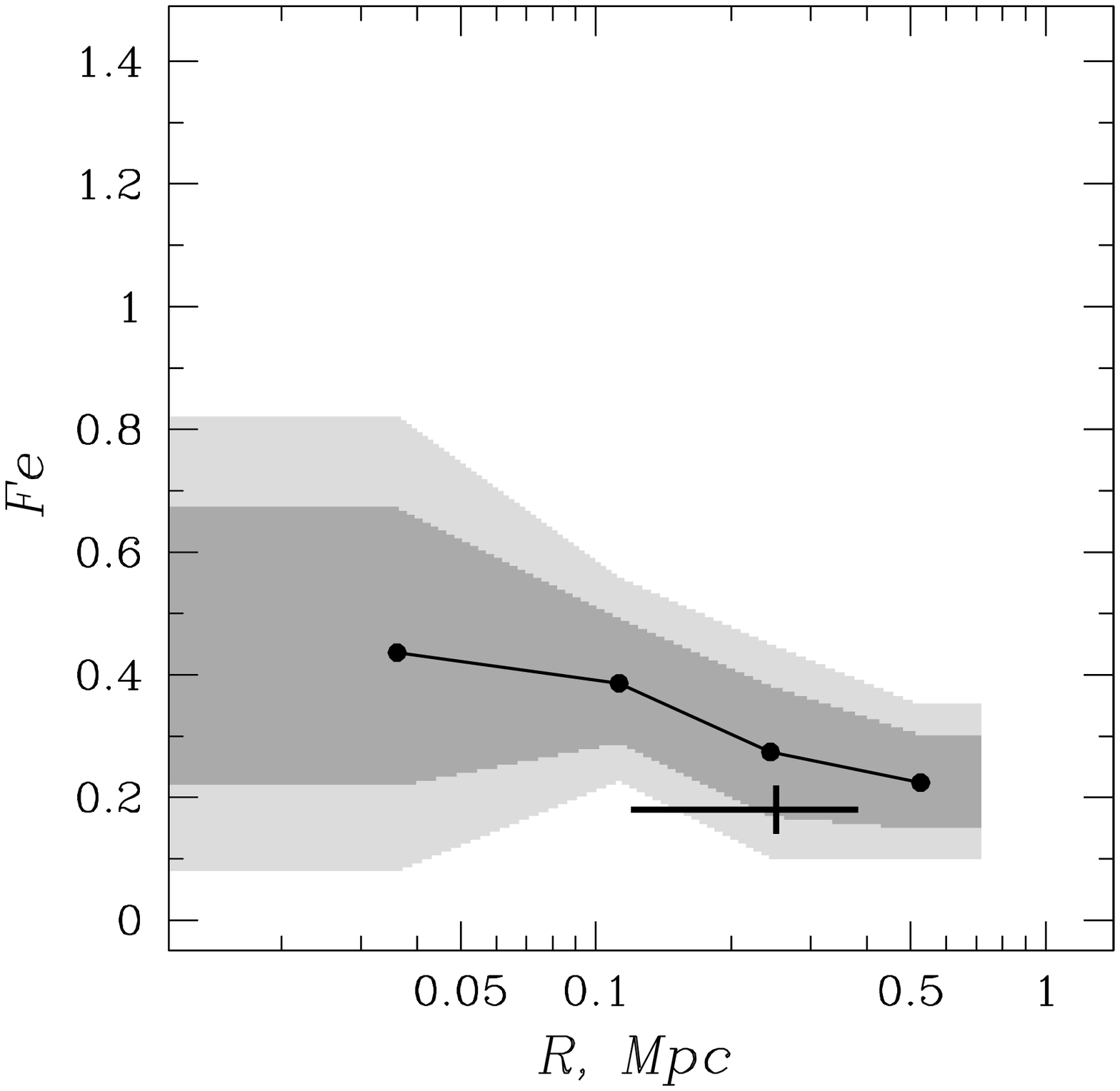} \hfill 
  \includegraphics[width=1.6in]{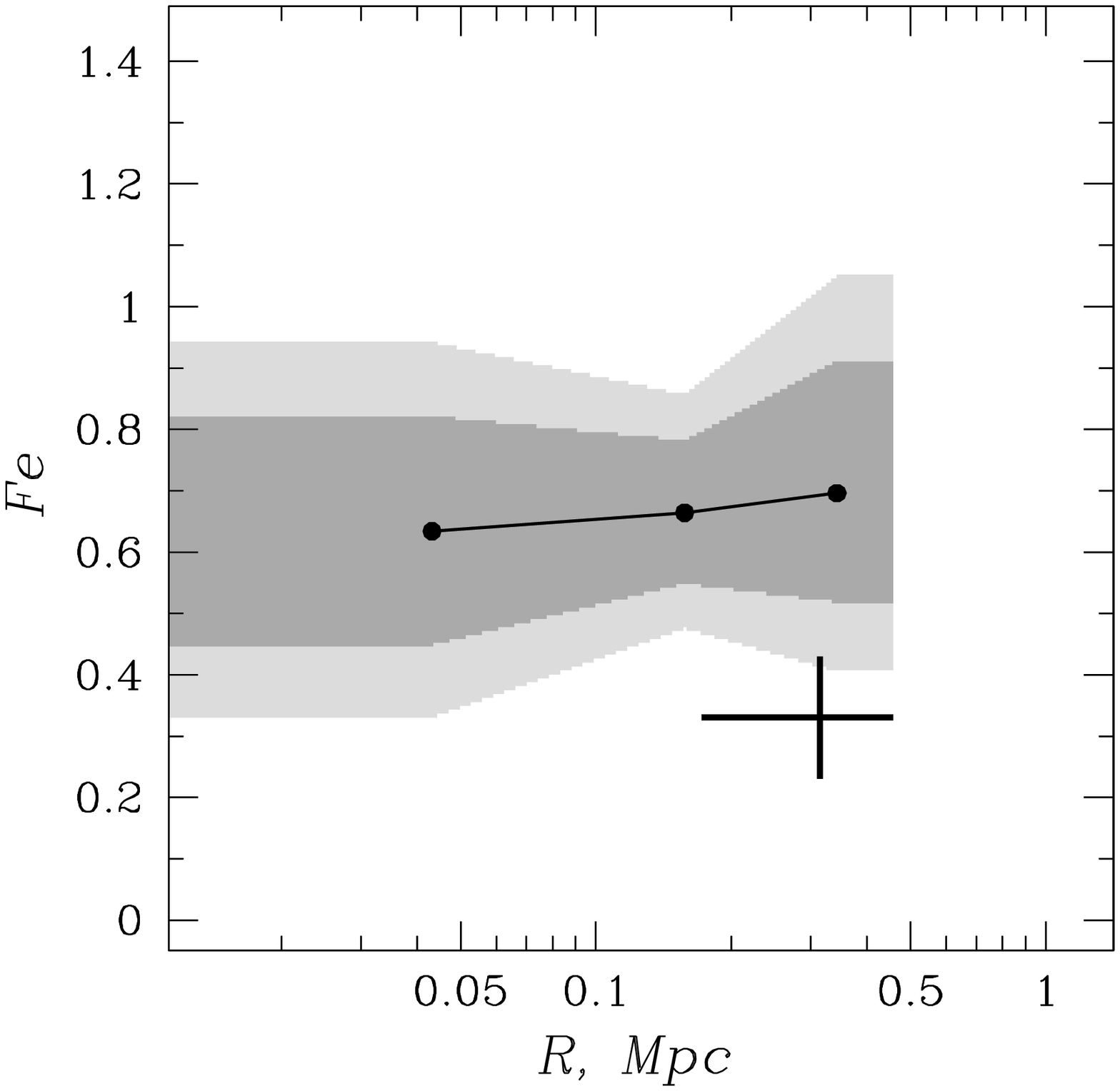} \hfill 
  \includegraphics[width=1.6in]{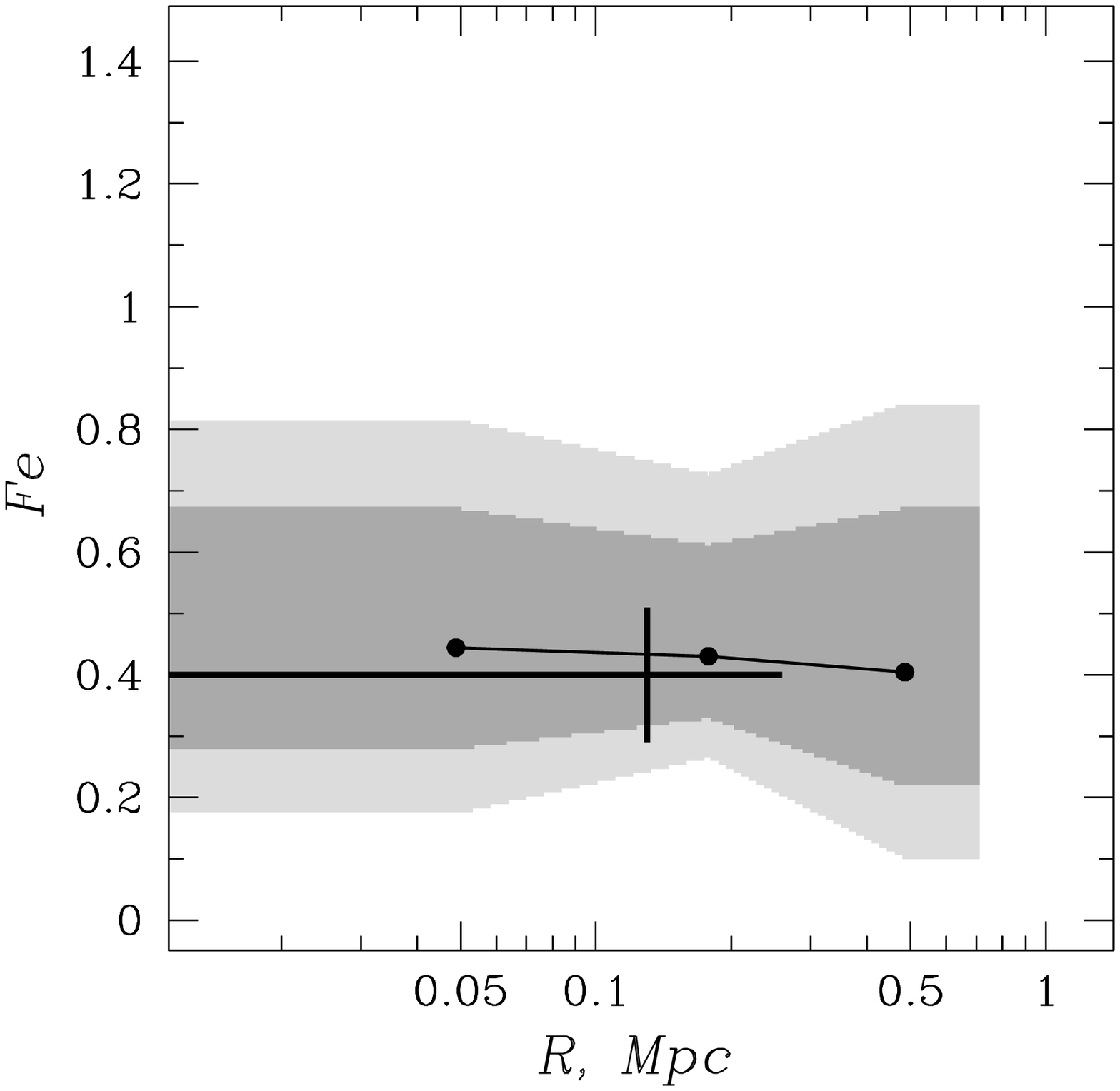}

   \includegraphics[width=1.6in]{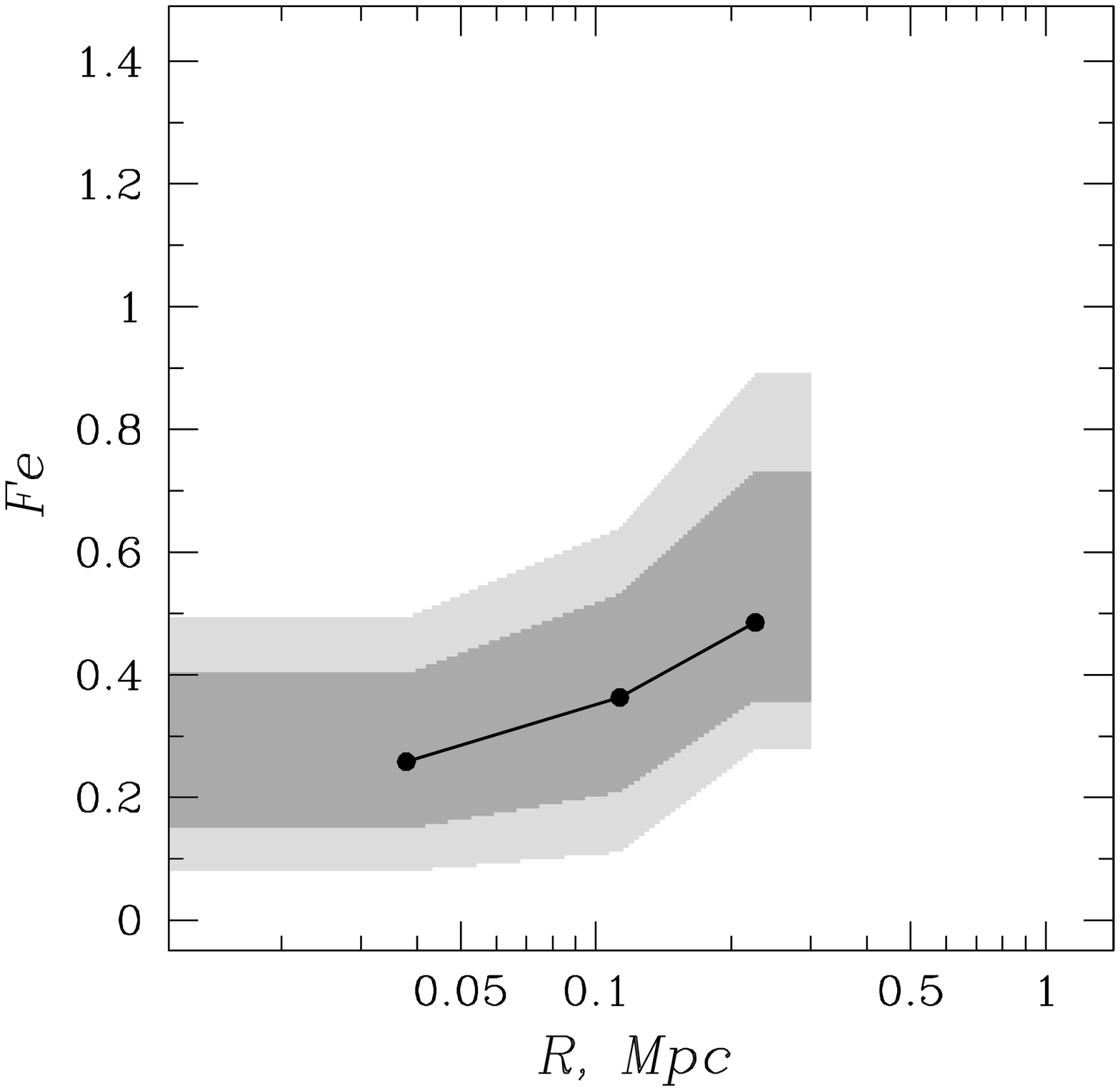}  \hfill 
  \includegraphics[width=1.6in]{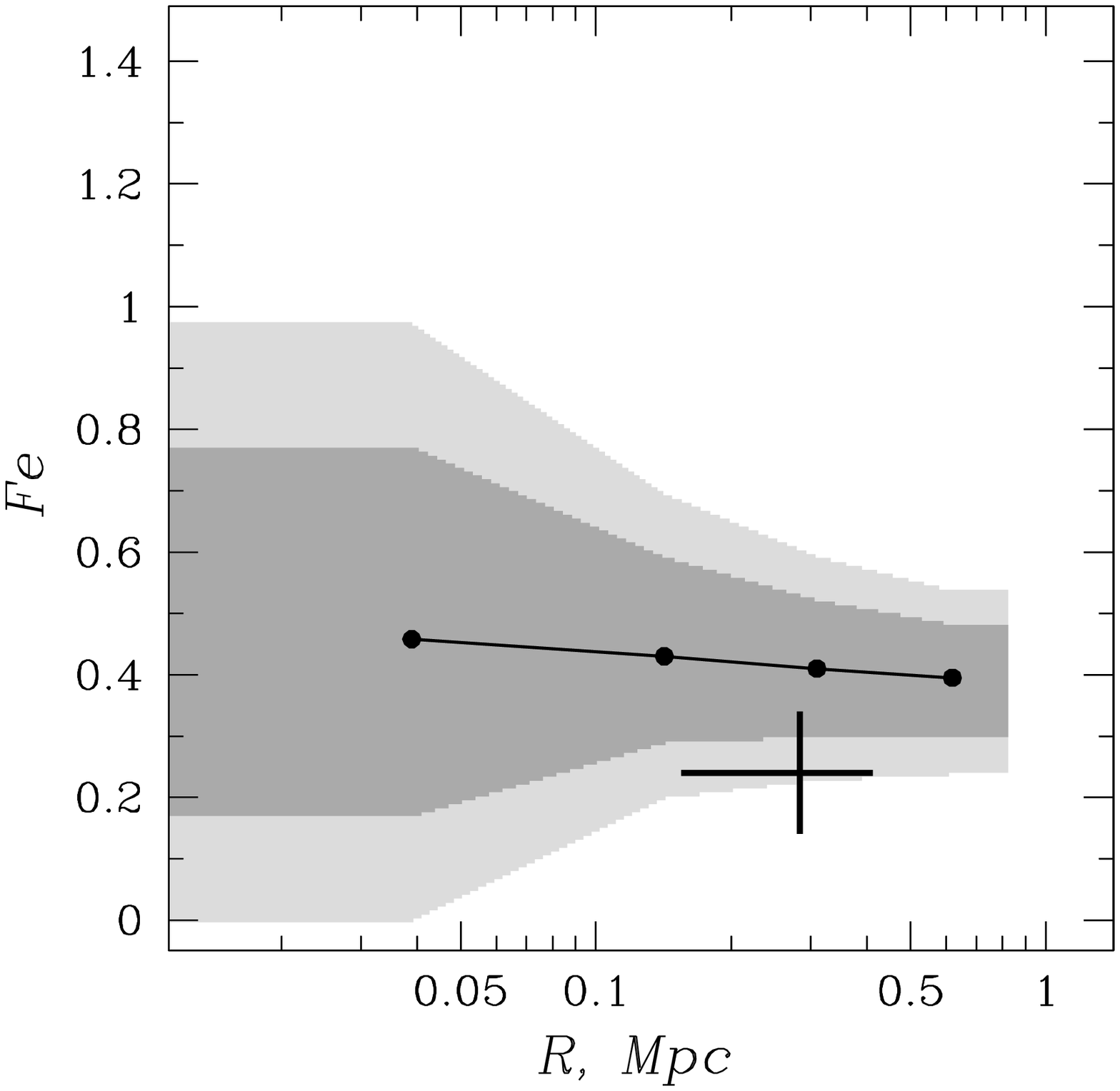} \hfill 
 \includegraphics[width=1.6in]{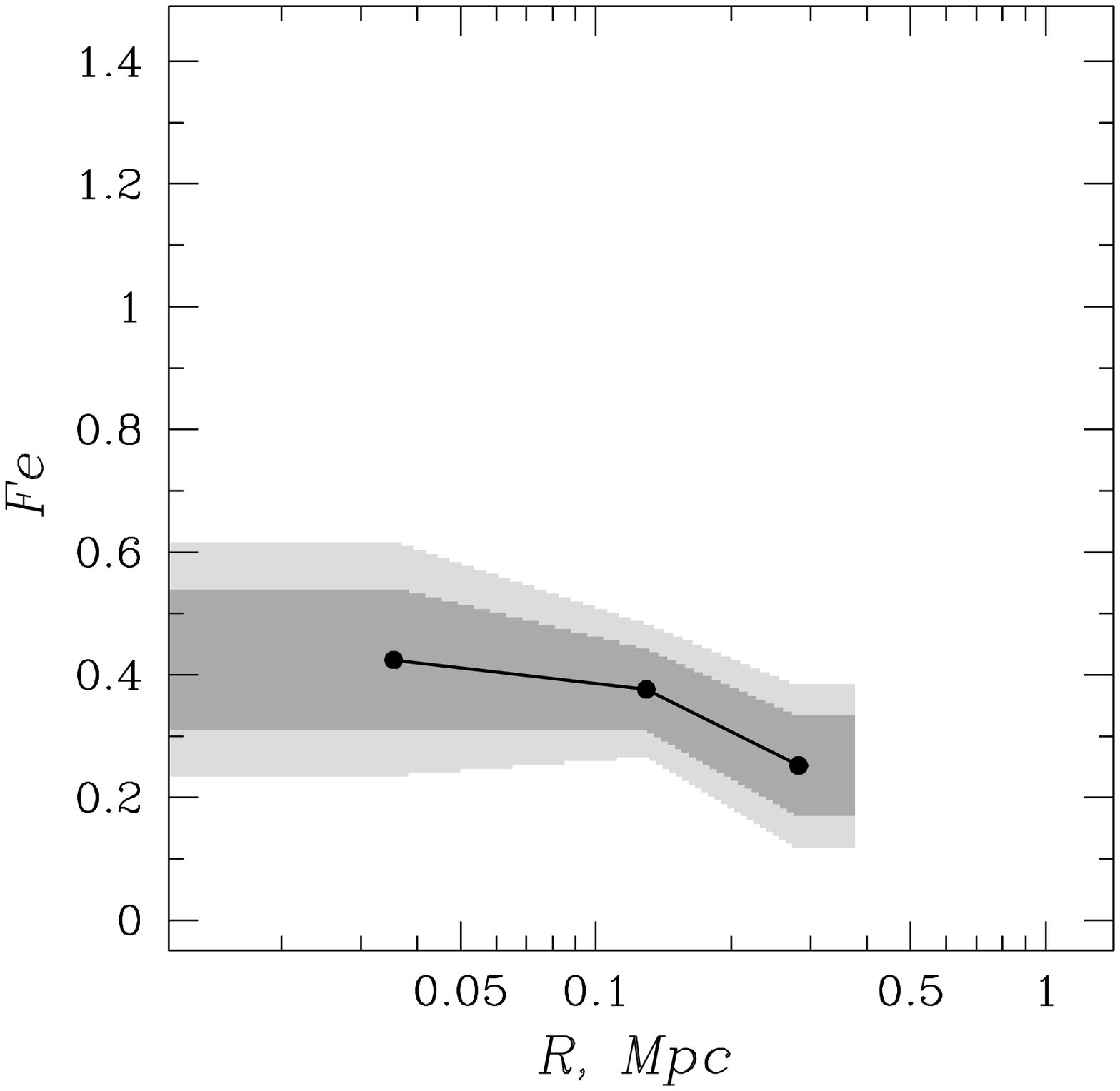} \hfill 
  \includegraphics[width=1.6in]{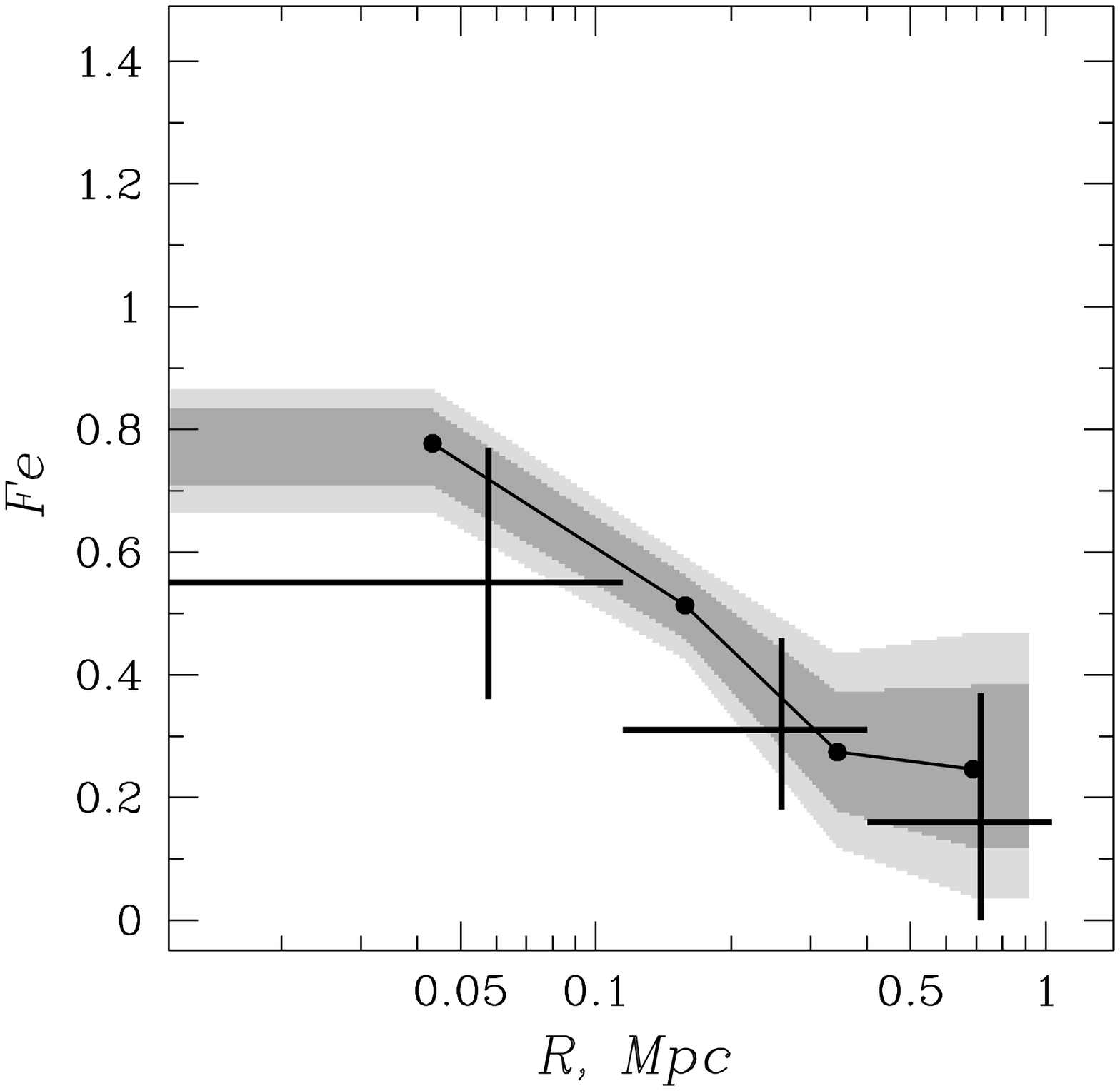} 

   \includegraphics[width=1.6in]{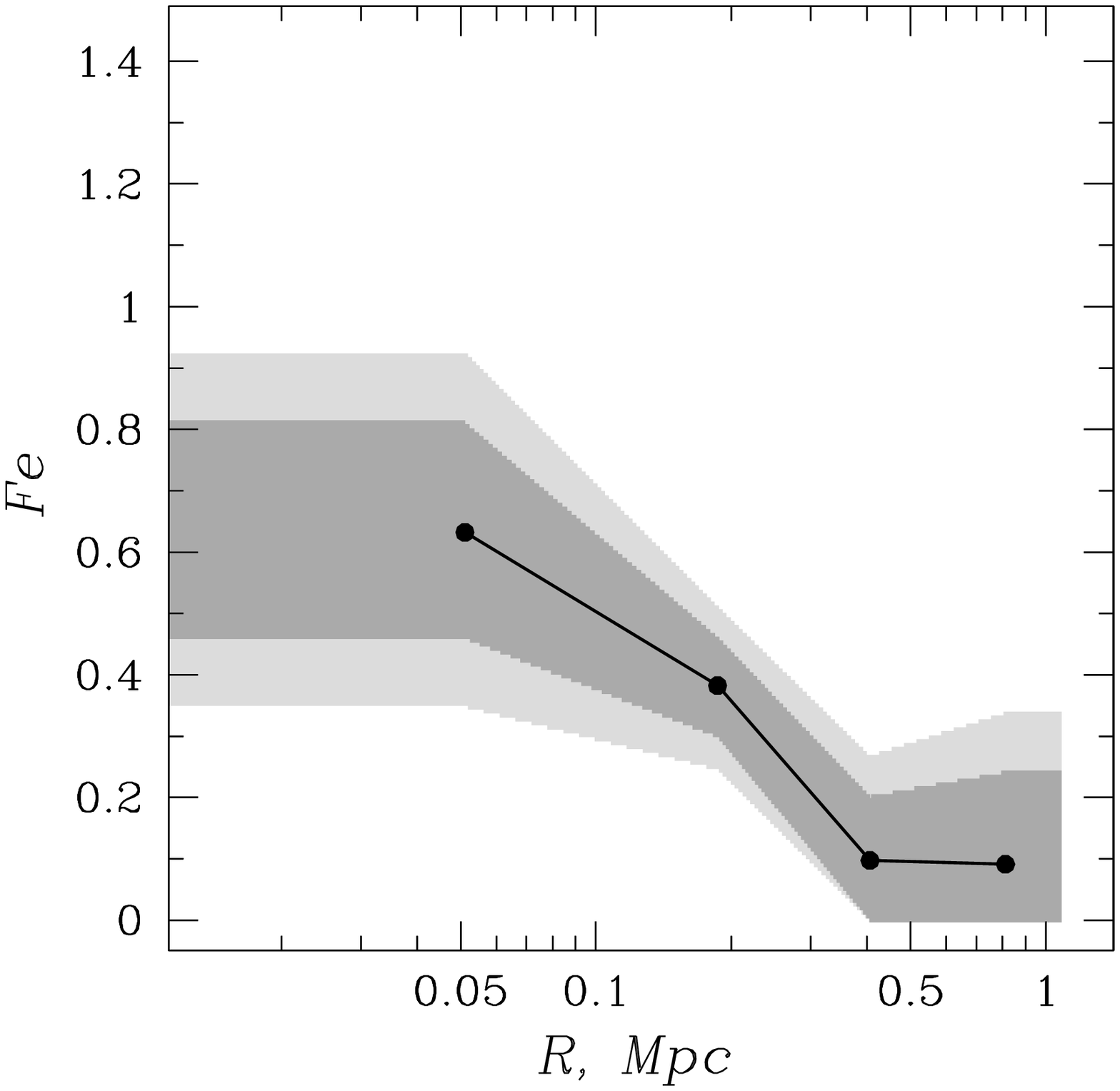} \hfill 
   \includegraphics[width=1.6in]{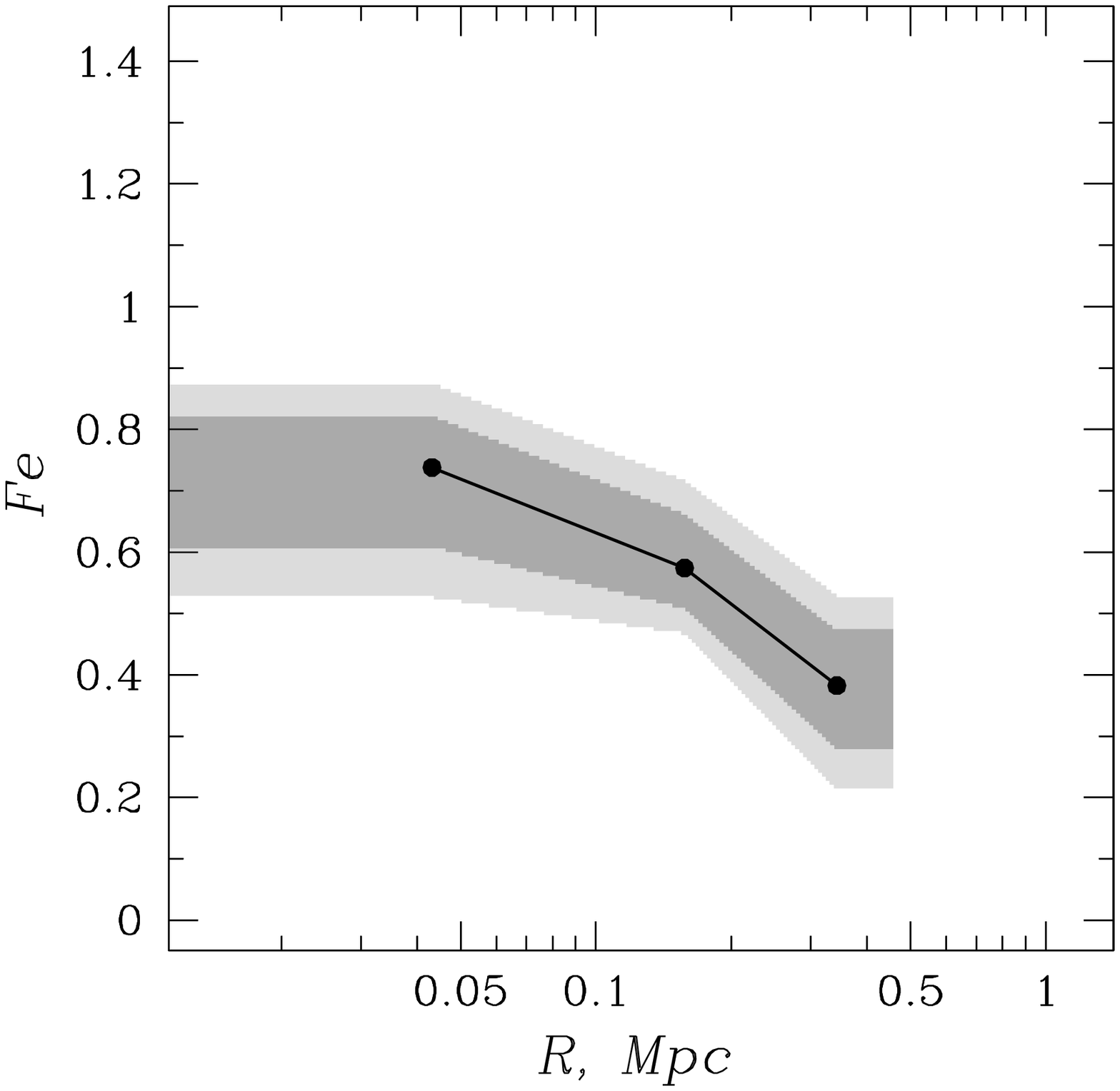} \hfill 
  \includegraphics[width=1.6in]{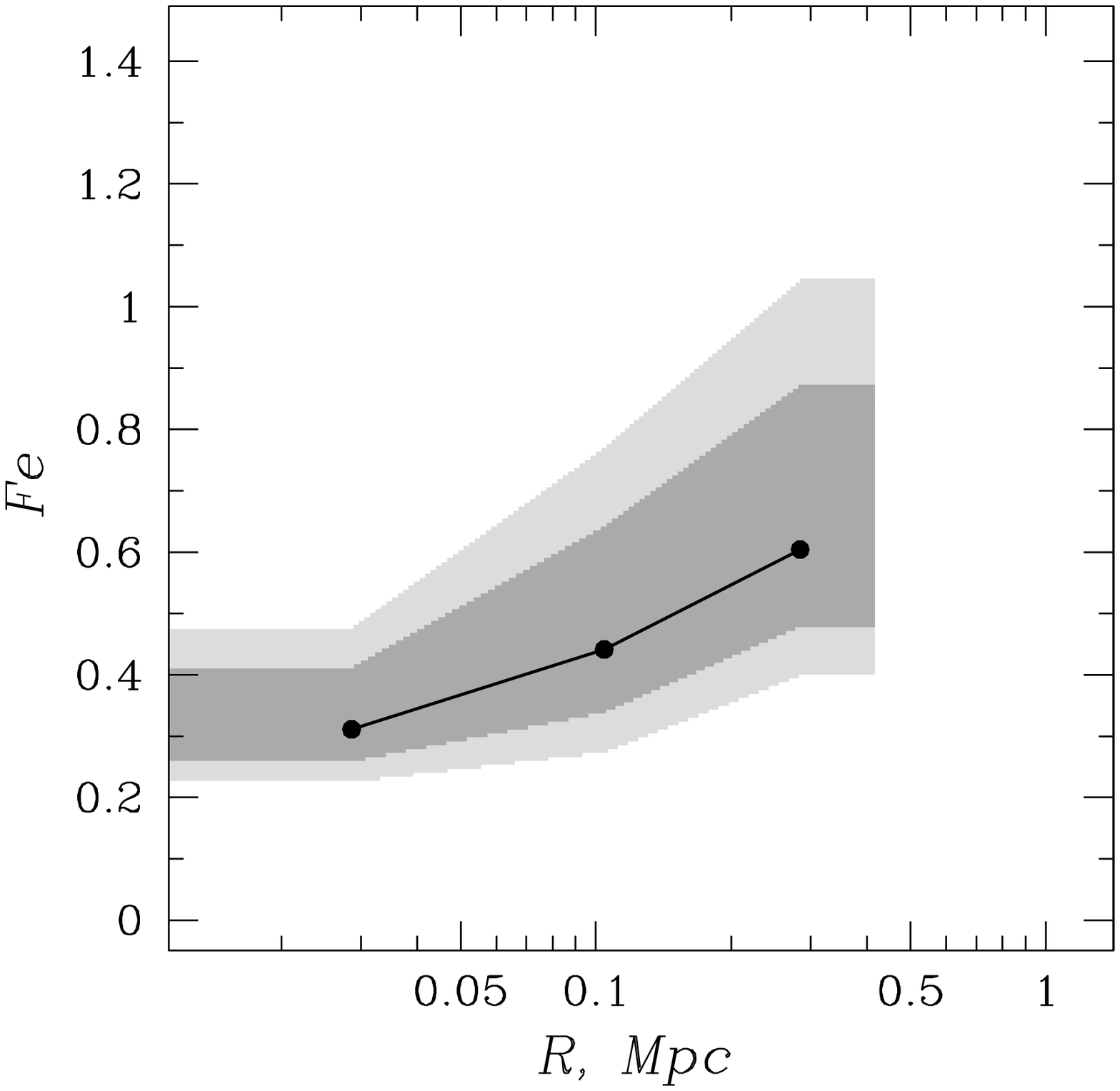} \hfill 
  \includegraphics[width=1.6in]{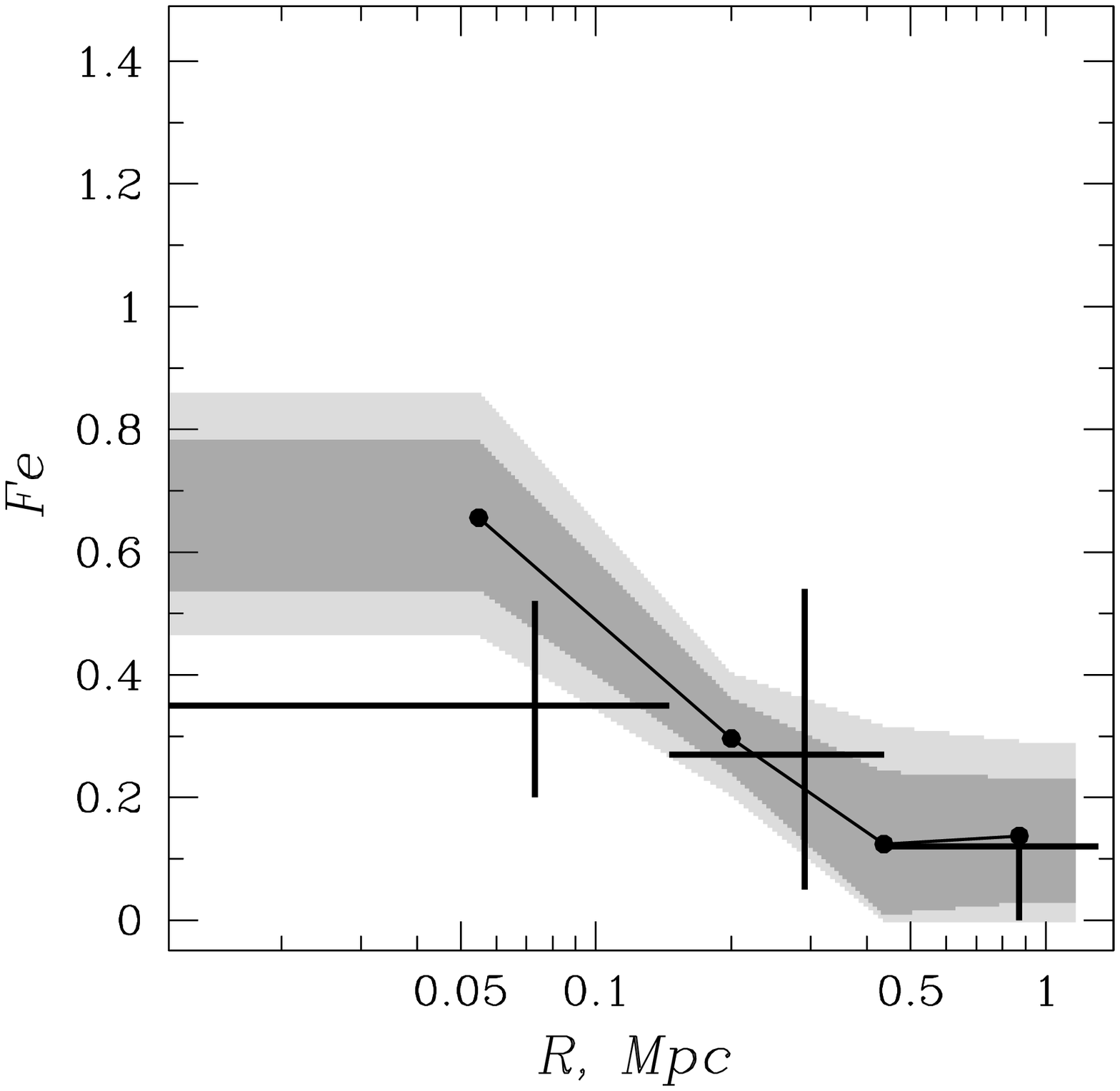} 

    \includegraphics[width=1.6in]{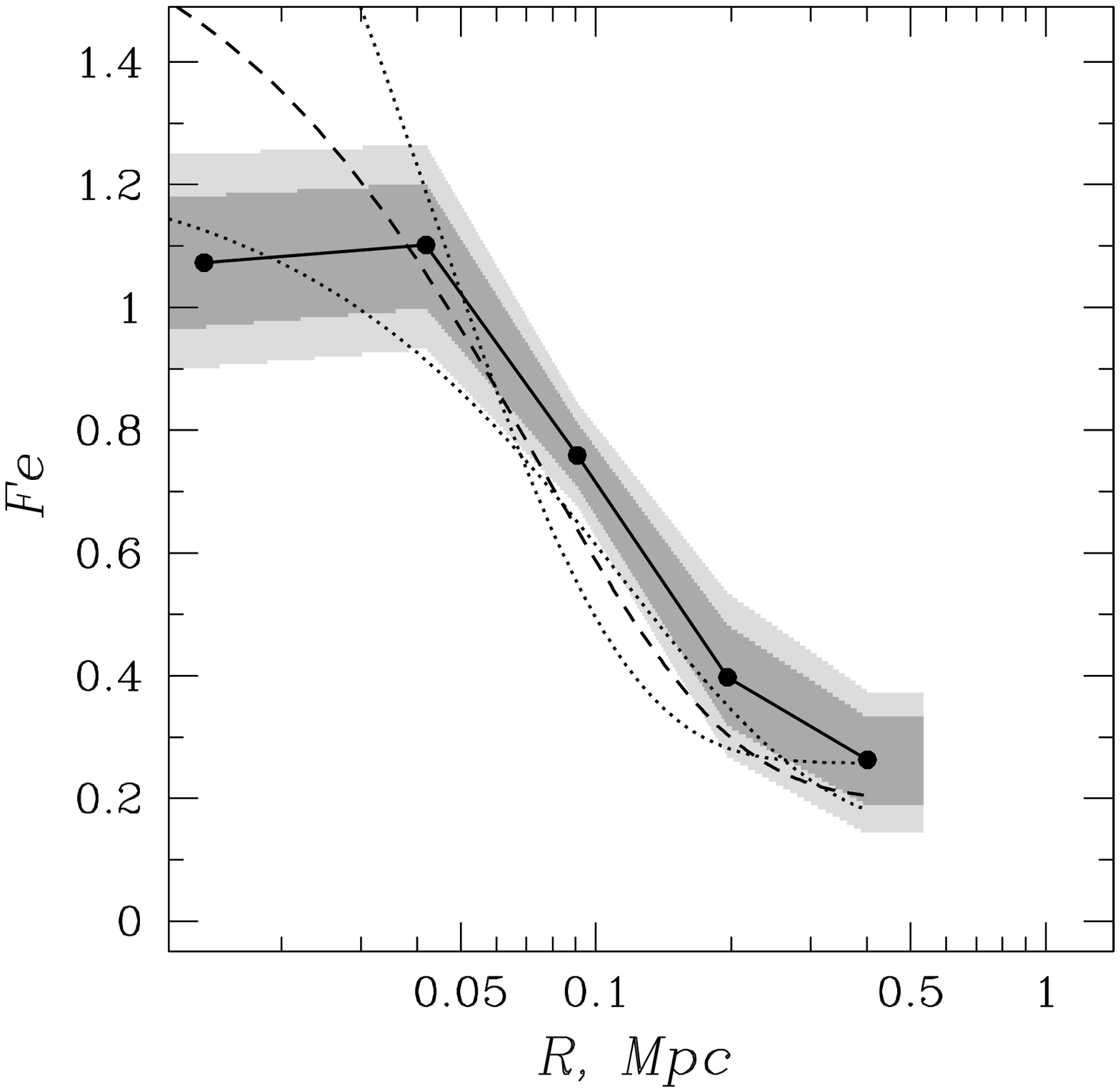}  \hfill 
  \includegraphics[width=1.6in]{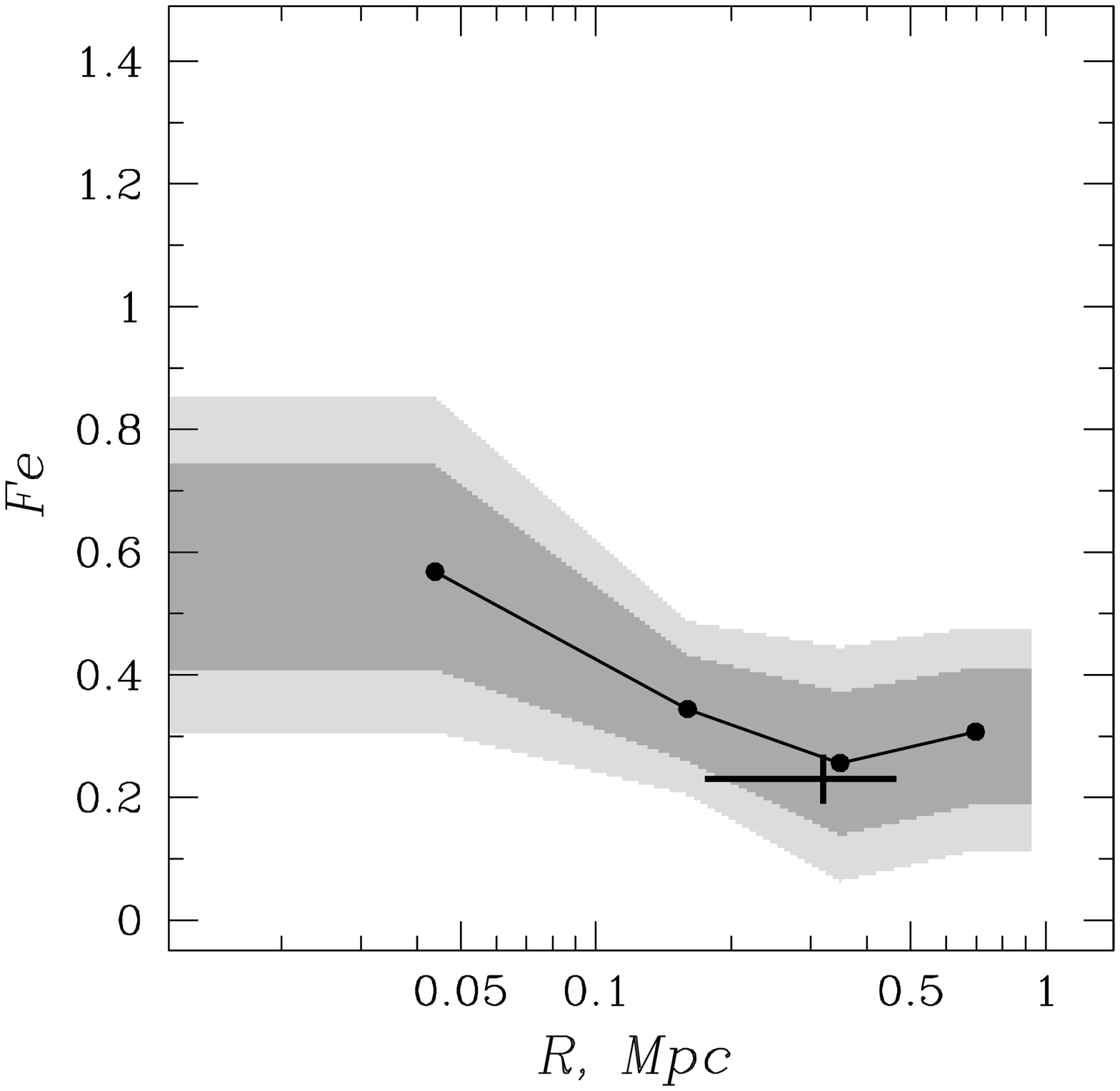} 

\figcaption{Derived Fe abundances (in units of 4.68 $10^{-5}$ for iron
number abundances relative to H). The solid lines correspond to the best-fit
Fe abundances derived from the ASCA data. The filled circles indicate the
spatial binning used in the analysis. Dark and light shaded zones around the
best fit curves denote the 68 and 90 per cent confidence areas.  Crosses on
A400, A194, A262, MKW4S, A539, AWM4, MKW9, A2634 and A2063 panels show the
results from Fukazawa \etal (1998) with radii of measurement from Fukazawa
(private communication). Crosses on the 2A0335 and MKW3S plots denote the
results of modeling of GIS data from Kikuchi \etal (1999) and similarly
lines on Cen plot from Ikebe \etal (1999).
\label{fe-fig}}
\vspace*{-21.1cm}

{\it \hspace*{2.4cm} A2197E \hspace*{3.9cm} A400 \hspace*{3.9cm} A194 \hspace*{3.9cm} A262}

\vspace*{3.55cm}

{\it \hspace*{2.6cm} MKW4S \hspace*{3.6cm} A539 \hspace*{3.8cm} AWM4 \hspace*{3.8cm} MKW9}

\vspace*{3.55cm}

{\it \hspace*{2.4cm} A2197W \hspace*{3.7cm} A2634 \hspace*{3.6cm} A4038 \hspace*{3.5cm} 2A0335}

\vspace*{3.55cm}

{\it \hspace*{2.6cm} HCG94 \hspace*{3.5cm} A2052 \hspace*{3.8cm} A779 \hspace*{3.5cm} MKW3S}

\vspace*{3.55cm}
{\it \hspace*{2.8cm} CEN \hspace*{13.3cm} A2063}

\vspace*{3.7cm}

\end{figure*}

\begin{figure*}

   \includegraphics[width=1.6in]{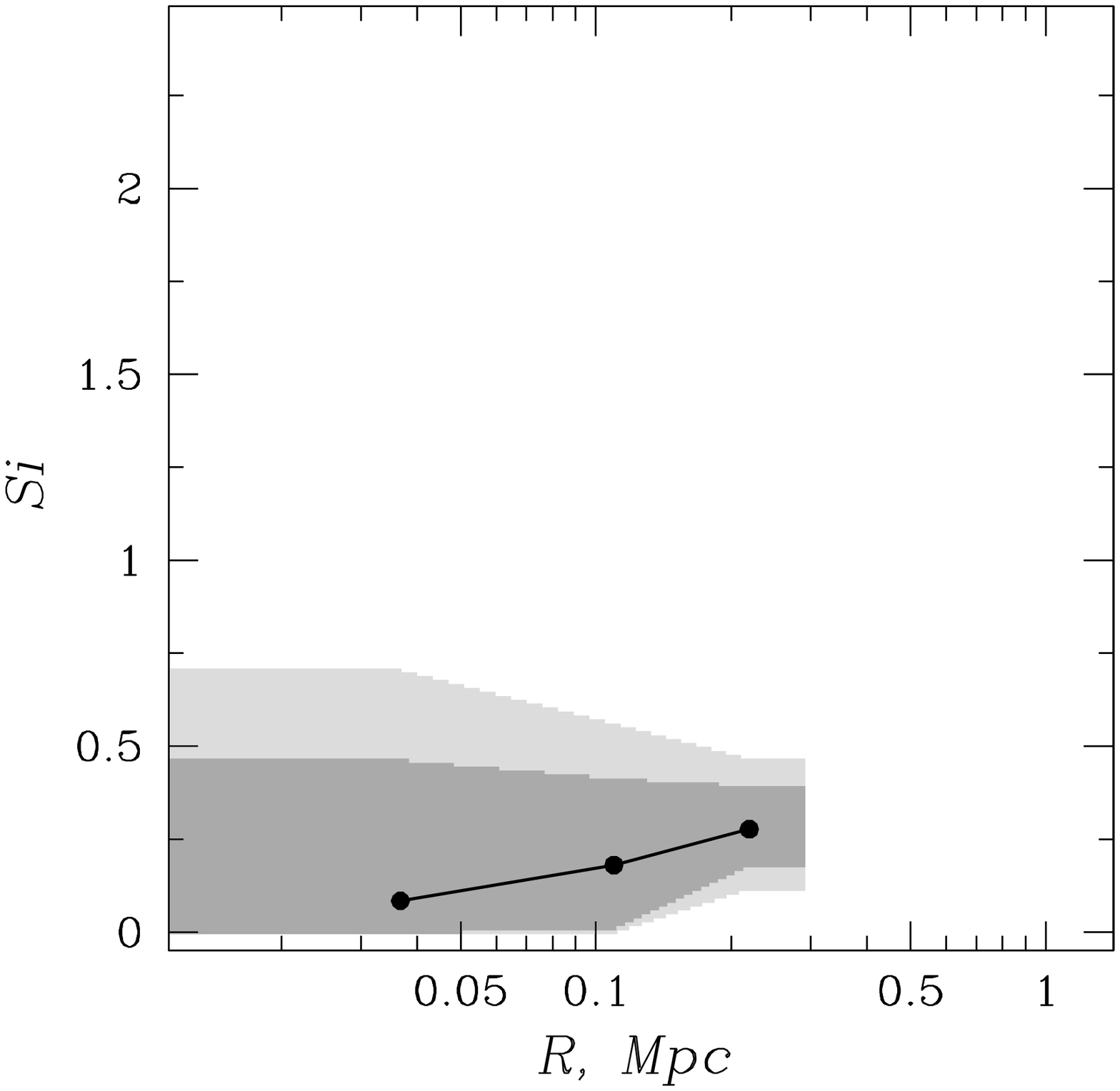} \hfill
   \includegraphics[width=1.6in]{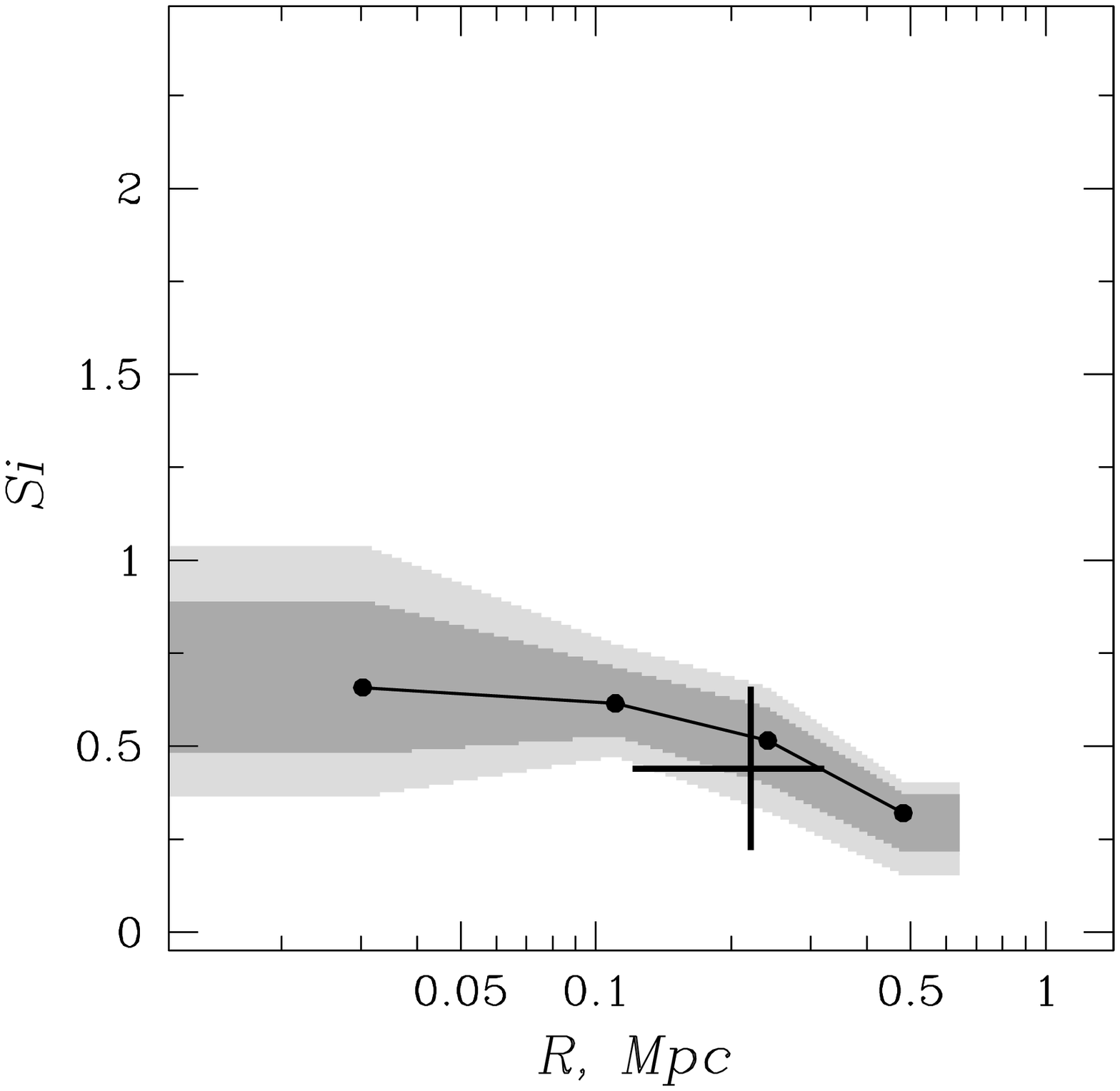} \hfill
   \includegraphics[width=1.6in]{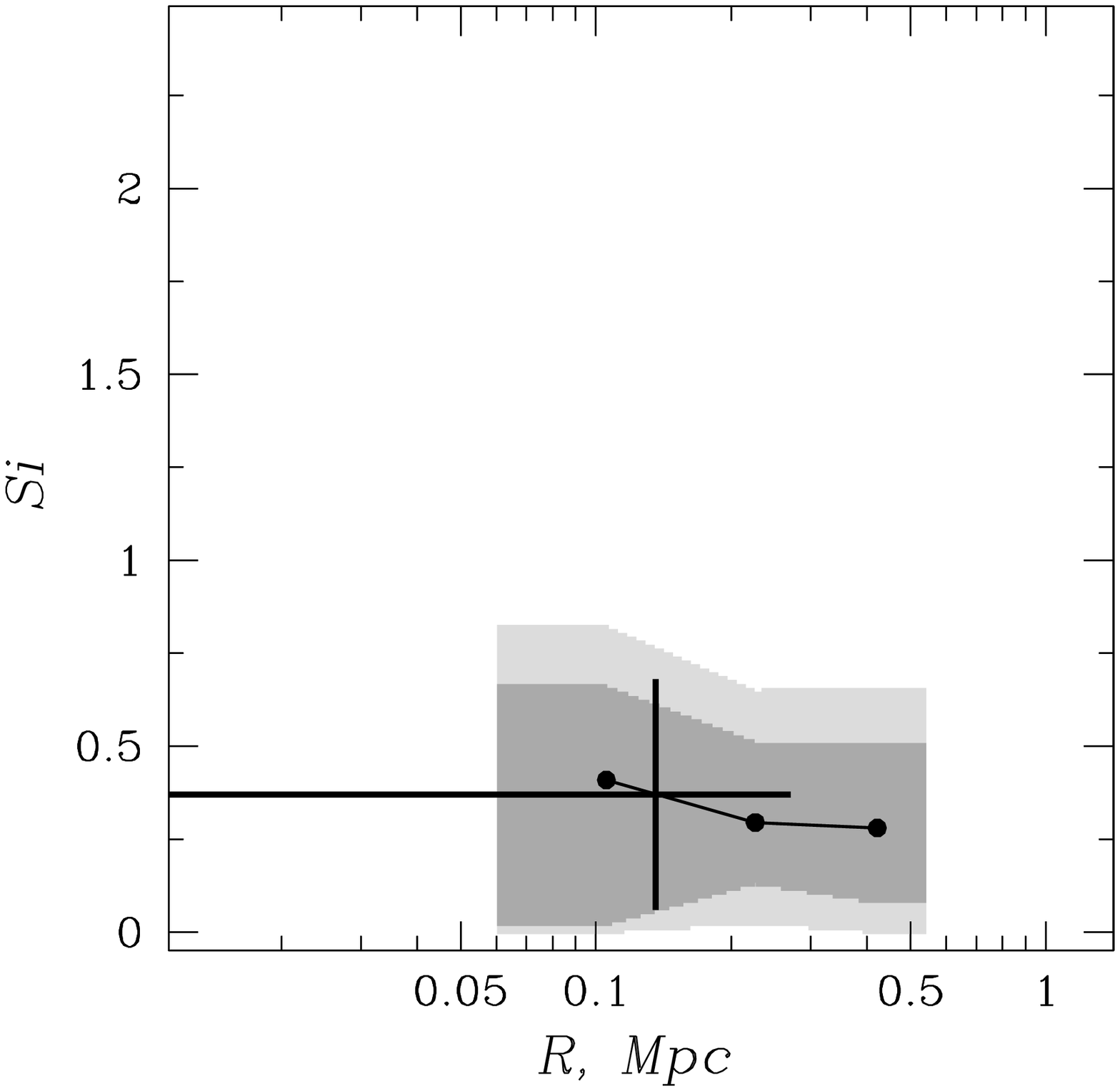} \hfill
   \includegraphics[width=1.6in]{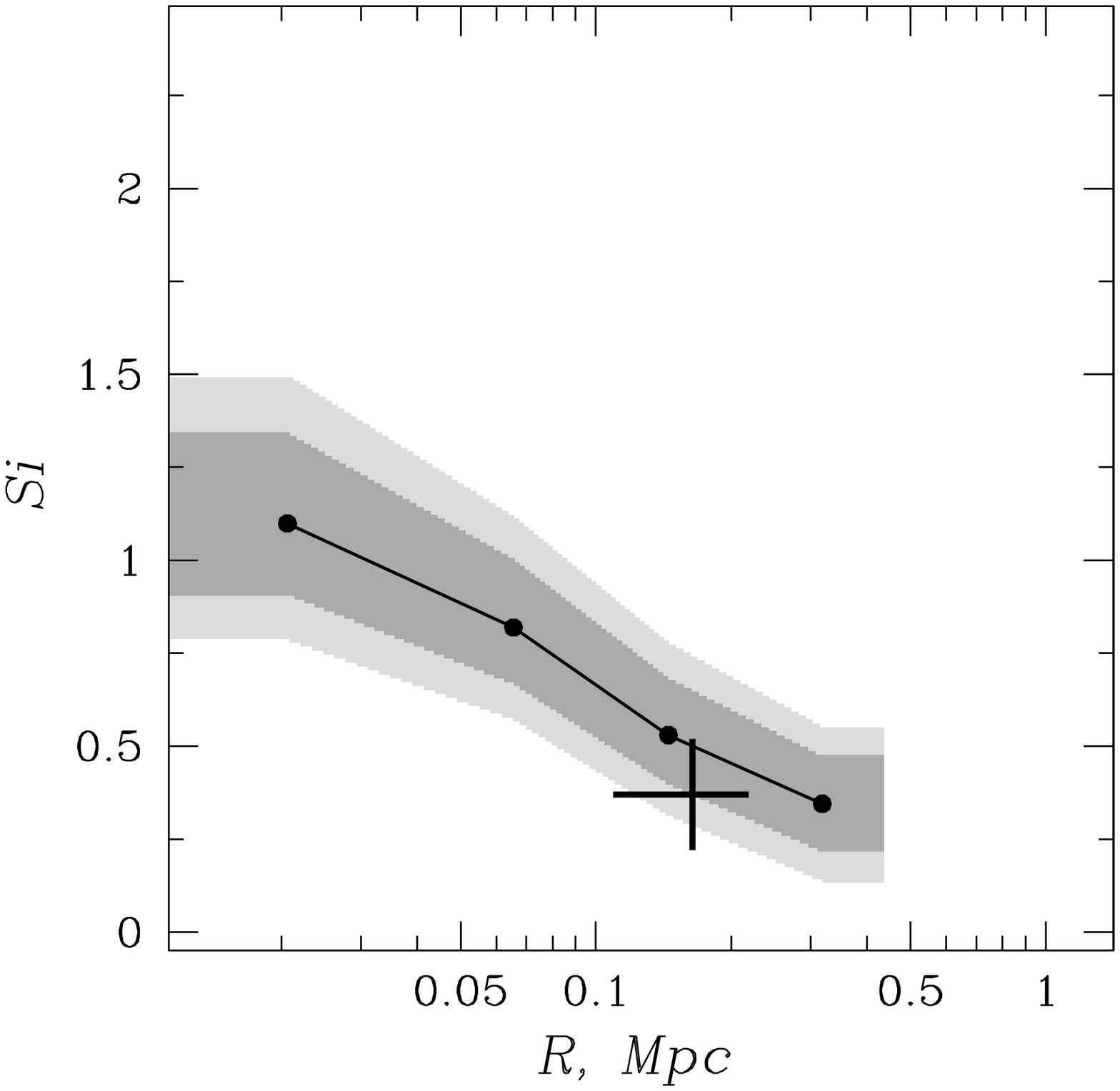} 

  \includegraphics[width=1.6in]{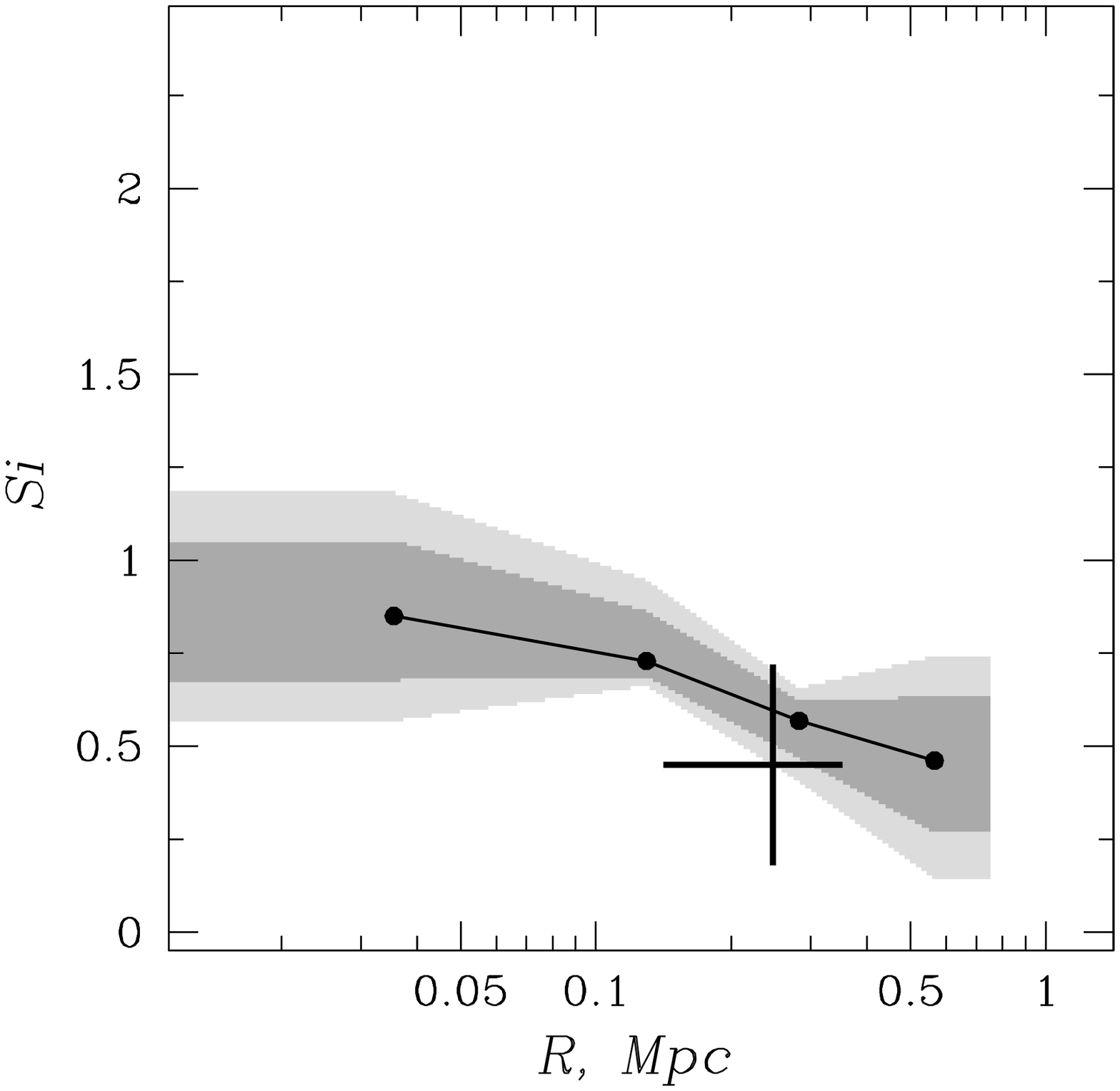} \hfill  
   \includegraphics[width=1.6in]{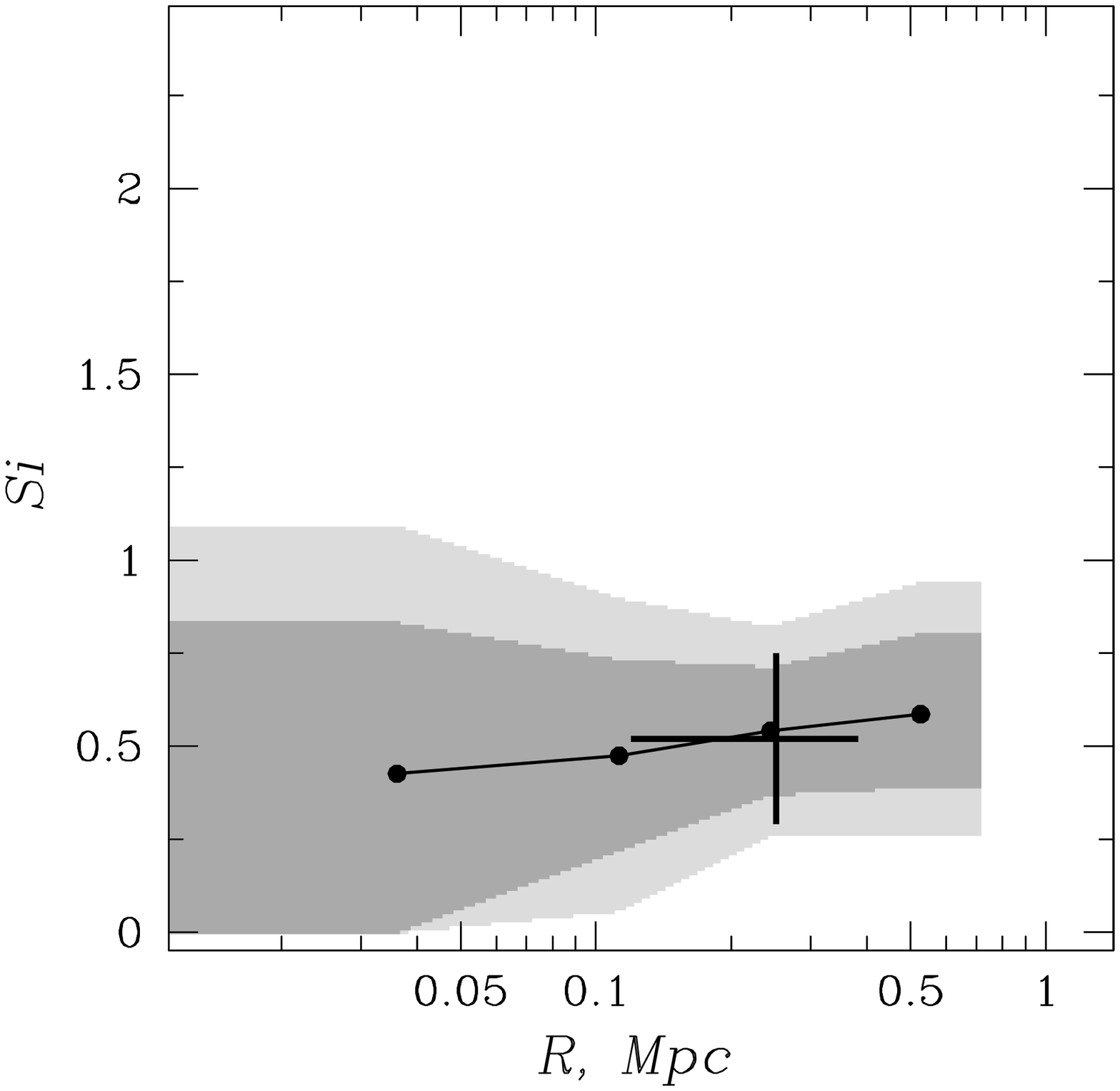} \hfill 
  \includegraphics[width=1.6in]{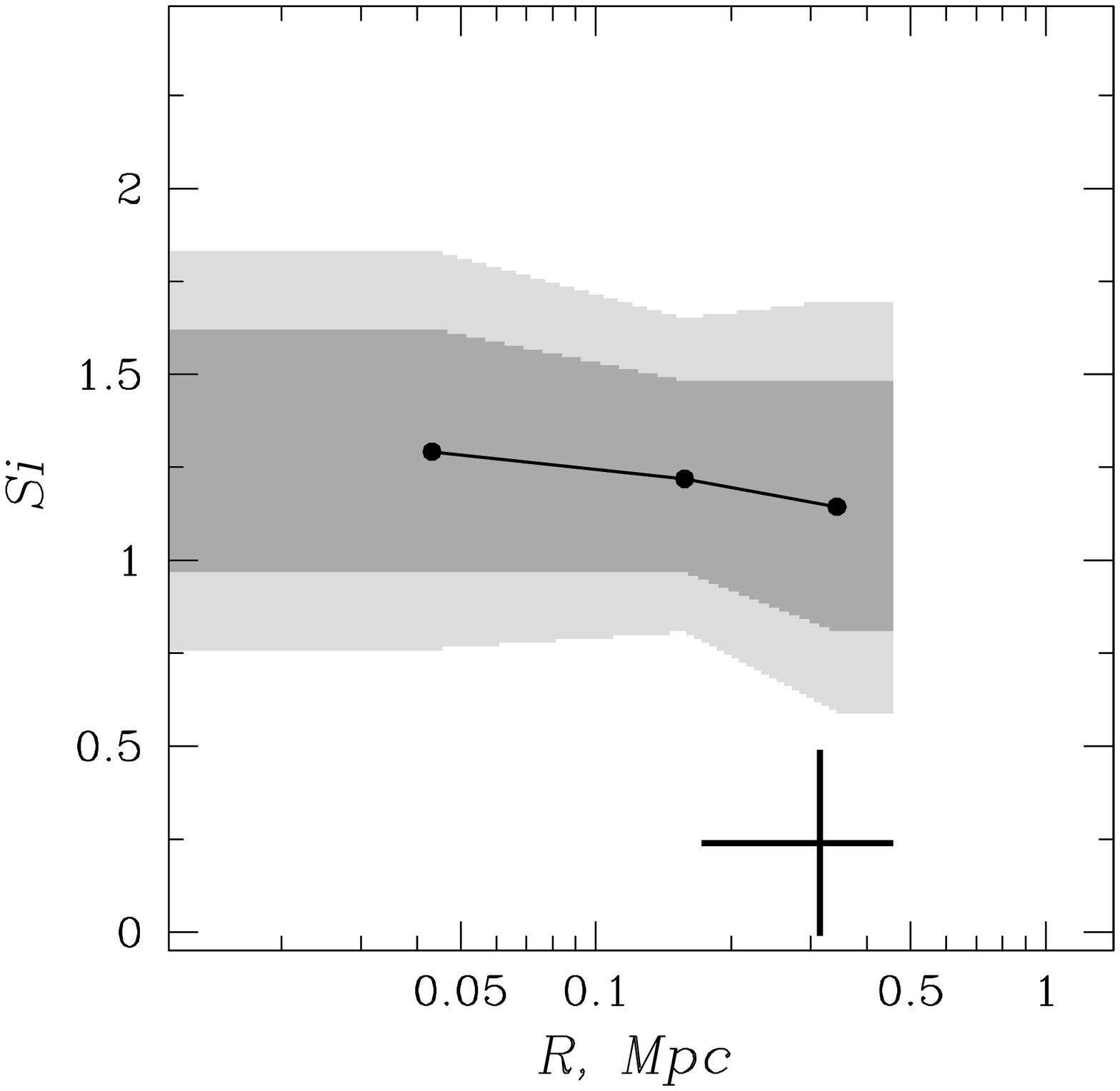} \hfill 
  \includegraphics[width=1.6in]{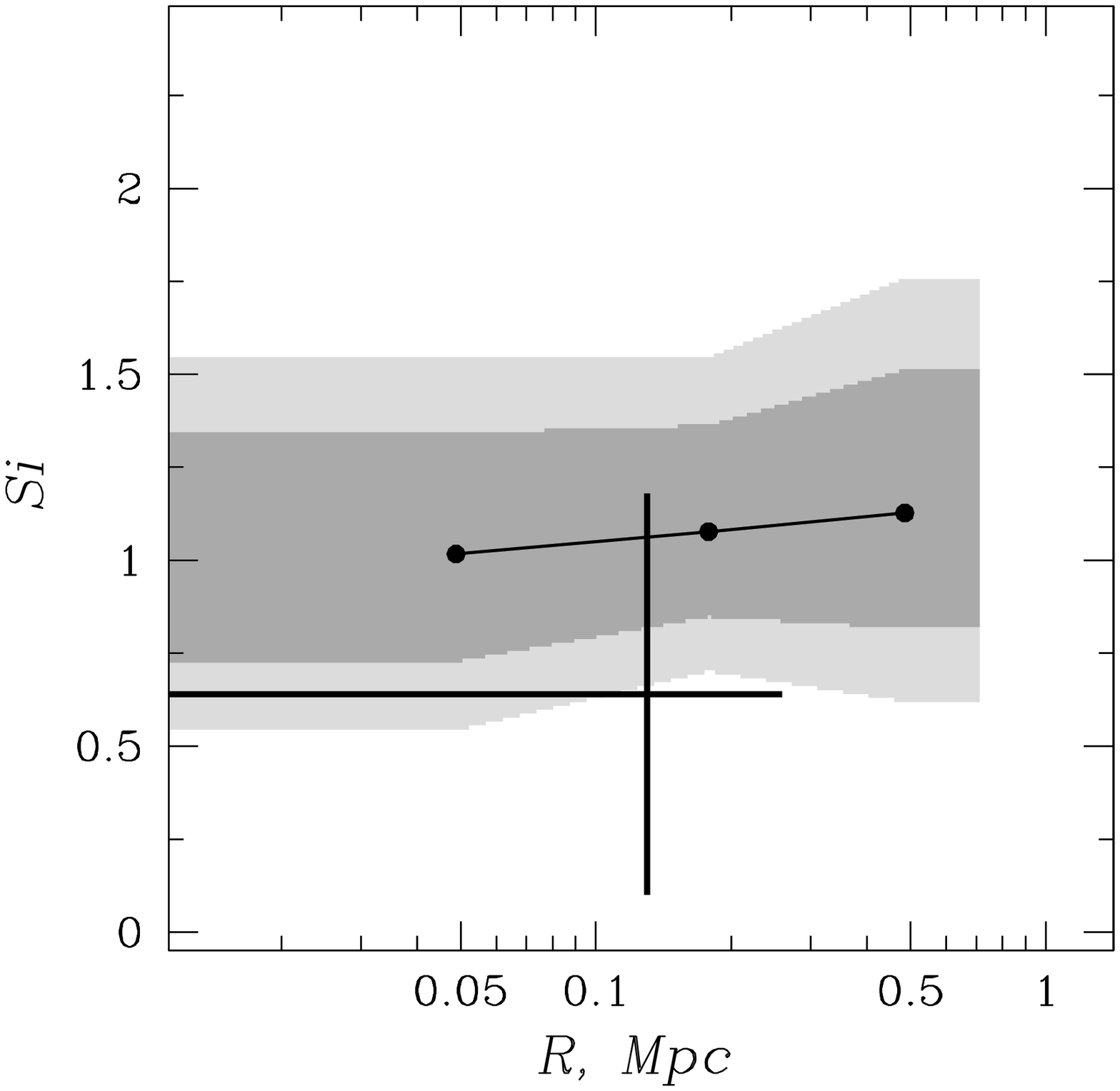}

   \includegraphics[width=1.6in]{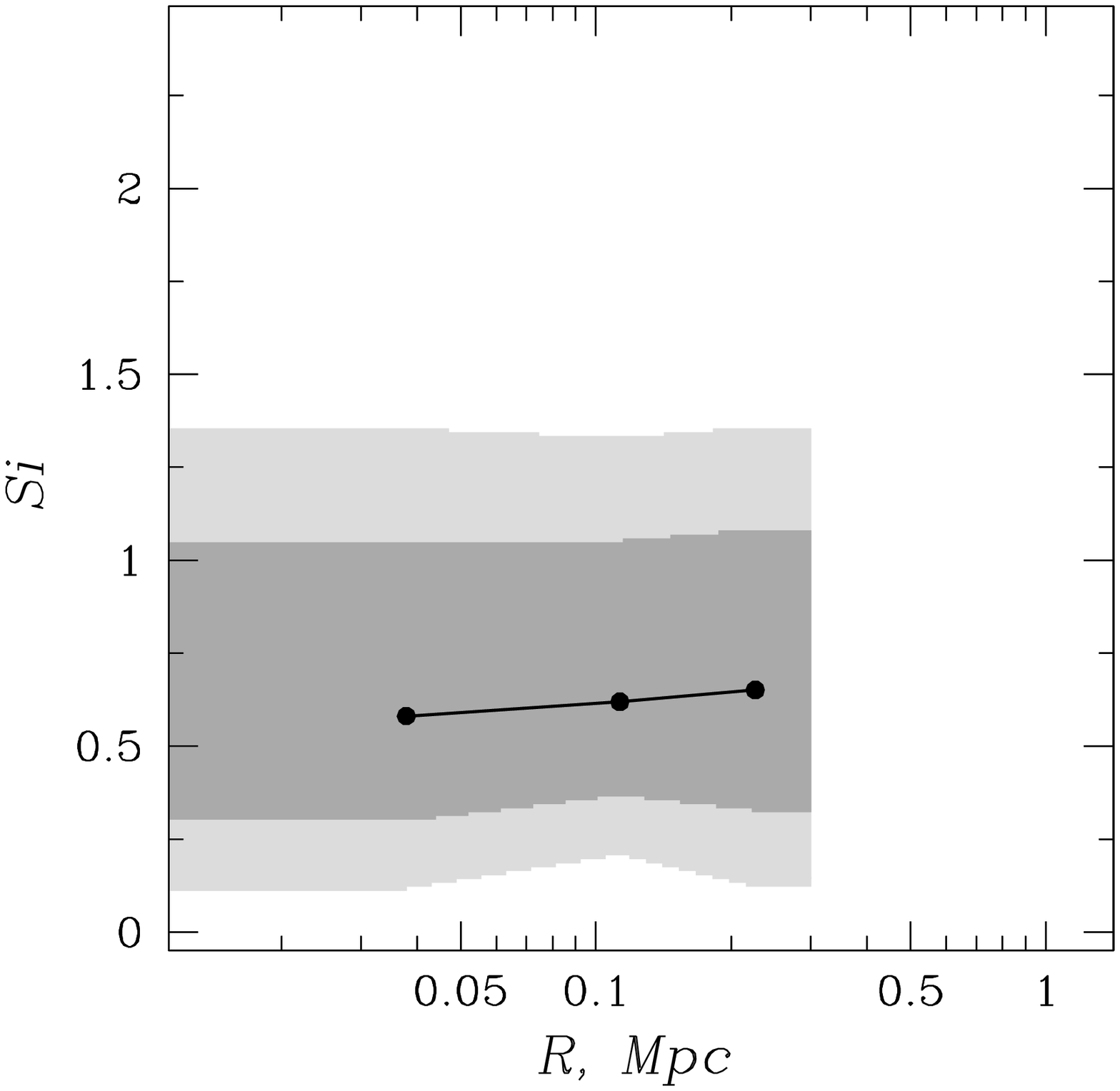}  \hfill 
  \includegraphics[width=1.6in]{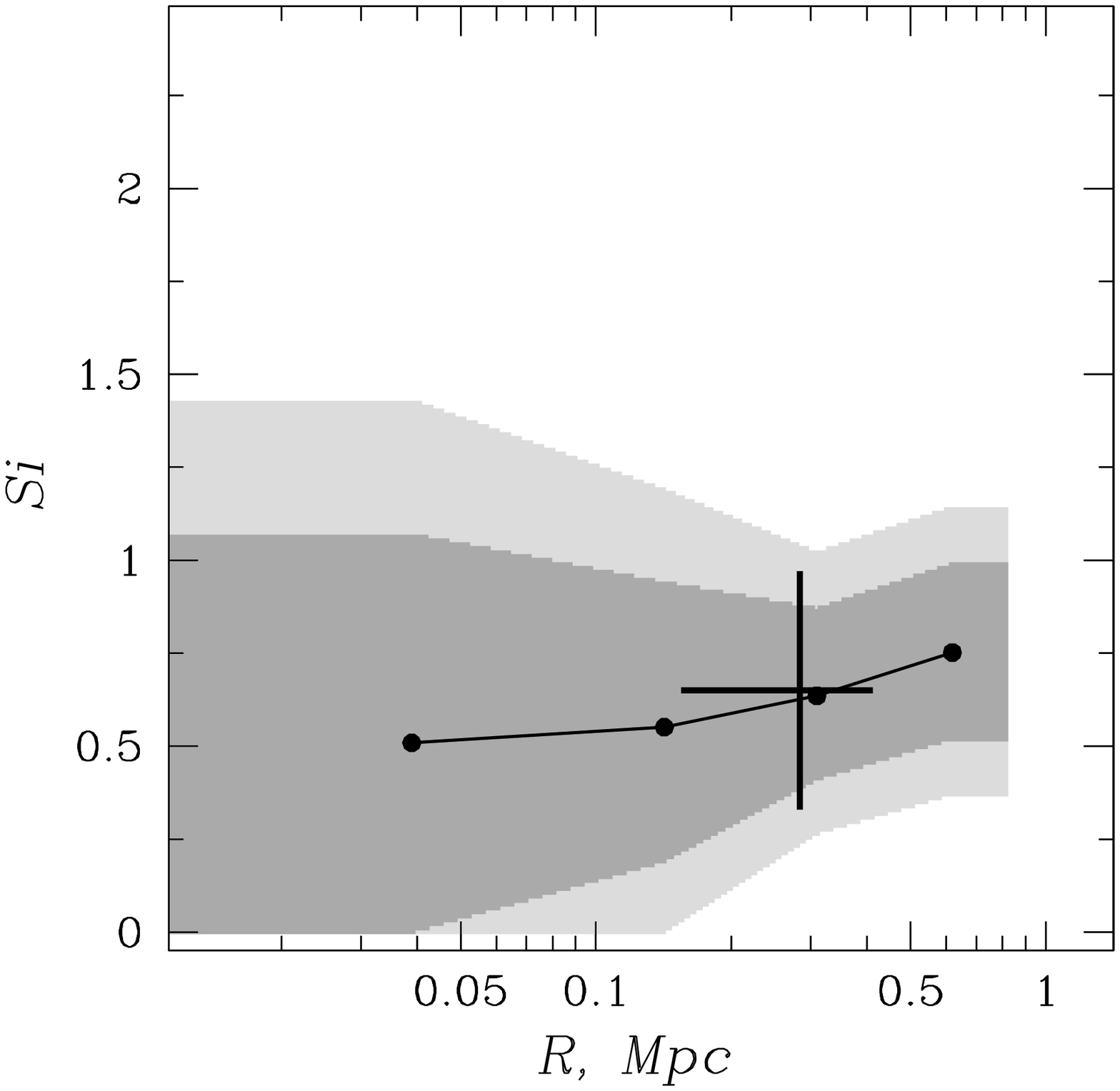} \hfill 
 \includegraphics[width=1.6in]{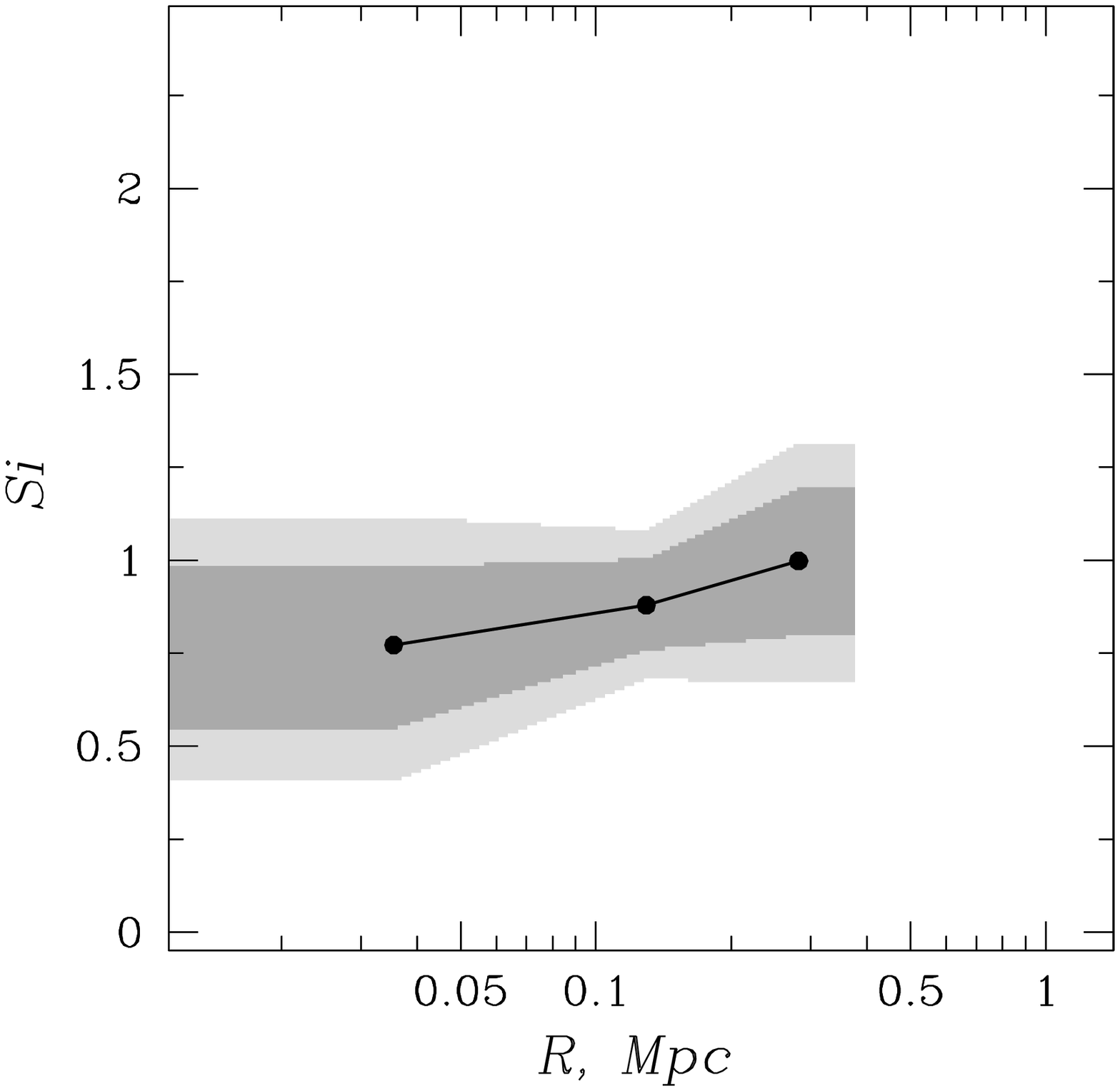} \hfill 
  \includegraphics[width=1.6in]{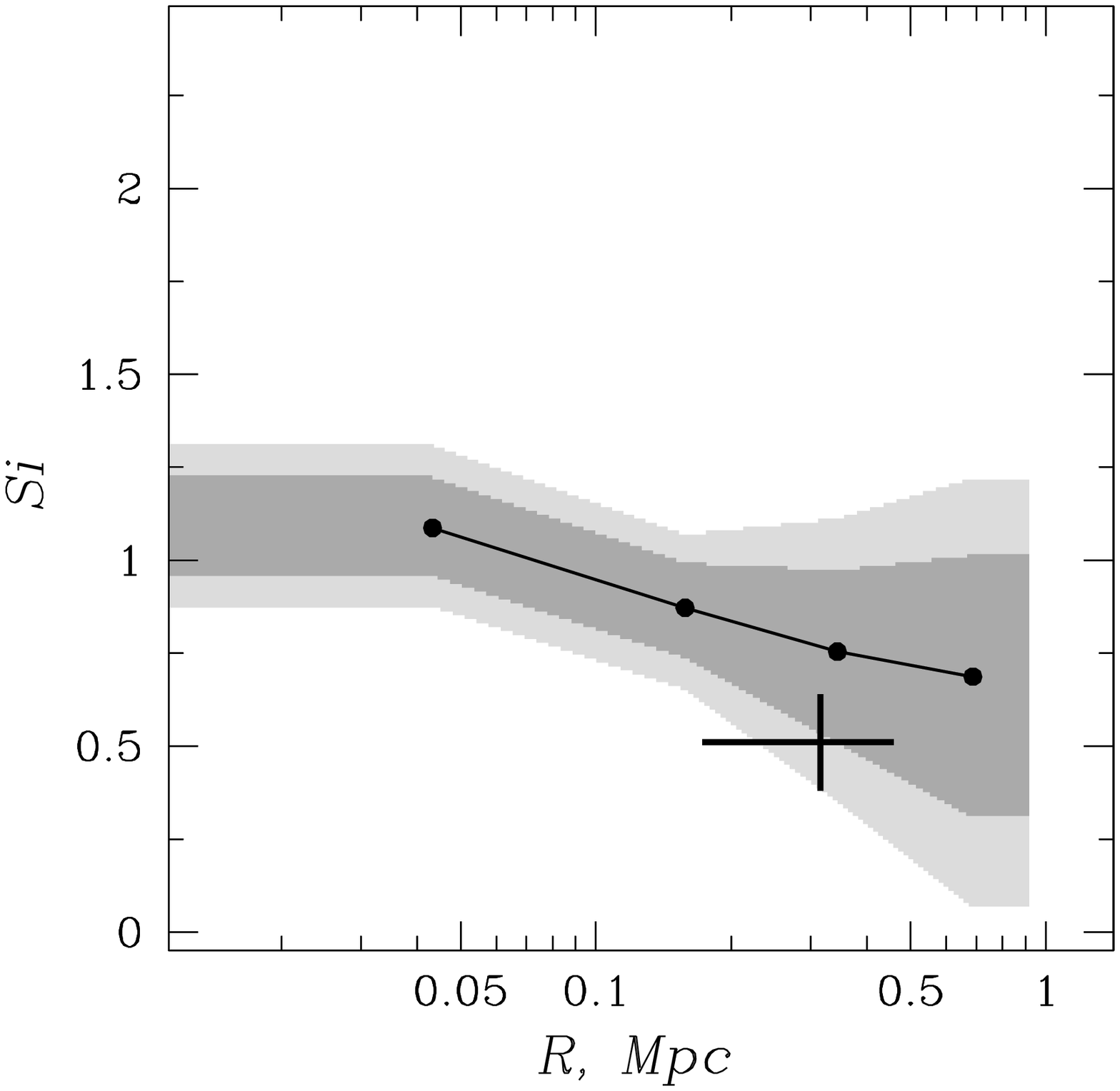} 

   \includegraphics[width=1.6in]{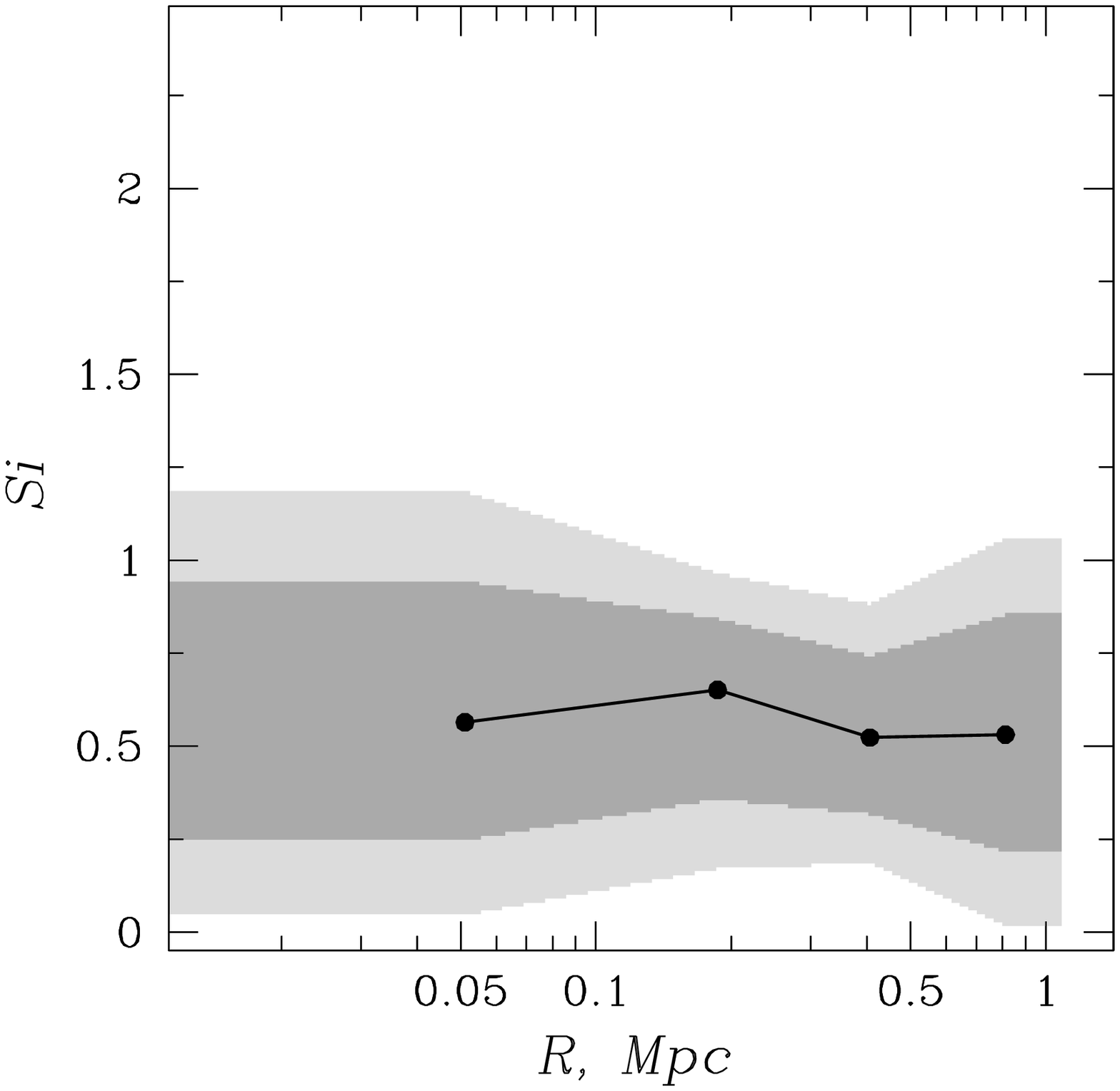} \hfill 
   \includegraphics[width=1.6in]{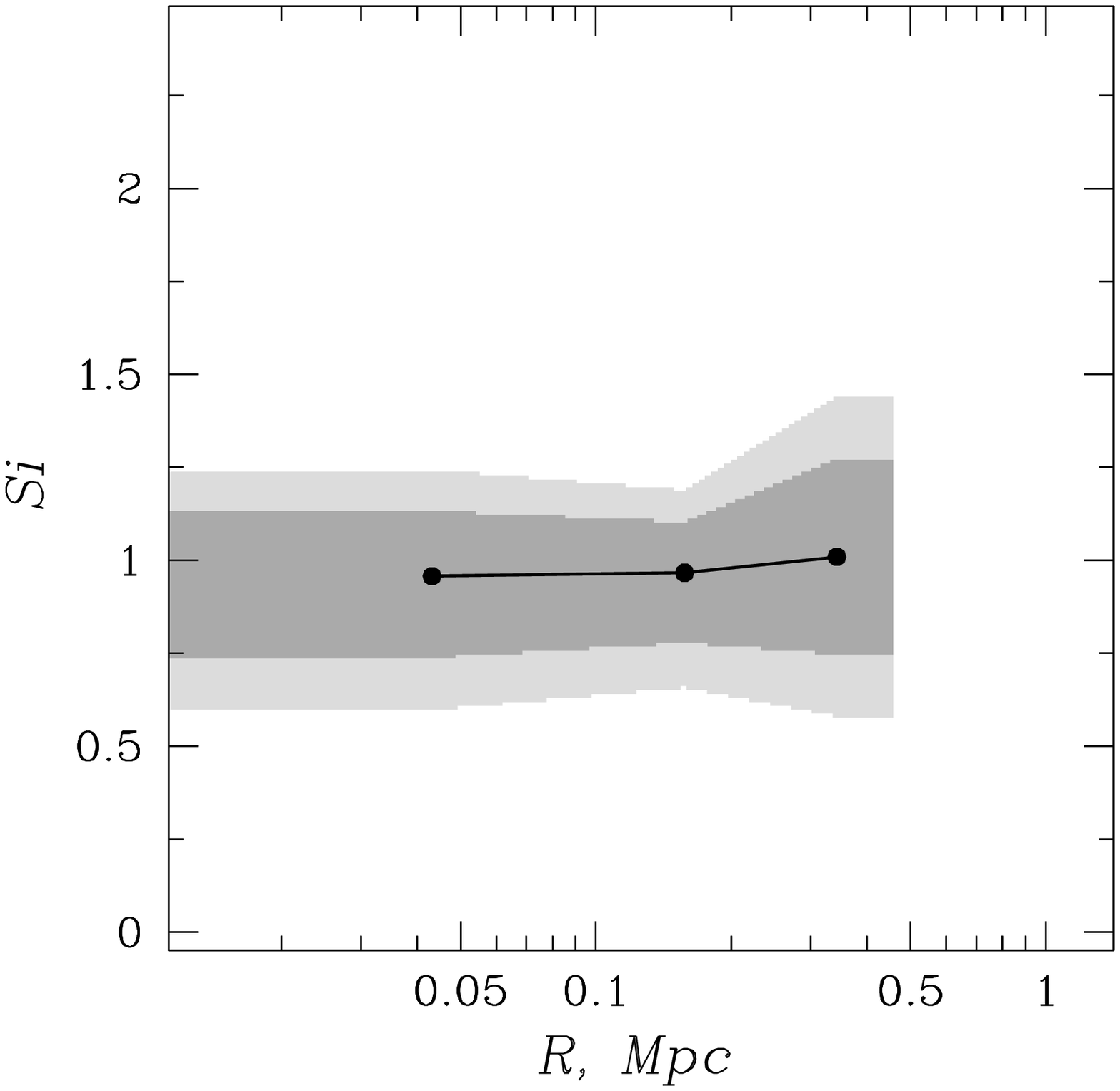} \hfill 
  \includegraphics[width=1.6in]{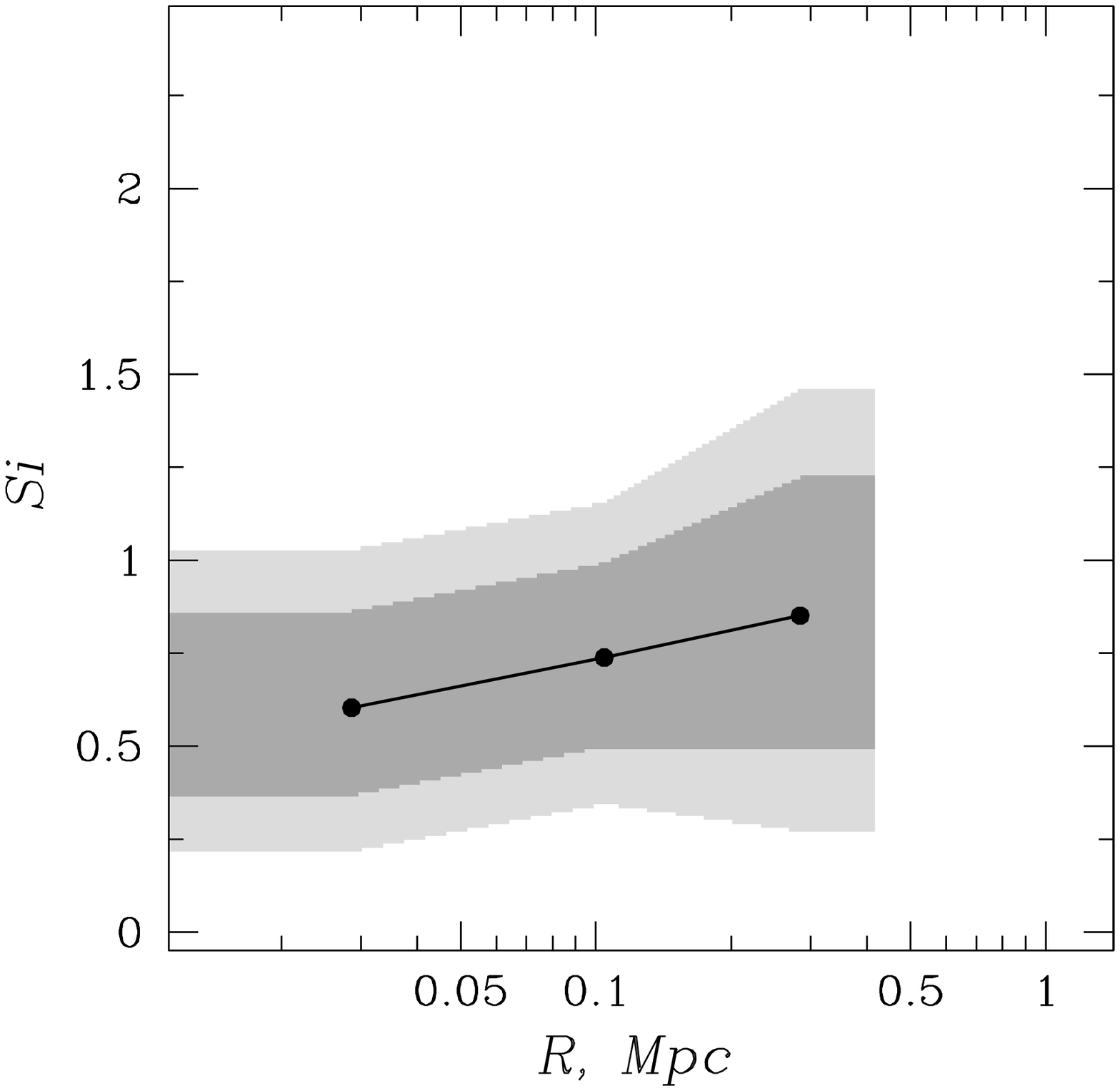} \hfill 
  \includegraphics[width=1.6in]{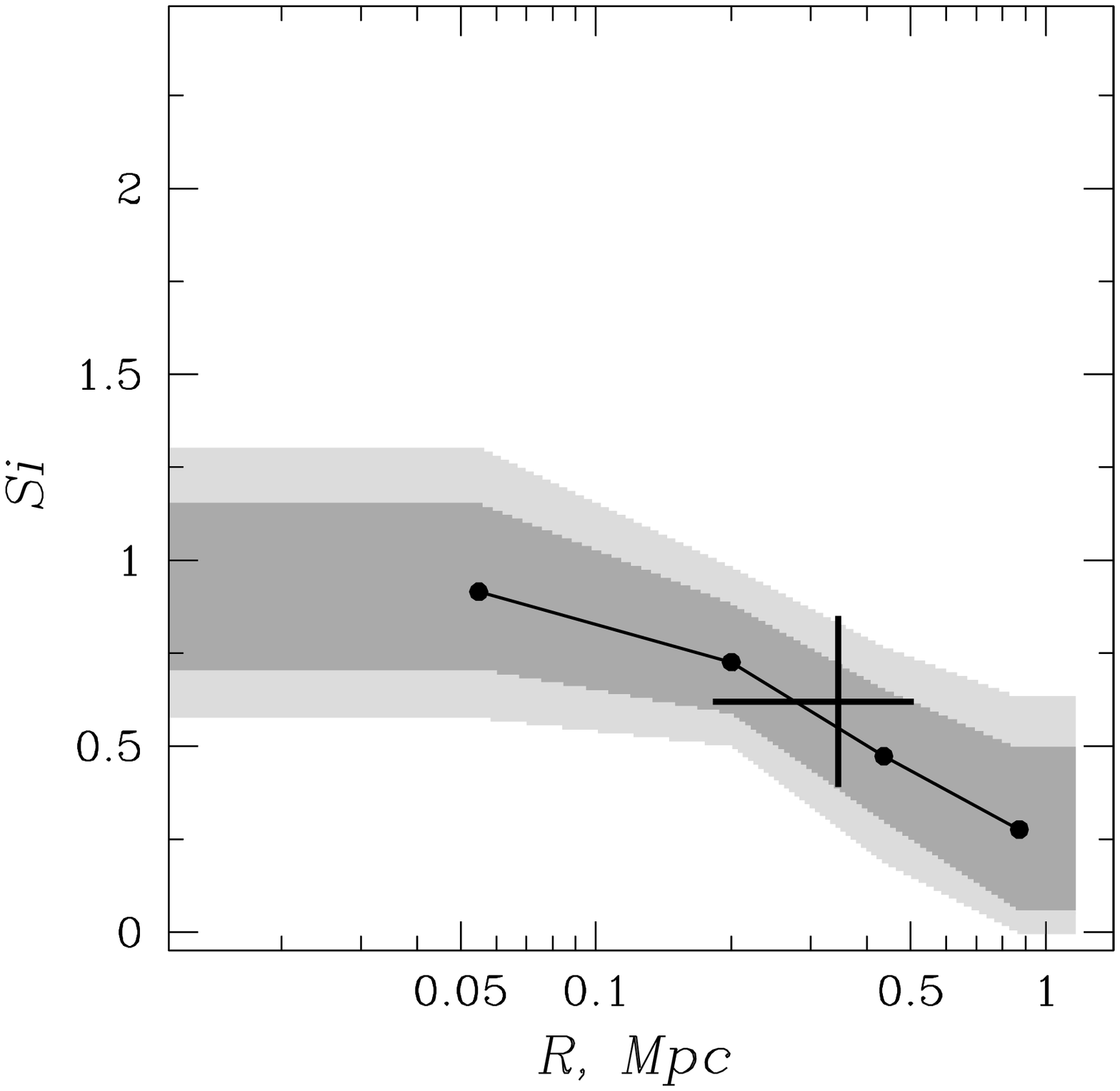} 

    \includegraphics[width=1.6in]{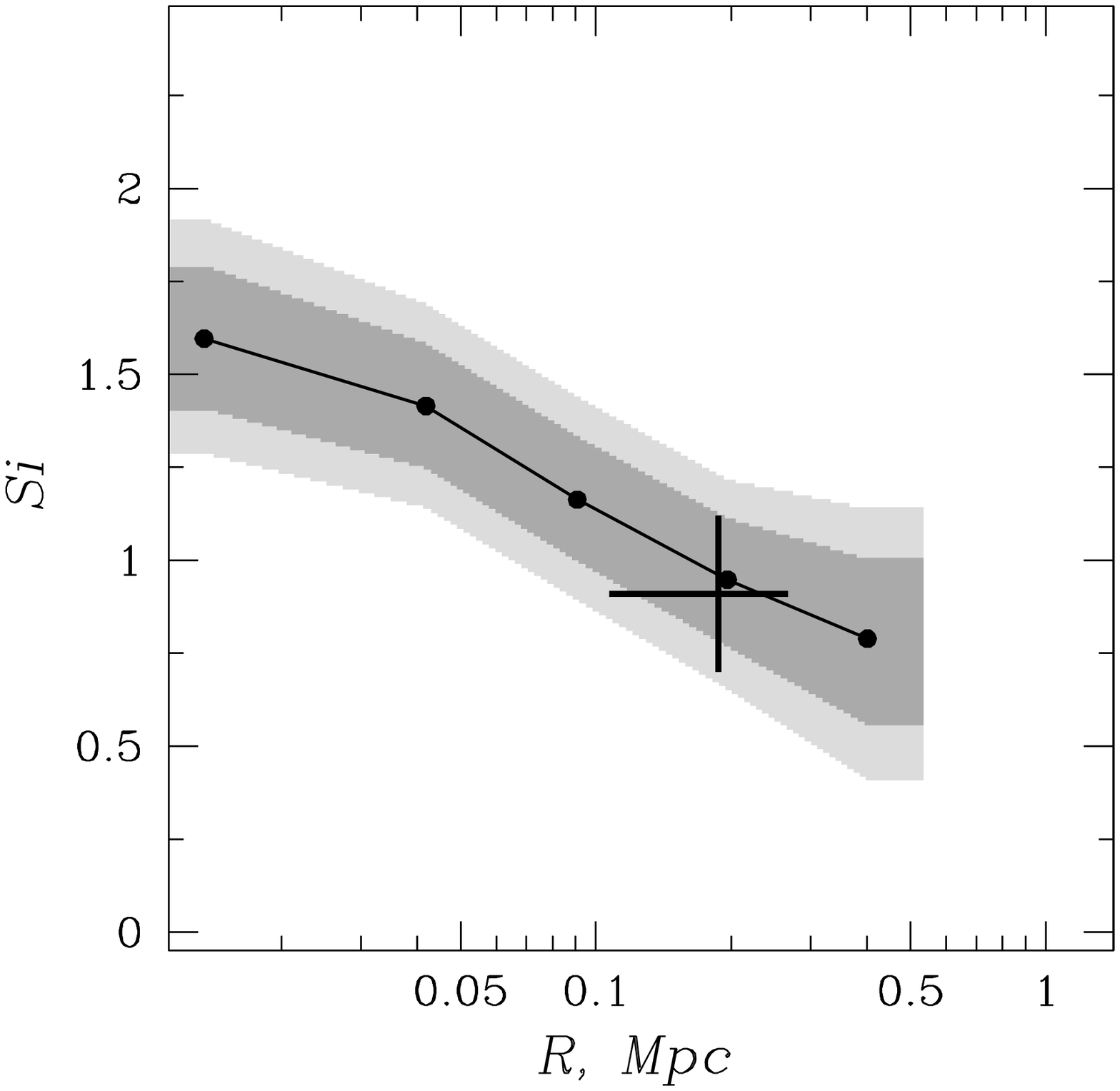}  \hfill 
  \includegraphics[width=1.6in]{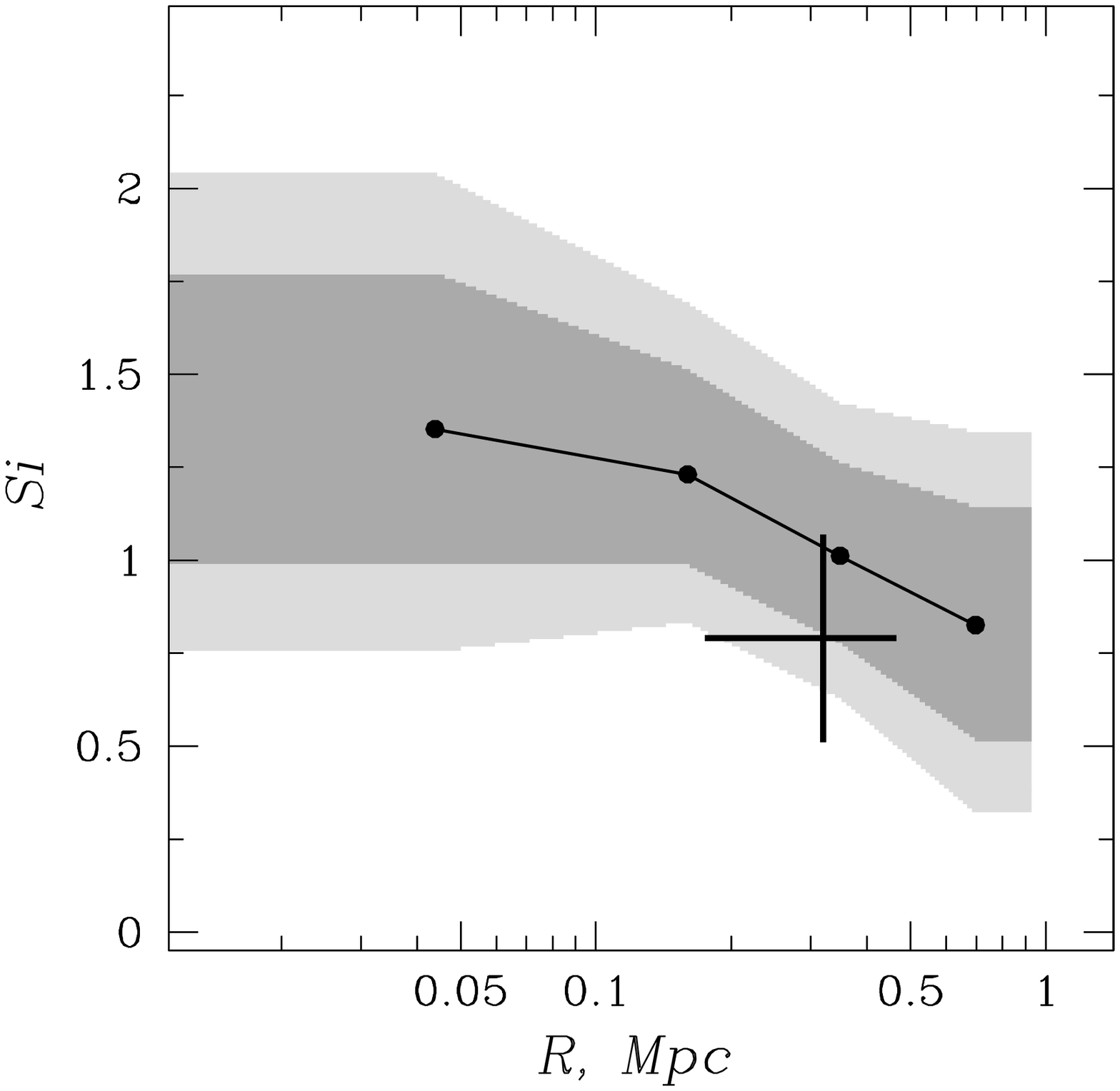} 

\figcaption{Derived Si abundances. The solid lines correspond to the
best-fit Si abundances derived from the ASCA data. The filled circles
indicate the spatial binning used in the analysis. Dark and light shaded
zones around the best fit curves denote the 68 and 90 per cent confidence
areas. Crosses on A400, A194, A262, MKW4S, A539, AWM4, MKW9, A2634, Cen and
A2063 panels show the results from Fukazawa \etal (1998) with radii of
measurement from Fukazawa (private communication).
\label{si-fig}}
\vspace*{-20.7cm}

{\it \hspace*{2.4cm} A2197E \hspace*{3.9cm} A400 \hspace*{3.9cm} A194 \hspace*{3.9cm} A262}

\vspace*{3.55cm}

{\it \hspace*{2.6cm} MKW4S \hspace*{3.6cm} A539 \hspace*{3.8cm} AWM4 \hspace*{3.8cm} MKW9}

\vspace*{3.55cm}

{\it \hspace*{2.4cm} A2197W \hspace*{3.7cm} A2634 \hspace*{3.6cm} A4038 \hspace*{3.5cm} 2A0335}

\vspace*{3.55cm}

{\it \hspace*{2.6cm} HCG94 \hspace*{3.5cm} A2052 \hspace*{3.8cm} A779 \hspace*{3.5cm} MKW3S}

\vspace*{3.55cm}
{\it \hspace*{2.8cm} CEN \hspace*{13.3cm} A2063}

\vspace*{3.7cm}

\end{figure*}

\begin{figure*}

   \includegraphics[width=1.6in]{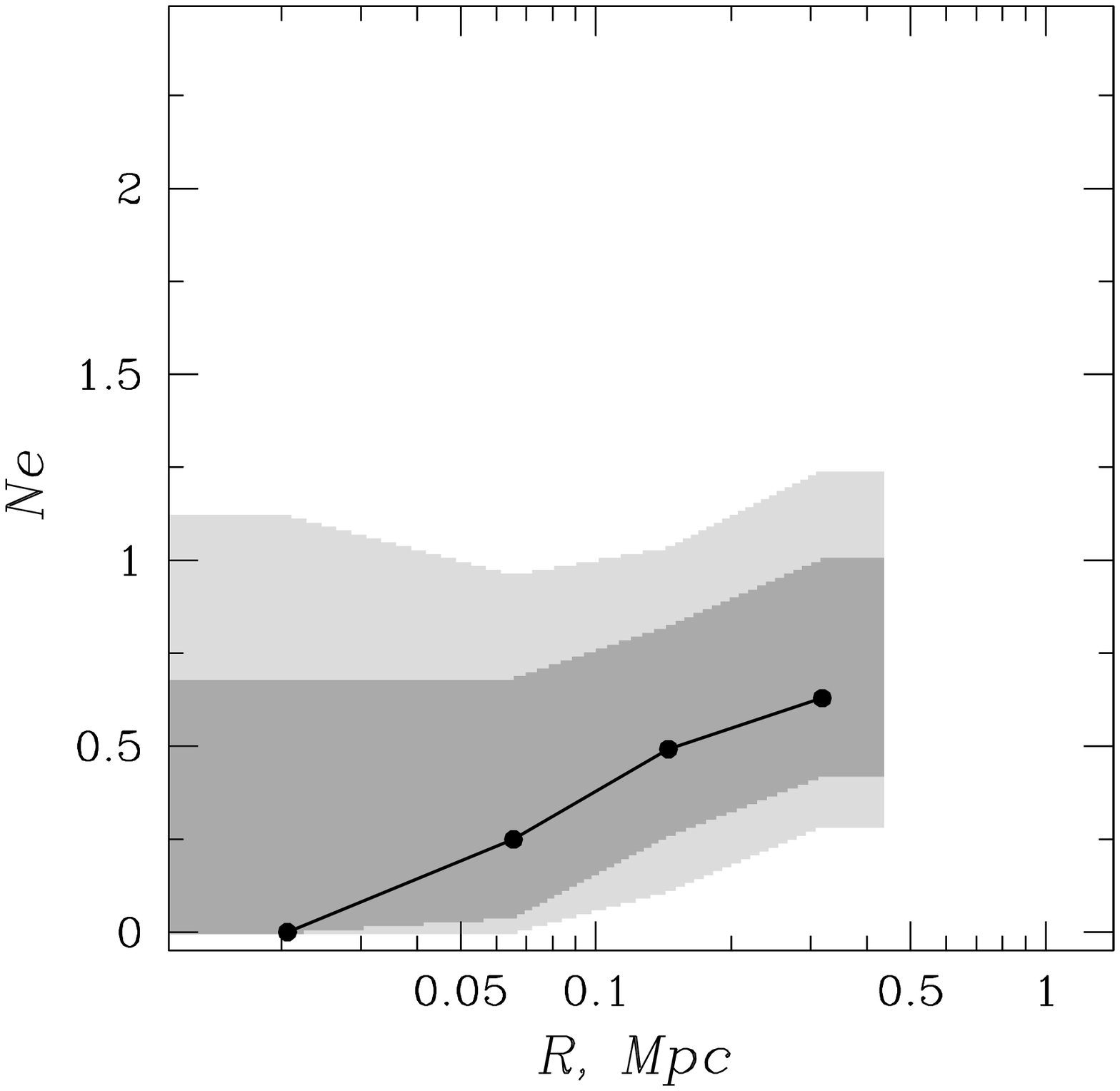} \hfill 
  \includegraphics[width=1.6in]{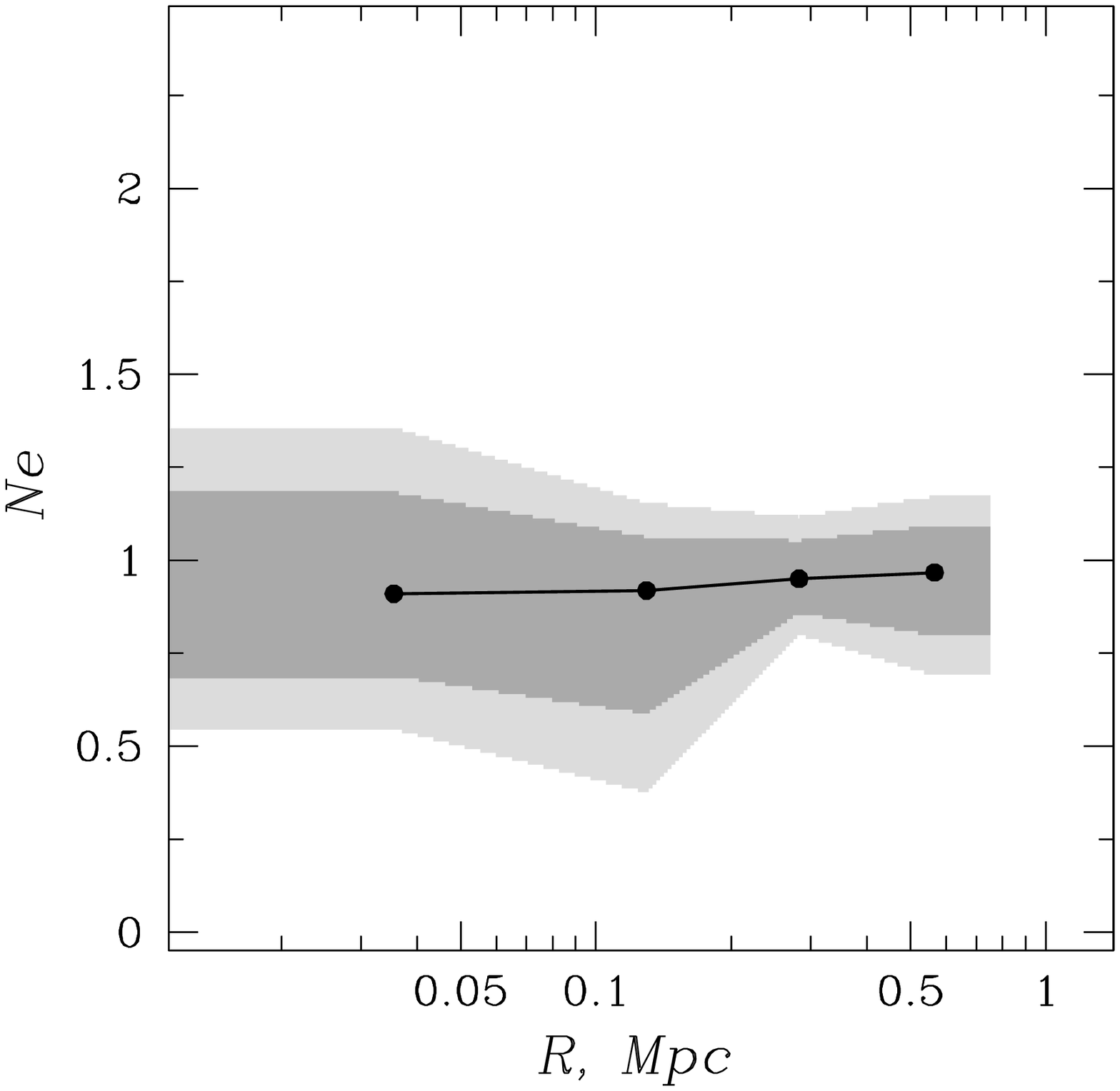}  \hfill
  \includegraphics[width=1.6in]{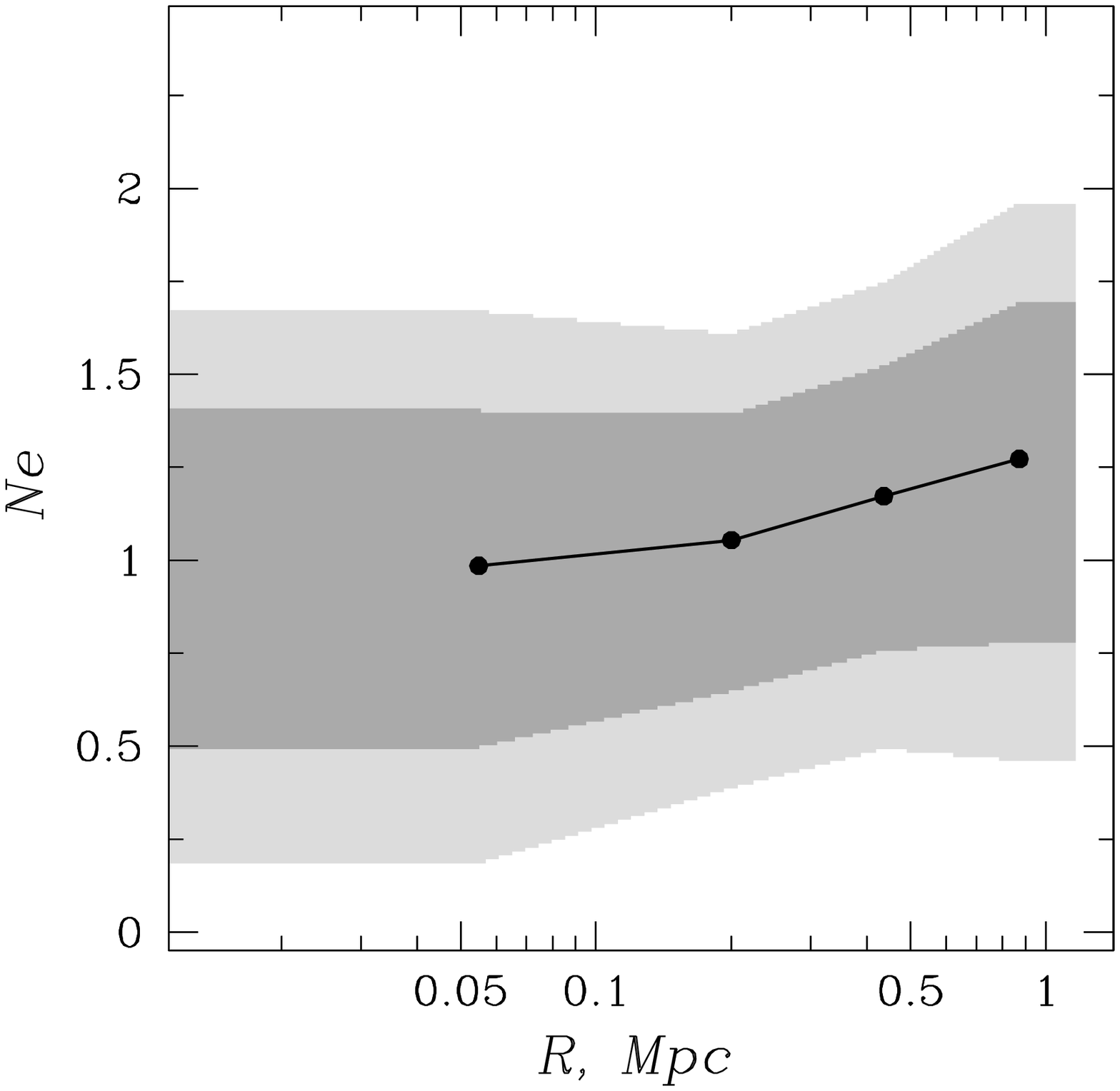} \hfill 
    \includegraphics[width=1.6in]{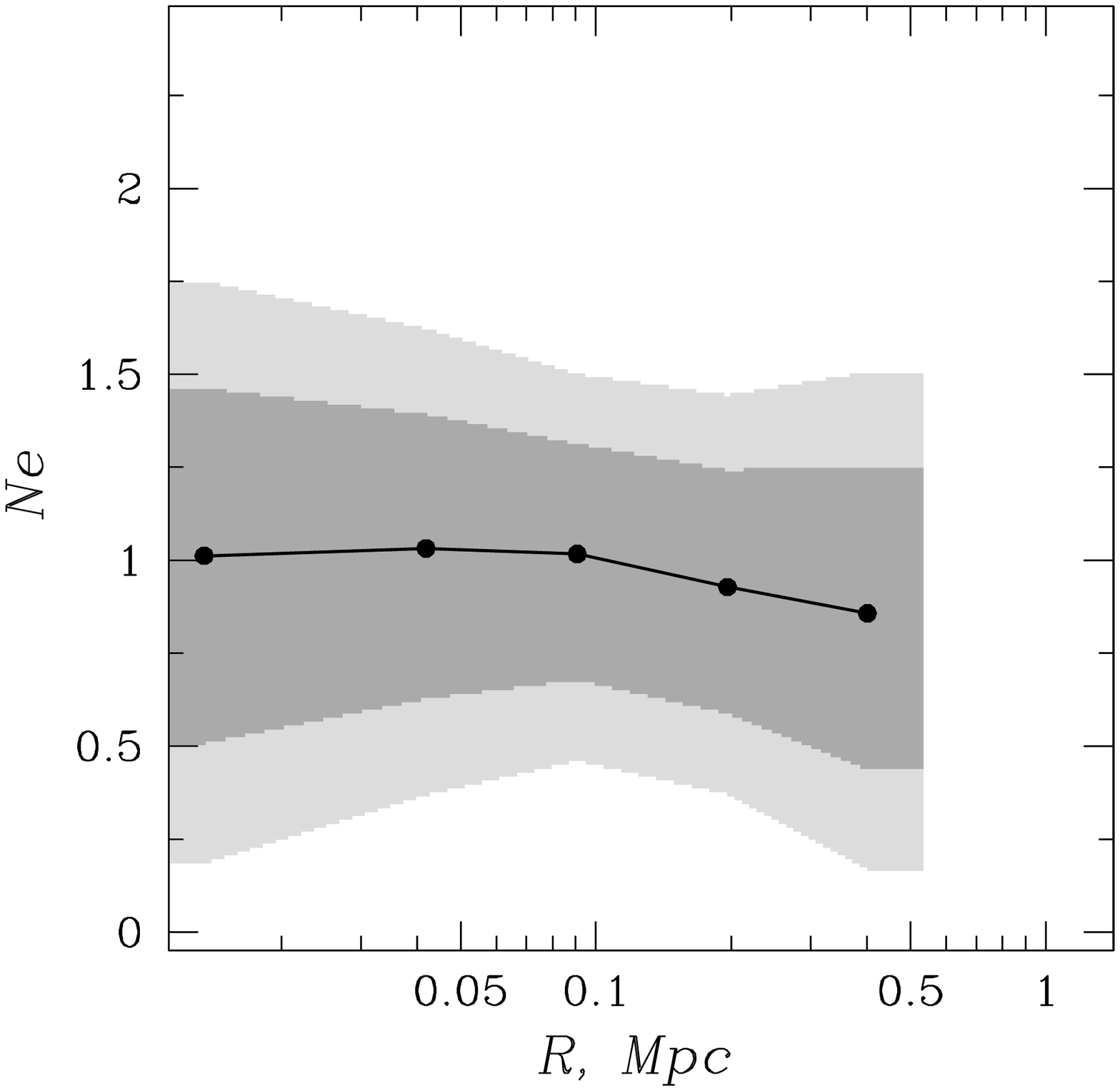} 

\figcaption{Derived Ne abundances. The solid lines correspond to the
best-fit Ne abundances derived from the ASCA data. The filled circles
indicate the spatial binning used in the analysis. Dark and light shaded
zones around the best fit curves denote the 68 and 90 per cent confidence
areas.
\label{ne-fig}}
\vspace*{-4.5cm}

{\it \hspace*{2.8cm} A262 \hspace*{3.6cm} MKW4S  \hspace*{3.5cm} MKW3S \hspace*{3.7cm} CEN}

\vspace*{4.3cm}

   \includegraphics[width=1.6in]{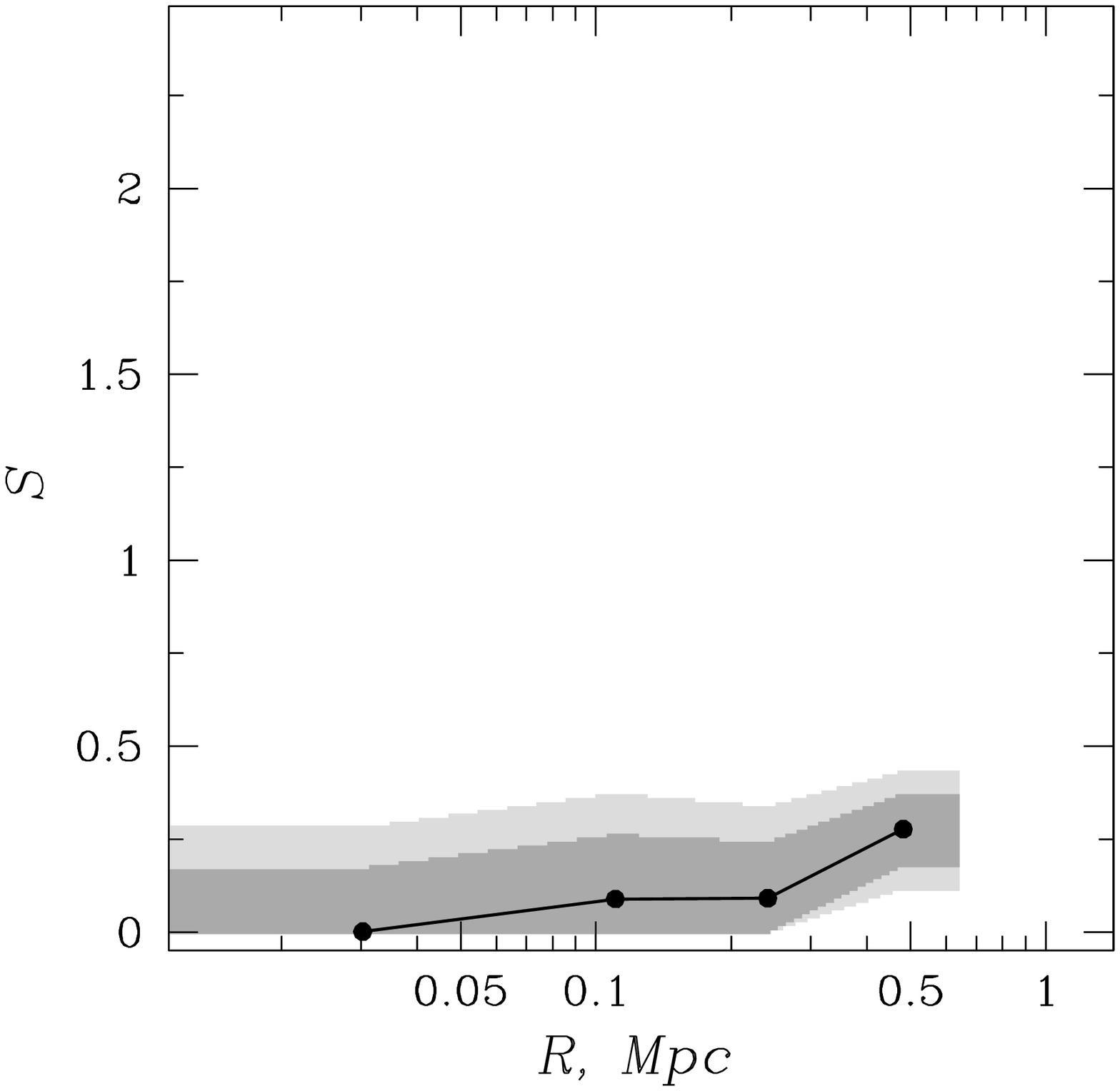} \hfill
   \includegraphics[width=1.6in]{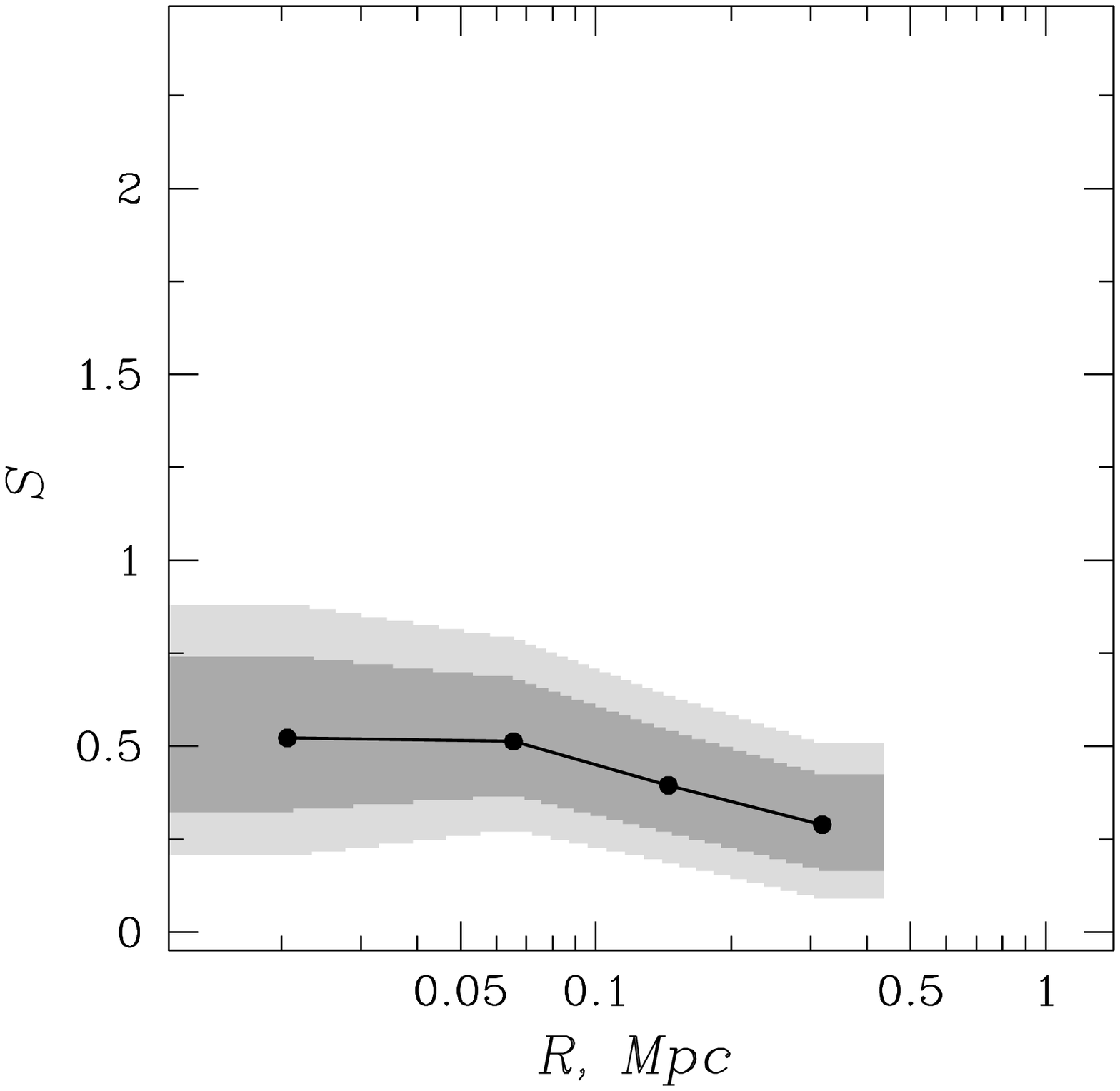} \hfill 
  \includegraphics[width=1.6in]{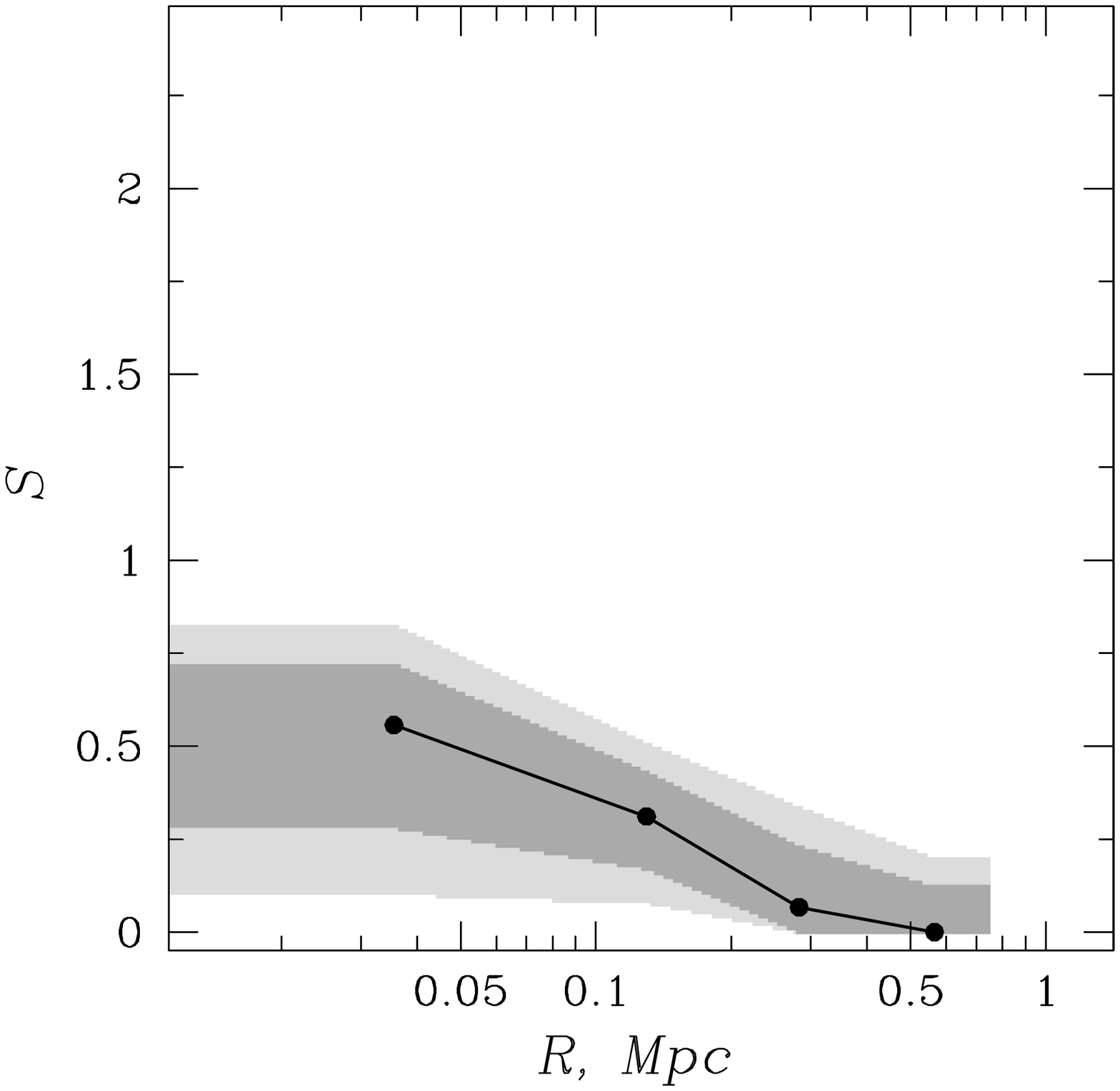}  \hfill
  \includegraphics[width=1.6in]{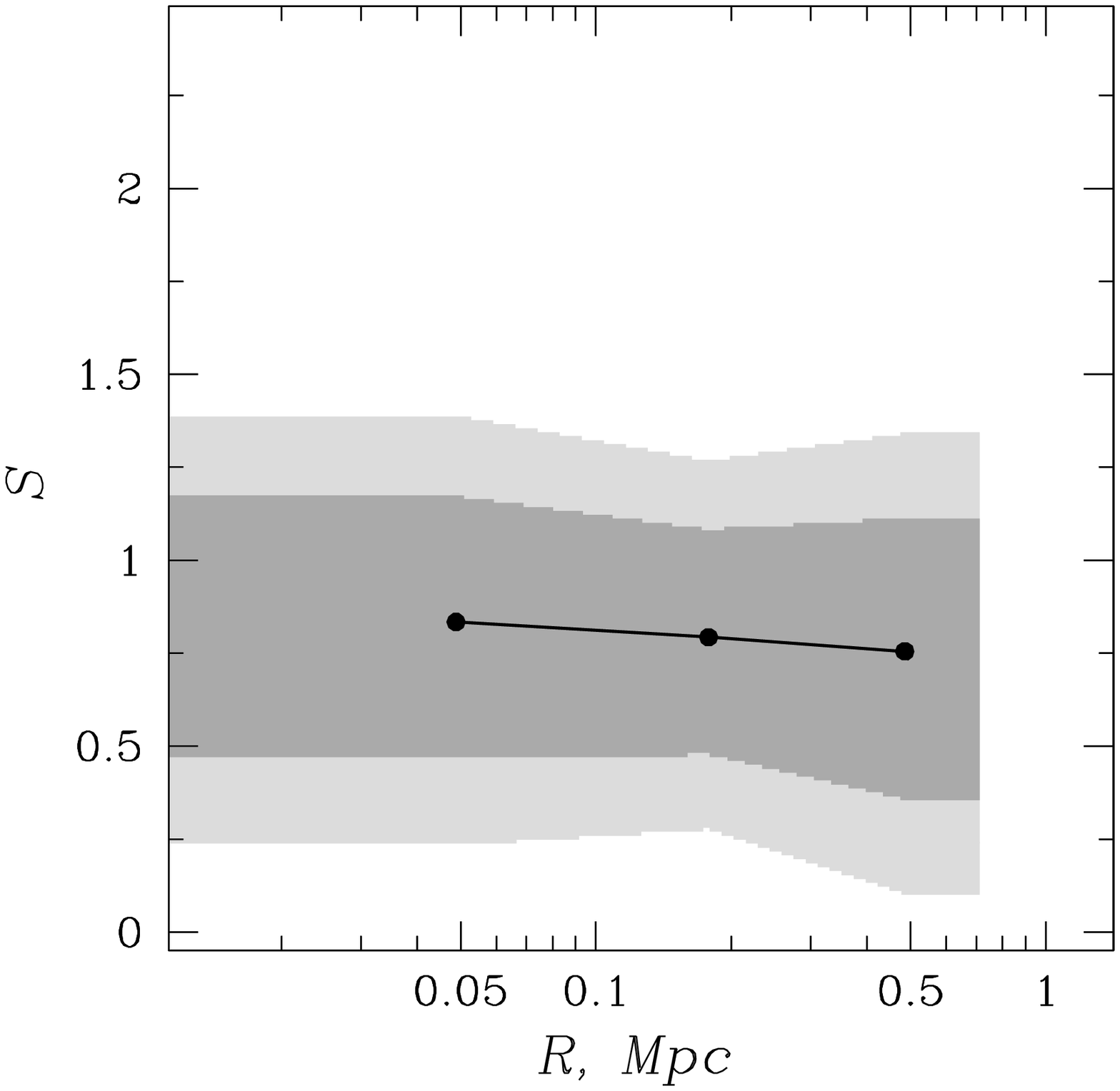}

  \includegraphics[width=1.6in]{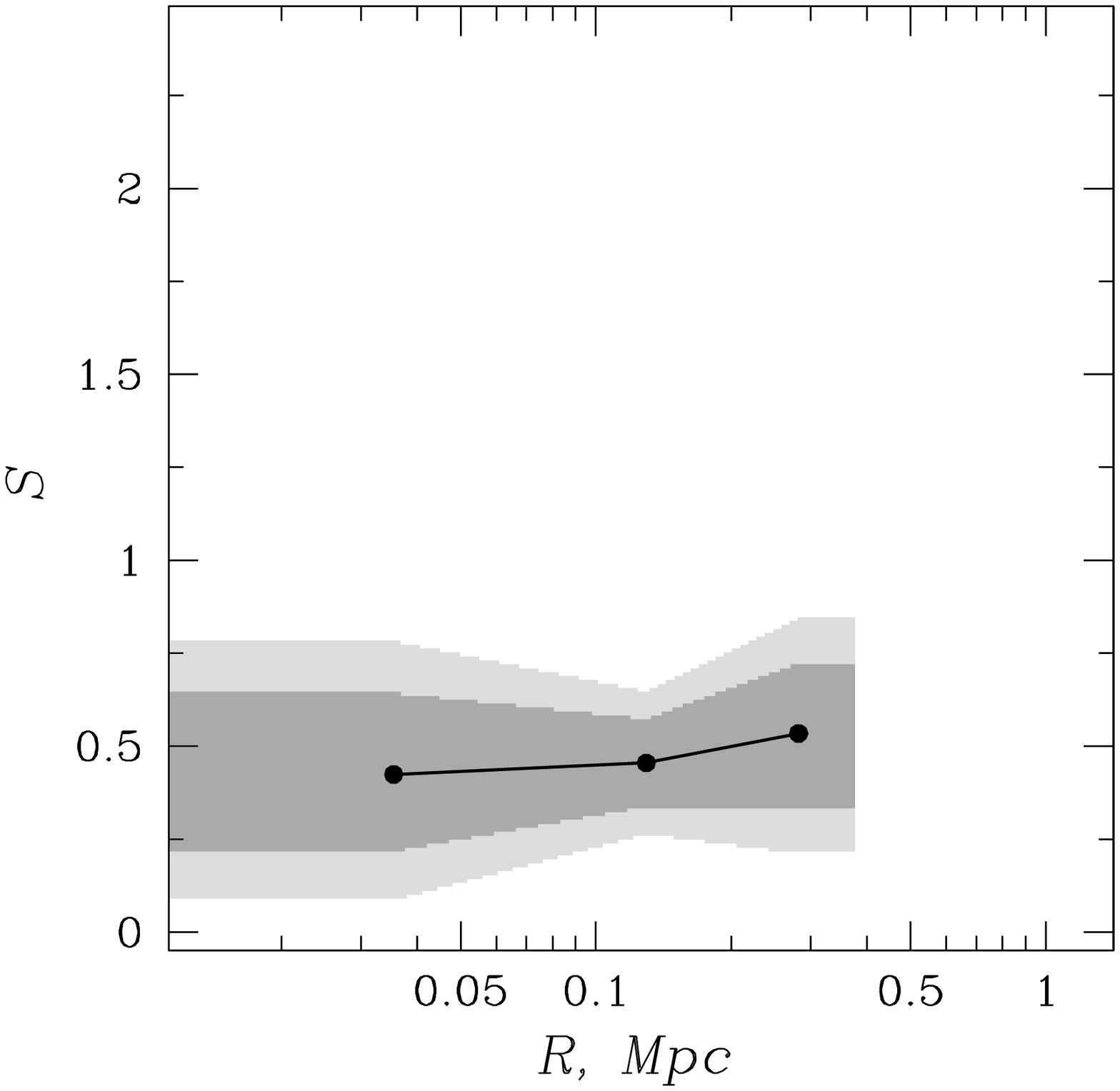} \hfill 
 \includegraphics[width=1.6in]{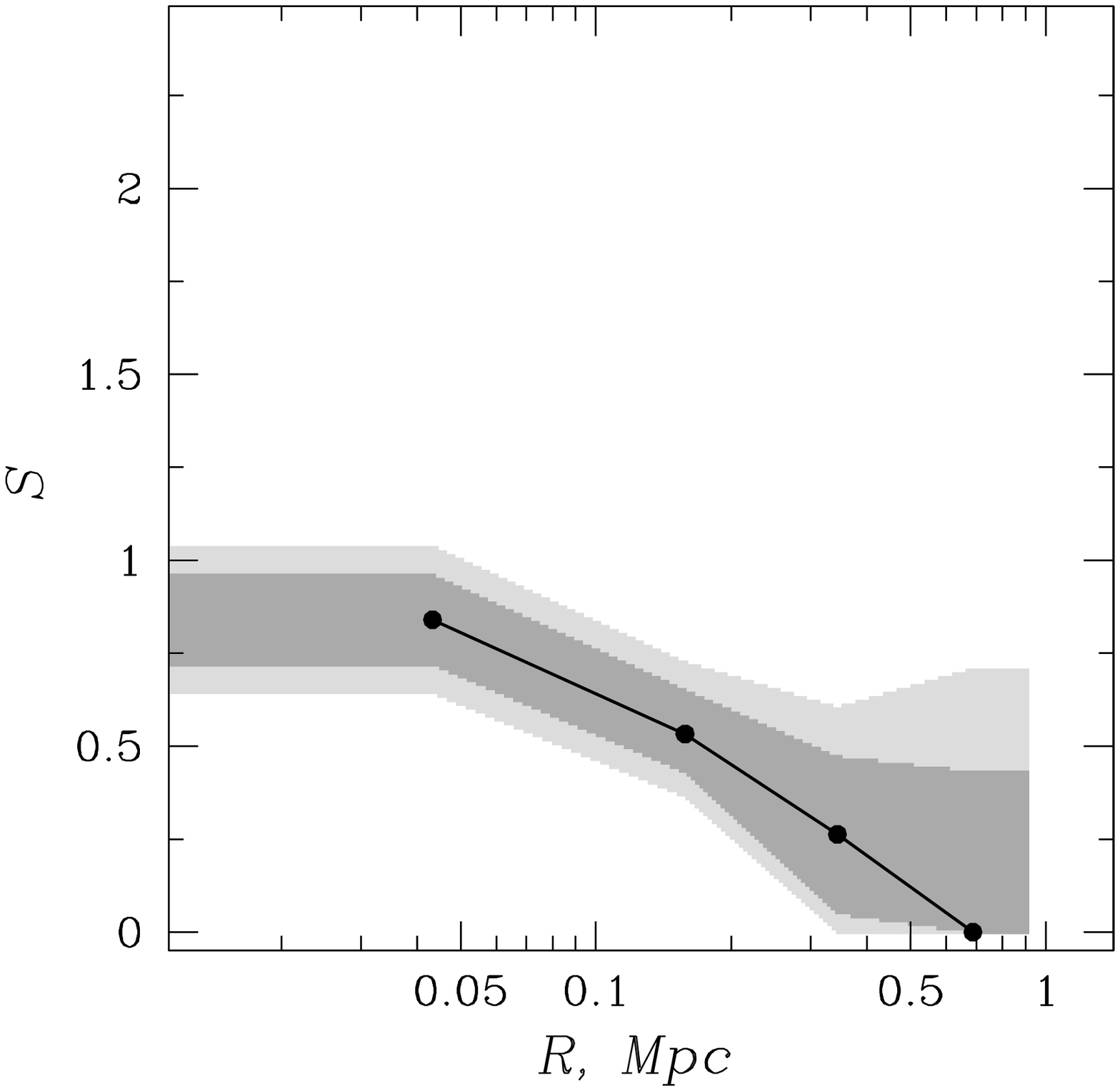} \hfill 
   \includegraphics[width=1.6in]{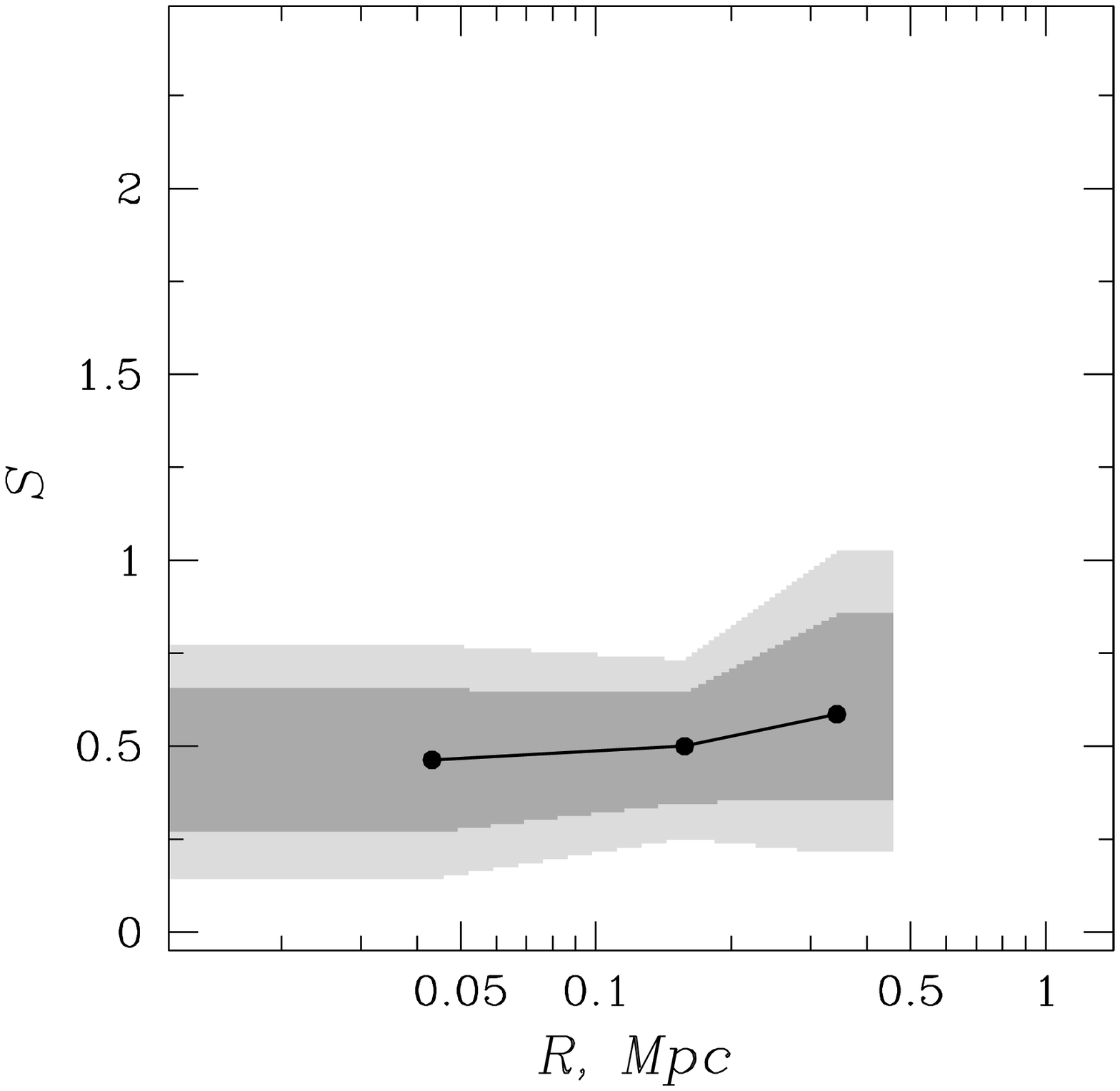} \hfill 
    \includegraphics[width=1.6in]{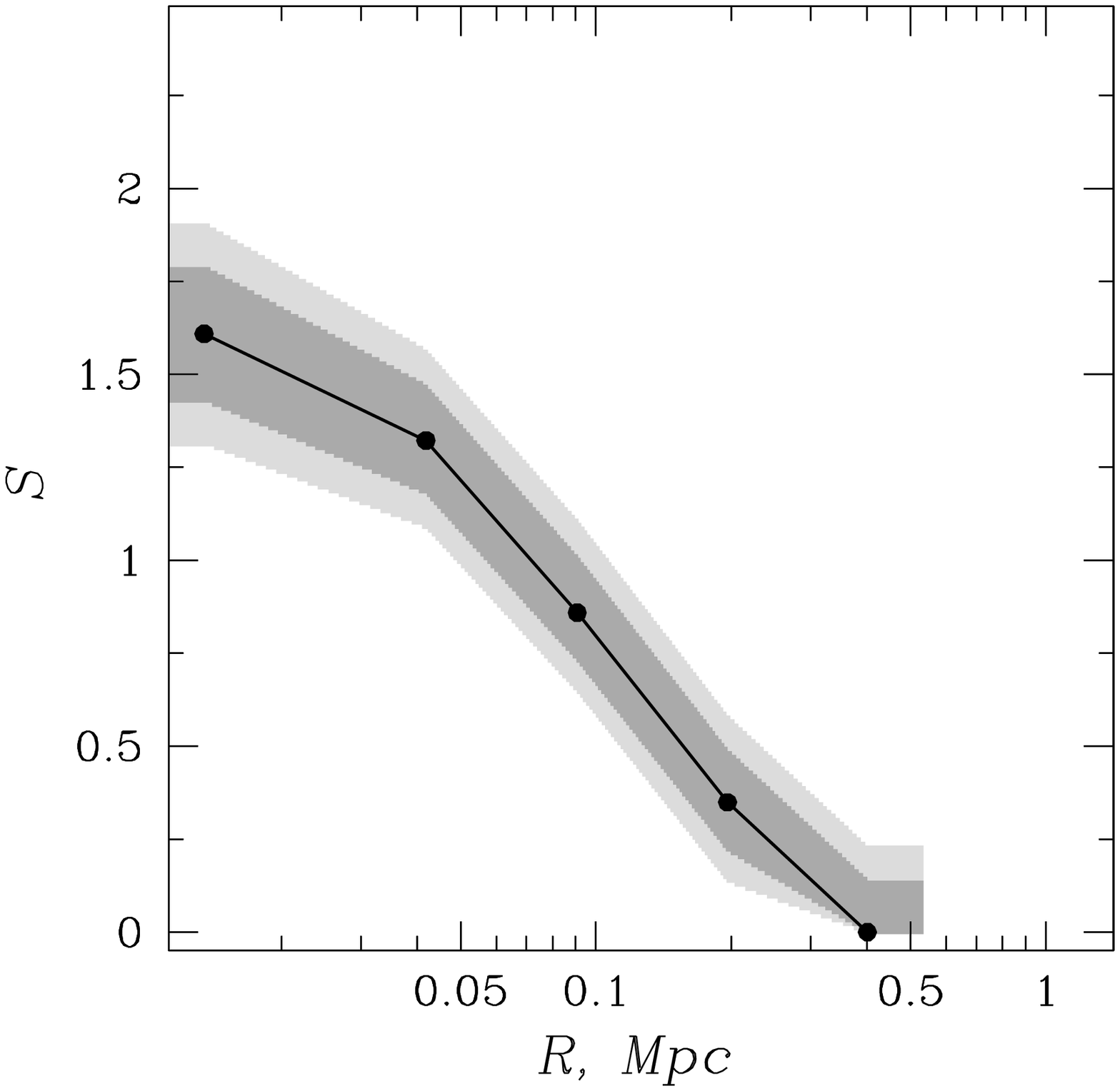} 

\figcaption{Derived S abundances. The solid lines correspond to the
best-fit S abundances derived from the ASCA data. The filled circles
indicate the spatial binning used in the analysis. Dark and light shaded
zones around the best fit curves denote the 68 and 90 per cent confidence
areas.
\label{s-fig}}
\vspace*{-8.5cm}

{\it \hspace*{2.8cm} A400 \hspace*{3.9cm} A262 \hspace*{3.6cm} MKW4S \hspace*{3.6cm} MKW9}

\vspace*{3.55cm}

{\it \hspace*{2.6cm} A4038 \hspace*{3.6cm} 2A0335 \hspace*{3.8cm} A2052 \hspace*{3.9cm} CEN}

\vspace*{3.55cm}

\end{figure*}

%\clearpage
\begin{figure*}

\hspace*{-0.2cm}\includegraphics[width=4.2cm]{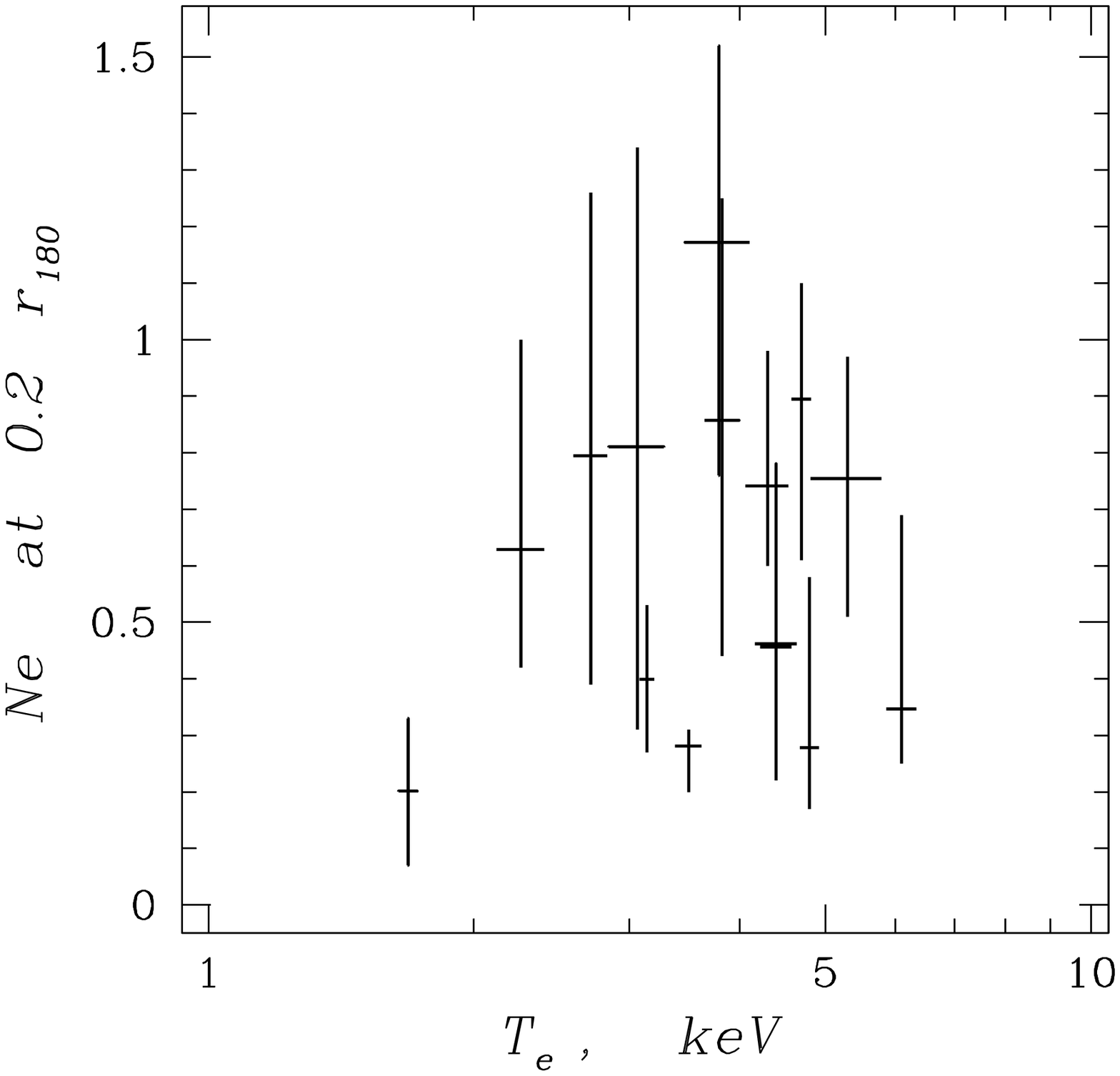}
\hspace{0.2cm}\includegraphics[width=4.2cm]{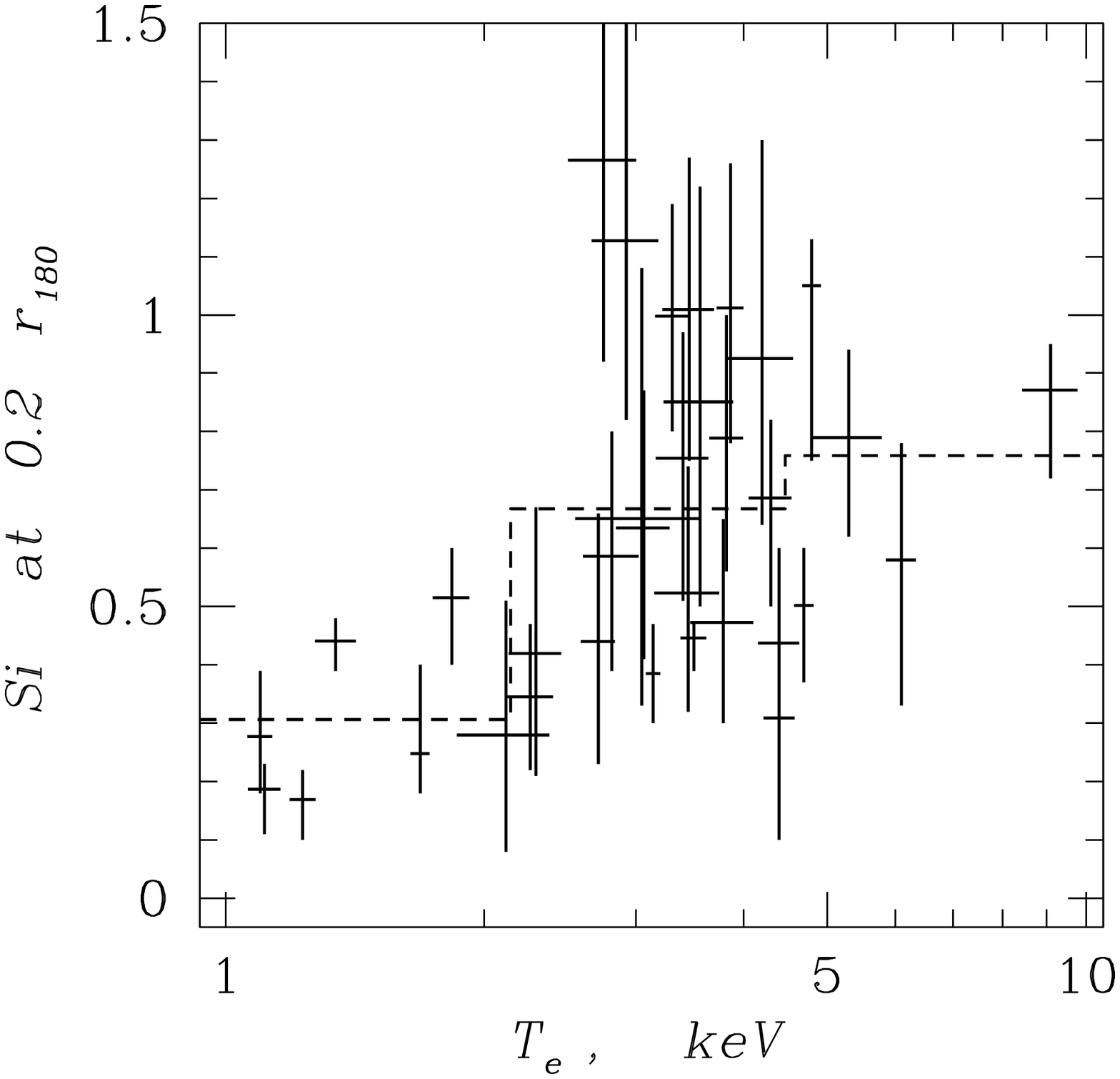}
\hspace{0.2cm}\includegraphics[width=4.2cm]{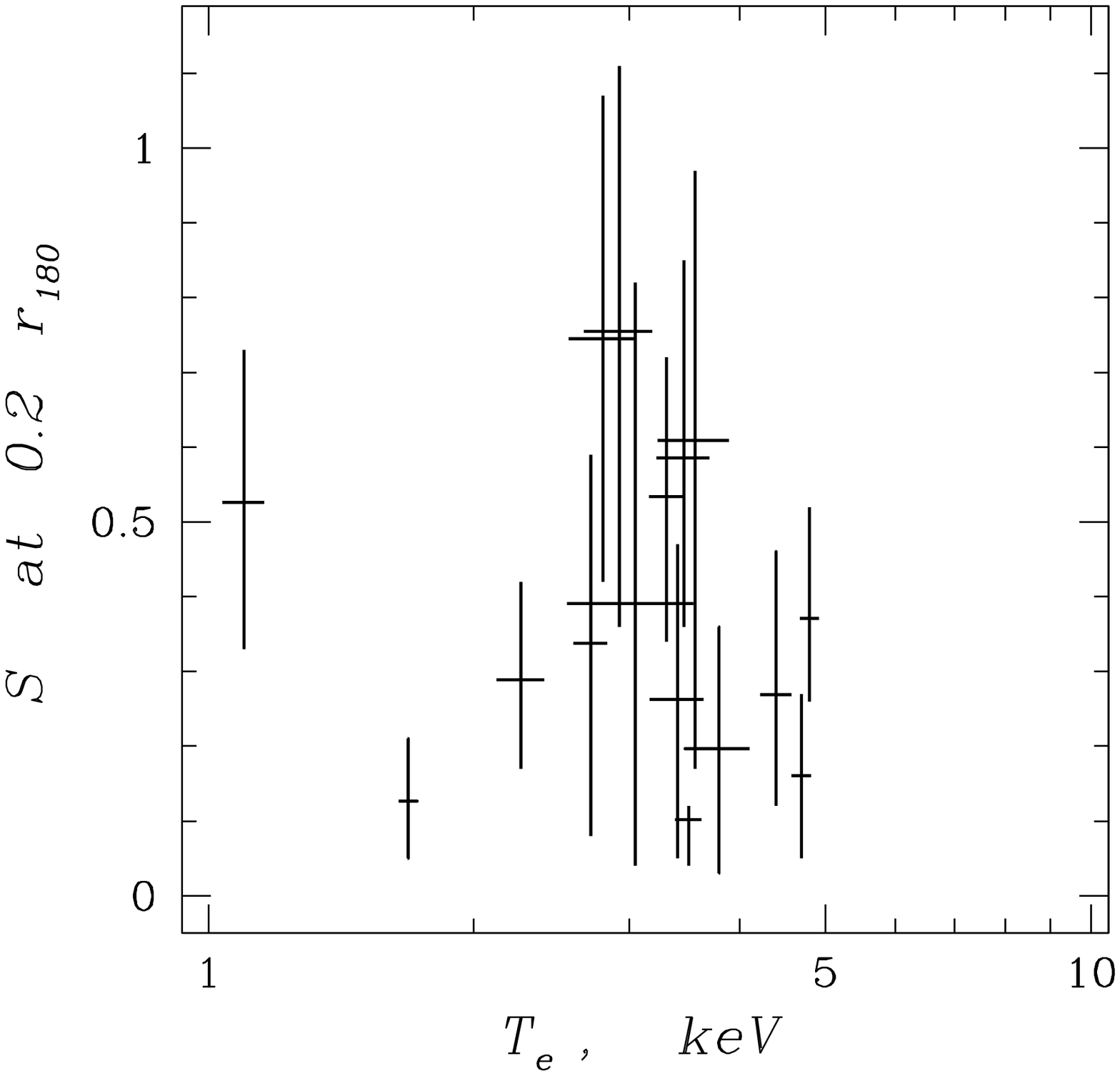}
\hspace{0.2cm}\includegraphics[width=4.2cm]{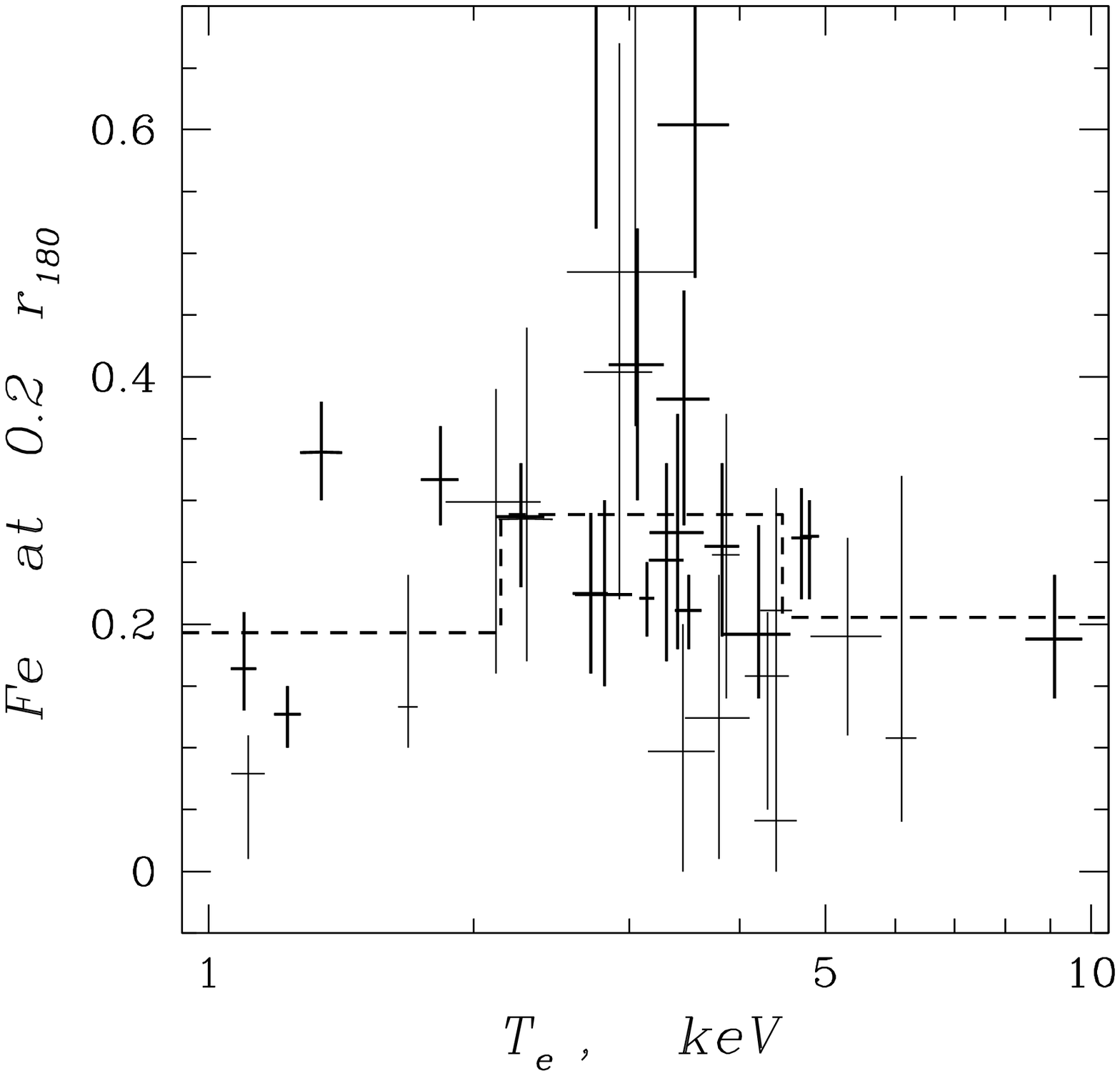}

\hspace*{-0.2cm}\includegraphics[width=4.2cm]{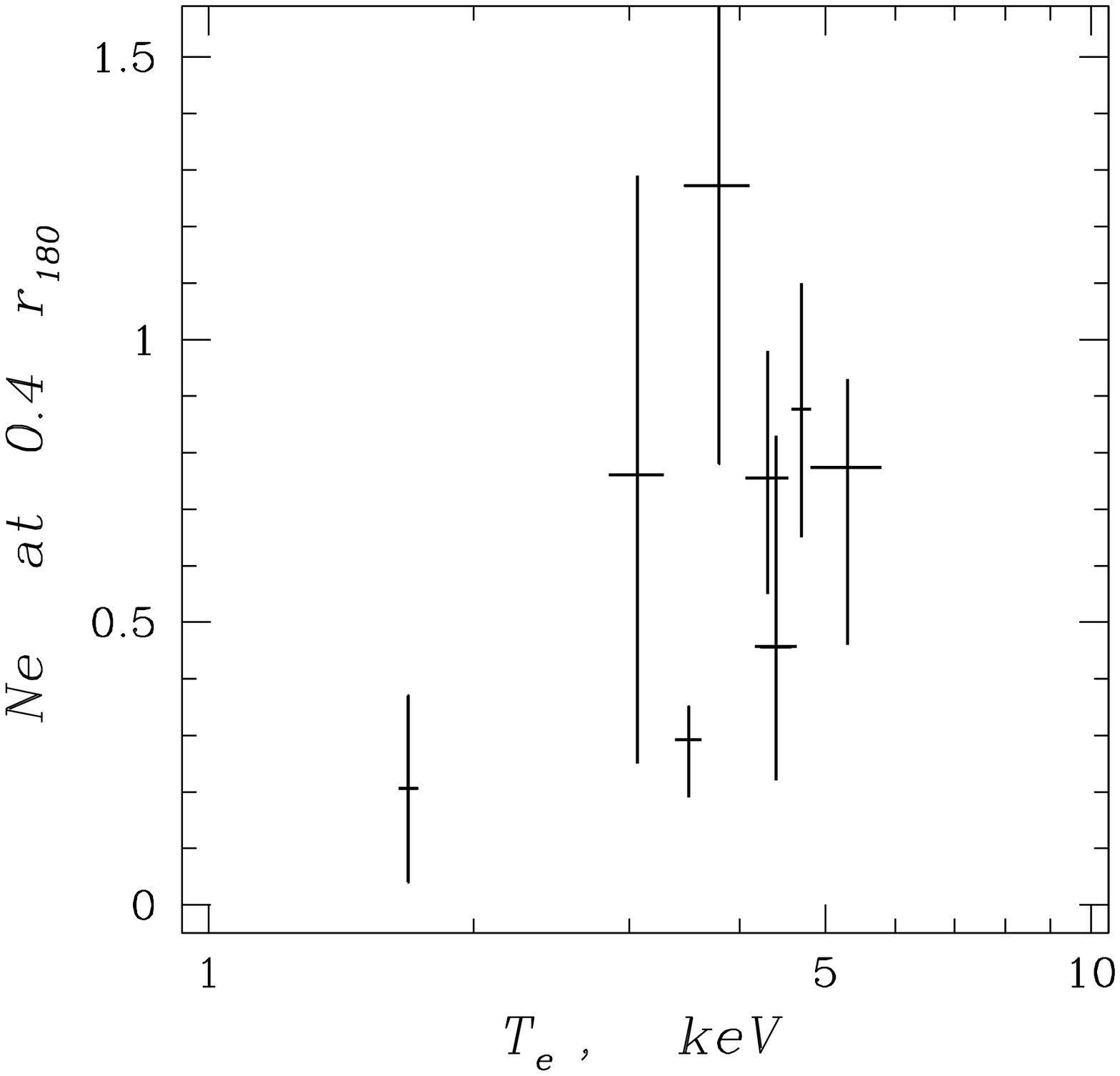}
\hspace{0.2cm}\includegraphics[width=4.2cm]{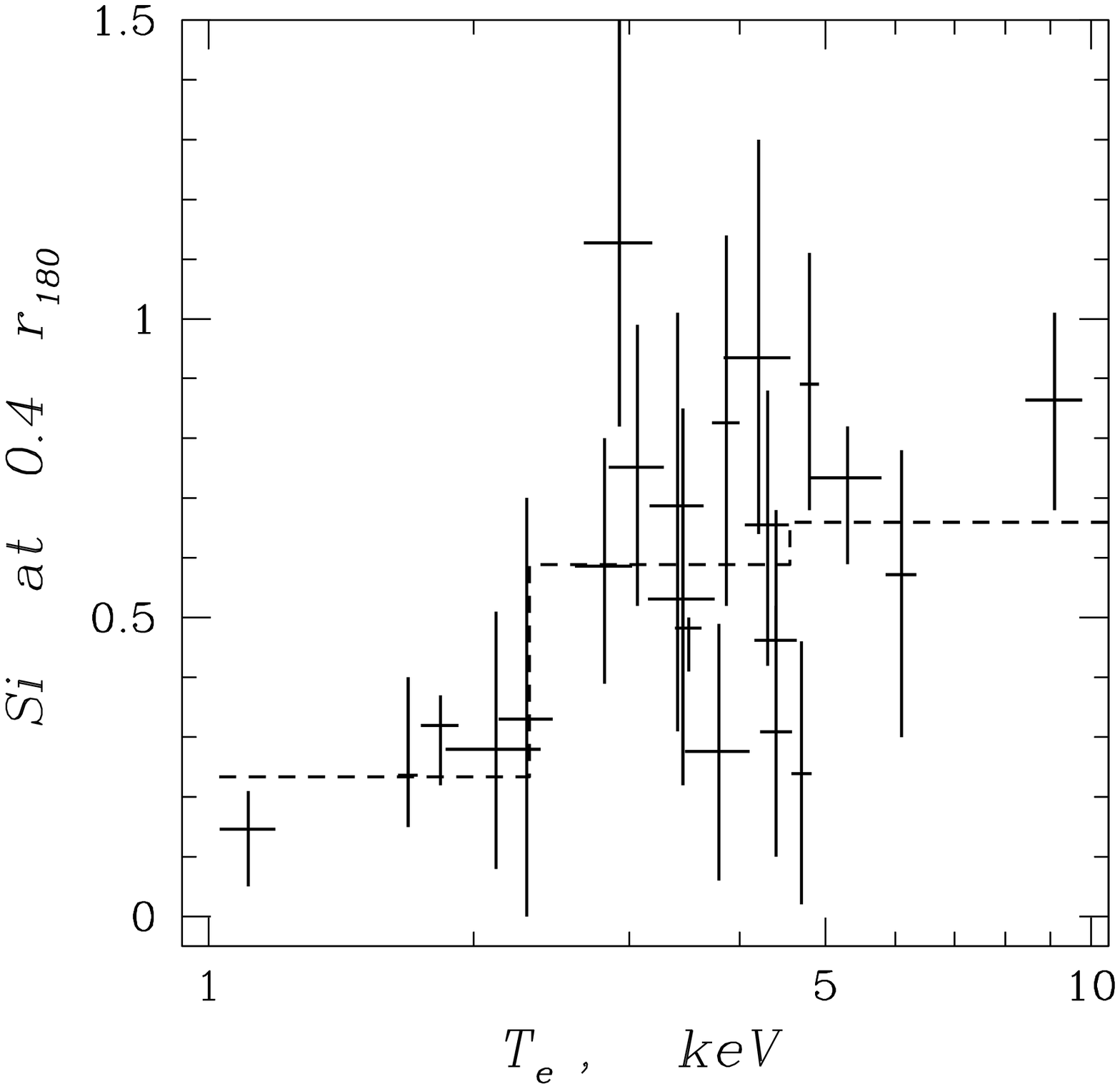}
\hspace{0.2cm}\includegraphics[width=4.2cm]{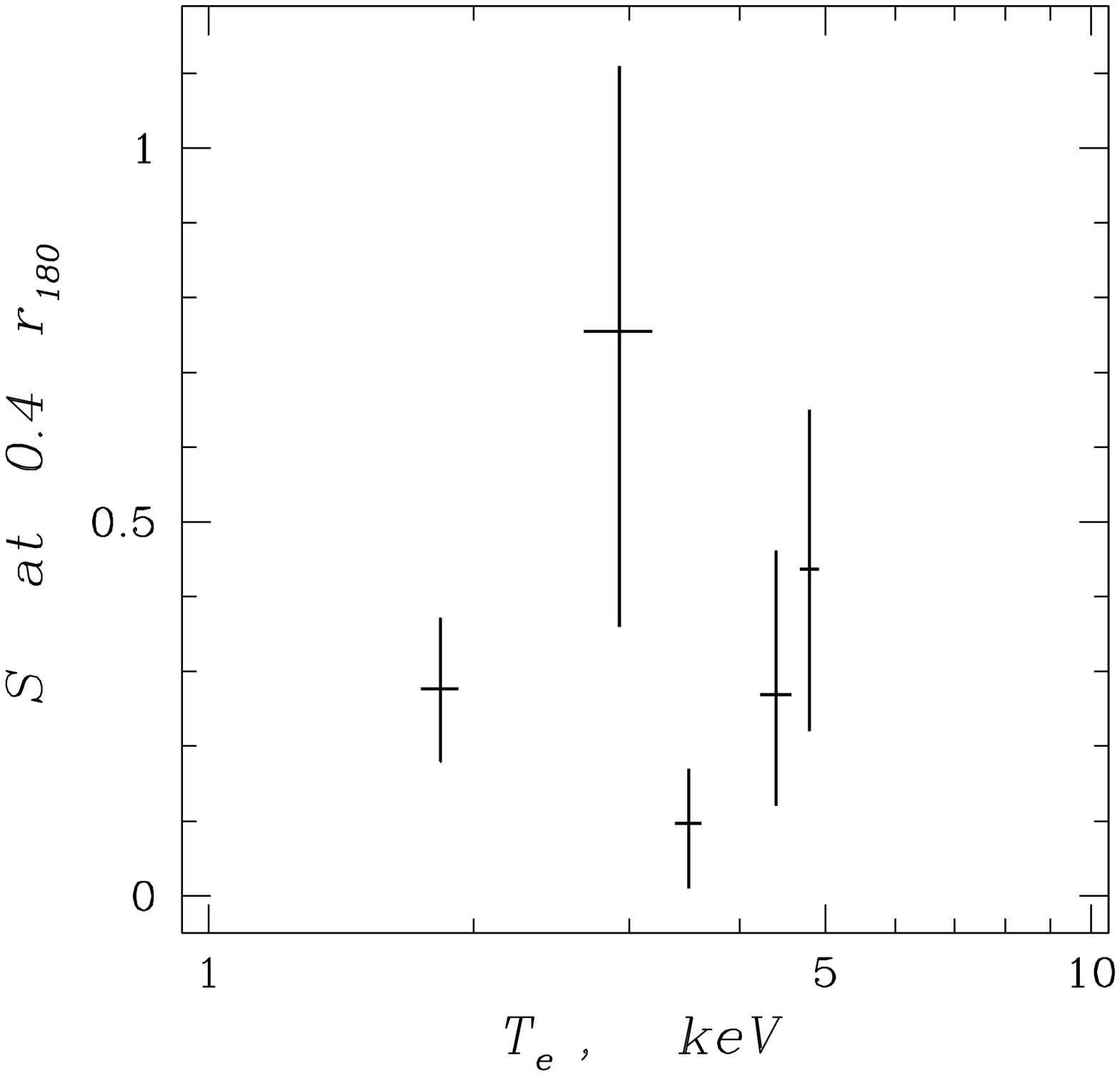}
\hspace{0.2cm}\includegraphics[width=4.2cm]{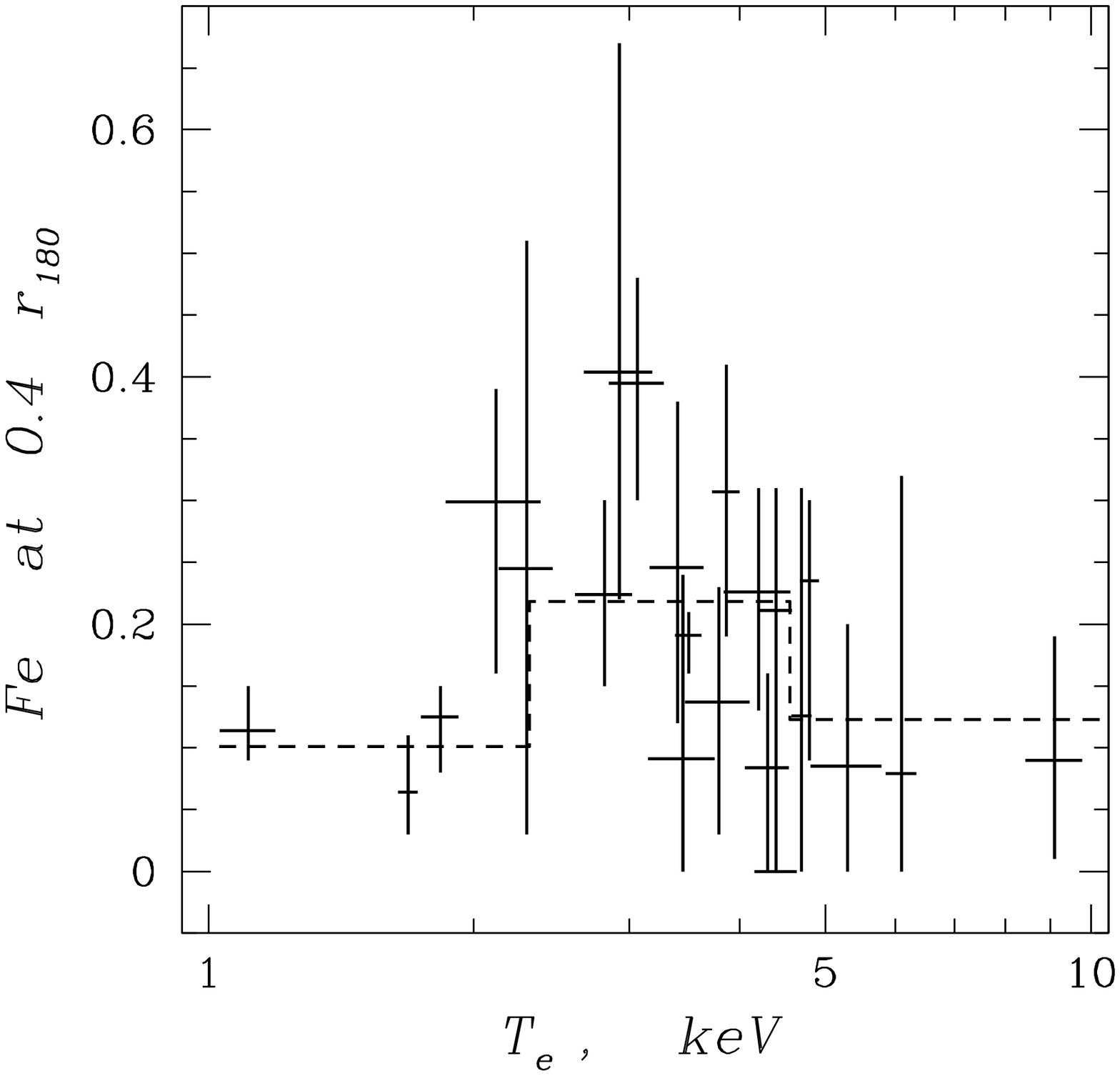}

\figcaption{Ne, Si, S and Fe abundance determined at a given (0.2 and 0.4)
fraction of the virial radius. Dashed lines in the Si and Fe figures denote
the change of the averaged value.
\label{ab-vir}}
\end{figure*}

\section{Radial Distribution of Heavy Elements}\label{sec:ab}

\subsection{Iron}
  
With a statistical threshold of 90\%, an iron abundance gradient is observed
in A400, A262, MKW4S, 2A0335+096, HCG94, MKW3S and Cen. Among these
clusters, an Fe gradient outside the central 100 kpc is seen in A400,
2A0335+096, and Cen. A comparison with other ASCA measurements produces good
agreement, except possibly for AWM4, where the presence of a strong point
source complicates the analysis. The analytical fit of the Fe abundance in
Cen from Ikebe \etal (1999) appears to be slightly more centrally
concentrated. However, one should take into account the spatial width of our
bins. We ascribe the abundance values to the center of the spatial bin, but
the abundance values in each bin are dominated by the emission closer to the
cluster center, which can explain our broader Fe distribution. Overall, good
agreement between our results and Ikebe \etal (1999) serves as an argument
against the suggestion of Buote (2000) about the underestimation of
elemental abundances in the center of clusters due to 'over-regularization'.

\subsection{Silicon, Neon and Sulfur}

With a statistical threshold of 90\%, Si abundance gradients are observed in
A400, A262, and Cen. Only in A400 is there a significant decrease in the Si
abundance beyond the central 100 kpc. A comparison with the results of
Fukazawa \etal (1998) shows very good agreement, except possibly for AWM4
(our value is $2\sigma$ higher).

No Ne abundance gradients are detected in this sample. Ne abundance are only
significantly constrained in A262, MKW4S, MKW3S and Cen. A sulfur abundance
gradient is detected in Cen excluding the central 100 kpc from
consideration. The S/Fe ratio changes by much less than the Si/Fe ratio with
radius, which was also noted by Fukazawa (1997). Central enhancements of
Sulfur are marginally seen in 2A0335+096 and MKW4S and S abundance is
constrained at any radius in A2197E, A400, A262, MKW9, A4038 and A2052.

\section{Heavy Elements at similar overdensity.}\label{sec:vir}

The derived elemental abundances in our sample of groups and clusters of
galaxies span the range from 1/10 to a few times solar, even within a single
system (\eg\ Cen cluster). To compare the results between different systems,
% we need to choose an appropriate physical scale. We use a radius
% corresponding to the same overdensity as proposed in the simulations by Cen
% and Ostriker (1999).  **Their suggestion to use the radius of similar
% overdensity results from its correspondence to similar concentration of
% galaxies, providing a similar metal enrichment (Cen and Ostriker 1999). Our
% choice of equal overdensity also implies, according to Dressler (1980) a
% similar morphological content of clusters at the radius of
% comparison.**[RQ-4,6a]
% 
we need to choose an appropriate physical scale. The natural physical scale
of a cluster is its virial radius and we thus compare data at given
fractions of $r_{180}$. This is equivalent to comparing data at similar
overdensity.
%(with respect to the density of the universe within which the
%cluster is embedded ), if, as suggested by numerical simulation and simple
%spherical collapse models, clusters form at constant overall overdensity
%(about 180) and their internal structures are self-similar. 

It is worth noting that the overdensity does appear to be a fundamental
parameter for the metal enrichment, as indicated by the simulations of Cen
and Ostriker (1999). Furthermore, the cluster morphological content, another
potentially important factor for chemical enrichment, should be the same at
the radius of comparison, since it depends on the local galaxy density
(Dressler 1980).

In Tab.\ref{tab:opt} we show $r_{180}$ for all the systems in our sample.
Comparisons between systems are made at 1/5 and 2/5 of $r_{180}$
(corresponding to overdensities of 8600 and 1600).  We choose an outer radius
of $0.4r_{180}$ due to the limited extent of our observations. The 
inner radius of 1/5 $r_{180}$ is chosen to avoid the central 200
kpc. Among the elements presented in Fig.\ref{ab-vir}, Ne and S do not reveal
any distinct trend, due to large measurement errors.

A major difference between our work and some earlier results (\eg\ Renzini
\etal 1993) is the dependence of the Fe abundance on gas temperature at
lower temperatures. We do not find an increase in the Fe abundance near 1
keV as previously reported.  Instead, the Fe abundances at $0.2r_{180}$ are
similar among groups and clusters with a tendency for the Fe abundance to
decrease at lower overdensities (compare the values at $0.2r_{180}$ and
$0.4r_{180}$).
% It seems that there is also a larger scatter in the Fe
% abundances among groups and that the highest Fe abundance are found in 3 keV
% clusters. 
The average trends in the Fe abundance at $0.2r_{180}$ with temperature
plotted in Fig.\ref{ab-vir}, are in good agreement with the findings 
of Fukazawa \etal (1998).  There is a 
peak in the Fe abundance of 0.3 solar for cool clusters, 
with hotter clusters having average iron abundances of 0.2 solar.

In contrast, the Si abundances increase strongly with gas temperature,
starting from 1/5 solar in groups and reaching solar values in hot clusters.
This agrees with the previous findings of Fukazawa \etal (1998). This
drastic difference between the Si and Fe abundances reflects the level of SN
II retainment indicated by the Si abundance, while the nearly constant Fe
abundance reflects the equal role of SN Ia in groups and clusters of
galaxies (e.g. FDP). To determine the relative enrichment from different
supernovae types, we adopt the yields in FDP, given by, $y_{Si}=0.133$\msun,
$y_{Fe}=0.07$\msun\ for SN II yields and $y_{Si}=0.158$\msun,
$y_{Fe}=0.744$\msun\ for SN Ia.

In Fig.\ref{m2l-vir} we show the Fe and Si mass accumulated within
1/5 and 2/5 of the $r_{180}$ expressed in a form useful for the
study of element production, i.e., the M/L$_B$ ratio
($h_{50}^{-1/2}$\msun/\lsun).  The light and gas within the central 
200 kpc are not included in this calculation for systems with cD galaxies. 
In Fig.\ref{m2l-vir} we also show the IMLR for each SN type.
As can be seen in Fig.\ref{m2l-vir}, both the Si M/L and the SN II Fe M/L
increase by a factor of 10 between groups and clusters of galaxies. In
contrast, the difference in Fe M/L and SN Ia Fe M/L is less prominent,
especially at $0.4r_{180}$. For comparison, we also plot in
Fig.\ref{m2l-vir} the corresponding gas mass fractions. Between 0.2 and 0.4
$r_{180}$ there is a significant change in $f_{gas}$, with clusters
tending to have similar gaseous fractions at large radii. These results can
be compared with the low scatter in gas fractions found at smaller
overdensities by Ettori \& Fabian (1999) and Vikhlinin \etal (1999).

\subsection{SN II and the Preferential Infall Scenario}\label{sec:infall}

The main goal of this section is to explain the different observed Si
abundances in these systems. In particular, we are interested
in the dependence of the Si {\it abundance} on the gas 
retainment (or accretion) in these systems. To
illustrate this idea, consider an outflow of material from a group, caused
by some form of heating. Such a scenario can explain the reduced 
gas fractions and mass in elements inside groups, but such a scenario 
cannot change the Si {\it abundance}.

In FDP, we proposed that the absence of strong gradients in alpha-elements
implies that SN II enrichment occurred prior to cluster collapse.  We also
note that the Si abundance does not vary significantly 
between $0.2r_{180}$ and $0.4r_{180}$ (see Figs.\ref{ab-vir} 
and \ref{m2l-vir}) even though there is a significant change in $f_{gas}$ 
between these radii.  This result also requires that the enrichment of 
the ICM with Si occur before the gas distribution in these systems
has been established. 
Thus, we propose a {\it preferential infall scenario} to explain
the different levels of alpha-elements in these systems. In this scenario,
intergalactic gas, after being enriched with SN II ejecta, becomes too hot 
(or has too high an entropy) to accrete into the shallow potential wells of 
cold systems, so only the low-entropy metal-poor gas is accreted onto groups.
The more strongly enriched, higher-entropy gas can only be accreted onto
rich clusters.

\medskip
\begin{minipage}[-ht]{8.8cm}

\hspace*{-0.2cm}\includegraphics[width=4.2cm]{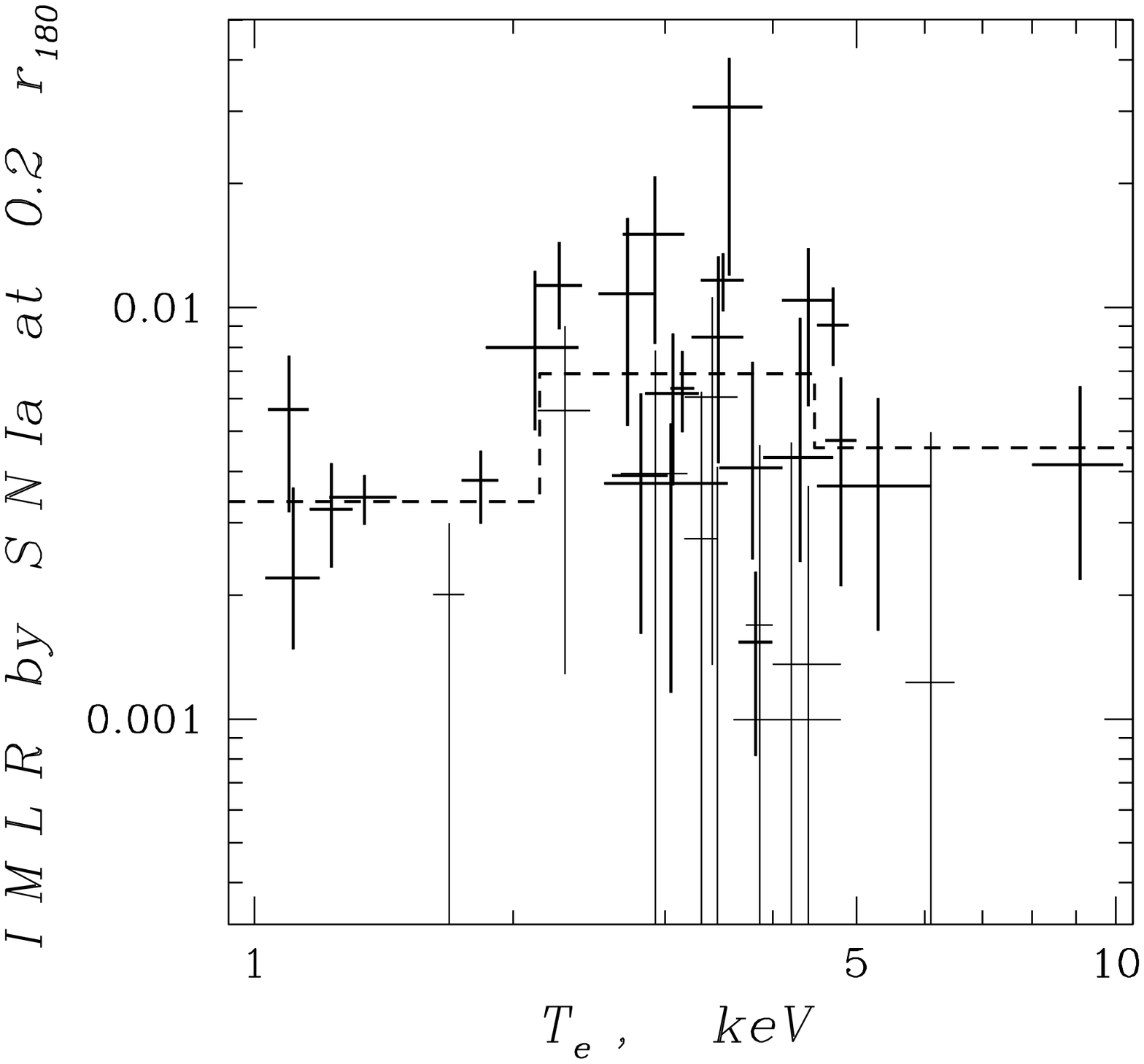} \hspace{0.2cm}
\includegraphics[width=4.2cm]{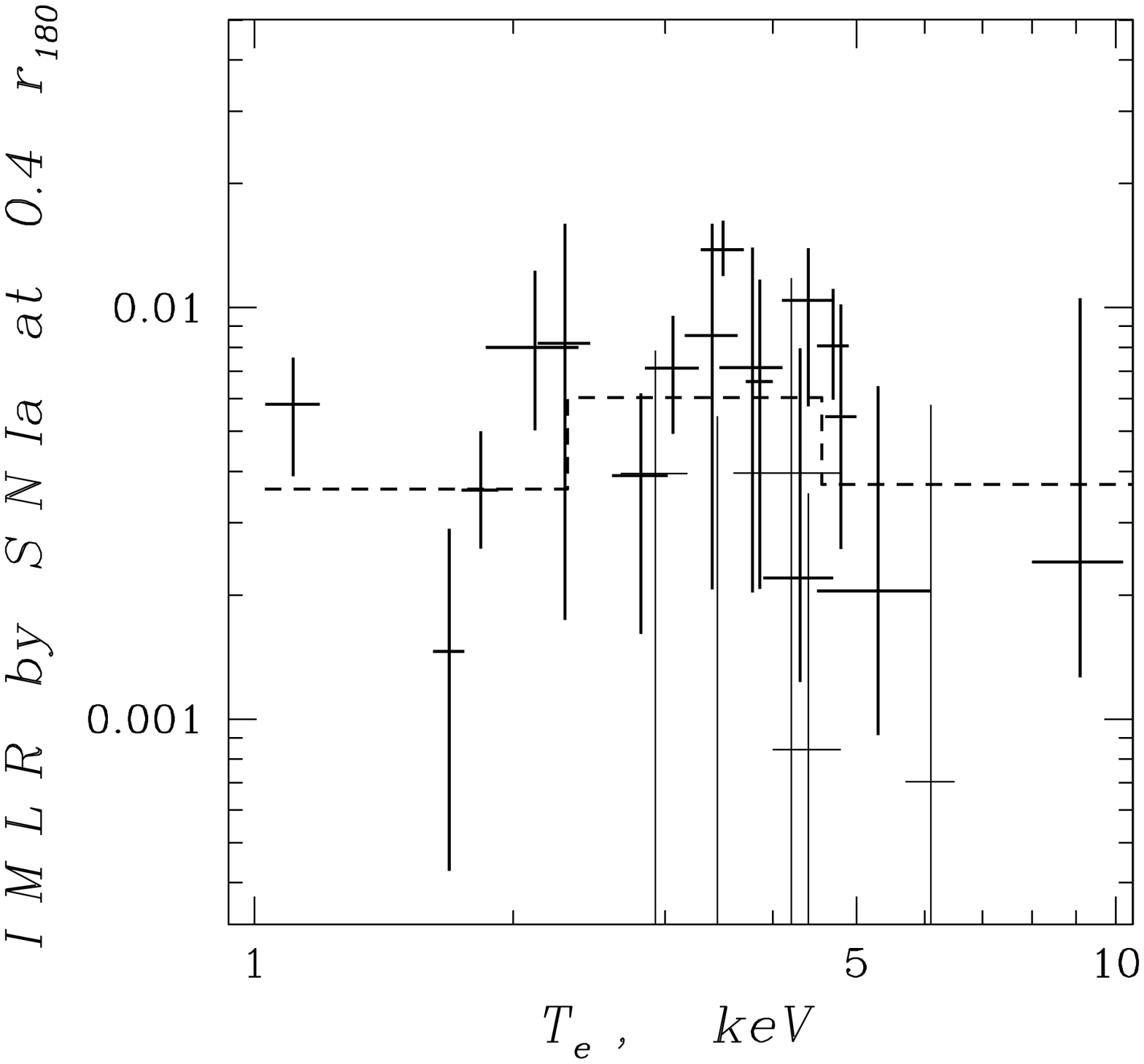}

\hspace*{-0.2cm}\includegraphics[width=4.2cm]{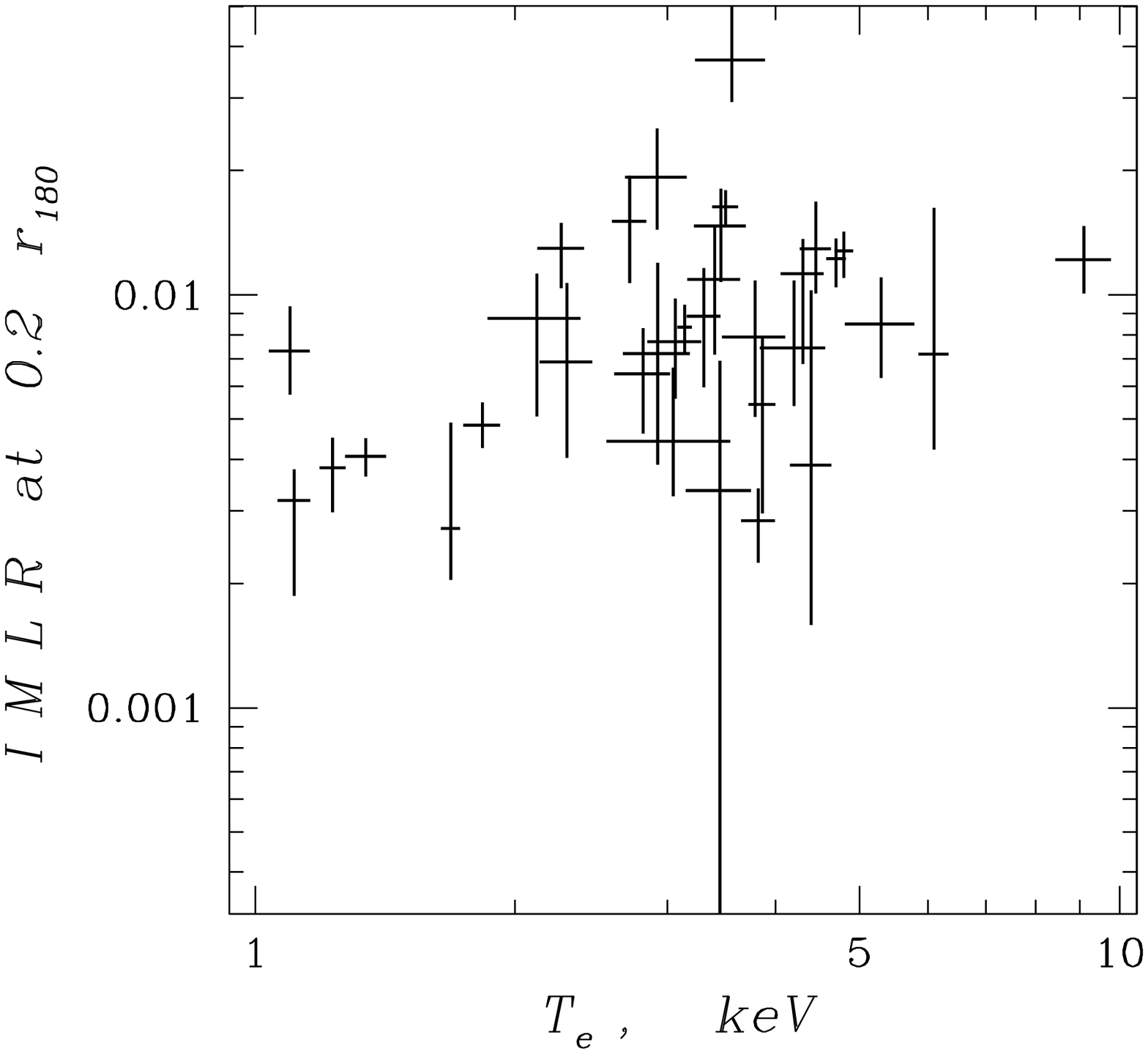} \hspace{0.2cm}
\includegraphics[width=4.2cm]{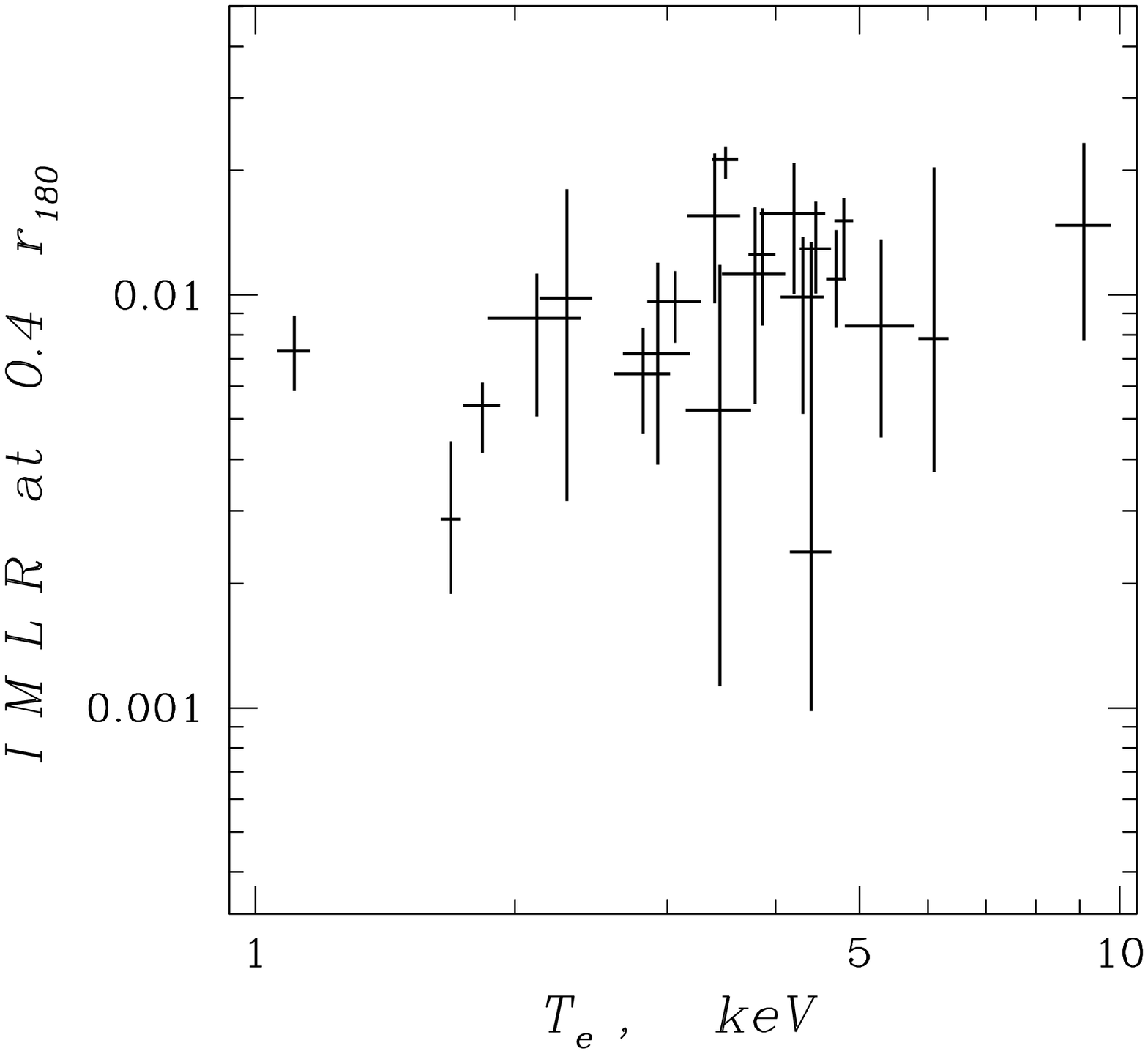}

\hspace*{-0.2cm}\includegraphics[width=4.2cm]{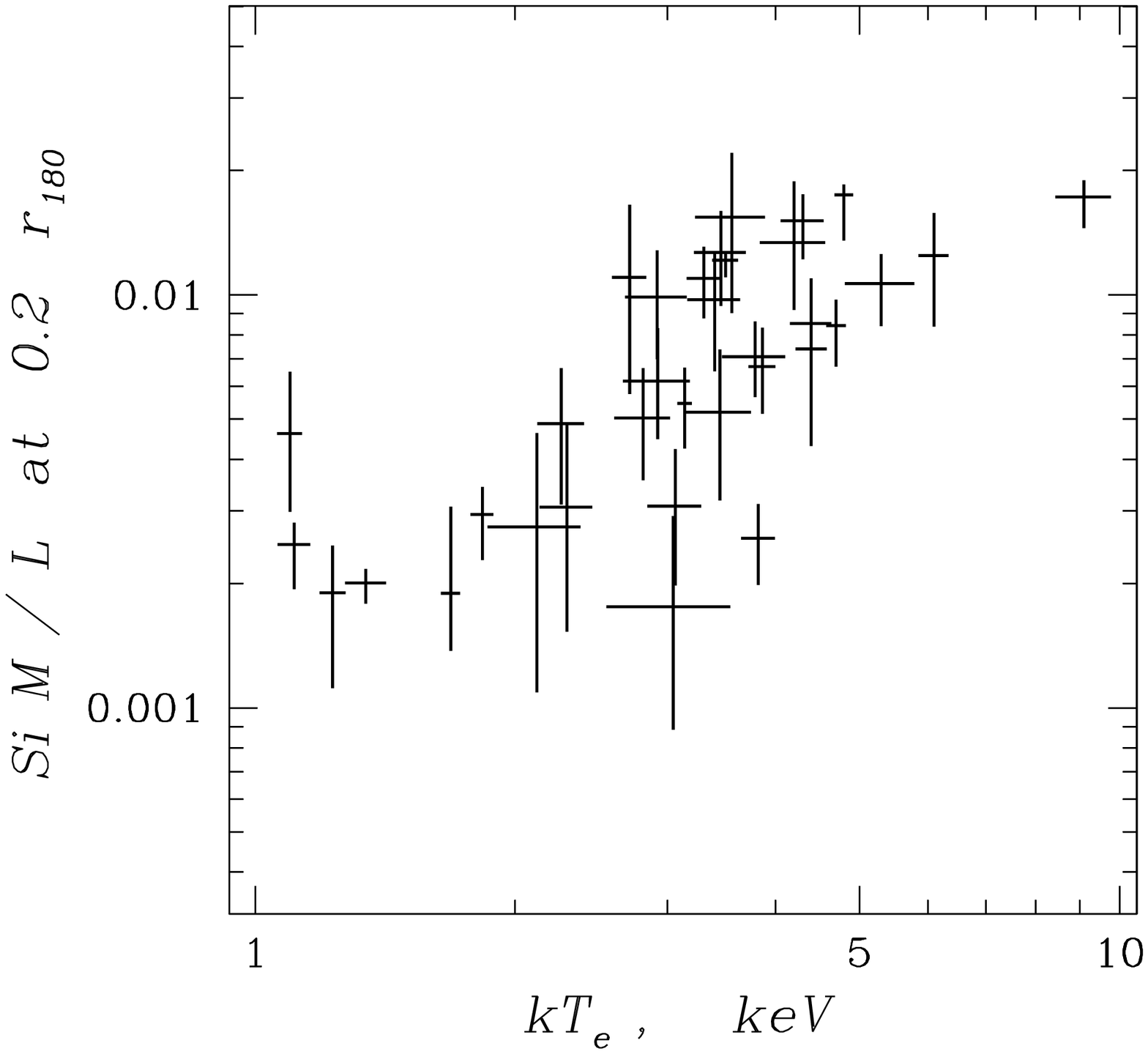} \hspace{0.2cm}
\includegraphics[width=4.2cm]{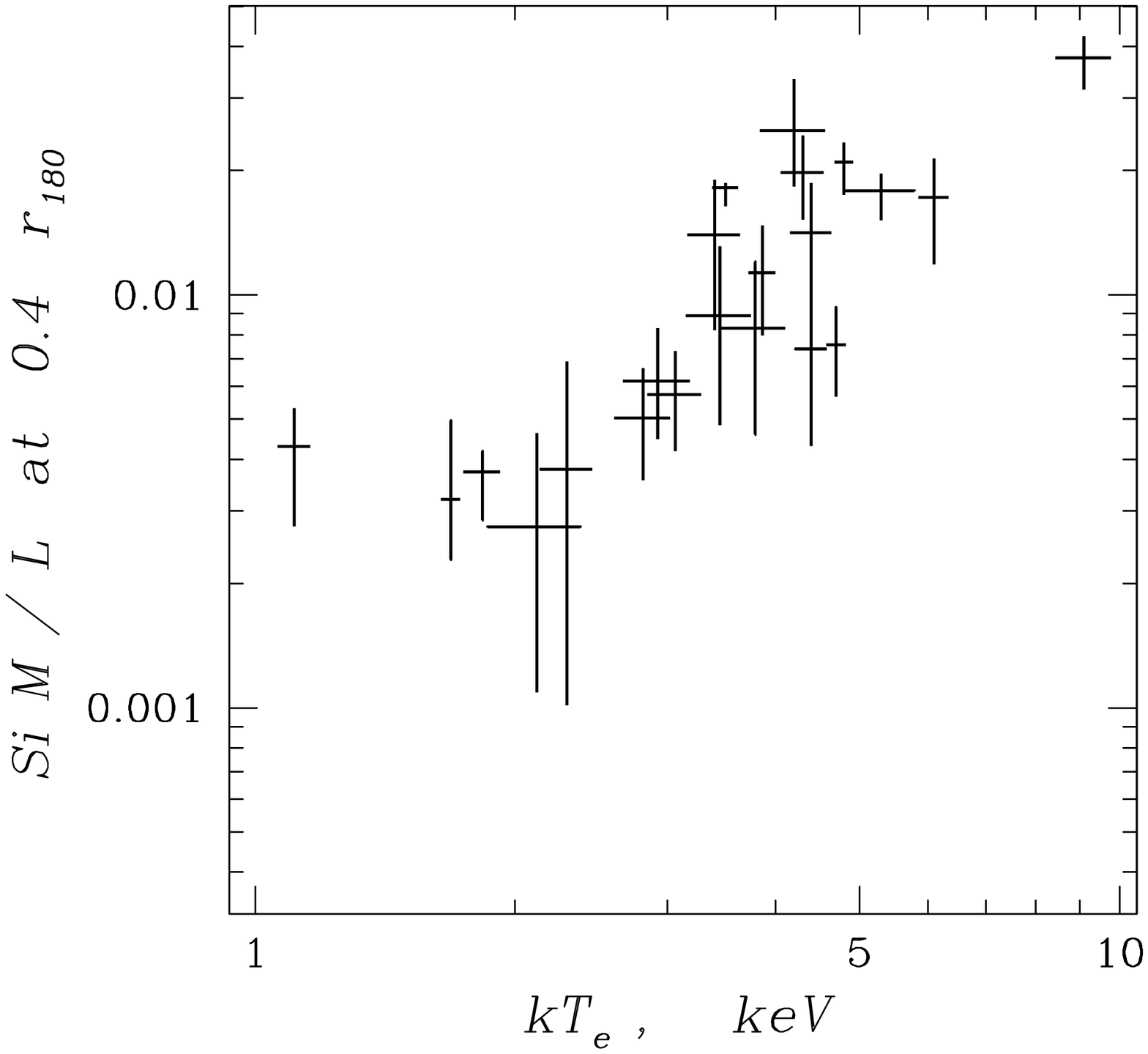}

\hspace*{-0.2cm}\includegraphics[width=4.2cm]{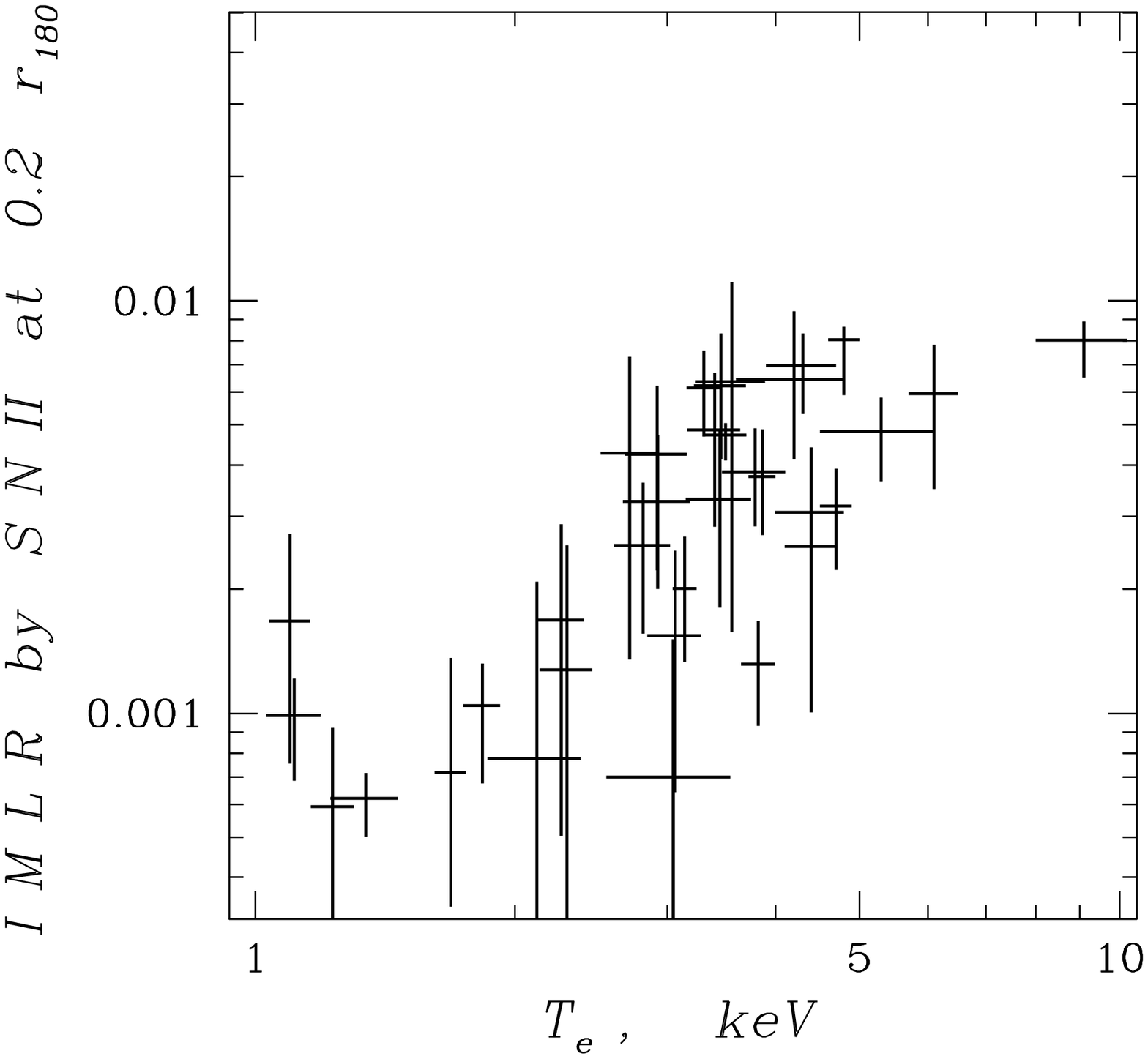} \hspace{0.2cm}
\includegraphics[width=4.2cm]{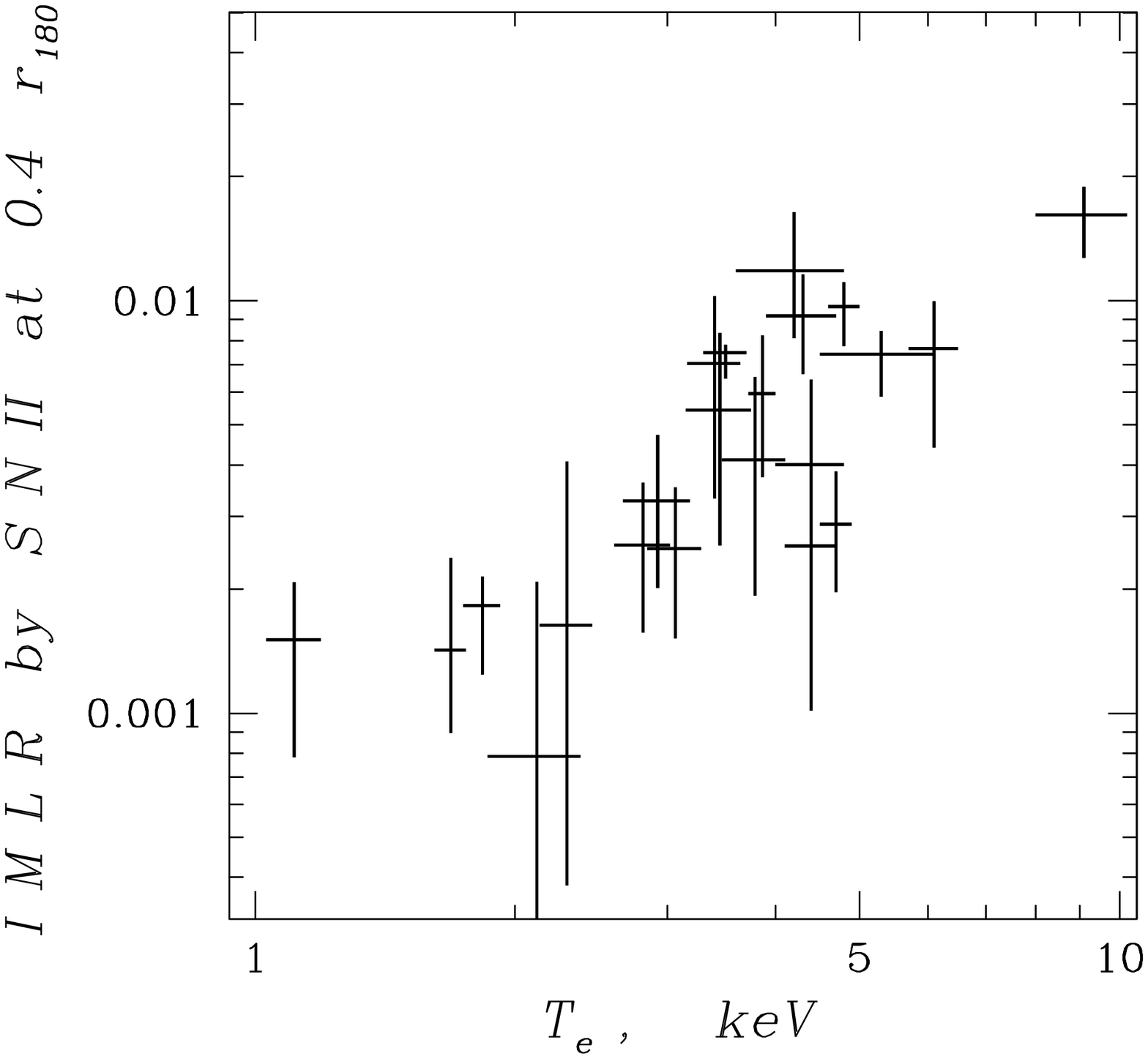}

\hspace*{-0.2cm}\includegraphics[width=4.2cm]{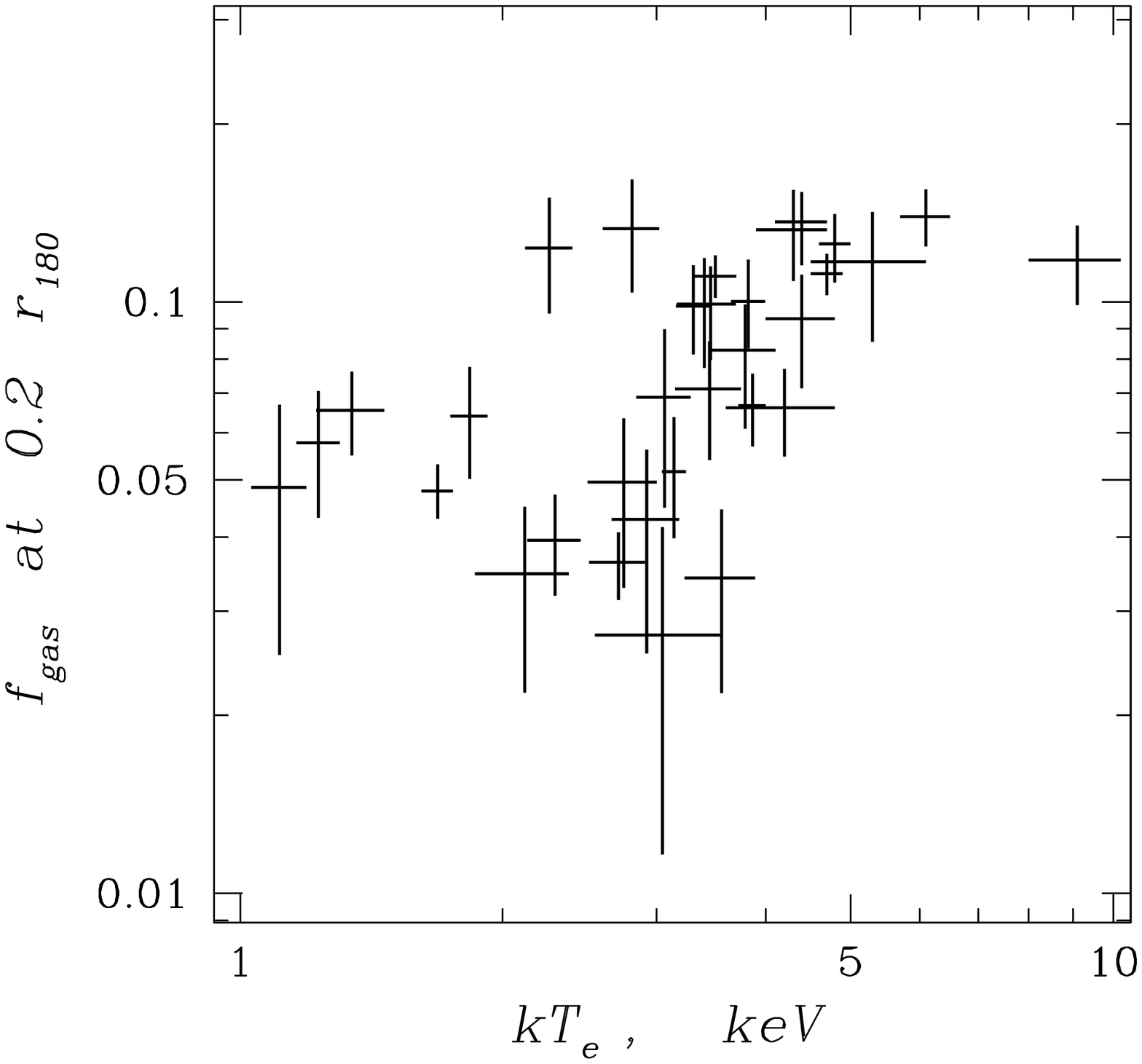} \hspace{0.2cm}
\includegraphics[width=4.2cm]{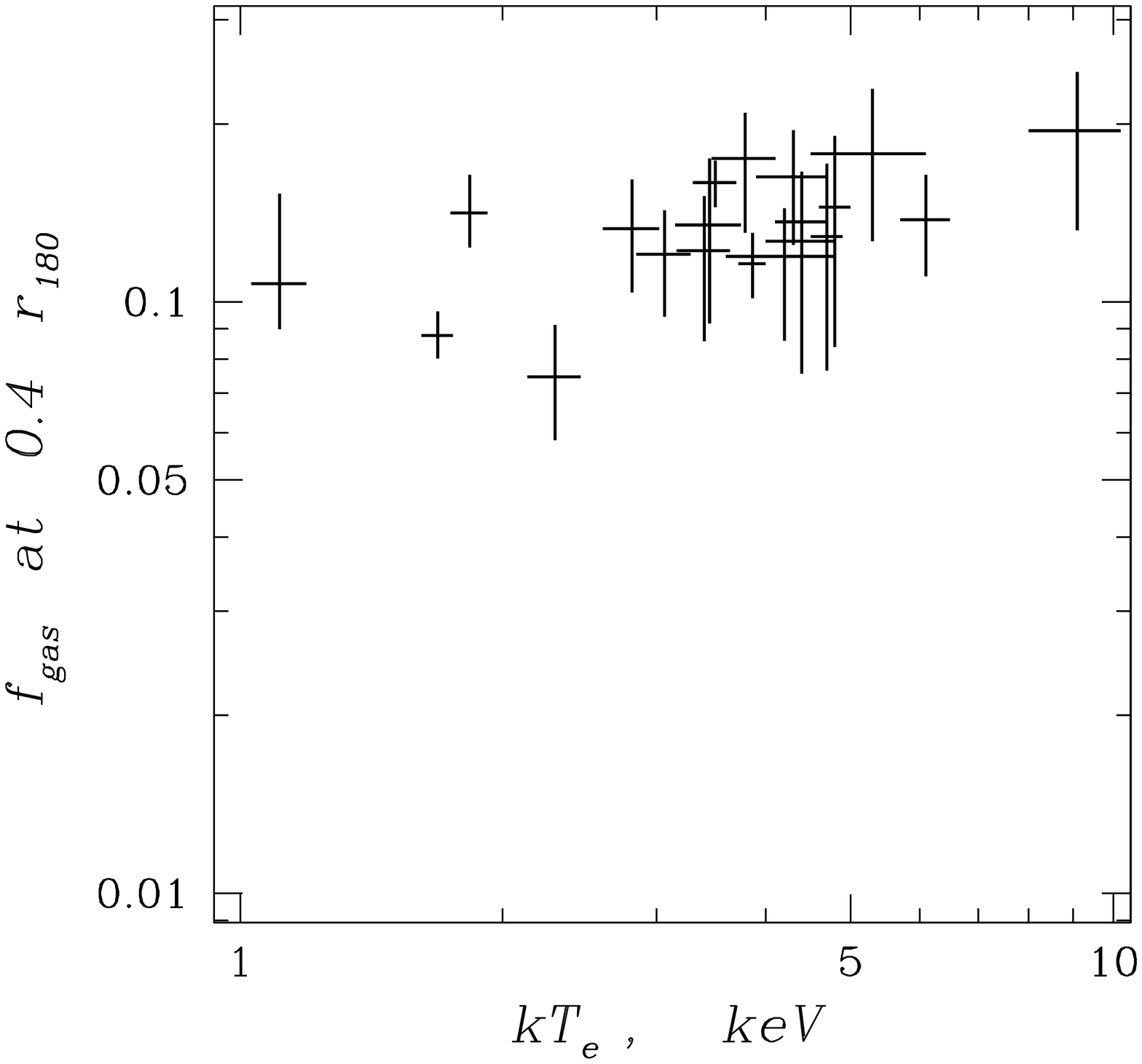}

\figcaption{$M/L_B$ ratios for Fe, Si and SN type II and Ia contribution to
Fe $M/L_B$. Behavior of gaseous fraction is shown at the bottom. Values are
displayed at 0.2 and 0.4 $r_{180}.$ Dashed lines in the SN Ia figures denote
the change of the averaged value.
\label{m2l-vir}}

\end{minipage}

In order for this scenario to work, the excess entropy of the SN-preheated
gas should exceed the entropy increase produced during the accretion and
shock heating of the gas in groups.  It is worth noting that the entropy
increase due to shocks and the entropy increase due to preheating exhibit
different scaling relations depending on the cluster formation redshift,
\eg\ Ponman, Cannon, Navarro (1999). This provides a possible test of the
above scenario, through future observations of the Si abundance levels in a
large sample of groups.  A large scatter in formation epochs for groups, as
suggested by simulations, may be a reflection of the relative importance of
preheating vs accretion shock heating, and may allow a more precise
determination of the actual energy released by SNe II in the form of
preheating.

Since between the 0.2 and 0.4 $r_{180}$ Si abundance is constant, as well as
$M_{tot}/L_B$ ratio, the observed differences in SN II Fe M/L ratio at the
0.2 and 0.4 $r_{180}$ simply follow the changes in $f_{gas}$.

One possible explanations for the reduced gas fractions in 
cold clusters is that the baryons are contained in the form of stars and 
the baryon fraction is actually a constant. However, such a scenario 
cannot reproduce the observed behavior in the Si abundance, since the 
consumption of any gas will change the mass of elements in the ICM, but 
will not alter the abundances.

In the scenario presented above, we propose that the trend in the observed
Si abundance is produced by varying degrees of SN II retainment.  In the
following, we discuss an alternative 'closed-box' scenario, in which the
increase in the Si abundance with gas temperature results from a dependence
of the star formation on the gravitational potential of the system.  This
requires that SN II are more common in hotter clusters, either because their
galaxies are more massive (Diaferio 1999) and more metals per given light
are released into the ICM, or the IMF is {\it top-heavy} in the host
galaxies of hotter clusters (Larson 1998).  In such a case, SN II are
favored in massive systems, resulting in higher [Si/H] and Si M/L ratios.
In terms of the slope of the IMF, the observed Si abundances require an IMF
slightly steeper then the Scalo IMF for systems cooler than 3 keV, and an
IMF that is slightly more top-heavy than a Salpeter IMF in hot clusters
(using the calculations of the SN II contribution to the IMLR from Renzini
\etal 1993, who adopt $y_{Fe}$ from SN II consistent with our definition;
$IMLR(x=1.7)=0.003 \Rightarrow IMLR(x=1.35)=0.009 \Rightarrow
IMLR(x=0.9)=0.035$.) We note that only the measurement for A2029 at 2/5
$r_{180}$ exceeds the amount of IMLR predicted for a Salpeter IMF.

\subsection{SN energy input and the scaling relations.}\label{sec:en-sn}

Preheating of the intergalactic medium by supernovae has long been
considered a possible explanation for the observed deviation of cluster
scaling laws from theoretical expectations based on simple gravitational
collapse.  More recently, the amount of preheating was estimated to be 1--3
keV per particle in order to explain the entropy floor in cool systems
(Ponman \etal 1999; Loewenstein 2000; Wu \etal 1999). This amount of energy
has been considered too high to be produced by SNe heating alone and heating
by AGNs has been suggested as an alternative solution to the problem (Wu
\etal 1999).

Our observations provide for the possibility of determining the
amount of energy associated with SNe in a robust way using the measured Si
abundances. The advantage of this method consists in the similar Si yields
for different SNe types, so a separation between SN Ia and SN II is not 
required to calculate the energy released.

In Fig.\ref{sn-en} we show our estimate for the energy associated with SN
feedback, assuming that each SN provides 0.14\msun\ of Si ($y_{Si}$) and
releases $E_{SN}=10^{51}$ ergs of mechanical energy. Based on these
assumptions a solar abundance of Si ($A_{Si}$) corresponds to 1.6
keV/particle ($ E_{SN} A_{Si} \mu m_p / y_{Si}$, where $\mu=0.6$, $m_p$ --
proton mass).  As can be seen from Fig.\ref{sn-en}, the data show a
flattening in the SN energy input around the cluster temperature of 3
keV. This means that for systems hotter than 3 keV, SN ejecta makes little
difference on the gas thermodynamics and that these systems should be
self-similar.  In fact, this is exactly what is observed (Ponman \etal
1999). If AGNs were responsible for preheating the gas in clusters, this
trend with the Si {\it abundance} would not be reproduced.

\includegraphics[width=3.2in]{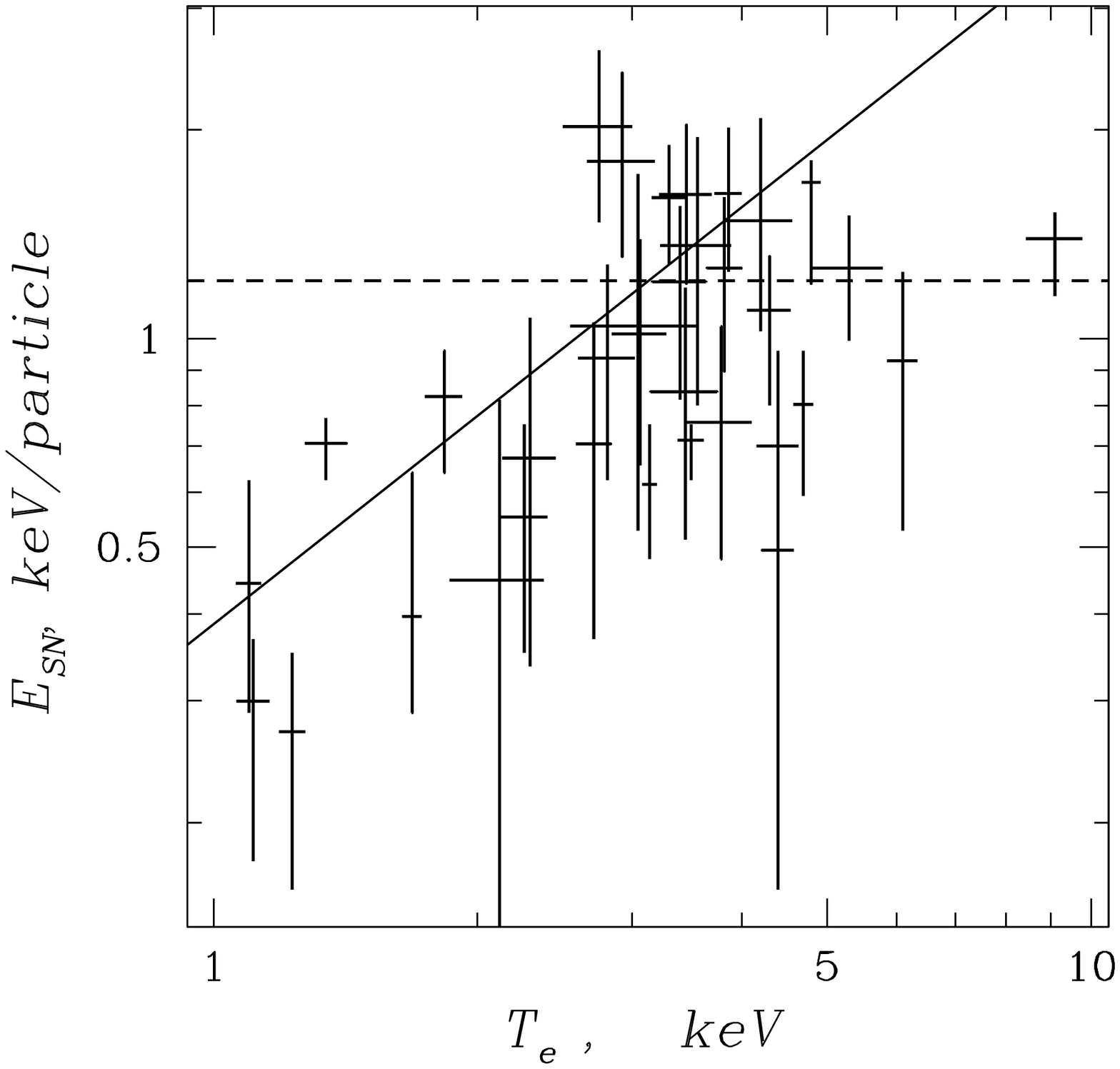} 

\figcaption{Thermal energy supplied by both types of SNe associated with
  measured Si abundance. Under assumption that each SN provides 0.14\msun\ 
  of Si and releases $10^{51}$ ergs in a form of thermal energy, a solar
  abundance of Si correspond to energy release of 1.6 keV per particle.
 The data is flat above 3 keV temperature with mean SN energy of 1.2 keV per 
 particle. 
\label{sn-en}
}

There exists a significant discrepancy in the normalizations of the $M-T$
relation between simulations (\eg\ Evrard \etal 1996) and observations
(Horner \etal 1999). One possible explanation is that the observed X-ray
temperatures are $\sim40$\% higher compared to that expected from pure
gravitational heating. The solid line in the figure shows the SN feedback
needed to reconcile the normalizations of the $M-T$ relations. As can be
seen from the figure, this fit matches our measurements below 3 keV.
However, for a similar SN-heating scenario to hold for a 10 keV cluster, the
observed Si abundance should reach 2 times solar, which is not observed
(Fukazawa \etal 1998). As a result, mass estimates for clusters hotter then
$\sim6$ keV are expected to get progressively closer to simulations of
Evrard \etal (1996), as $\Delta T \over T$ decreases with T. Application of
this scenario to the observed $M-T$ relation is further discussed in
Finoguenov, Reiprich, Boehringer (2001).

%   As
% simulations of Metzler \& Evrard (1994) show, SN driven winds have little
% effect on the ICM temperature (and thus on the M-T relation), when they
% occur after cluster collapse because most of the energy is consumed as work
% in lifting the gas, not to heat it up. Therefore, to explain the $M-T$ relation
% for low mass systems, and also taking into account the findings of Ponman \etal
% (1999), the feedback from SN must occur before cluster collapse.  However,
% for a similar SN-heating scenario to hold for a 10 keV cluster, the observed
% Si abundance should reach 2 times solar, which is not observed (Fukazawa
% \etal 1998). As a result, mass estimates for clusters hotter then $\sim6$
% keV are expected to get progressively closer to simulations of Evrard \etal
% (1996), as $\Delta T \over T$ decreases with T. 

% However, in addition to the normalization, the observed $M-T$ relation has
% also a steeper slope, than is expected from gravitational effects
% (Finoguenov, Reiprich, Boehringer 2001), so in the above consideration,
% groups do not receive the required energy {\it unless} the gas is heated
% before group collapse, in which case the temperature jump associated with
% SNe heating will be increased by a factor of 4 due to the collapse by a
% factor of two. As is noted by several authors (\eg\ Lloyd-Davies \etal
% 2000), heating is most efficient at turn around.  Simulations confirm the
% importance of even a small amount of SN energy (Tozzi \& Norman
% 2001). 

We note that the preferential infall scenario naturally supports the two
levels of entropy ($250\pm100$ and $500\pm200$ keV cm$^2$), that are
introduced in Tozzi \& Norman (2001) to explain the $L-T$ relation. It is
remarkable that at 0.4$r_{180}$ the observed entropy in groups begins to
exceed the adopted preheating level, there is no drastic difference in the
gas fraction among these systems (see Fig.\ref{m2l-vir}). Yet, the Si
abundance does not increase in groups, which implies that there is no
substantial enrichment at later times and that previously enriched gas
escaped the group's surroundings. This escaped gas can enrich a cluster of
galaxies siting on the same filament and create enhanced Si
abundance. Fukugita \etal (1998) estimated the total amount of baryons in
groups maybe similar to that in clusters. Therefore, hot clusters may
accrete a substantial amount of alpha elements (up to 50\%), but not (yet?)
the galaxies.

The largest uncertainty in estimating the mechanical heating from SNe comes
from radiative losses. Under the assumption of a thermal wind, where the
wind velocity is equal to the escape velocity of a galaxy, Renzini (2000)
obtained a value for SN energy input that is 10\% of the initial energy
released. We argue, that this is a lower limit on SN heating. Wind
velocities up to a $\sim1000$ km/sec implied by the assumption of low
radiative losses are justified in Lloyd-Davies \etal (2000). If metals are
ejected via thermal winds, as assumed by Renzini (2000), then most of the
metals will remain bound to groups. The observed entropy at ($0.2r_{180}$)
in groups is approximately 300 keV cm$^2$. At the corresponding formation
epoch the preheated gas entropy is $500 \Delta E$ keV cm$^2$ (following
Lloyd-Davies \etal 2000), where $\Delta E$ is in keV per particle. If the
assumption of thermal winds is correct, the preheating entropy is then only
50 keV cm$^2$, which cannot produce a significant loss of enriched gas from
group potentials. Also, no infall avoidance of enriched gas is seen in
hotter systems, where the entropy exceeds 700 keV cm$^2$, leaving little
room for alternatives.  Therefore, we conclude that our estimate of the
energy input from SNe is justified. One important detail in calculating the
energy released by SNe is that we only detect in clusters of galaxies the
elements that successfully escaped from the galaxies (which gives a rough
estimate of $f_{ej}$ of 2/3 for modeling of the starbursts, just from
observing two thirds of metals in ICM). Some of the material that
experiences large radiative losses will inevitably remain bound to the
galaxy (\eg\ Larson \& Dinerstein 1975).

\subsection{SN Ia products as an indicator of the cluster formation age.}\label{sec:sn1a}

The retainment of SN Ia products shown in Fig.\ref{m2l-vir} reveals a
significant differences between systems. On average, cool clusters ($T=2-4$
keV) have an SN Ia IMLR that is 1.5 times higher compared to hot clusters.
In explaining the observed variations in the SN II products in the ICM, we
speculated on the preferential accretion of low-entropy low-abundance
gas. Differences in the SN Ia products cannot be explained in this way. In
fact, as we noted above, clusters that are rich in SN Ia products are not
necessarily the hottest ones. Furthermore, the lower Fe abundance at higher
gas mass fraction suggests that Fe enrichment occur mostly after the gas
density distribution was established, and not before cluster collapse. In
this section we discuss possible mechanisms that may lead to variations in
the amount of SNe Ia ejecta in the ICM.

In the classical scenario for the formation of elliptical galaxies, most of
the stars form in a short burst of star formation at early times. SN Ia
(long lived secondary star) will continue to explode and the SN Ia products
will not be recycled because of the lack of star formation, but will
continuously build up over the lifetime of the galaxy.  Thus, in a given
galaxy, the SN Ia products will remain in the ISM and their abundance will
increase with time, until the host galaxies fall into the dense ICM
environment. If striping is continuous after the galaxy has entered the
cluster the SN Ia Fe M/L will be constant.  Since this is not observed, we
need to modify this standard scenario.

Possible assumptions are

\begin{itemize}

\item The galaxy recycles its elements through star formation while outside
the cluster (may be due to ellipticals being not the only source of SN Ia
products in clusters of galaxies)

\item The halos of infalling groups/galaxies are stripped at very large
distances

\end{itemize}

Both scenarios can be characterized as attempts to hide the SN Ia products
before the galaxy's infall, either in the stellar population of the host
galaxies, or by spreading them through the outskirts of clusters. There are
certain observational and theoretical grounds for the above
modifications. First, it is found in studies of star formation in galaxies,
that galaxies probably significantly reduce their star formation after
falling into clusters (Balogh \etal 1998; Poggianti \etal 1999). Second,
in hydrodynamical simulations (Klypin, private communication), the gaseous
content of infalling groups of galaxies is segregated (stripped) from the
infalling galaxies at the outskirts of clusters.  Third, observed
galaxies entering clusters (\eg\ M86) show that the amount of
metals associated with their halos is $\sim10^{-4}$ in terms of Fe M/L
ratio, which is only a few percent of the ICM value.

What are the observational consequences of the proposed modifications?  Only
SN Ia explosions occurring after galaxies infall will contribute to the SN~Ia
ICM enrichment. First of all, if clusters form from inside out by the
gradually accretion of galaxies (\eg\ Klypin \etal 1999), then one 
expects that the central parts of clusters should be exposed to SN Ia
enrichment for the longest time. This scenario reproduces the Fe M/L
ratio gradient discussed in FDP. The rising Fe M/L ratios detected in
compact groups (FDP) can be understood within the context of this scenario
if the galaxies have collapsed only recently onto the center of groups (in
agreement with a statement that compact groups are dynamically young).
Second, the Fe M/L ratio should be correlated with the formation of the
galactic components of clusters, since clusters that form earlier will retain
more ejecta. Within this scenario, we can easily explain the scatter in the
Fe M/L as well as the peak in Fe M/L for cold clusters in view of direct
similarity to Butcher-Oemler effect (\eg\ Kauffmann 1995). We can therefore
test our scenario by comparing with simulations of large-scale structure.

To derive the redshift of cluster formation from the observed mass of
heavy elements, we have to assume a present-day SN Ia rate and the redshift
dependence of the SN Ia rate.  Soon (with the advent of the NGST) this will
be a nicely determined function, at present, in addition to the local
measurements (Cappellaro \etal 1997) there is only one point at $z=0.4$ from
SN Ia Cosmology Project (Pain \etal 1996) obtained for field galaxies.
Following a suggestion of Renzini \etal (1993), we assume a power law
dependence of SN Ia rate with time, using the two measured points.
We adopt for this calculation $h=0.75$ (and rescale
our IMLRs), local $R_{SN Ia}=0.2$ SNu (1 SN per century per
$10^{10}L_{B\odot}$; adopting this value we consider all the galaxies as SN
Ia producers), $s=2$ (see Eq.7 in Renzini \etal 1993). The use of $s=2$ (which
implies a greater SN Ia rate in the past) is also justified by the large amount
of Fe attributed to SN Ia in our measurements. For example, a time-independent
rate of SN Ia can only provide about $\sim10$\% of the measured value
(this consideration is similar to the estimate of Renzini \etal 1993).
We also consider two different cosmologies: $\Omega=1$ and
$\Lambda$CDM with $\Omega_m=0.3$.

The calculation itself could be expressed as

\[ M_{Fe, SN Ia} = \int_{t(z=0)}^{t(z=\infty)} L_{z=0} \theta (t -
t_{formation}) A(t) dt \]

where A(t) is the SN Ia metal production rate per unit luminosity. We
approximate the evolution of the galactic components of clusters by $L_{z=0}
\theta (t - t_{formation})$, \ie\ by observing the present-day light for
some period of time ($\theta$ is equal to 1 after cluster collapse epoch and
0 before). The results of this calculation are shown in Fig.\ref{z-cl}. We
choose a redshift binning of 0.5 to improve the statistical significance of
the points and yet retain the most useful information. Every cluster is
added as a Gaussian of width corresponding to the uncertainty of the
measurement. In order for these plots to be directly compared with
observations, we should correct for the incompleteness of our sample. To
accomplish this we used the cluster number function from Henry \& Arnaud
(1991), counting systems colder then 2 keV as 2 keV clusters.

In the plot we also show the analytical and numerical results on the
formation of clusters (Lacey \& Cole 1993), following the formulae presented
in Balogh \etal (1999) for $\Omega_m=0.3$. We set the mass fraction of
clusters whose formation we want to trace to 40\%, representing the ratio of
the mass at 1/5 of virial radius (chosen for comparison with the SN Ia
products) to the total virialized mass of the cluster and do the calculation
for two masses (2 and $5\times10^{14}$\msun), characterizing two subsamples,
shown in Fig.\ref{z-cl}. In making comparisons with the simulations, we
implicitly assume that there is no segregation between the clustering of
mass and light. As can be seen from the figure, formation of the central
40\% of the mass in the nearby clusters is shifted to earlier epochs.

\includegraphics[width=3.2in]{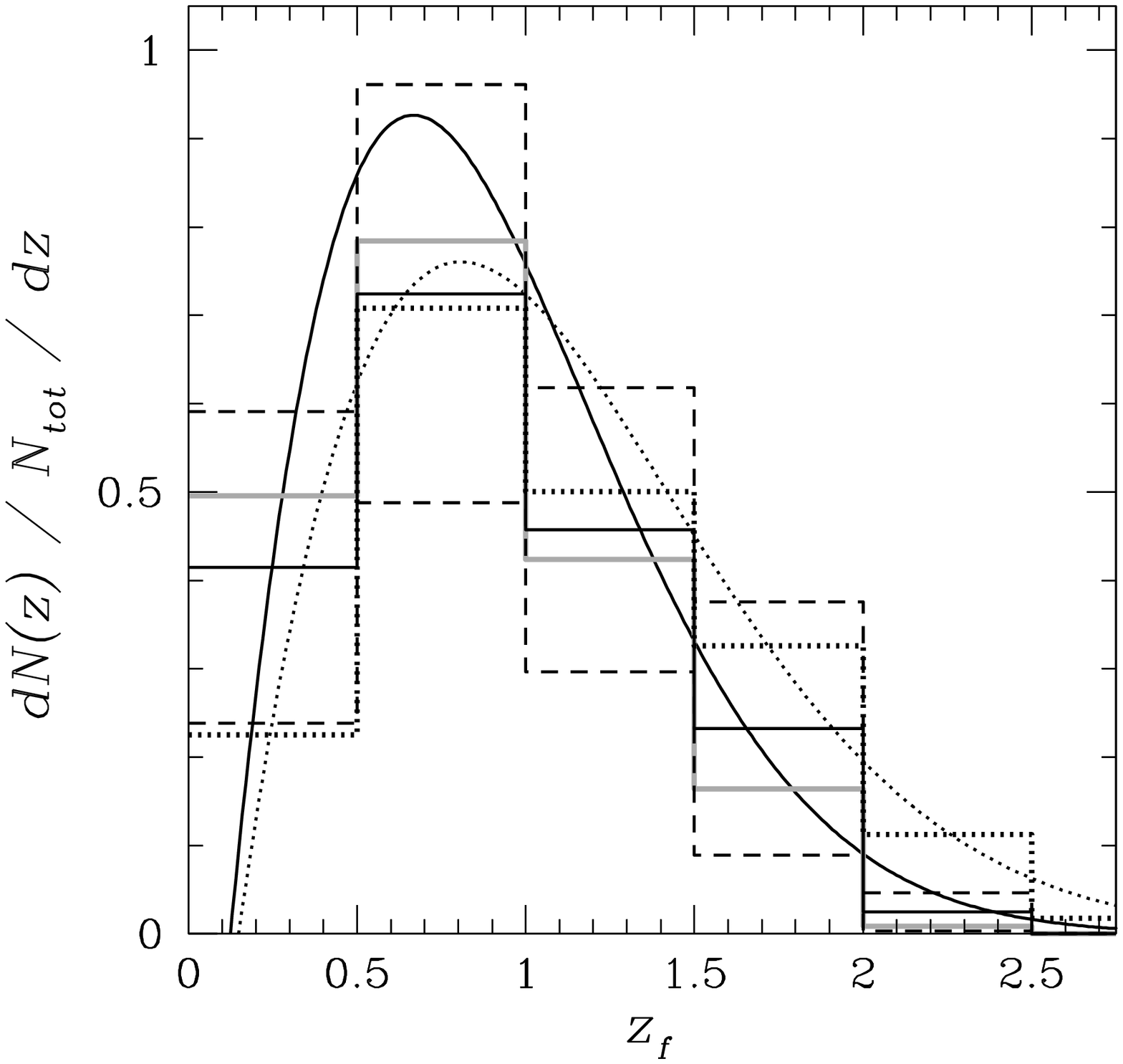}

\figcaption{Redshifts of cluster formation deduced from Fe M/L ratio
attributed to SN Ia. We correct for incompleteness of our sample using
number function of Henry \& Arnaud (1991). Black solid histogram shows
calculation for the bright end of the sample ($kT>4$ keV), calculated for
$\Lambda$CDM, with dashed histograms indicate the number count errors at
68\% confidence level. Grey histogram shows calculation for
$\Omega=1$. Dotted histogram shows the calculation for the whole sample,
assuming $\Lambda$CDM.  Predictions from numerical simulations of structure
formation (Lacey \& Cole 1993), calculated using the formulae in Balogh (1999)
for $\Omega_M=0.3$, are shown for formation of first 40\% of
$M=5\times10^{14}$\msun\ (solid line) and $M=2\times10^{14}$\msun\ (dotted
line), corresponding to the averaged mass of two samples shown.
\label{z-cl}}

The remarkable similarity between the distribution of formation redshifts
derived here with the calculations of Lacey \& Cole (1993) lends support to
our explanation for the observed scatter in SN Ia Fe M/L ratio being due to
differences in the epoch of cluster formation. The exact meaning of
Fig.\ref{z-cl} is the formation of a certain part of the studied clusters
($0.2r_{180}$). Similar plots can be done for any overdensity and can thus
provide an effective mechanism for tracing the formation of large-scale
structure.

\section{Conclusions}\label{sec:sum}

By combining a new analysis of 18 cool clusters with our previous
work we have a sample of clusters that span a factor of 10 in temperature.
We derive the following conclusions based on abundance measurements at
$0.2r_{180}$ and $0.4r_{180}$. 

\begin{itemize}
  
\item Groups and cool clusters preferentially accreted low-entropy 
low-abundance gas as best illustrated by the strong correlation between
Si abundance and
emission-weighted gas temperature of clusters. This supports the claim that 
energetic SNe II winds, driven at earlier epochs by star-burst galaxies, are
responsible for preheating the gas.
  
\item We detect a drastic change in the behavior of the gas mass fractions
between $0.2r_{180}$ and $0.4r_{180}$ which are not accompanied by changes
in the Si abundance. This requires that the enrichment by Si (SN II) occurs
before the observed gas distribution was formed. Lower Fe abundances at
higher gas mass fractions suggests that the SN Ia enrichment also occurred
after gas density distribution was established. Accounting for the baryons
in stars cannot solve the problem of the low gas mass fractions in low-mass
systems.
  
\item The energy per particle associated with SN explosions, when plotted
against the emission-weighted system temperature, exhibits a break at 3 keV.
At comparable temperatures many cluster scaling relation start to deviate from
self-similarity (\eg\ Ponman \etal 1999). This proves the importance of SN
feedback on the formation of the gaseous component of clusters.
  
\item We observe a significant scatter in the amount of SN Ia products
between systems, with cool clusters among the most SN Ia rich systems.
We propose an explanation based on the distribution of cluster
formation redshifts. A comparison of the predicted distribution of formation
redshifts of our sample with analytical formulation by Lacey \& Cole (1993)
for $\Omega_m=0.3$ demonstrates good agreement.

\end{itemize}

\section*{Acknowledgments}

MA thanks S. Maurogordato for fruitful discussions on cluster photometry and
for providing her background subtraction code.  MA thanks E. Vangioni-Flam
and M. Cass\'e for fruitful discussion.  AF acknowledges useful discussions
with S. Borgani, S. Gottloeber, A. Klypin, C. Metzler, J. Muecket,
V. Mueller and P. Tozzi. Authors thank the referee for useful comments on
the manuscript. AF acknowledges support from Alexander von Humboldt Stiftung
during preparation of this work. The authors acknowledge the devoted work of
the ASCA operation and calibration teams, without which this paper would not
be possible. This work was partially supported by NASA grant NAG5-3064.

%\clearpage

%\endrefs

\end{document}